\documentclass[letterpaper, 11pt]{article}

\usepackage{mathtools}
\usepackage{array}
\usepackage{multirow}
\usepackage[T1]{fontenc}
\usepackage[utf8]{inputenc}
\usepackage{helvet}
\usepackage{enumitem}
\usepackage{amsmath,amsfonts,amssymb,bm}
\usepackage{titlesec}
\usepackage{pgfgantt,rotating}
\usepackage{float}
\usepackage{graphicx}
\usepackage{sidecap}
\usepackage{wrapfig}
\usepackage{bold-extra}
\usepackage{authblk}
\usepackage{appendix}
\usepackage{amsfonts}
\usepackage{amsmath}
\usepackage{upgreek}
\usepackage{newtxmath}
\usepackage[utf8]{inputenc}
\usepackage{color}
\usepackage{rotating}
\usepackage{cite}
\usepackage{bm}
\usepackage{lipsum}
\usepackage{mwe}

\usepackage[margin=1in]{geometry}

\usepackage[linesnumbered,ruled]{algorithm2e}
\usepackage{algpseudocode} 
\algnewcommand{\Initialize}[1]{%
  \State \textbf{Initialize:}
  \Statex \hspace*{\algorithmicindent}\parbox[t]{.8\linewidth}{\raggedright #1}
}

\newcommand{\beq}{\begin{equation}}
\newcommand{\eeq}{\end{equation}}
\newcommand{\bgqar}{\begin{eqnarray}}
\newcommand{\enqar}{\end{eqnarray}}
\newcommand{\bgqarn}{\begin{eqnarray*}}
\newcommand{\enqarn}{\end{eqnarray*}}
\newcommand{\bgary}{\begin{array}}
\newcommand{\enary}{\end{array}}

\newcommand{\etal}{{\it et al. }}

\graphicspath{{figures/}}

\title{Data-Driven Structural State Estimation via Multi-Fidelity Gaussian Process Models}

\author{Yiming Fan}
\author{Fotis Kopsaftopoulos\footnote{Corresponding author.}}

\affil{\small Intelligent Structural Systems Laboratory (ISSL) \\ Department of Mechanical, Aerospace and Nuclear Engineering \\ Rensselaer Polytechnic Institute, Troy, NY, USA \\ Email: \{fany5,kopsaf\}@rpi.edu}

\date{\today}

\begin{document}

\maketitle

%----------------------------------------------------------------------------------------------------
% Abstract
%----------------------------------------------------------------------------------------------------

\begin{abstract}

Guided wave-based techniques have been used extensively in Structural Health Monitoring (SHM). Models using guided waves can provide information from both time and frequency domains to make themselves accurate and robust. Probabilistic SHM models, which have the ability to account for uncertainties, are developed when decision confidence intervals are of interest. Most active-sensing guided-wave methods rely on the assumption that a large dataset can be collected, making them impractical when data collection is constrained by time or environmental factors. Meanwhile, although simulation results may lack the accuracy of real-world data, they are easier to obtain. In this context, models that integrate data from multiple sources have the potential to combine the accuracy of experimental data with the convenience of simulated data, without requiring large and potentially costly experimental datasets. The goal of this work is to introduce and assess a probabilistic multi-fidelity Gaussian process regression framework for damage state estimation via the use of both experimental and simulated guided waves. The main differences from previous works include the integration of damage-sensitive features (damage indices, DIs) extracted from both experimental and numerical sources, as well as the use of a relatively small amount of real-world data. The proposed model was validated by two test cases where multiple data sources exist. For each test case, experimental data were collected from a piezoelectric sensor network attached to an aluminum plate with various structural conditions, while simulated data were generated using either multiphysics finite element model (FEM) or physics-based signal reconstruction approaches under the same conditions.

\end{abstract}

\newpage\pagebreak 

\tableofcontents 

%----------------------------------------------------------------------------------------------------
% Introduction
%----------------------------------------------------------------------------------------------------
\section{Introduction} \label{sec:intro}

Structural safety, a primary concern in engineering, is actively studied within the mechanical, aerospace, and civil engineering communities \cite{farrar2007introduction,diamanti2010structural}. Conventional inspections involve applications of non-destructive evaluation techniques with scheduled procedures under predetermined inspection time intervals \cite{Jordan-etal18,Frangopol-Maute03}. To enhance the efficiency and robustness of inspections, various Structural Health Monitoring (SHM) techniques with autonomous operation capabilities have been employed \cite{Cawley18,Amer-Kopsaftopoulos19a,Amer-Kopsaftopoulos20}. A fully developed SHM process includes damage detection, localization, quantification and remaining-useful-life estimation \cite{Janapati-etal16}. Active sensing, an SHM technique that utilizes piezoelectric transducers as both actuators and receivers, has been widely used in such processes. Once inputs are provided, features sensitive to various state factors can be generated for damage identification. Typical features from both time and frequency domains such as normalized amplitude, phase change and spectral quantity can be extracted from the first-arrival wave packet of transmitted signals \cite{2016-9,2017-5}. 

However, when structural changes are minimal, detecting causal differences using conventional features from signals can be challenging. To address this issue, a metric known as the Damage Index (DI) has been developed and widely applied in the field of SHM for aerospace structures. Typical DIs are obtained by comparing signals from unknown states with those from healthy cases. The values vary with respect to signal amplitude, phase, and energy, providing a useful tool for detecting damage states \cite{Jin-etal18,Xu-etal13,Janapati-etal16,Giurgiutiu11,Nasrollahi-etal18} and quantification \cite{Lim-etal11,Soman-etal18} using limited features. Researchers have explored various types of DI to develop more robust damage quantification metrics. For instance, Vanniamparambil \etal developed a data fusion technique that considers the guided waves, acoustics and digital images as inputs for crack size quantification \cite{Vanniamparambil-etal12}. The major drawback of DIs, however, is their deterministic nature with the inability in taking uncertainties of operations and environments into account \cite{farrar2007introduction,Amer-Kopsaftopoulos19a,ahmed2019uncertainty,Kopsaftopoulos-etal-MSSP18}. In addition, to capture features of stochastic time-varying responses, a large amount of data is often required to reach certain model accuracy and robustness, especially when damage states exhibit complex dynamics. Obtaining reliable signal responses can be time-consuming, requiring user expertise and domain knowledge for data generation. Additionally, collecting data from damaged states often involves manually inducing damage to structures, further increasing costs.

Various strategies have been developed to reduce the data costs associated with structural health monitoring. For instance, computational models using numerical methods, such as the finite element method (FEM), have been created to provide accurate results in complex dynamic systems. An example could be accurate modeling the effects of temperature perturbations \cite{ahmed2019uncertainty}. Yet the structural sensitivity of such models to variations can increase the computational cost significantly. 

Metamodeling, also known as surrogate modeling, includes techniques like kernel-based methods \cite{cortes1995support,cristianini2000introduction} and neural networks \cite{zhu2019physics,tripathy2018deep}, could be another option. Kernel methods are effective at building non-linear classifiers by transforming linearly inseparable data, but their complexity becomes prohibitive when the data size is large. While metamodeling approaches are explored typically to reduce computational costs, some methods, like neural networks, often require substantial data to achieve accuracy. Autoencoders, which learn a latent representation of the input data, are increasingly used in nonlinear reduced-order modeling. When combined with Long Short-Term Memory (LSTM) networks, these models are capable of recreating dynamic temporal responses \cite{simpson2021machine,nakamura2021convolutional}. While neural networks can achieve high accuracy, they typically require large training datasets and extended training times due to their complex structures. Both kernel-based and neural network methods are deterministic and lack interpretability, meaning the confidence in their predictions is unclear, and the underlying model structures or parameters are difficult to explain.
% PCA:
% Another strategy to require less computation cost involves reducing input dimension in the input space. Researchers have developed and applied principle component analysis methods \cite{tibaduiza2016structural,pearson1901liii,ma2011kernel} to compress the space by keeping the major components that is the most informative while rejecting the minor ones. 
% PCA ends

Gaussian Process \cite{raissi2018numerical,chen2015uncertainty,bilionis2013multi}, a non-deterministic and interpretable approach within metamodeling, allow the quantification of posterior distributions at target points using priors, even with limited training data. Given their ability to account for experimental uncertainty and provide confidence bounds in predictions, Gaussian Process Regression Models (GPRMs) have gained popularity in the active-sensing SHM community \cite{rasmussen2003gaussian}. For example, in online damage detection, where SHM must remain robust against various uncertainty sources during flight operations \cite{Dutta20}, the probabilistic nature of GPRMs are crucial for monitoring system safety levels. Additionally, GPRMs serve as valuable tools for parametric load estimation when direct measurements are costly \cite{fuentes2014aircraft}. Recently, GPRMs have been applied for probabilistic damage quantification in active-sensing SHM \cite{Amer-Kopsaftopoulos19a,Amer-Kopsaftopoulos19b}, addressing the limitations of deterministic DIs by using them as inputs for probabilistic regression models. While this approach allows for high-accuracy damage state estimation, it cannot incorporate input data from sources with varying fidelity levels.

% However, it is quite common that accurate formula or codes with high fidelity that describe the dynamic systems are not feasible and this framework would be prohibited. Even though highly-qualified descriptions are limited, it is still possible to reach accurate data regardless of inner information of stochastic systems.

To address this, an approach that integrates less accurate but low-cost data sources with informative, high-quality data would be highly promising. In fact, models that enhance accuracy by leveraging secondary correlated data have already been developed and shown to be efficient and effective. For example, Lewis \etal calibrated low-fidelity model parameters using high-fidelity simulations, guided by mutual information to optimally select high-fidelity evaluations for subsequent iterations \cite{lewis2016information}. However, this framework requires repeated low-fidelity model evaluations, which can become costly as the model structure increases in complexity.

To improve data efficiency, multi-fidelity Gaussian Process Regression Models (multi-fidelity GPRMs) have been proposed. These models not only retain the advantages of standard GPRMs but also combine data from different fidelity levels. Kennedy and O’Hagan \etal initially developed auto-regressive stochastic co-kriging models using computational simulations of varying costs. Their pioneering work introduced a framework that simplifies computation by decomposing simulations into an approximation component and a discrepancy component. The main objective function is then learned by prioritizing cheaper functions in regions of interest, assuming that expensive data is limited but ample low-cost data is available \cite{forrester2007multi}.
Gratiet \etal extended the method from Kennedy and O’Hagan by decoupling the multi-level auto-regressive problem into independent kriging problems, further reducing matrix inversion computation \cite{le2014recursive}. This approach was recognized for providing a rigorous and tractable workflow, making it feasible to apply multi-fidelity methods to physical models \cite{perdikaris2015multi}. Recently, variations and applications of these models have gained popularity. % Bonfiglio \etal \cite{bonfiglio2018improving} demonstrates the application of the multi-fidelity framework on discovering the optimal small waterplane area twin hull shapes for seakeeping. 
Gattiker \etal \cite{gattiker2006combining} combined experimental data and simulations to enhance predictions for a flyer plate. Their model accounted for unknown calibration parameters as well as the discrepancy between the simulator and reality. However, the use of a high-dimensional Gaussian Process (GP) was necessary, and the limited number of spatial dependence parameters hindered model performance. To address these computational challenges, Higdon \etal \cite{higdon2008computer} reduced the dimensionality of the framework and accelerated computations, particularly in cases where field data and simulator outputs were highly multivariate.

Multi-fidelity models have also demonstrated the ability to handle discontinuities and solve ordinary and partial differential equations (ODEs and PDEs) when combined with neural networks \cite{raissi2016deep,raissi2017inferring,raissi2017machine,meng2021multi}. Recently, researchers have begun applying multi-fidelity models in the Structural Health Monitoring (SHM) domain \cite{ torzoni2023deep,fan2023damage}.

% Towards this end, to develop a solver that incorporates less accurate, yet low-cost data sources, the co-kriging \cite{cressie2015statistics} approach was developed. Based on the pioneering work of Kennedy and O’Hagan \cite{kennedy2000predicting}, this approach is suitable for multi-fidelity simulations. Aiming at using data in a more efficient manner, multi-fidelity Gaussian process models (multi-fidelity GPRMs), which not only embrace the properties of standard Gaussian process models but also have the ability to combine data with different fidelity levels, have been proposed \cite{raissi2016deep }. Such models have been proved with the capability of handling discontinuities and solving ODEs \cite{raissi2017inferring,raissi2017machine}.

The objective of this work is to explore multi-fidelity GPRMs that optimally combine data from both experimental and computational SHM sources. This approach aims to reduce the reliance on costly and time-consuming experiments for model construction. Rather than using complex neural network models that often require large datasets and longer training times, a more concise framework, inspired by \cite{perdikaris2015multi,raissi2017inferring,raissi2017machine} has been implemented. Roy \etal \cite{roy2015load, roy2014novel} proposed a guided wave compensation model that accounts for varying loads and the effects of operational and environmental changes. This physics-based model enables the analysis of in-situ strain in the structure, which can be used to reconstruct signals at target states or compensate for underlying effects \cite{amer2021gaussian}. The second test case in this study implemented such methods to create a simulation dataset as the data source. The main contributions of this work can be summarized as follows: 

\begin{itemize}
\item Applied multi-fidelity GPRMs in the SHM domain for damage state quantification across various real-world tasks; 
% \item Compared with standard GPRM to demonstrate its accuracy on regression estimation; 
\item Extracted guided wave features (Damage Indices; DIs) sensitive to damage growth and severity level, using these as model inputs to enable uncertainty assessments, contrasting with traditional DI implementations;
% \item Tuning optimization process using experimental data-based constraints when minimizing the negative log maximum likelihood function;
\item Optimized the tuning process by incorporating experimental data-based constraints on estimated uncertainties to reflect internal data variance and prevent overconfidence;
\item Integrated multi-fidelity GPRMs with active learning using diverse criteria and acquisition functions to enhance data efficiency.
\end{itemize}

The structure of this work is organized as follows: Section 2 presents the methodology and background of the applied models, including the multi-fidelity Gaussian Process Regression Model (Section 2.1), the DIs formulations (Section 2.2), and the strategy for integrating active learning with criteria for selecting the next sampling point (Section 2.3). Section 3 illustrates the application of the proposed model with real-world data. The models were applied and compared in the first test case, which involved varying damage sizes for two tasks: (1) increasing the amount of simulated data while keeping the experimental data constant, and (2) filling data gaps in regions lacking experimental data using data from other sources. In Section 4, the same approach was applied to a different dataset under varying loading conditions to generalize the findings. Additionally, the proposed model was combined with active learning in a third task to further enhance data efficiency. Results from both test cases are discussed in detail, with final conclusions presented in Section 6.

\begin{figure}[h]
\centering
\includegraphics[scale=0.5]{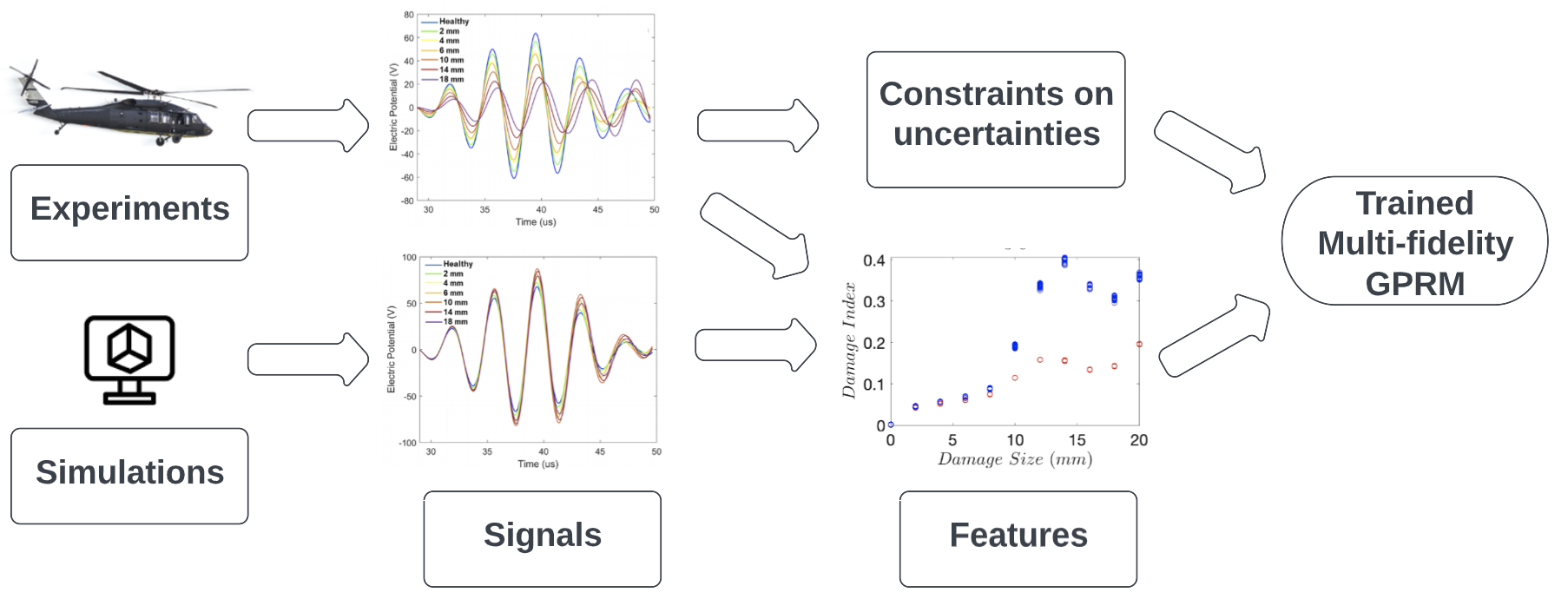}
\caption{The flowchart that demonstrates the main steps of guided-wave based multi-fidelity GPRM training procedure.}.
\label{fig:flowchart} 
% \vspace{10pt}
\end{figure}

\section{Methodology and Theoretical Background} \label{Sec:background}

This practical study aims to provide a comprehensive Gaussian Process-based continuous estimation of feature values across the domain of damage states. Assuming that the training and testing data share the same form and that features can be extracted using a consistent method, a continuous mapping between the damage state domain and the feature domain is explored for state estimation. These models require relatively few inputs while offering both mean and variance estimation at target states.  Researchers like Ahmad \cite{amer2021gaussian} have demonstrated the feasibility of state quantification using regression models by predicting state probabilities through the Cumulative Distribution Function (CDF) of GPRM targets. In light of this, the present work focuses on constructing and comparing effective regression models as the foundation for damage state quantification tasks.

Figure \ref{fig:flowchart} schematically depicts the framework of training process of the proposed model using data from multiple sources, i.e., experiments and simulations. 
% Data preprocessing is applied on signals from various sources first to reduce the noise effect. 
About $75\%$ of the data is designated as the training set, from which features are extracted and input into the regression models. Additionally, uncertainty constraints based on experimental data are applied to ensure that the confidence bounds of the model predictions are not overly narrow, which would conflict with real-world conditions. For the remaining $25\%$ of the data, DIs extracted from high-fidelity experimental signals are used as the testing set. It is important to note that standard GPRMs only accept data from a single source, whereas multi-fidelity GPRMs incorporate data from multiple sources. As a result, the two types of models have different training sets but share the same testing set.

\subsection{Multi-fidelity Gaussian Process Regression Model} \label{Sec:model}

The Bayesian approach provides event probabilities based on prior knowledge of the data, a property that can be seamlessly integrated into Gaussian Process Regression \cite{rasmussen2003gaussian}. Multi-fidelity GPRMs, a variant of conventional GPRM, also retain this valuable Bayesian property. In addition to estimating the target value, these models can predict the confidence interval of the estimation, offering the potential for further refinement through techniques such as active learning.

In this paper, a two-level fidelity model is applied, as the dataset originates from two different sources, i.e., simulation and experiments. The ground truth mapping of model inputs (damage states) to outputs (DI values) of the two sources are represented by $f_1(\mathbf{x})$ and $f_2(\mathbf{x})$ respectively. The basic assumption is data from $f_2(\mathbf{x})$ is less since obtaining high-fidelity data can be hard. 

\subsubsection{Formulation} 

In this section, a concise overview of multi-fidelity GPRMs is provided. For a more comprehensive treatment, readers are referred to \cite{raissi2017inferring}. The general formulation of a two-level fidelity model is introduced, utilizing the squared exponential (SE) kernel throughout the formulation:
\begin{equation}
 k(\mathbf{x},\mathbf{x'})=\sigma^2\exp(-\frac{1}{2}\frac{(\mathbf{x}-\mathbf{x'})^T(\mathbf{x}-\mathbf{x'})}{l^2})
 \label{eq:SE_kernel}
\end{equation}
In equation (\ref{eq:SE_kernel}), $\sigma^2$ is the output variance, and $l$ is the lengthscale parameter. For convenience, we will use $\bm{\theta} = (\sigma^2,l^2)$ to represent the hyperparameters in the kernel. The bold letter herein indicates a vector representation.

The initial goal is to use the mapping from simulations $f_1(\mathbf{x})$ to represent the mapping from experiments $f_2(\mathbf{x})$. Both of the mappings are considered noiseless and the noisy formulation will be introduced later. Assuming two independent Gaussian processes which are $f_1(\mathbf{x}) \sim \mathcal{GP}(0,g_1(\mathbf{x},\mathbf{x'};\bm{\bm{\theta_1}}))$ and $\delta(\mathbf{x}) \sim \mathcal{GP}(0,h(\mathbf{x},\mathbf{x'};\bm{\theta_d}))$, where $f_1(\mathbf{x})$ is the ground truth mapping from simulations, i.e., the first-level fidelity data, $ \delta(\mathbf{x})$ is the additive correlation surrogate, $g_1(\mathbf{x},\mathbf{x'};\bm{\bm{\theta_1}})$, $h(\mathbf{x},\mathbf{x'};\bm{\theta_d})$ are covariance functions with the same form introduced in equation (\ref{eq:SE_kernel}) and $\bm{\bm{\theta_1}}, \bm{\theta_d}$ are hyperparameters related to the corresponding kernel functions. The mapping from experiments, i.e., the second-level fidelity function, is then formulated by: 
\begin{equation}
    f_2(\mathbf{x})=\rho
    f_1(\mathbf{x})+\delta(\mathbf{x}) 
    \label{eq:SGPRM}
\end{equation}
where $\rho$ is the cross-correlation parameter which links the two fidelity mappings. Though the actual value can change at different locations, the results show that such an assumption can still give better predictions compared to the conventional Gaussian process estimation. Since a linear combination of Gaussian processes remains a Gaussian process, it can be shown that:
\begin{equation}
    f_2(\mathbf{x})\sim\mathcal{GP}(0,g_2(\mathbf{x},\mathbf{x}';\bm{\theta_2})) 
    \label{eq:SGPRM}
\end{equation}
where $g_2(\mathbf{x},\mathbf{x}';\bm{\theta_2})=\rho^2
g_1(\mathbf{x},\mathbf{x}';\bm{\theta_1})+h(\mathbf{x},\mathbf{x'};\bm{\theta_d}), \bm{\theta_2}$ represents for the hyperparameter vector of the kernel, which leads to:
%\bm{\theta_2}=(\bm{\theta_1},\bm{\theta_d},\rho)$
\begin{equation}
 \left[\begin{array}{c} f_1(\mathbf{x}) \\ f_2(\mathbf{x}) \end{array} \right]\sim\mathcal{GP}
 \left(\mathbf{0},\begin{array}{cc} \mathbf{K}_{11}(\mathbf{x},\mathbf{x}';\bm{\theta_1}) & \mathbf{K}_{12}(\mathbf{x},\mathbf{x}';\bm{\theta_1},\rho) \\
 \mathbf{K}_{21}(\mathbf{x},\mathbf{x}';\bm{\theta_1},\rho) & \mathbf{K}_{22}(\mathbf{x},\mathbf{x}';\bm{\theta_1},\bm{\theta_d},\rho) \end{array} \right)
\end{equation}
where
\begin{subequations}
\begin{eqnarray}
\mathbf{K}_{11}(\mathbf{x},\mathbf{x}';\bm{\theta_1})=g_1(\mathbf{x},\mathbf{x}';\bm{\theta_1})
\\
\mathbf{K}_{12}(\mathbf{x},\mathbf{x}';\bm{\theta_1},\rho)=\mathbf{K}_{21}(\mathbf{x},\mathbf{x}';\bm{\theta_1},\rho)=\rho
g_1(\mathbf{x},\mathbf{x}';\bm{\theta_1})
\\
\mathbf{K}_{22}(\mathbf{x},\mathbf{x}';\bm{\theta_1},\bm{\theta_d},\rho)=\rho^2
g_1(\mathbf{x},\mathbf{x}',\bm{\theta_1})+h(\mathbf{x},\mathbf{x'};\bm{\theta_d})
\end{eqnarray}
\end{subequations}
Given a training dataset $\mathcal{D}$ that contains data with two fidelity levels, $\{\mathbf{x}_{L1},\mathbf{y}_{L1}\}$, $\{\mathbf{x}_{L2},\mathbf{y}_{L2}\}$, the multi-fidelity data with noise can be expressed as:
\begin{subequations}
\begin{eqnarray}
    \mathbf{y_{L1}}=f_1(\mathbf{x}_{L1})+\epsilon_1, \quad \epsilon_1 \sim iid \, \mathcal{N}(0,\sigma_1^2) \\
    \mathbf{y_{L2}}=f_2(\mathbf{x}_{L2})+\epsilon_2, \quad \epsilon_2 \sim iid \, \mathcal{N}(0,\sigma_2^2)
    \label{eq:SGPRM}
\end{eqnarray}
\end{subequations}
% \begin{equation}
% f_{1}(\mathbf{x}) \sim \mathcal{GP}(m(\mathbf{x}),k(\mathbf{x},\mathbf{x'})),
% \end{equation}
In the above equations,  $\mathbf{\epsilon_i}$ (i=1,2) is the white noise term. The joint distribution of $\mathbf{y}_{L1}$ and $\mathbf{y}_{L2}$ also follows a Gaussian distribution:
\begin{equation}
\mathbf{y} \sim \mathcal{N}(0,\mathbf{K})
\end{equation}
where
\begin{equation}
\mathbf{y} = \left[\begin{array}{c} \mathbf{y_{L1}} \\ \mathbf{y_{L2}} \end{array} \right], \quad \mathbf{K}=\left(\begin{array}{cc} \mathbf{K}_{11}(\mathbf{x},\mathbf{x}';\bm{\theta_1})+\sigma_1^2\mathbb{I}
& \mathbf{K}_{12}(\mathbf{x},\mathbf{x}';\bm{\theta_1},\rho) \\
 \mathbf{K}_{21}(\mathbf{x},\mathbf{x}';\bm{\theta_1},\rho) & \mathbf{K}_{22}(\mathbf{x},\mathbf{x}';\bm{\theta_1},\bm{\theta_d},\rho)+\sigma_2^2\mathbb{I} \end{array} \right)
\end{equation}
\subsubsection{Training} 

To find the optimized hyperparameters, the marginal likelihood is maximized, or equivalently, minimizing the negative log marginal likelihood (NLML) \cite{rasmussen2003gaussian,raissi2017inferring}: 
\begin{equation}
\mathcal{NLML}(\mathbf{x},\mathbf{x}';\bm{\theta_1},\bm{\theta_d},\rho)
= -\frac{1}{2}\mathbf{y}^{T}(\mathbf{K})^{-1}\mathbf{y}-\frac{1}{2}\log |\mathbf{K}|-\frac{n}{2}\log2\pi
%\end{eqnarray}
\label{eq:nlml}
\end{equation}
where ${n}$ denotes the total number of the training data pairs. The hyperparameters need training are $\bm{\theta_{total}}=[\bm{\theta_1},\bm{\theta_d},\rho,\sigma_1^2,\sigma_2^2]$. A constraint was added during the training process: the variance at the target value must be no less than the variance of the high-fidelity data, i.e., $\sigma_\ast^2$ $\geq$ $\sigma_2^2$. This integrates the preliminary distribution information into the model, and various boundaries of data uncertainties have been examined.

%
% \vspace{-6pt}
\subsubsection{Prediction} 

The joint distribution of the observed target values and the function values at test positions under the priors can be derived as:
\begin{equation}
\left[\begin{array}{c} y_\ast \\ \mathbf{y} \end{array} \right]  \sim  \mathcal{N}\left(\mathbf{0},\begin{array}{cc} \mathbf{K}_{22}(\mathbf{x_\ast},\mathbf{x_\ast};\bm{\theta_1},\bm{\theta_d},\rho) & \mathbf{q}^T \\ \mathbf{q}
&
\mathbf{K} \end{array} \right)
\end{equation}
where
\begin{equation}
\mathbf{q}^T = \left[ \begin{array}{c}
\mathbf{K}_{21}(\mathbf{x_\ast},\mathbf{x}_{L1};\bm{\theta_1},\rho) \quad \mathbf{K}_{22}(\mathbf{x_\ast},\mathbf{x}_{L2};\bm{\theta_1},\bm{\theta_d},\rho)
\end{array}\right]
\end{equation}
from which, the predictive equation for multi-fidelity Gaussian process regression can be shown as:
\begin{equation}
y_\ast|(\mathbf{x}_\ast,X,\mathbf{y})
\sim \mathcal{N}(\mathbf{q}^{T}\mathbf{K}^{-1}\mathbf{y}, \mathbf{K}_{22}(\mathbf{x_\ast},\mathbf{x_\ast};\bm{\theta_1},\bm{\theta_d},\rho)-\mathbf{q}^{T}\mathbf{K}^{-1}\mathbf{q})
\label{eq:predict}
\end{equation}

\subsection{Damage Indices} \label{Sec:SGPRM}

% \subsubsection{Formulation}

In this work, two different methods for DI calculation are demonstrated. Since the application is mainly on the damage state identification, DI that is sensitive for various damage states would be appropriate. The first DI type chosen herein was derived by Janapati et al \cite{Janapati-etal16}. In their work, the authors verified that this formulation is highly sensitive to damage growth (damage size and orientation) while less sensitive to other properties such as structural material and variation in adhesive thickness. The DI is defined as follows:
\begin{eqnarray}
    Y_{u}^n[t] =
    \frac{y_u[t]}{\sqrt{\sum_{t=1}^{N}{y^2_u[t]}}}\nonumber \\
    % \quad
    Y_{0}^n[t]=\frac{\sum_{t=1}^{N}{(y_0[t]\cdot Y_{u}^n[t])}}{\sum_{t=1}^{N}{y_0^2[t]}}y_0[t]
    \nonumber \\ DI_{Janapati}=\sum_{t=1}^{N}{(Y^n_{u}[t] - Y^n_{0}[t])^2}
\label{eq:janapati}
\end{eqnarray}
where $y_0[t]$ and $y_u[t]$ are normalized signals from the healthy and unknown states of the system, indexed with normalized discrete time $t$ ($t=1,\ldots, N$, where $N$ is the data length in samples). 
% ; in this work, after a preliminary investigation of different data lengths, $N$ was selected as 2500, which includes the first $100 \ \mu s$ of the wave propagation signals). 
%
The second method is called Root-Mean-Square Deviation (RMSD):
\begin{equation}
DI_{RMSD} = \sqrt{\frac{\sum_{t=1}^{N}{(y_0[t]-y_u[t])^2}}{N}}
\label{eq:rmsd}
\end{equation}

\begin{algorithm} [!t]
    \SetKwInOut{Input}{Input}
    \SetKwInOut{Output}{Output}
    % \SetKwProg{Init}{Init}{}{}
    \SetKwInput{kwInit}{Init}
    
    % \underline{function Euclid} $(a,b)$\;
    \Input{Experimental data $\{\mathbf{x}_{L1},\mathbf{y}_{L1}\}$ and simulated data $\{\mathbf{x}_{L2},\mathbf{y}_{L2}\}$}
    \Output{DI regression estimation}
    \kwInit{
    Hyperparameters $\bm{\theta_{total_o}}=[\bm{\theta_{1_o}},\bm{\theta_{d_o}},\rho_o,\sigma_{1_o}^2,\sigma_{2_o}^2]$, iteration number $n$}

    % \For{$i = 1$ to ${n_1}$}
    % \While{$itr \leq n$}{
    Extract feature (DI) from signals to form dataset $\mathcal{D}$ by (\ref{eq:janapati}) or (\ref{eq:rmsd})\;
    % Cut out $\mathcal{D}_o$ to $\mathcal{D}$ according to real-world applications, e.g., in this work, $\mathcal{D}$ contains all available simulated DIs as well as experimental DIs only at 0 and 20 kN when combined with active learning\;
    Divide dataset into training and testing sets
    % $\bm{D_train}$
    $\mathcal{D}=[\mathcal{D}_{train}, \mathcal{D}_{test}]$, where $\mathcal{D}_{train}=[\mathcal{D}_{train_{exp}}, \mathcal{D}_{train_{sim}}], \mathcal{D}_{test}=\mathcal{D}_{test_{exp}}$, initially $\mathcal{D}_{train_{sim}}=Null$\;
    Set the bounds for $[\sigma_{1}^2,\sigma_{2}^2]$ according to $\mathcal{D}_{train_{exp}}$\;
    
    Train standard GPRM with $\mathcal{D}_{train_{exp}}$ as baseline\;
    
    \For{$itr\leftarrow 1$ \KwTo $n$}{
        Find the simulated data that is the closest to the target location and update $\mathcal{D}_{train_{exp}}, \mathcal{D}_{train_{sim}}$\;
        Train Multi-fidelity GPRM with experimental and simulated DIs in $\mathcal{D}_{train}$ by (\ref{eq:nlml})\;  
            % return Euclid$(b,a\mod b)$\;
          % }
        Estimate regression along the domain by (\ref{eq:predict})\;
        Find the target location according to the applied acquisition function\;
        % Find the simulated data that is the closest to the target location and update $\mathcal{D}_{train_{exp}}, \mathcal{D}_{train_{sim}}$\;
    } 
    Compare model performance to GPRM using $RMSE$ and ${R}^2$
    % \EndFor
    % next line
    \caption{Multi-fidelity Gaussian Process Regression Model (GPRM) Combined with Active Learning}
    \label{alg:alg1}
\end{algorithm}

\subsection{Active Learning} \label{Sec:AL}
%
%Intro to active learning
When combined with active learning, selecting the location of the next sampling point can be critical on the rate of decreasing the error. After each iteration, extra data which is the closest to the selected location from the lower fidelity dataset will be added to the inputs for another training round. This process is repeated until a certain number of iteration is reached. The algorithm of implementing the framework is shown in Algorithm \ref{alg:alg1}.
% acquisition function intro starts
To locate the next sampling point, the available simulation inputs will be looped and the one that gives the highest function value will be chosen. Four criteria are applied herein to select the next sampling point. The first one is $L2$ loss defined by:

\begin{equation}
% L(x,y,f(x))=(y-f(x))^2=\sum_{i=1}^{N}{(y_i - f(x_i))^2}
L(x,y,f(x))=\sum_{i=1}^{N}{(y_i - f(x_i))^2}
\end{equation}
where $i$ is the index of load at which the functions are evaluated, $y$ is the base function value and $f(x)$ is the Multi-fidelty GPRM output. One thing to be noted is, since only 5 experimental sets are available in this test case, we regressed GPRM using all these 5 sets to obtain the base function values along the domain. 
% In practice it is not always the case to have an accurate base function, but we want to demonstrate the convergence can be achieved with this approach.

The second method is max variance which is trivial since the variance can be directly obtained from the regression results.

The third one is upper confidence bound (UCB) contains explicit exploitation $\mu({x})$ and exploration terms $\sigma({x})$:

\begin{equation}
\alpha(x;\lambda)=\mu({x})+\lambda\sigma({x})
    \label{eq:ucb}
\end{equation}
The forth one is expected improvement (EI) under the GP model can be analytically evaluated as:
\begin{equation}
EI(x) = \begin{cases}
    (\mu(x)-f(x^+)-\xi)\Phi(Z)+\sigma(x)\phi(Z),& \text{if } \sigma(x)\geq 0\\
    0,              & \text{if } \sigma(x)=0
\end{cases}
\label{eq:ei}
\end{equation}
where
\begin{equation}
Z = \begin{cases}
    \frac{\mu(x)-f(x^+)-\xi}{\sigma(x)},& \text{if } \sigma(x)\geq 0\\
    0,              & \text{if } \sigma(x)=0
\end{cases}
\end{equation}
where $\mu(x)$ and $\sigma(x)$ in equations \ref{eq:ucb} and \ref{eq:ei} are the mean and standard deviation of the GP posterior prediction at $x$, respectively. $\Phi$ and $\phi$ are the CDF and PDF of the standard normal distribution. $\lambda$ and $\xi$ are parameters that determine the degree of exploration and exploitation.
\section{First Test Case: Al Coupon Under Varying Damage Sizes}

\begin{table}[t!]
%\vspace{-16pt}
\centering
\small
\caption{Summary of material properties of the Al plate}\label{tab:plate_prop}
\renewcommand{\arraystretch}{1.3}
{\footnotesize\begin{tabular}{|c|c|c|c|c|c|c|}
\hline
Elastic  & Maximum & Density & Initial & Poisson & Energy per \\ 
% \cline{2-3}
% \multicolumn{7}{c|}{Replaced Number of dataset} \\ \cline{2-7}
Modulus & Principal Stress &  & Crack Length & Ratio & Unit Area \\
(GPa) & (MPa) & (g/cm3) & (m) & (g/cm3) & (J/m2) \\
\hline
68.9 & 242 & 2.7 & 0.0944 & 0.33 & 12206.095 \\
\hline
\end{tabular}} 
\end{table}
\begin{table}[t!]
%\vspace{-16pt}
\centering
\small
\caption{Summary of Al plate dimensions}\label{tab:plate_dim}
\renewcommand{\arraystretch}{1.3}
{\footnotesize\begin{tabular}{|c|c|c|c|c|c|c|}
\hline
Thickness  & Hole & Sensor & Sensor & Width & Length \\ 
% \cline{2-3}
% \multicolumn{7}{c|}{Replaced Number of dataset} \\ \cline{2-7}
 & Diameter & Diameter & Thickness & of Plate & of Plate \\
(mm) & (mm) & (mm) & (mm) & (mm) & (mm) \\
\hline
2.36 & 12.7 & 3.175 & 0.2 & 254 & 152.4 \\
\hline
\end{tabular}} 
\end{table}
% \subsection{{\color{blue}Experiment and Simulation Setup}} 

% \subsection{{\color{blue}Damage State Quantification}} \label{sec:prediction}
% 

\subsection{{Experiment Setup}} \label{sec:exp_setup}

\begin{figure}[!t]
   \begin{minipage}{0.48\textwidth}
     \centering
     \includegraphics[scale=0.42]{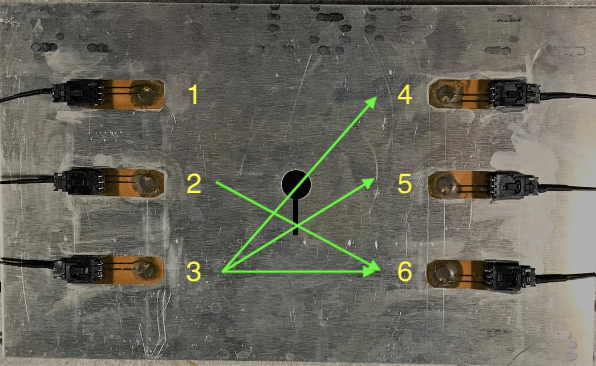}
     % \caption{The notched Al coupon used in this study with a 20-mm notch (largest damage size). Sensors 1-3 were used as activators and 4-6 were used as receivers. The arrows indicate the four paths analyzed in this test case.}\label{fig:Al_coupon}
   \end{minipage}\hfill
   \begin{minipage}{0.48\textwidth}
     \centering
     \includegraphics[scale=0.26]{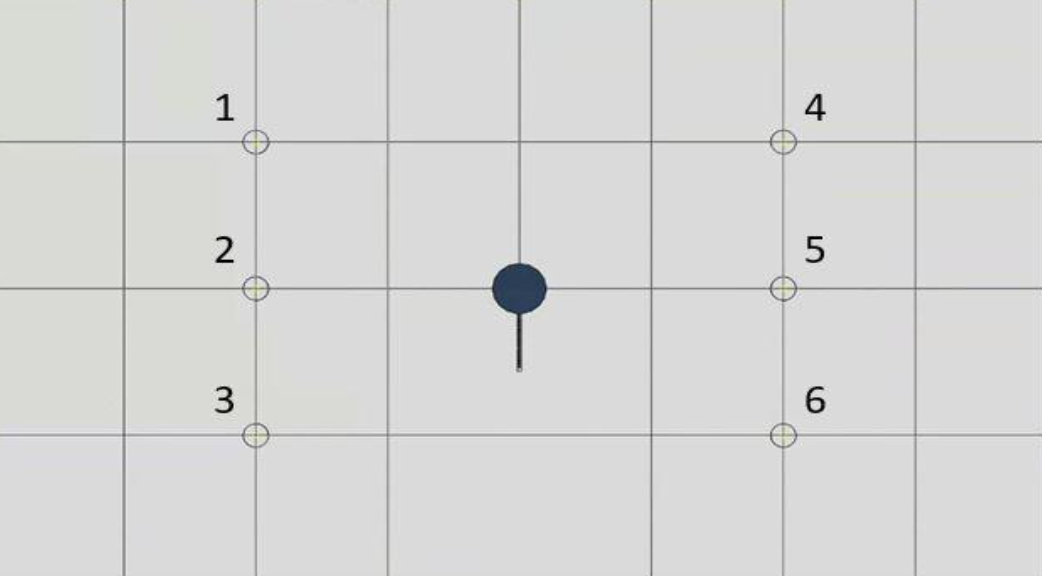}
     % \caption{The simulated Al coupon used in this study with a 20-mm notch (largest damage size). Sensors 1-3 were used as activators and 4-6 were used as receivers. The specifications of Al plate were kept the same as the experiments.}\label{fig:Al_coupon_sim}
   \end{minipage}
   \caption{Left: The notched aluminum coupon used in the first test case, featuring a 20-mm notch (representing the largest damage size). Sensors 1-3 were designated as activators, and sensors 4-6 functioned as receivers. The arrows indicate the four signal paths analyzed in this test case. Right: The simulated aluminum coupon used in this study, also with a 20-mm notch (largest damage size). The specifications of the aluminum plate were kept identical to those used in the experiments.}\label{fig:Al_coupon}
\end{figure}

In this study, the signals were acquired using active-sensing methods from both experiments and simulations on a notched Al plate. Table \ref{tab:plate_prop} presents the material properties, while Table \ref{tab:plate_dim} outlines the dimensions of the coupon. For the experiments, a notch was manually generated from a 12-mm (0.5-in) diameter hole at the center of a 6061 Aluminum plate whose dimension is 152.4 × 254 × 2.36 mm, serving as a simulated crack. Six piezoelectric sensors were attached to the Al plate, as shown on the left side of Figure \ref{fig:Al_coupon}.  During the experiments, crack sizes ranged from 0 mm to 20 mm, increasing in 2 mm increments. Sensors 1-3 were designated as actuators, while sensors 4-6 functioned as receivers. One sensor was actuated with 5-peak tone burst signals at a time, while the three receivers recorded the signals simultaneously. The actuating signals were Lamb waves, known for their efficient propagation through thin structures. The received signals were collected using a ScanGenie III data acquisition system (Acellent Technologies, Inc.) with a sampling frequency of 24 MHz.

%
% \begin{figure}[t!]
% \centering
% \includegraphics[ scale=0.6]{Figures/1sttest/al_coupon2.png} \hspace{0cm}\includegraphics[scale=0.26]{Figures/rep/simu_plate.png}
% \caption{(a) The notched Al coupon used in this study with a 20-mm notch (largest damage size). Sensors 1-3 were used as activators and 4-6 were used as receivers. The arrows indicate the four paths analyzed in this test case. (b) The simulated Al coupon. Sensors 1-3 were used as activators and 4-6 were used as receivers.}
% \label{fig:Al_coupon} 
% % \vspace{10pt}
% \end{figure}
% % \vspace{-12pt}
% %\vspace{-8pt}

% \begin{figure}[t!]
% \centering
% \includegraphics[scale=0.36]{Figures/rep/simu_plate.png}
% \caption{The simulated Al coupon. Sensors 1-3 were used as activators and 4-6 were used as receivers.}
% \label{fig:Al_coupon_sim} 
% % \vspace{10pt}
% \end{figure}
% \vspace{-12pt}
%\vspace{-8pt}

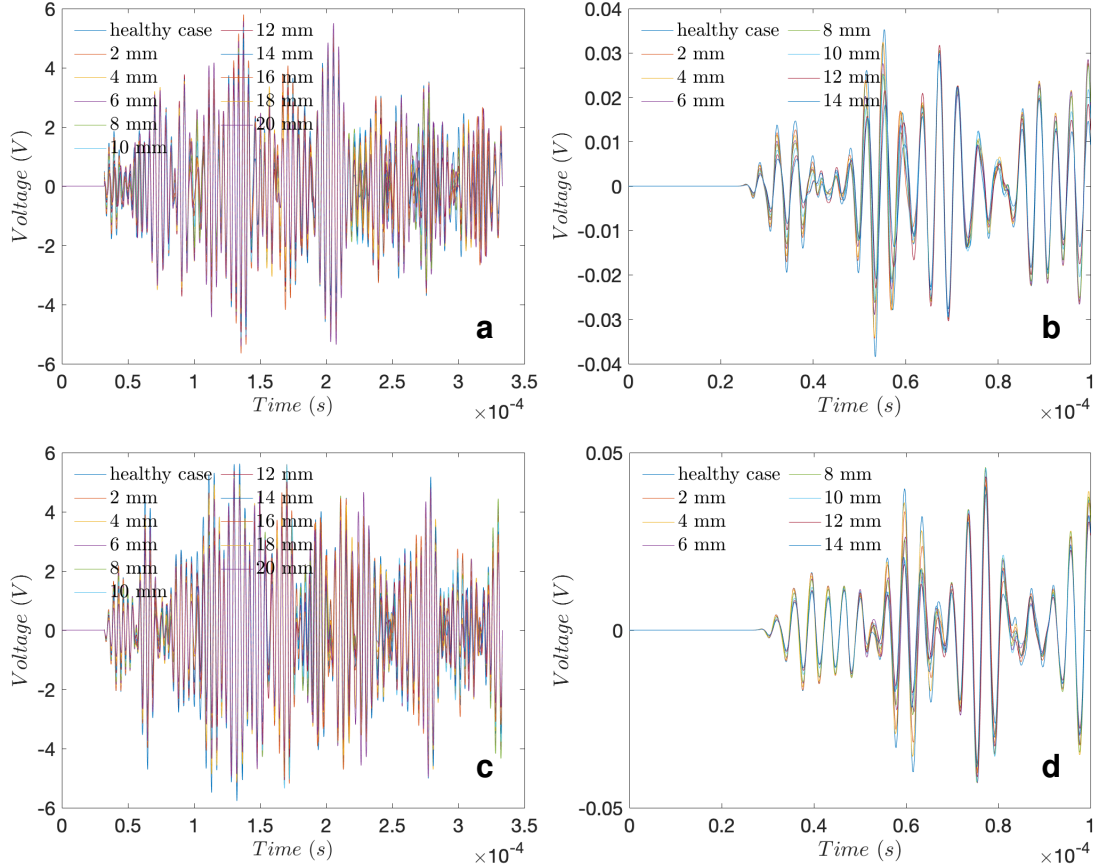
\begin{figure}[t!]
    % \centering
    \begin{picture}(500,328)
    \put(10,168){\includegraphics[width=0.48\textwidth]{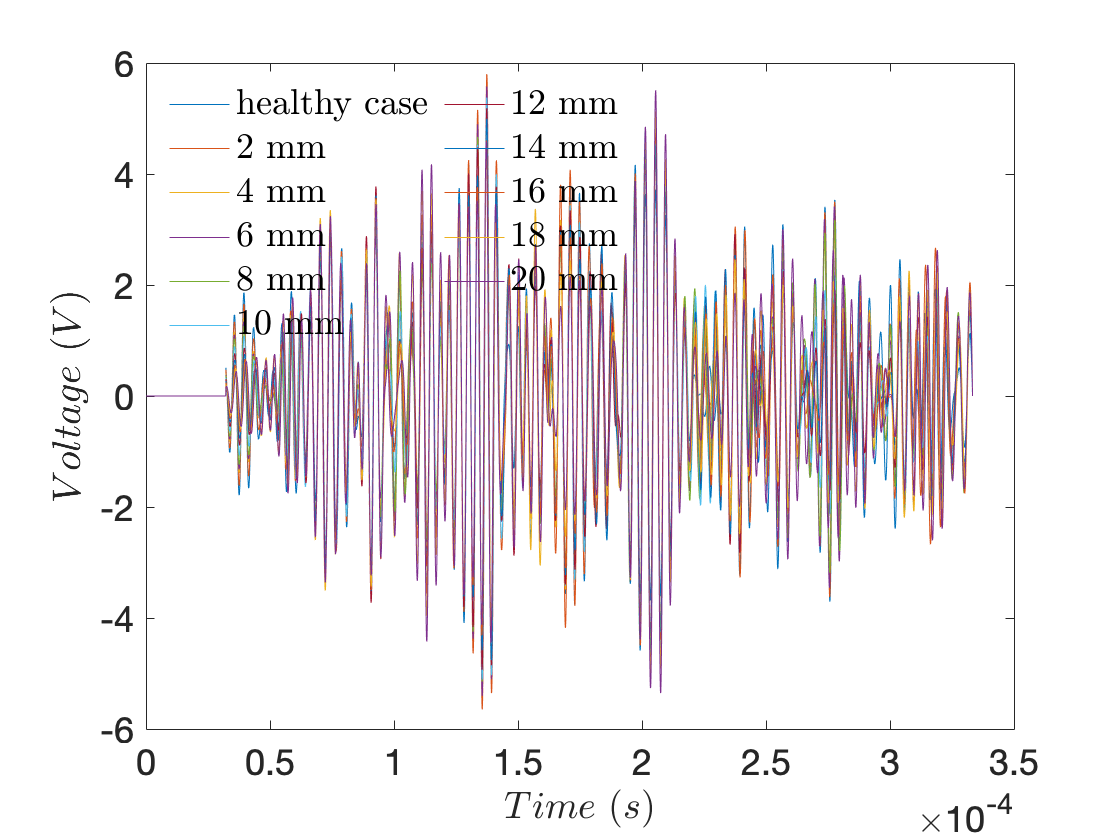}}
    \put(224,168){\includegraphics[width=0.48\textwidth]{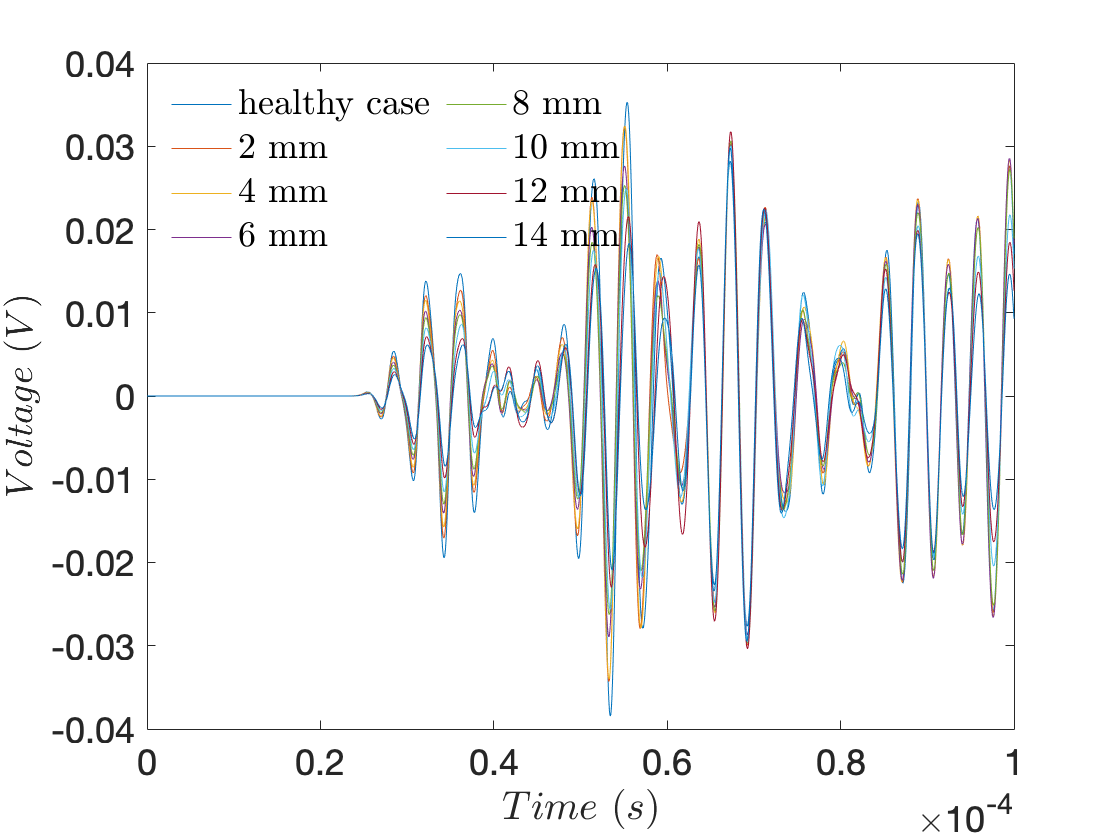}}
    \put(10,0){\includegraphics[width=0.48\textwidth]{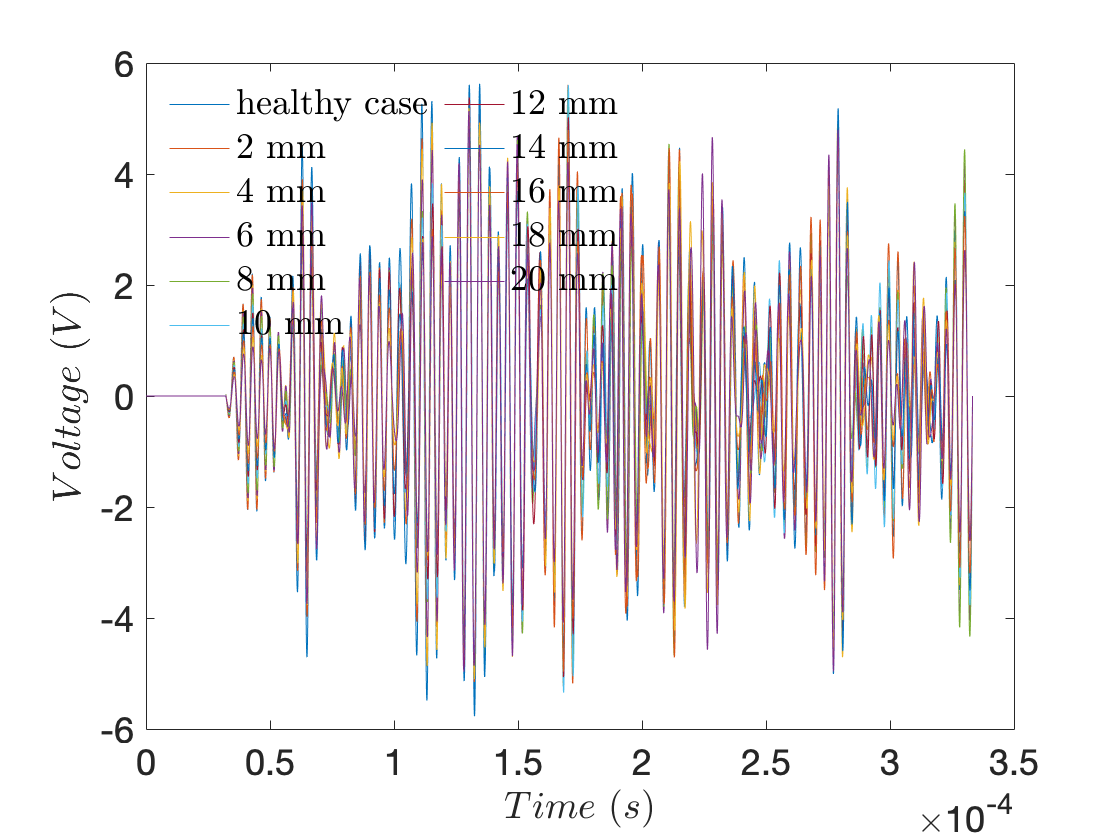}}
    \put(224,0){\includegraphics[width=0.48\textwidth]{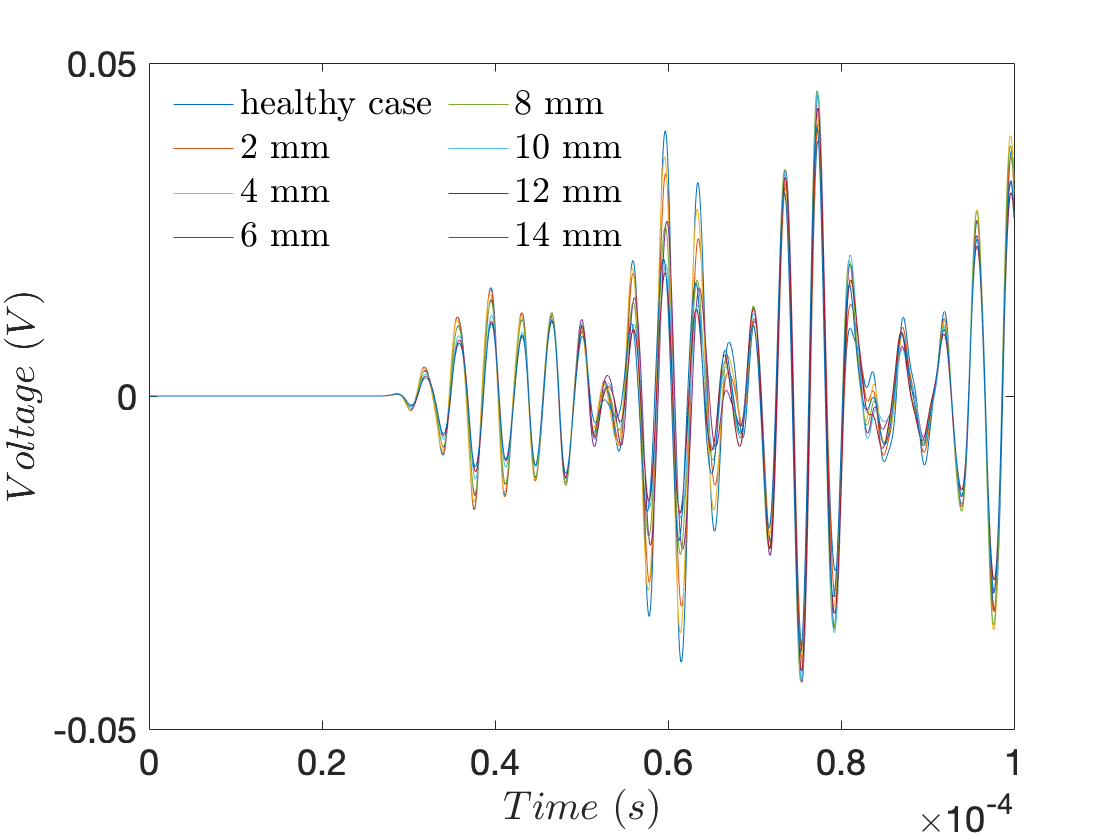}}
    \put(196,200){\color{black} \large {\fontfamily{phv}\selectfont \textbf{a}}}
    \put(410,200){\large {\fontfamily{phv}\selectfont \textbf{b}}}
   \put(196,34){\large {\fontfamily{phv}\selectfont \textbf{c}}} 
   \put(410,34){\large {\fontfamily{phv}\selectfont \textbf{d}}} 
    \end{picture} 
    \caption{
Sample signals collected in test case 1: Panel a and b display the experimental and simulated signals from path 2-6, respectively, while Panel c and d show the experimental and simulated signals from path 3-4, respectively.}
\label{fig:signals_DIs} 
\end{figure}

%%%% signals from 3-5 3-6
% \begin{figure}[t!]
%     % \centering
%     \begin{picture}(500,410)
%     \put(10,250){\includegraphics[width=0.48\textwidth]{Figures/rep/3-5exp copy.png}}
%     \put(224,250){\includegraphics[width=0.48\textwidth]{Figures/rep/3-5sim copy.png}}
%     \put(10,0){\includegraphics[width=0.48\textwidth]{Figures/rep/3-6exp copy.png}}
%     \put(224,0){\includegraphics[width=0.48\textwidth]{Figures/rep/3-6sim copy.png}}
%     \put(176,280){\color{black} \large {\fontfamily{phv}\selectfont \textbf{a}}}
%     \put(380,280){\large {\fontfamily{phv}\selectfont \textbf{b}}}
%    \put(176,114){\large {\fontfamily{phv}\selectfont \textbf{c}}} 
%    \put(380,114){\large {\fontfamily{phv}\selectfont \textbf{d}}} 
%     \end{picture} 
%     \vspace{-100pt}
%     \caption{Sample signals collected in test case 1. Panel a and b are experimental and simulated signals from path 3-5, respectively; Panel c and d are experimental and simulated signals from path 3-6, respectively}
% \label{fig:signals_DIs} 
% % \vspace{-12pt}
% \end{figure}

%% RMSD & Janapati
\begin{figure}[t!]
\centering
% \hspace{-0.5cm}
\includegraphics[scale=0.2]{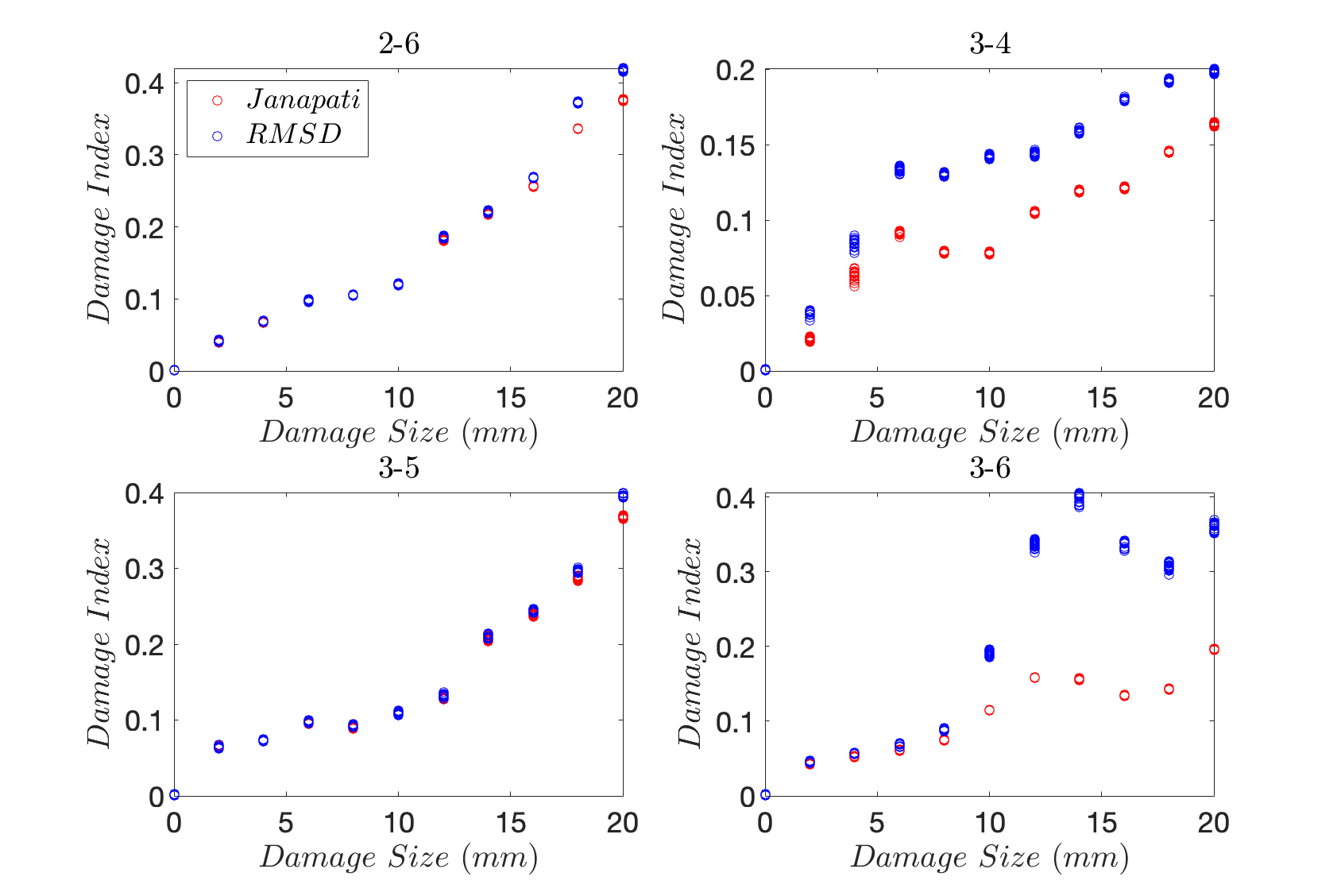}
% \includegraphics[scale=0.168]{Figures/rep/JvsR1.png}
% \vspace{-1.5cm}
\vspace{-8pt}\caption{The indicative evolution of two types of introduced Damage Indices (DIs) with respect to notch size for the four paths is shown. It can be observed that the RMSD DI exhibits a larger range and is more sensitive to variations in damage size compared to the other DI.}
\label{fig:DI_evo} 
% \vspace{2pt}
\end{figure}

%% RMSD
\begin{figure}[t!]
    % \centering
    \begin{picture}(500,300)
    \put(10,150){\includegraphics[width=0.42\textwidth]{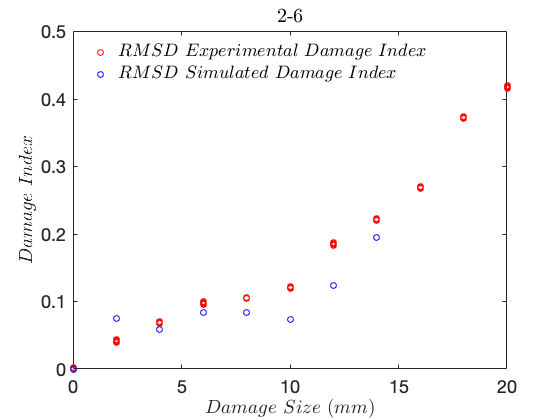}}
    \put(224,150){\includegraphics[width=0.42\textwidth]{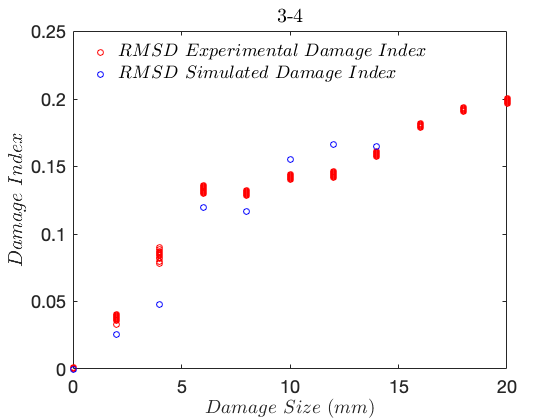}}
    \put(10,0){\includegraphics[width=0.42\textwidth]{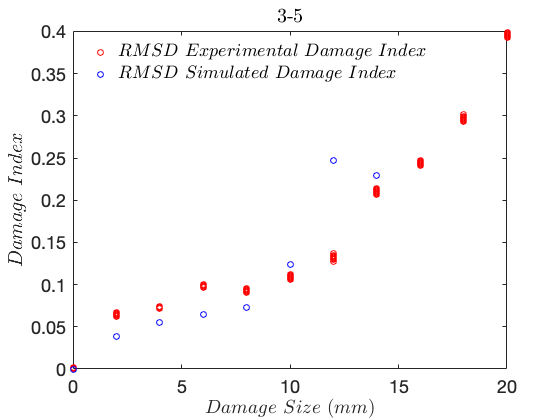}}
    \put(224,0){\includegraphics[width=0.42\textwidth]{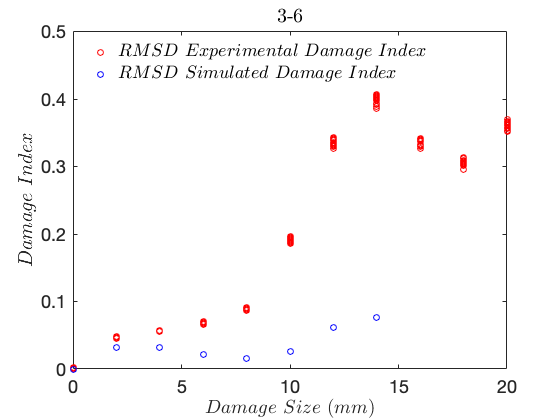}}
    \put(196,170){\color{black} \large {\fontfamily{phv}\selectfont \textbf{a}}}
    \put(410,170){\large {\fontfamily{phv}\selectfont \textbf{b}}}
   \put(196,16){\large {\fontfamily{phv}\selectfont \textbf{c}}} 
   \put(410,16){\large {\fontfamily{phv}\selectfont \textbf{d}}} 
    \end{picture} 
    % \vspace{-100pt}
    \caption{Indicative RMSD DI evolution of two data sources applied in this study with respect to notch size for four paths.}
\label{fig:signals_DIs} 
% \vspace{-12pt}
\end{figure}
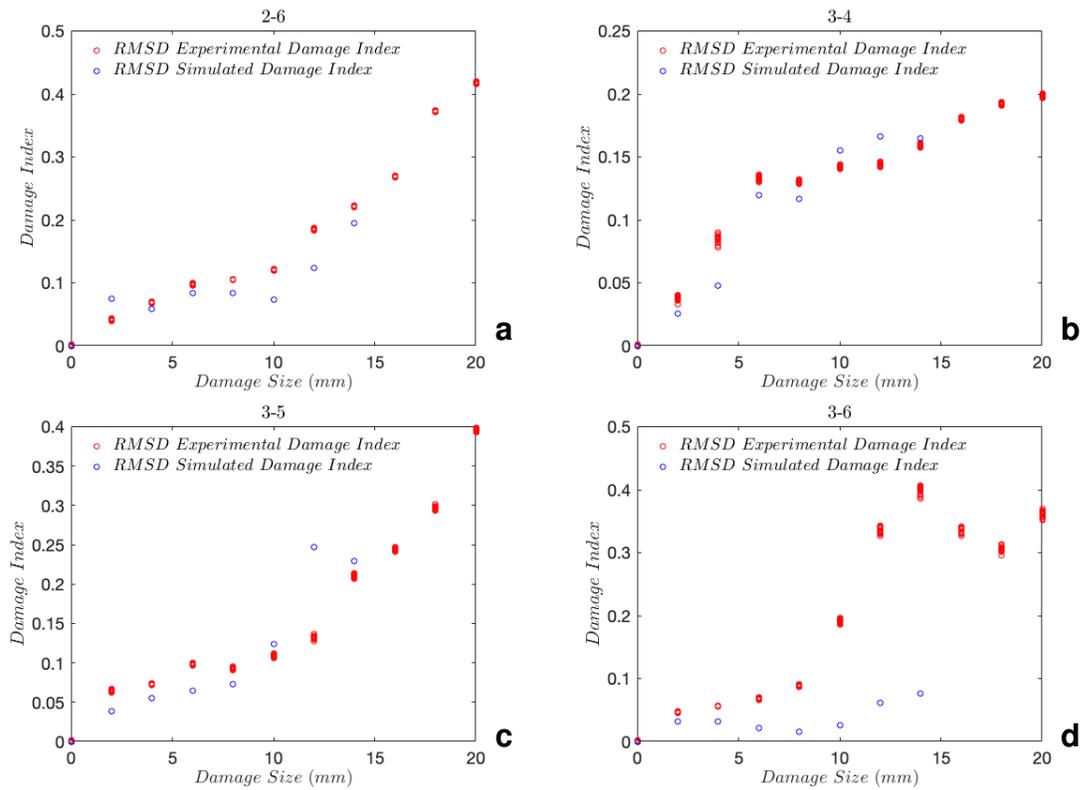
% \begin{figure}[t!]
%     \centering
%     \begin{picture}(400,300)
%     \put(0,40){\includegraphics[trim = 20 0 20 15,clip,scale=0.8]{Figures/DI_example_1.png}}
%     \put(0,-95){\includegraphics[trim = 20 0 20 15,clip,scale=0.8]{Figures/DI_example_2.png}}
%   \put(45,295){\color{black} \large {\fontfamily{phv}\selectfont \textbf{a}}}
%     \put(236,295){\large {\fontfamily{phv}\selectfont \textbf{b}}}
%   \put(45,160){\large {\fontfamily{phv}\selectfont \textbf{c}}} 
%   \put(236,160){\large {\fontfamily{phv}\selectfont \textbf{d}}} 
%     \end{picture} \vspace{-55pt}
%     \caption{Indicative DI plots from path 1-6 in the third test case in this study (Al coupon with simulated damage) showing the DI evolution with respect to damage size and load using both classes of reference signals: (a) evolution of DI values calculated using class 1 reference signals with damage size at a load of 5 kN; (b) evolution of DI values calculated using class 1 reference signals with load in the healthy state; (c) evolution of DI values calculated using class 2 reference signals with damage size at a load of 5 kN; (d) evolution of DI values calculated using class 2 reference signals with load in the healthy state. The red dots indicate the means of the DI values at every state.} 
% \label{fig:example_DI} \vspace{0pt}
% \end{figure}

\subsection{{Simulation Setup}} \label{sec:sim_setup}

For the SEM simulations, the specifications of Al plate were kept the same as the experiments as shown on the right side of Figure \ref{fig:Al_coupon}. The crack sizes ranged from 0 mm to 14 mm in 2 mm increments. Since the sampling frequency for the simulations was 250 MHz, which differs from that used in the experiments, the experimental data was clipped to ensure that the data from both sources covered the same time duration.

\subsection{Results and Discussion} \label{Sec:notched_Al}

\begin{figure}[b!]
    % \centering
    \begin{picture}(500,328)
    \put(10,168){\includegraphics[width=0.48\textwidth]{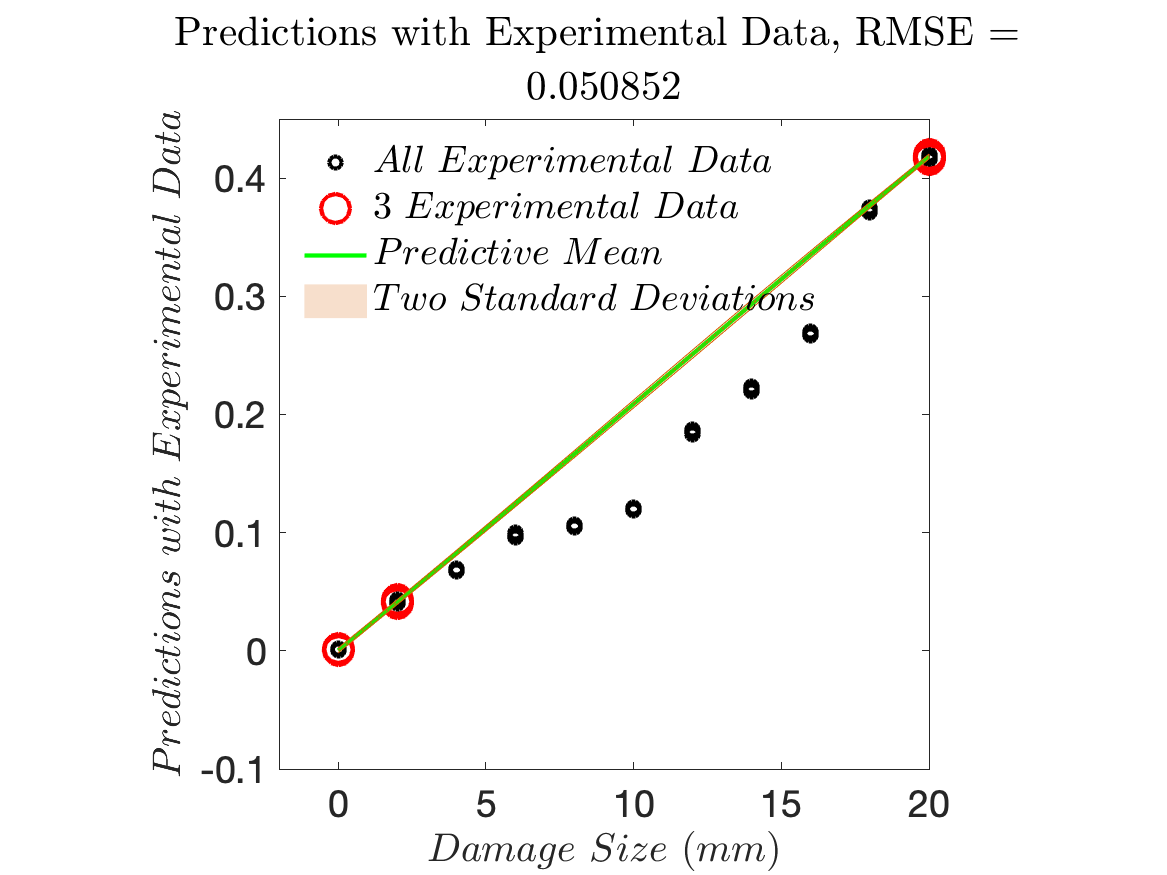}}
    \put(224,168){\includegraphics[width=0.48\textwidth]{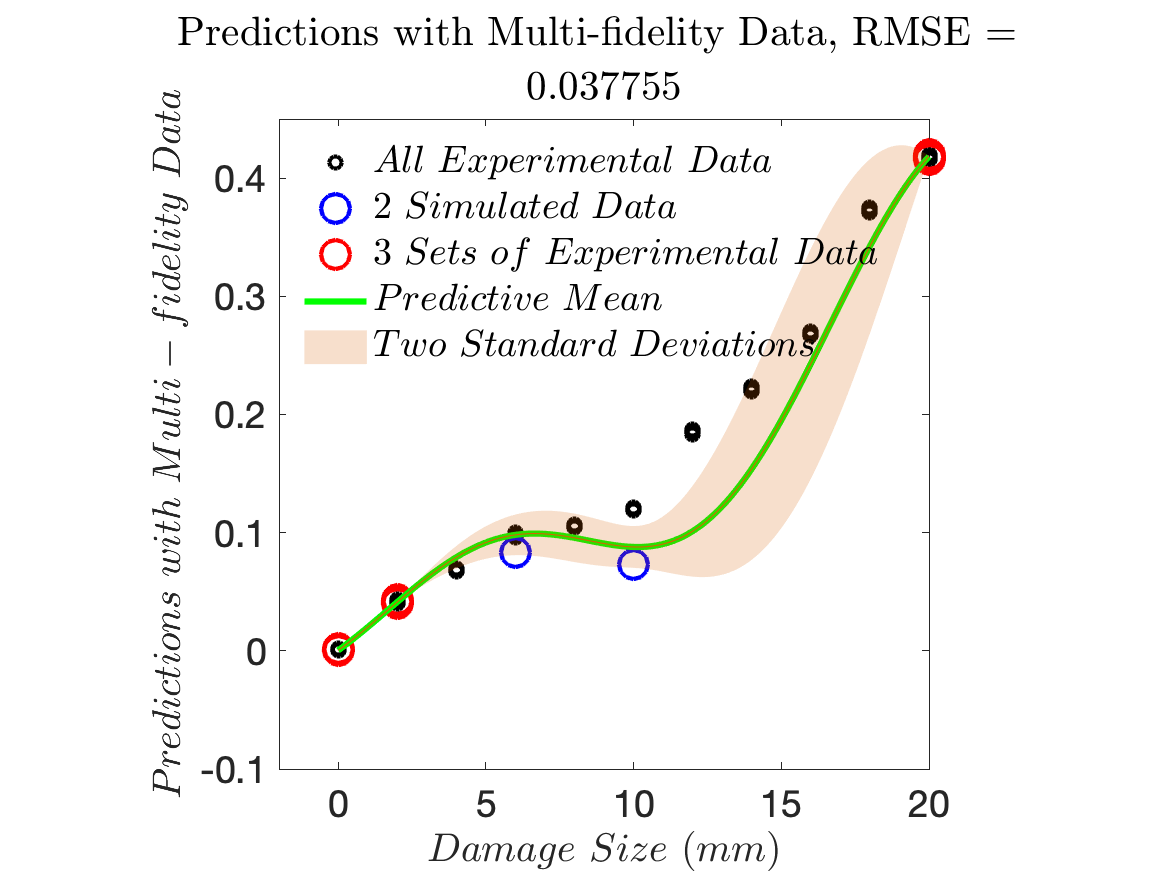}}
    \put(10,0){\includegraphics[width=0.48\textwidth]{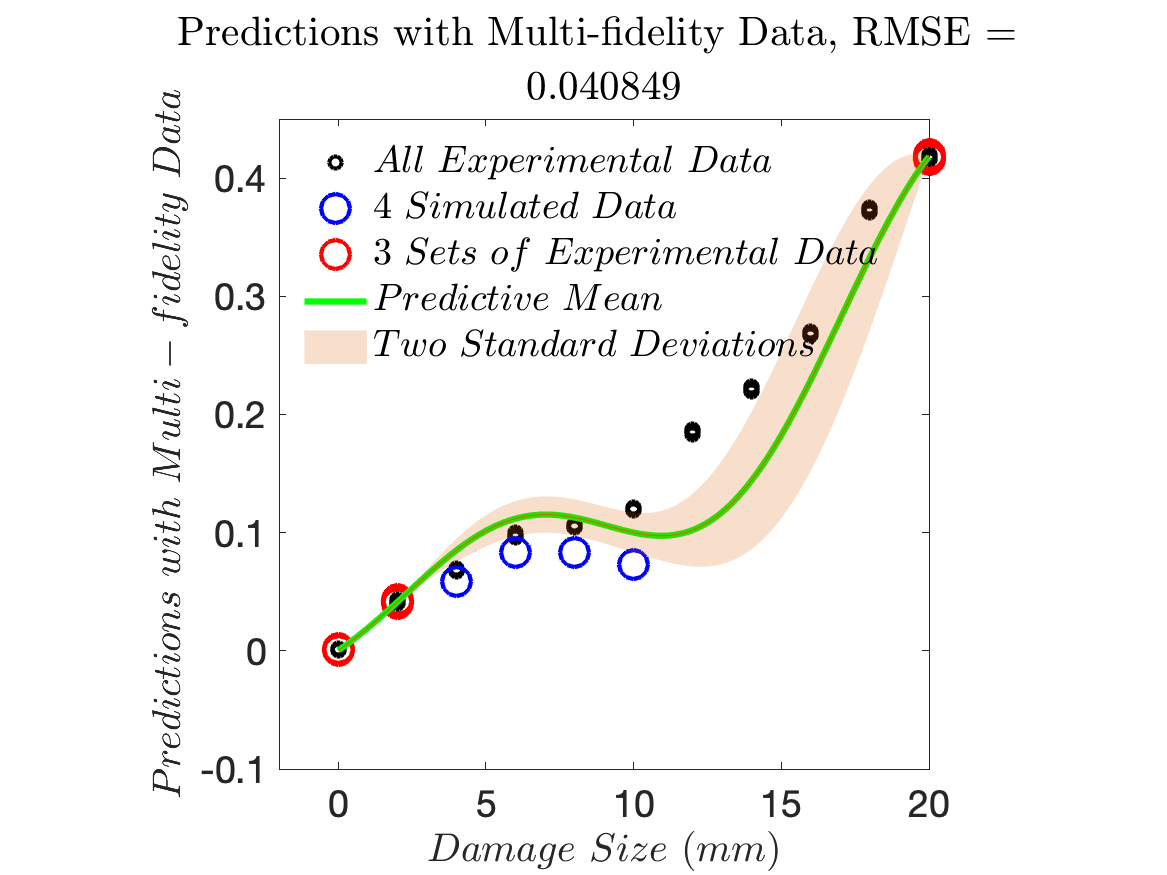}}
    \put(224,0){\includegraphics[width=0.48\textwidth]{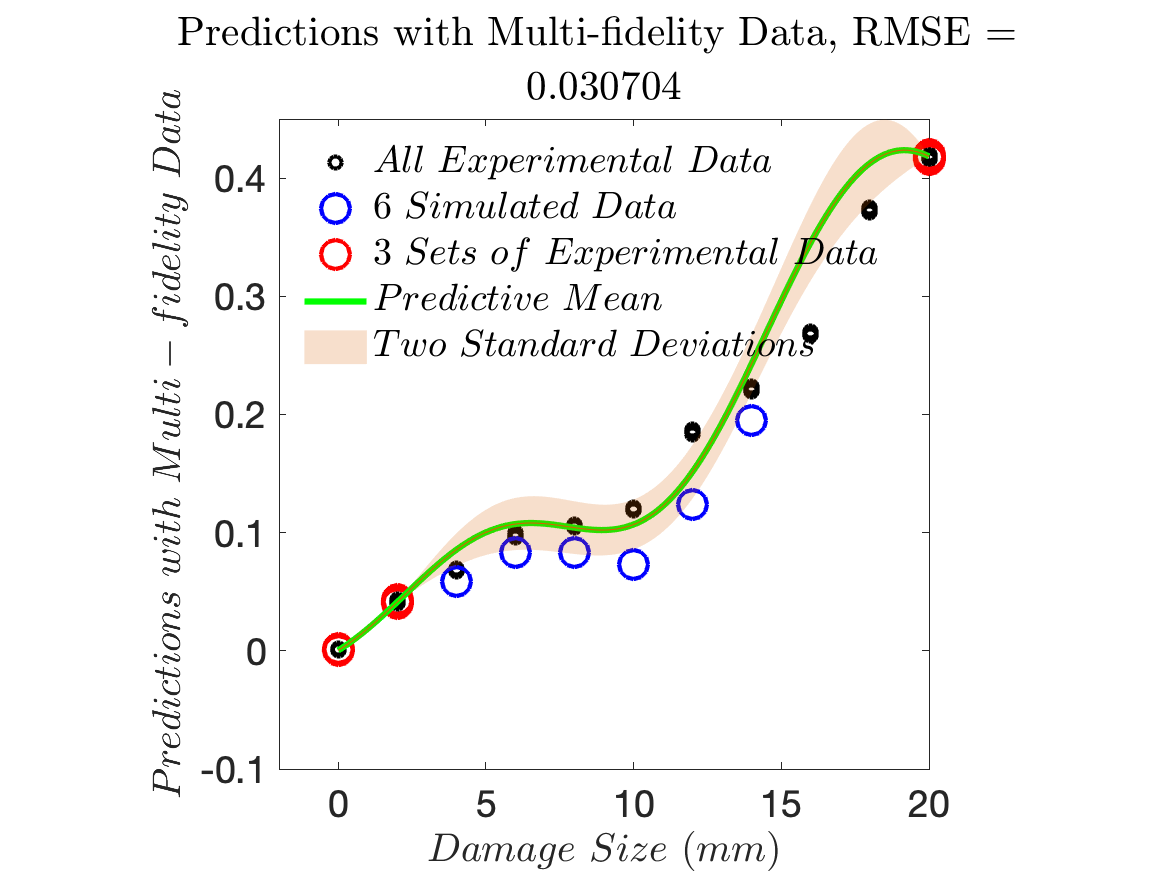}}
    \put(196,200){\color{black} \large {\fontfamily{phv}\selectfont \textbf{a}}}
    \put(410,200){\large {\fontfamily{phv}\selectfont \textbf{b}}}
   \put(196,34){\large {\fontfamily{phv}\selectfont \textbf{c}}} 
   \put(410,34){\large {\fontfamily{phv}\selectfont \textbf{d}}} 
    \end{picture}
    \caption{DI regression for path 2-6 from GPRM and multi-fidelity GPRM: (a) prediction  using 3 experimental sets at 0, 2 and 20 mm; (b) prediction  using 3 experimental sets and 2 simulated data points; (c) prediction  using 3 experimental sets and 4 simulated data points; (d) prediction  using 3 experimental sets and 6 simulated data points.}
\label{fig:test1_1} 
\end{figure}
% \subsubsection{Task 1: Fix the experimental data while increasing simulated data}
% To evaluate the performance of the multi-fidelity Gaussian Process Regression Model (GPRM) in active sensing SHM, the Damage Index (DI), a widely used method for damage detection, was employed after data collection. DIs were computed based on signals captured under both healthy and damaged conditions. 
A common characteristic of DI is its tendency to increase with increasing damage size.  Figure \ref{fig:DI_evo}  illustrates the DI evolution from the two aforementioned methods as a function of crack length for four sensor network paths. As shown, DIs progressively rise for damage-intersecting paths, such as 2-6 and 3-5, while showing greater fluctuations for damage-non-intersecting paths, such as 3-6. However, even in damage-intersecting cases, DIs do not increase uniformly, making it challenging to predict their evolution, especially for crack sizes where experimental data is unavailable. Conventional GPRMs address this issue by providing both mean predictions and confidence intervals, but they remain limited in practice due to the time-consuming experimental setup and the difficulty in obtaining high-accuracy data. To overcome these challenges, multi-fidelity GPRMs were applied in this study, integrating data with two levels of fidelity. It is observed that, although both methods show similar trends, the RMSD DI evolution displays a wider range compared to the DI proposed by Janapati et al. \cite{Janapati-etal16}, making it more sensitive to structural changes in this application. The root mean squared error (RMSE) was employed to evaluate the trained models by quantifying the RMSD DIs.
\begin{figure}[b!]
    % \centering
    \begin{picture}(500,328)
    \put(10,168){\includegraphics[width=0.48\textwidth]{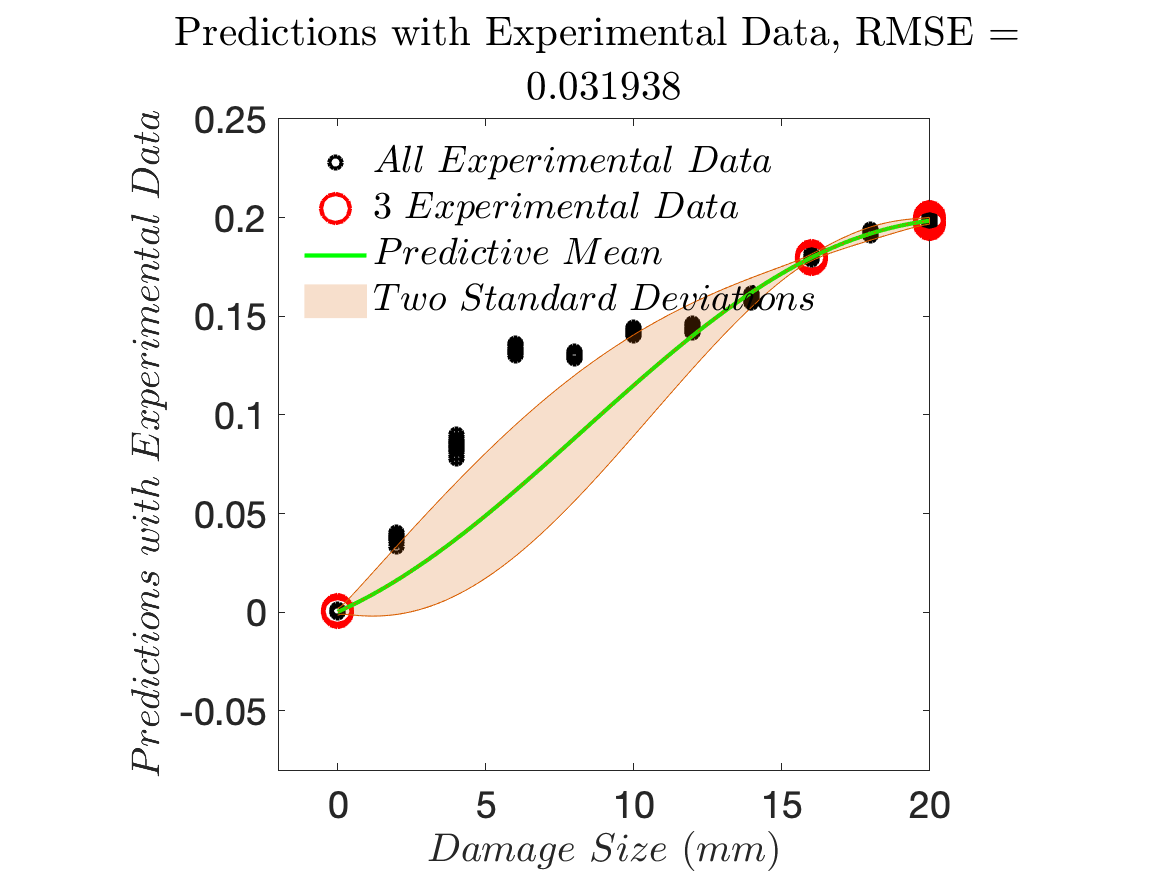}}
    \put(224,168){\includegraphics[width=0.48\textwidth]{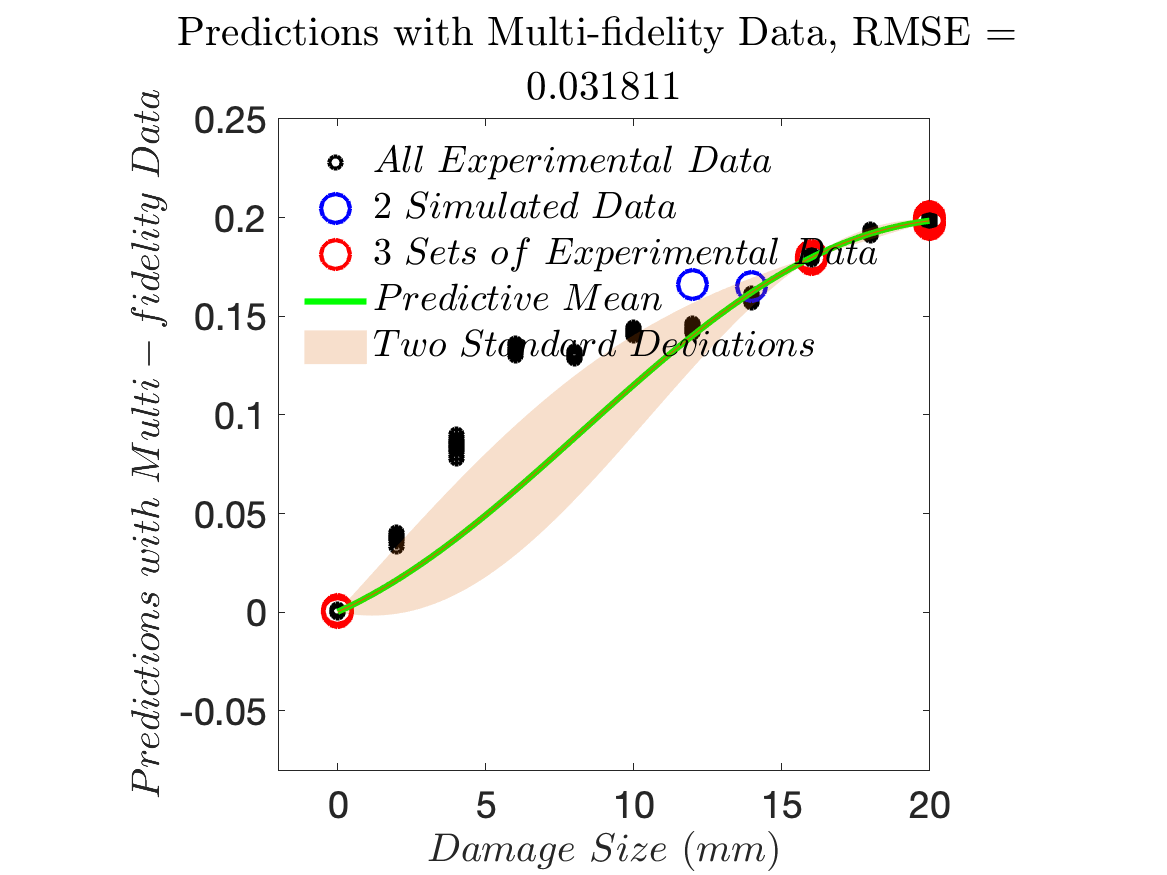}}
    \put(10,0){\includegraphics[width=0.48\textwidth]{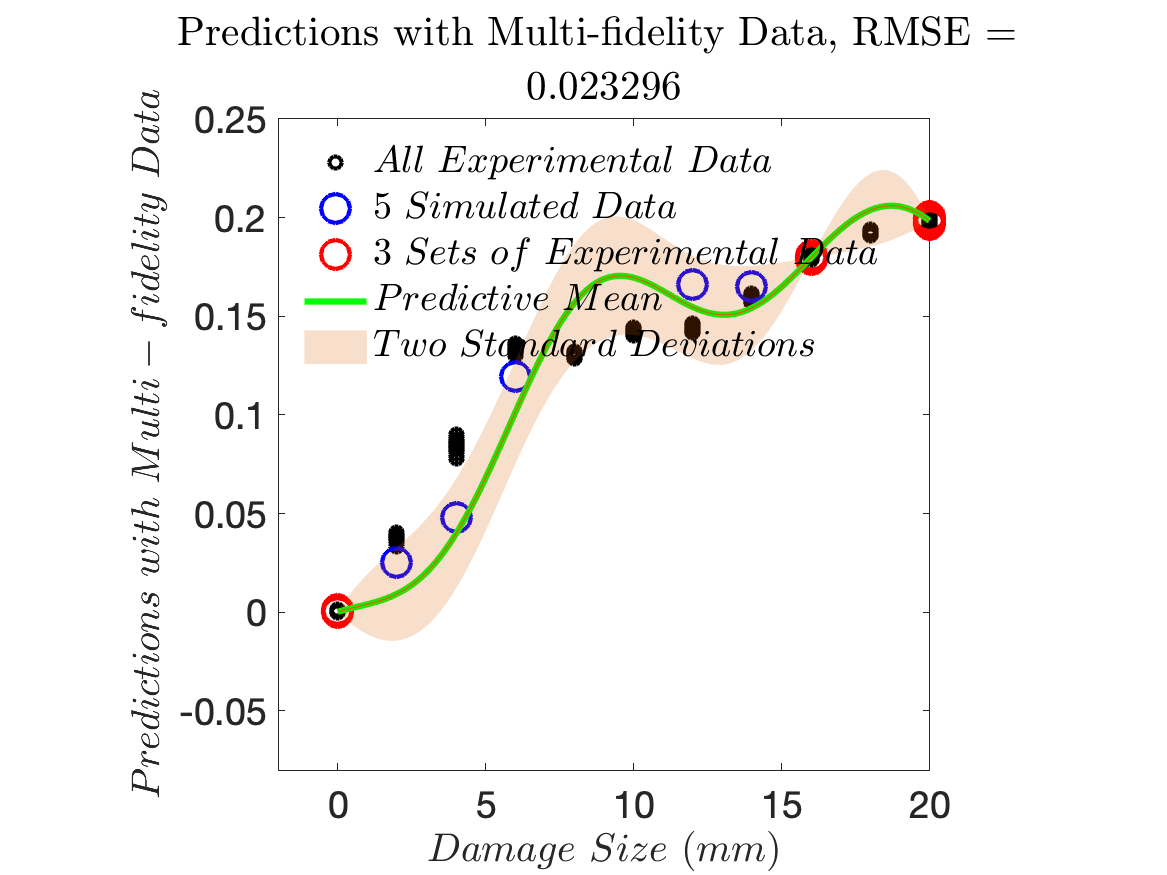}}
    \put(224,0){\includegraphics[width=0.48\textwidth]{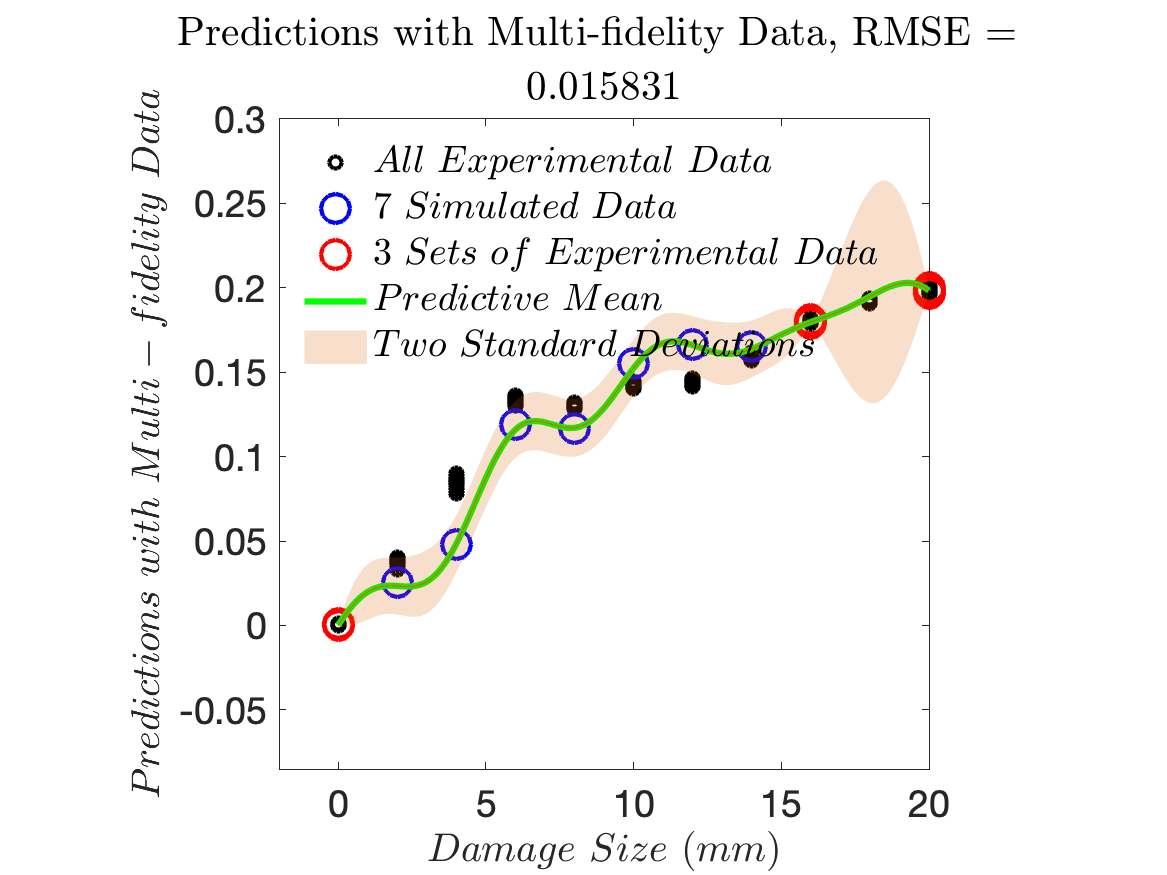}}
    \put(196,200){\color{black} \large {\fontfamily{phv}\selectfont \textbf{a}}}
    \put(410,200){\large {\fontfamily{phv}\selectfont \textbf{b}}}
   \put(196,34){\large {\fontfamily{phv}\selectfont \textbf{c}}} 
   \put(410,34){\large {\fontfamily{phv}\selectfont \textbf{d}}} 
    \end{picture}
    \caption{DI regression for path 3-4 from GPRM and multi-fidelity GPRM: (a) prediction  using 3 experimental sets at 0, 16 and 20 mm; (b) prediction  using 3 experimental sets and 2 simulated data points; (c) prediction  using 3 experimental sets and 5 simulated data points; (d) prediction  using 3 experimental sets and 7 simulated data points.}
\label{fig:test1_2} 
\end{figure}
\begin{figure}[t!]
    % \centering
    \begin{picture}(500,300)
    \put(10,150){\includegraphics[width=0.42\textwidth]{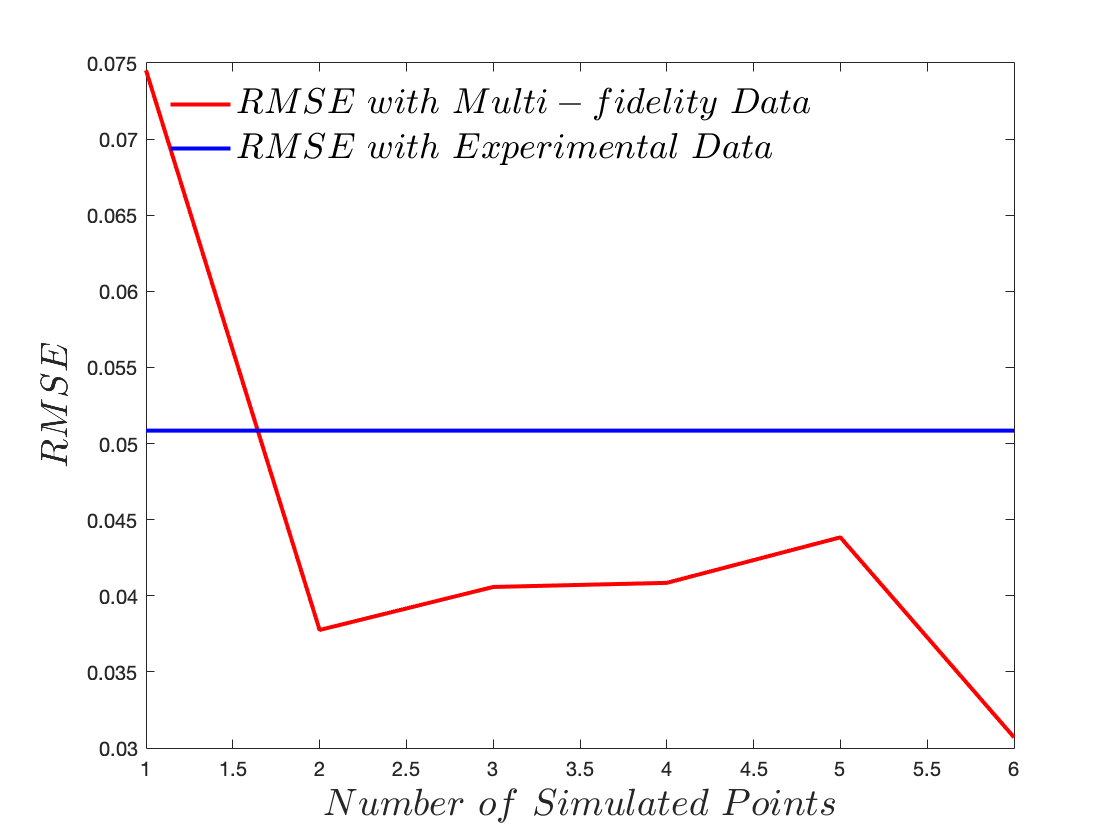}}
    \put(224,150){\includegraphics[width=0.42\textwidth]{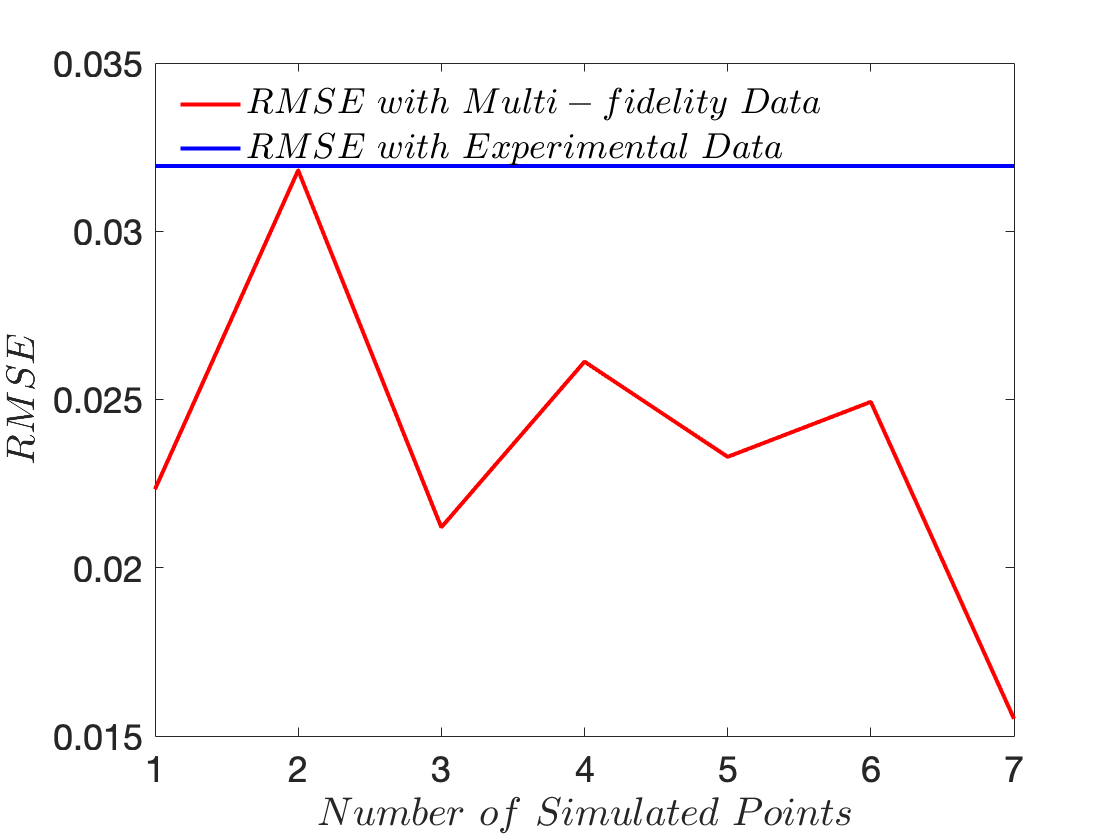}}
    \put(10,0){\includegraphics[width=0.42\textwidth]{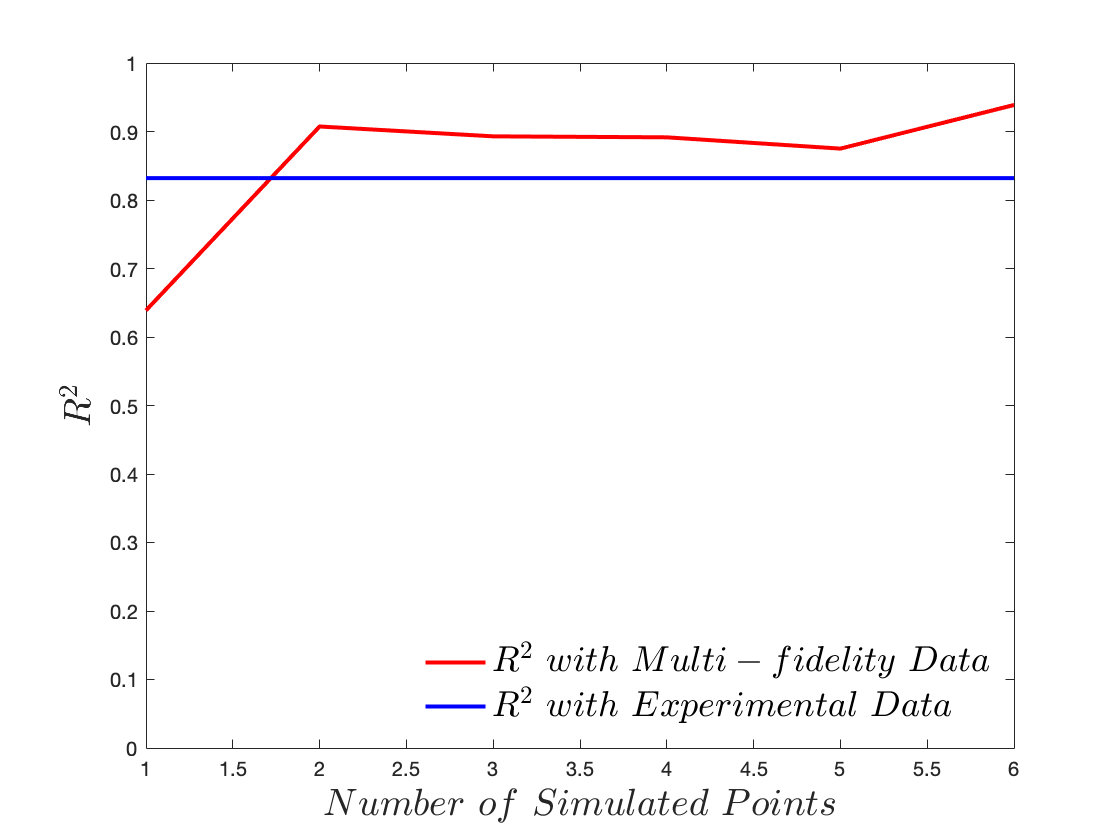}}
    \put(224,0){\includegraphics[width=0.42\textwidth]{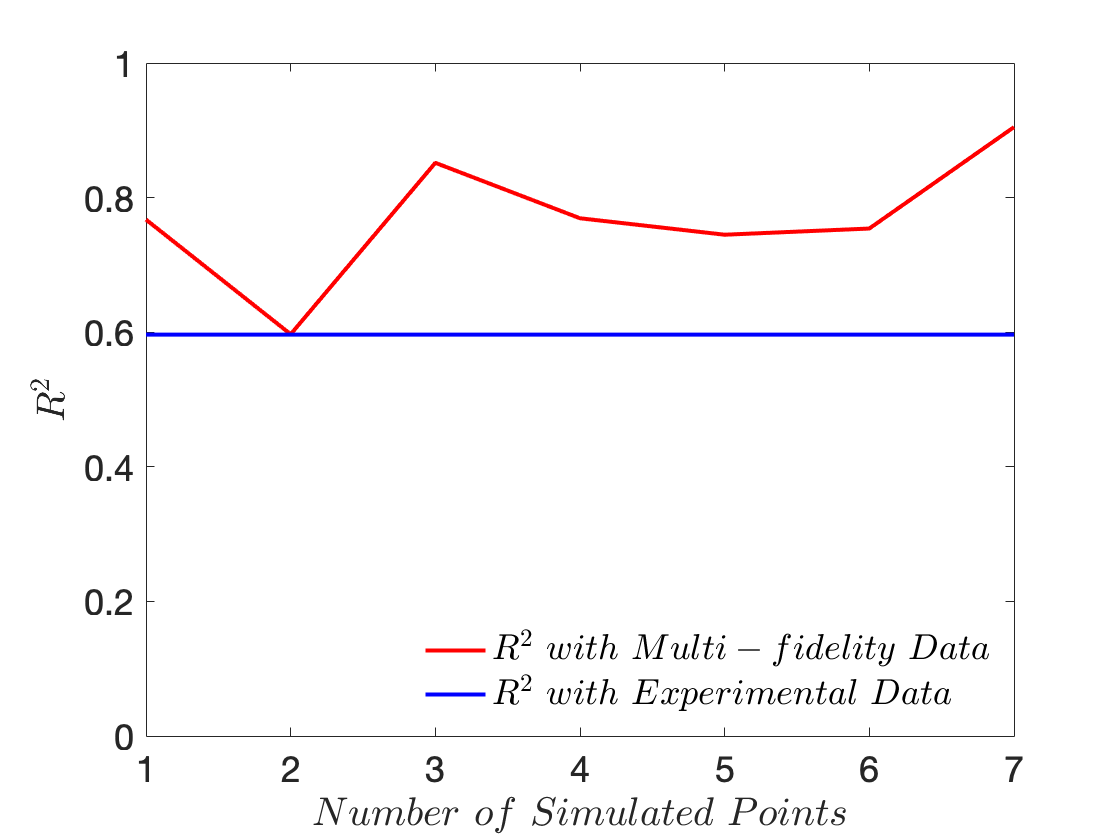} }
    \put(196,170){\color{black} \large {\fontfamily{phv}\selectfont \textbf{a}}}
    \put(410,170){\large {\fontfamily{phv}\selectfont \textbf{b}}}
   \put(196,16){\large {\fontfamily{phv}\selectfont \textbf{c}}} 
   \put(410,16){\large {\fontfamily{phv}\selectfont \textbf{d}}} 
    \end{picture} 
    \caption{RMSE and ${R}^2$ comparison between GPRM and multi-fidelity GPRM for path 2-6 and 3-4.}
\label{fig:test1_3} 
\end{figure}
%
% RMSE and ${R}^2$ comparison between GPRM and multi-fidelity GPRM for path 2-6 and 3-4: (a) RMSE comparison for path 2-6 with experimental sets at 0, 2 and 20 mm; (b) RMSE comparison for path 3-4 with experimental sets at 0, 16 and 20 mm; (c) ${R}^2$ comparison for path 2-6 with experimental sets at 0, 2 and 20 mm; (d) ${R}^2$ comparison for path 3-4 with experimental sets at 0, 16 and 20 mm.

% From the evolution shown in Figure \ref{fig:DI_evo}, paths 2-6 and 3-6 have been chosen as representatives in of 2 different cases, i.e., DIs grow steadily and DIs fluctuate, respectively. 
Figure \ref{fig:signals_DIs} panels (a) through (d) illustrate all available DIs from both experiments and simulations for each of the four studied paths. Experimental data, which has higher fidelity, serves as the ground truth for model assessment. With the extracted features, two real world application scenarios has been investigated, i.e., 1) Fixed High-Fidelity Data: The model is evaluated by adding lower fidelity data while keeping the high-fidelity data constant; 2) Constant Total Number of States: The total number of corresponding states remains unchanged, with lower fidelity data used to fill in data-sparse regions. Both standard GPRM and multi-fidelity GPRM were trained and evaluated to model the feature’s evolution for further comparison. 
\subsubsection{Task 1: Fixed High-Fidelity Data}
In Task 1, the experimental dataset at each state consists of 20 realizations to mitigate the influence of outliers, while the simulated set contains only one realization, as multiple simulations produce identical results. Of the 20 experimental realizations, 15 are randomly selected for training, while the remaining 5 are reserved for testing. Experimental data from the undamaged state and the state with the largest damage are always included as model inputs. This is based on the rationale that experiments on healthy structures are the most straightforward to conduct, and the data from both extremes of the regression curve can provide a general trend within the target region. After comparison, three times the largest variance of the experimental data was chosen as the lower constraint for both $\sigma_1^2$ and $\sigma_2^2$, as introduced in Section \ref{Sec:model}.  

Figure \ref{fig:test1_1} presents the results from two models applied to path 2-6, utilizing experimental datasets at 0, 2, and 20 mm. At each crack size, black circles represent all experimental DIs from 20 repeated realizations, while red and blue circles denote the experimental and simulated DIs used for training, respectively. Panel (a) illustrates an extreme case where only data at three states are available. The $95\%$ confidence interval represented by the yellow region produced by GPRM is quite narrow, as these three input clusters almost coincidentally align linearly. This counterintuitive behavior is alleviated by adding two simulated data points at 6 and 10 mm, as shown in panel (b). The green curve, representing the mean prediction, shifts downward in the central region, aligning more closely with the experimental dat (black dots). In panels (c) and (d), the addition of more data further improves the model's predictive accuracy, as the mean curve approaches the real trend and the root mean squared error (RMSE) decreases to around 0.0307.
% It is worthy to mention that even the DI evolution formed by simulated points gradually deviates from the one formed by experimental sets as the notch size increases, the positions where the tendency change can be captured relatively accurately. 
%

Figure \ref{fig:test1_2}, panel (a), displays the performance of the standard GPRM on path 3-4, using experimental datasets from three of the eleven damage states. Although this model provides a rough estimate of the mean curve, it lacks sufficient information at damage sizes between 2 and 14 mm, failing to capture the DI fluctuations within this range. In panel (b), a multi-fidelity GPRM is applied using the same experimental sets as in panel (a), supplemented with simulated points at 12 and 14 mm. Despite these additions, the RMSE, mean curve, and uncertainty remain almost unchanged, as the two new points fail to provide significant additional information. In panel (c), the inclusion of five additional points improves the model’s performance obviously comparing to panel (a), with the mean curve more accurately reflecting the underlying trend, particularly in the range where experimental data is sparse. Further data augmentation in panel (d) results in clearer DI evolution, reduced confidence intervals, and a significant decrease in RMSE from 0.032593 to 0.030838, approximately halving the original value from GPRM.
% may add explanation on why CI does not cover exp data here, more results in Appendix
% A side effect of this process was also noticed. As inputs cover most of the damage sizes, the model gets more confident about its prediction and the CI may not cover the true values. A comparison of results with different constraints on the lower bound of the model noise variance is provided in Appendix. From the comparison results, the lower bound was selected as three times the largest variance of experimental data at one damage size.

Figure \ref{fig:test1_3} illustrates the variation in RMSE and ${R}^2$ as the number of added simulated data points increases, while the experimental data remains fixed, for paths 2-6 and 3-4. A consistent trend is observed: although adding more simulated data initially introduces deviations from the ground truth, with further additions, the model eventually achieves a lower RMSE and a higher ${R}^2$ compared to the standard GPRMs. These results suggest that, with the inclusion of simulated data, the multi-fidelity model can produce more accurate mean estimates and reduce uncertainty. Based on this conclusion, a second test case, where simulated data is more abundant, was conducted and validated using the proposed framework in Section \ref{Sec:test2}, ensuring the results were not coincidental.
% 
% \subsection{Task 2: Replace the experimental data by simulated data}

\begin{figure}[t]
    % \centering
    \begin{picture}(500,328)
    \put(10,168){\includegraphics[width=0.48\textwidth]{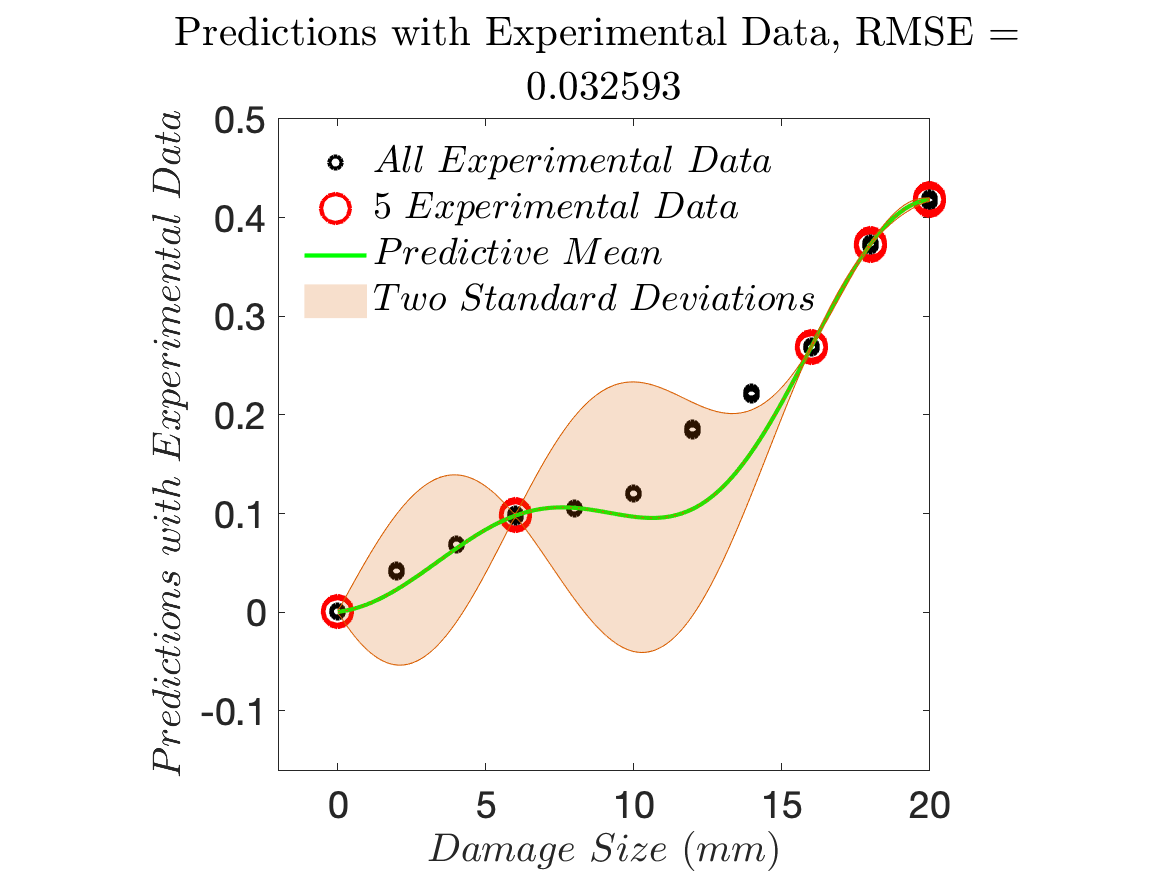}}
    \put(224,168){\includegraphics[width=0.48\textwidth]{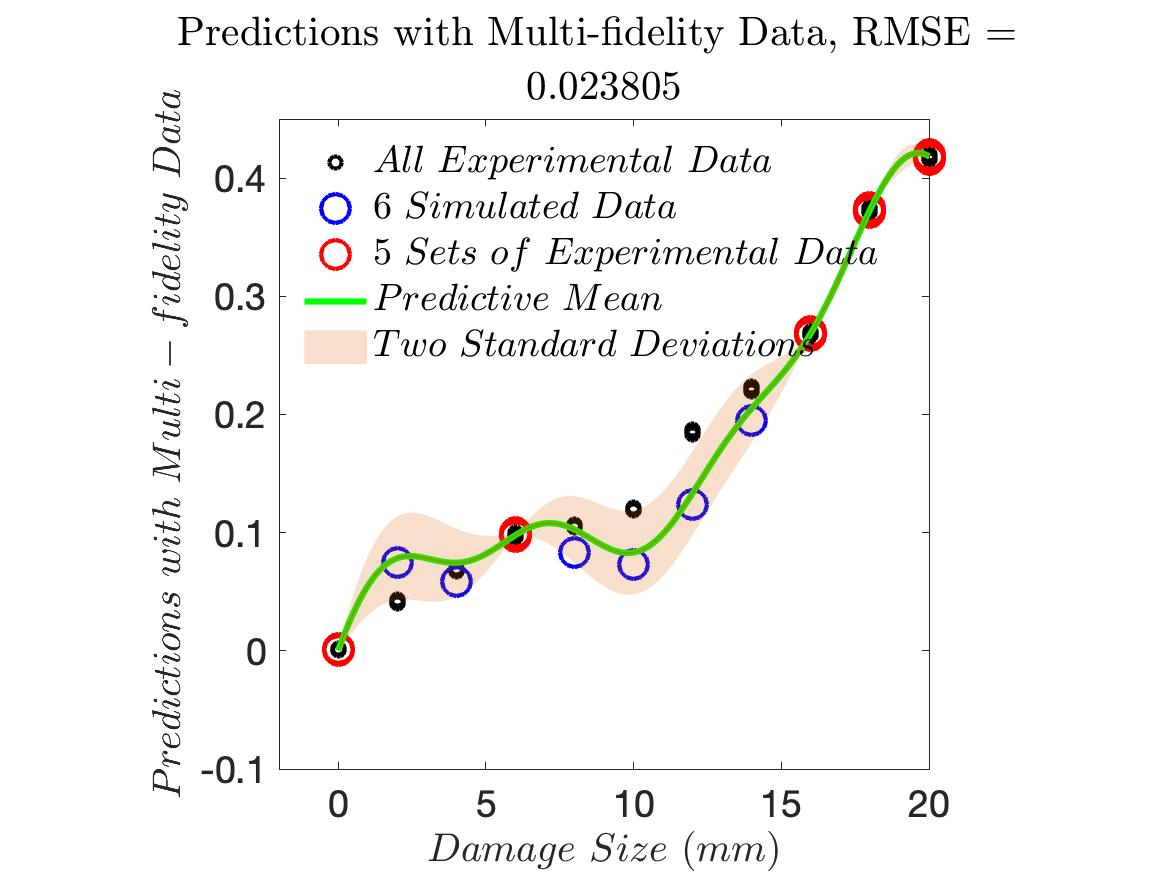}}
    \put(10,0){\includegraphics[width=0.48\textwidth]{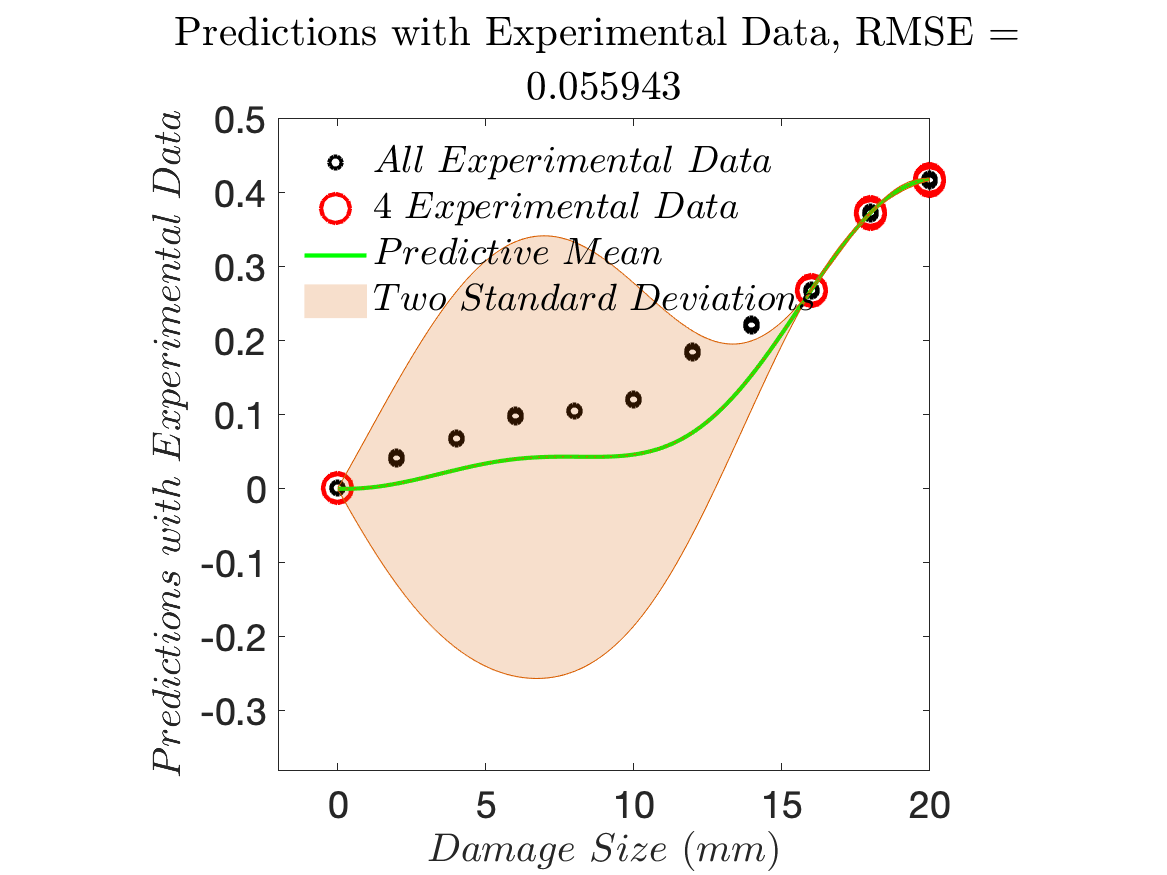}}
    \put(224,0){\includegraphics[width=0.48\textwidth]{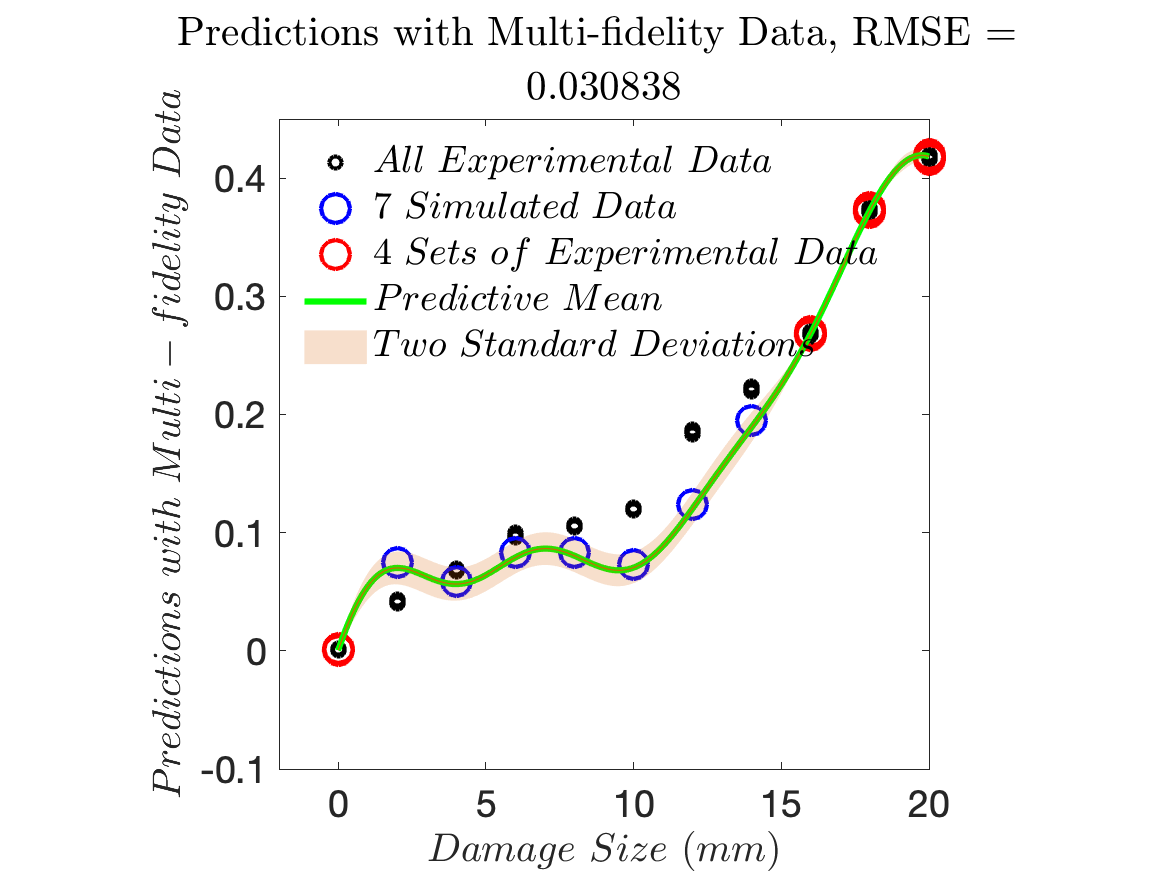}}
    \put(196,200){\color{black} \large {\fontfamily{phv}\selectfont \textbf{a}}}
    \put(410,200){\large {\fontfamily{phv}\selectfont \textbf{b}}}
   \put(196,34){\large {\fontfamily{phv}\selectfont \textbf{c}}} 
   \put(410,34){\large {\fontfamily{phv}\selectfont \textbf{d}}} 
    \end{picture} 
    \caption{DI regression for path 2-6 from GPRM and multi-fidelity GPRM: (a) prediction  using 5 experimental sets at 0, 6, 16, 18 and 20 mm; (b) prediction  using 5 experimental sets and 6 simulated data; (c) prediction  using 4 experimental sets at 0, 16, 18 and 20 mm; (d) prediction  using 4 experimental sets and 7 simulated data.}
\label{fig:test1_4} 
\end{figure}
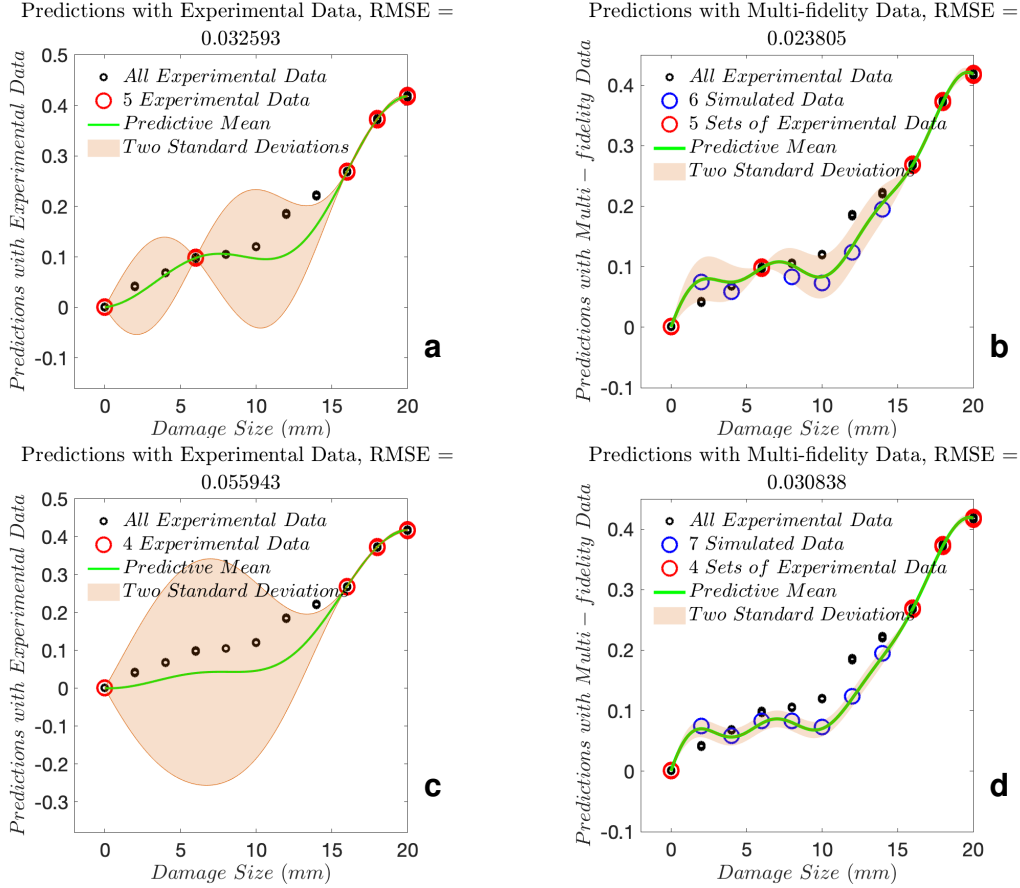
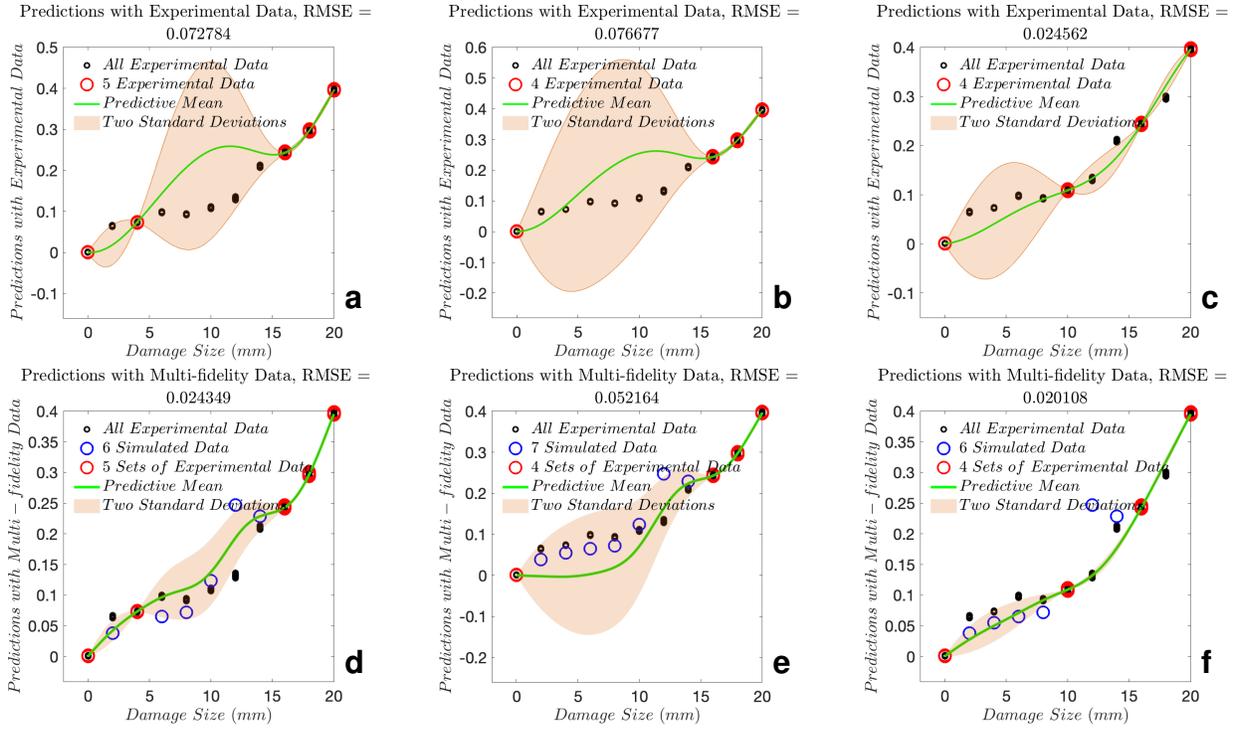
\begin{figure}[t!]
    % \centering
    \begin{picture}(500,280)
    \put(-20,138){\includegraphics[width=0.39\textwidth]{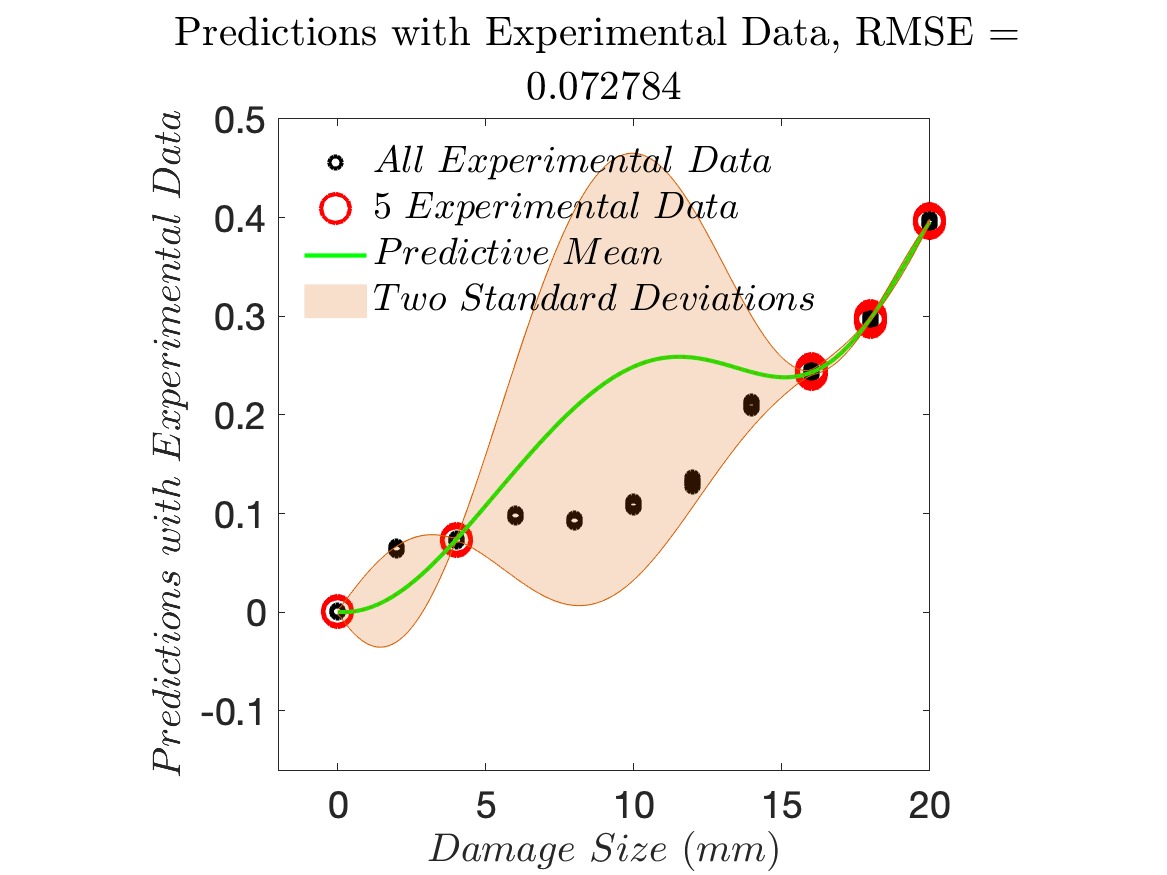}}
    \put(142,138){\includegraphics[width=0.39\textwidth]{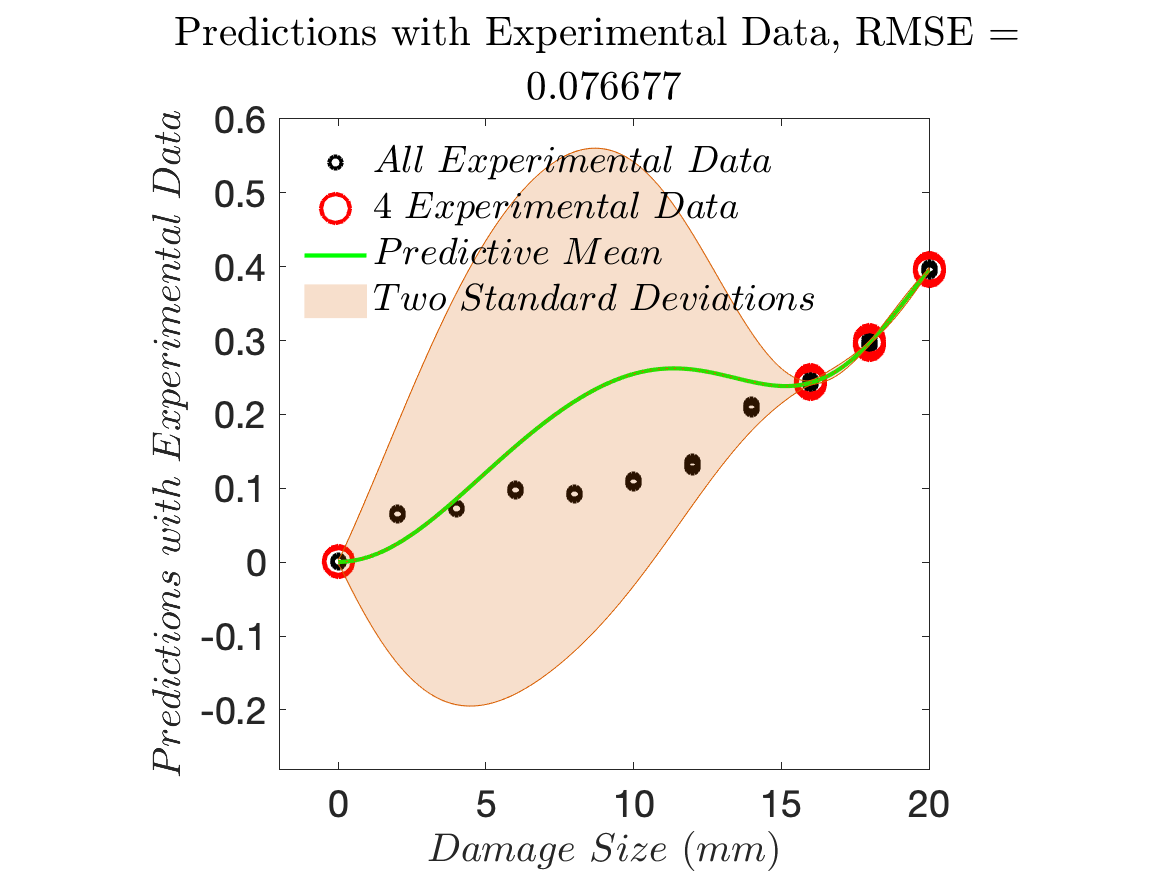}}
    \put(304,138){\includegraphics[width=0.39\textwidth]{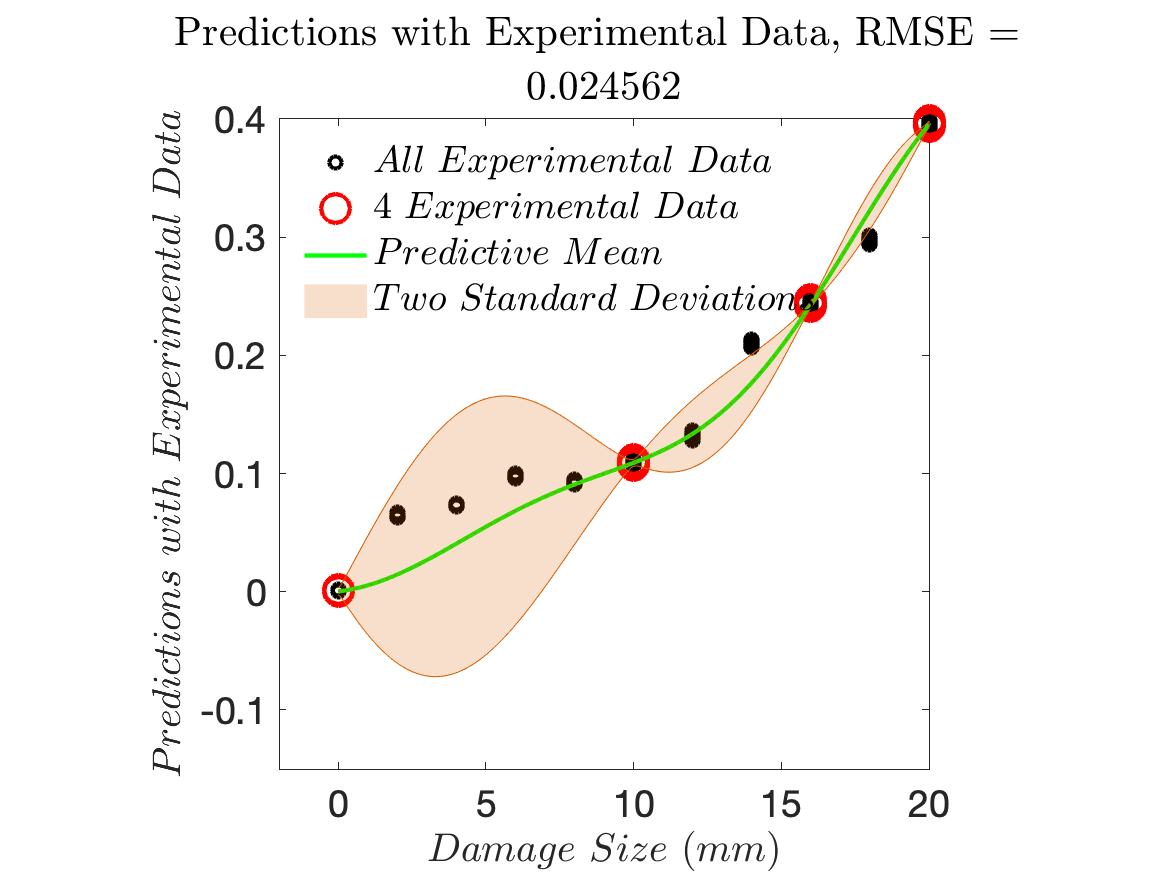}}
    \put(-20,0){\includegraphics[width=0.39\textwidth]{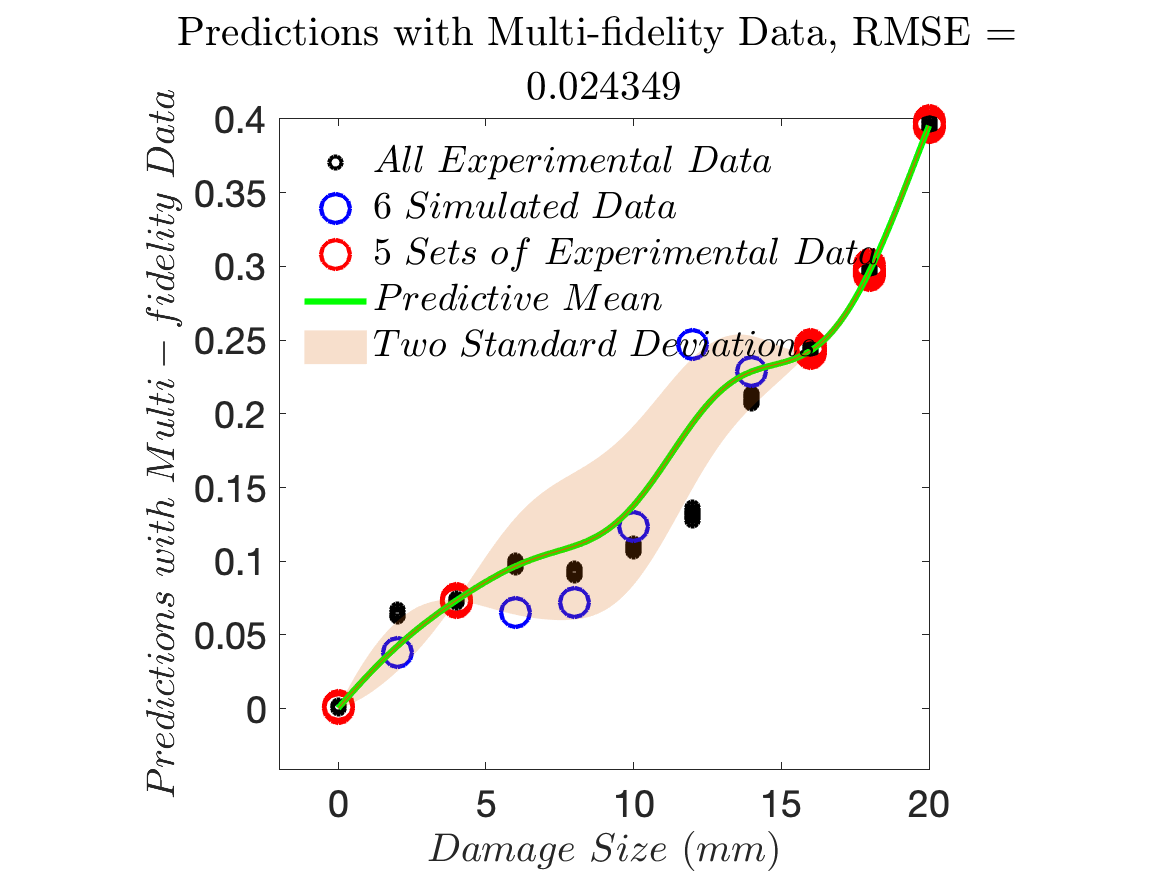}}
    \put(142,0){\includegraphics[width=0.39\textwidth]{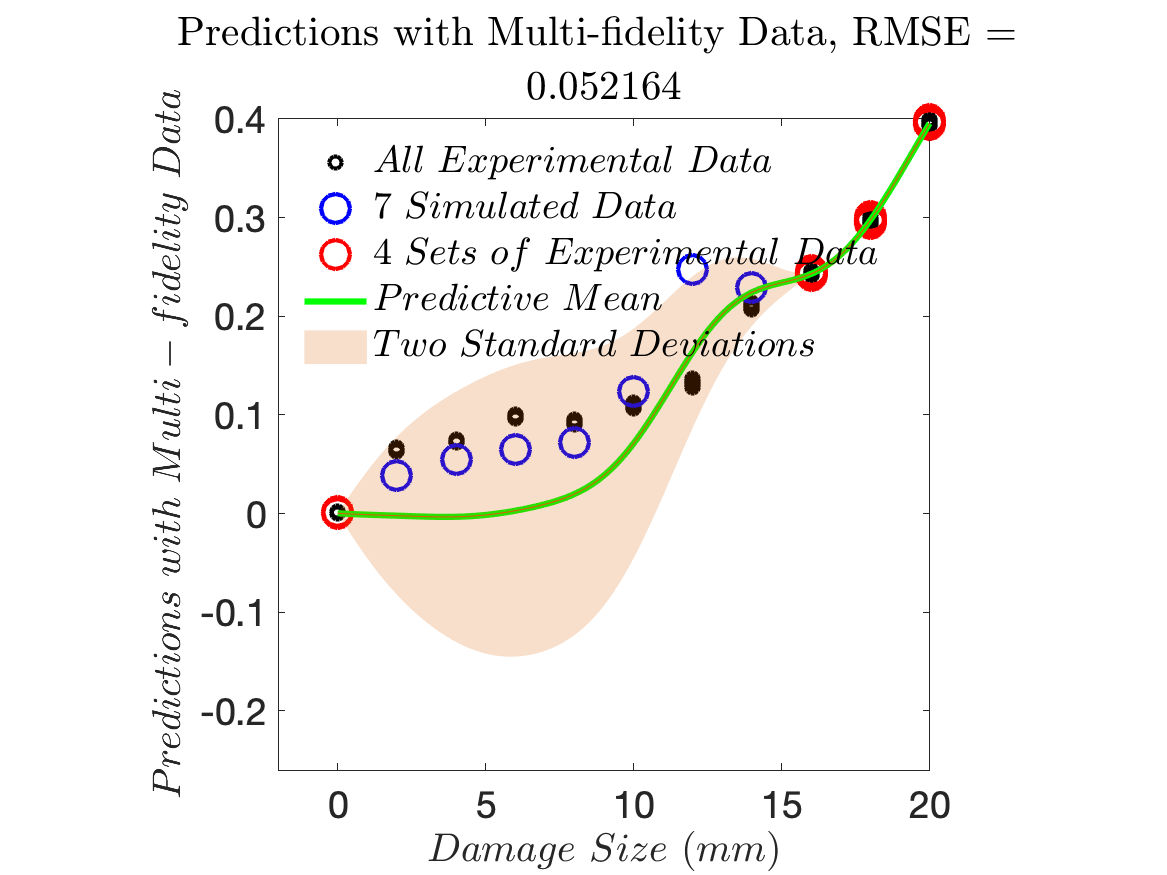}}
    \put(304,0){\includegraphics[width=0.39\textwidth]{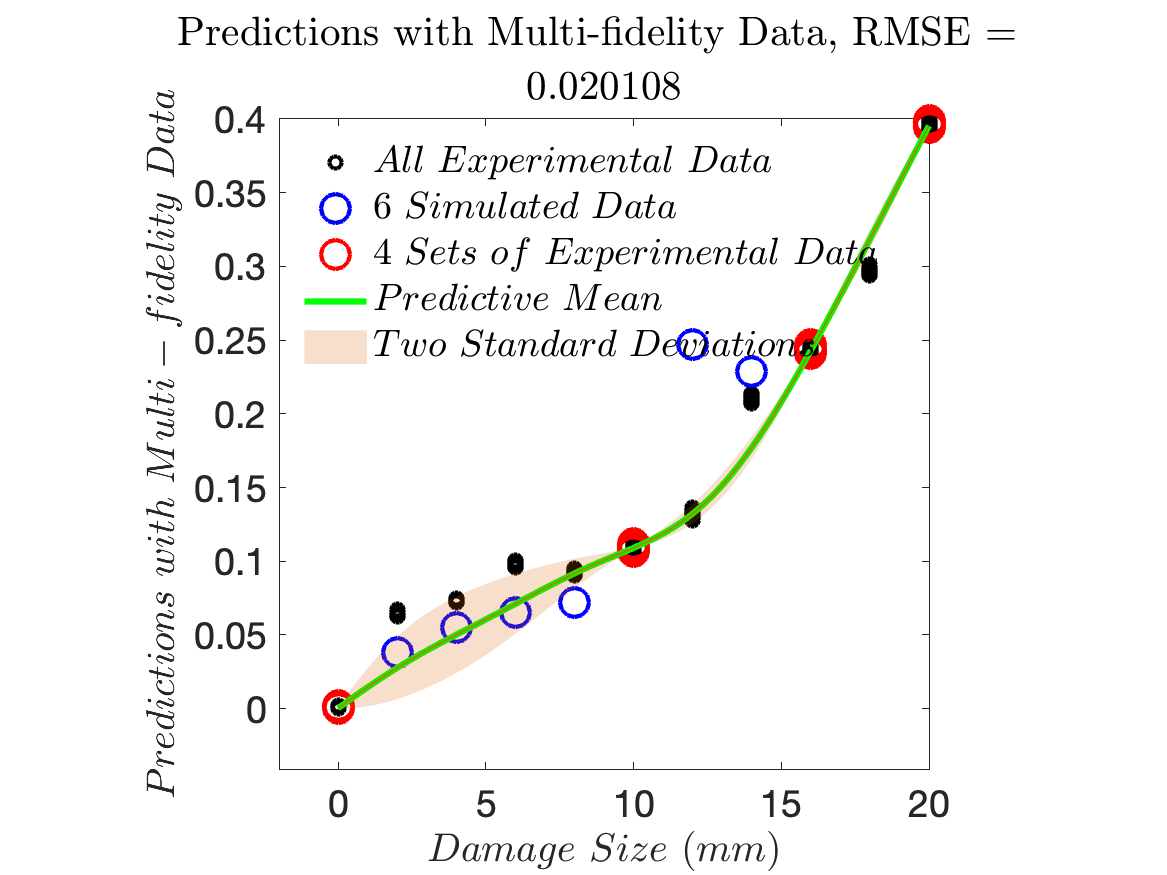}}
    
    \put(130,158){\color{black} \large {\fontfamily{phv}\selectfont \textbf{a}}}
    \put(292,158){\large {\fontfamily{phv}\selectfont \textbf{b}}} 
    \put(454,158){\large {\fontfamily{phv}\selectfont \textbf{c}}} 
    \put(130,20){\large {\fontfamily{phv}\selectfont \textbf{d}}}
   \put(292,20){\large {\fontfamily{phv}\selectfont \textbf{e}}} 
   \put(454,20){\large {\fontfamily{phv}\selectfont \textbf{f}}} 
    \end{picture} 
    \caption{DI regression for path 3-5 from GPRM and multi-fidelity GPRM: (a) to (c) predictions  using experimental data only; (d) to (f) predictions using additional simulated data to fill the gap.}
\label{fig:test1_5} 
% \vspace{-8pt}
\end{figure}

\begin{figure}[b!]
    % \centering
    \begin{picture}(500,240)
    \put(10,144){\includegraphics[width=0.41\textwidth]{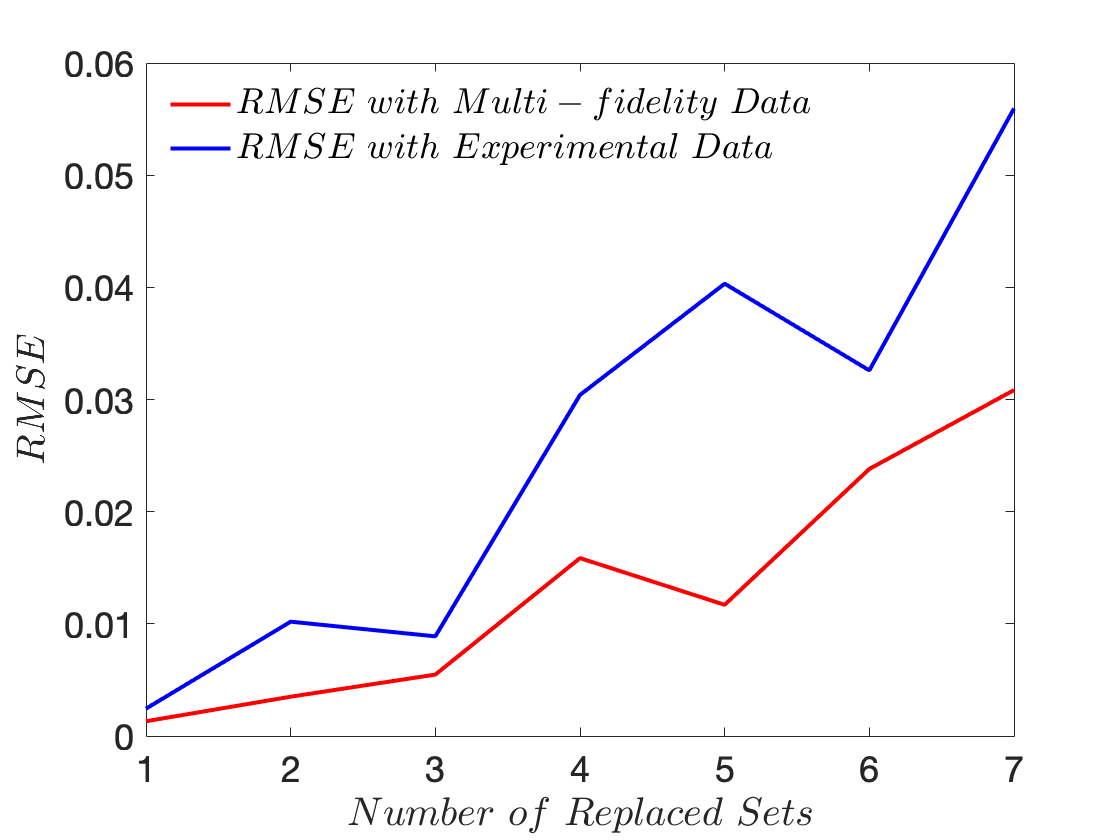}}
    \put(224,144){\includegraphics[width=0.41\textwidth]{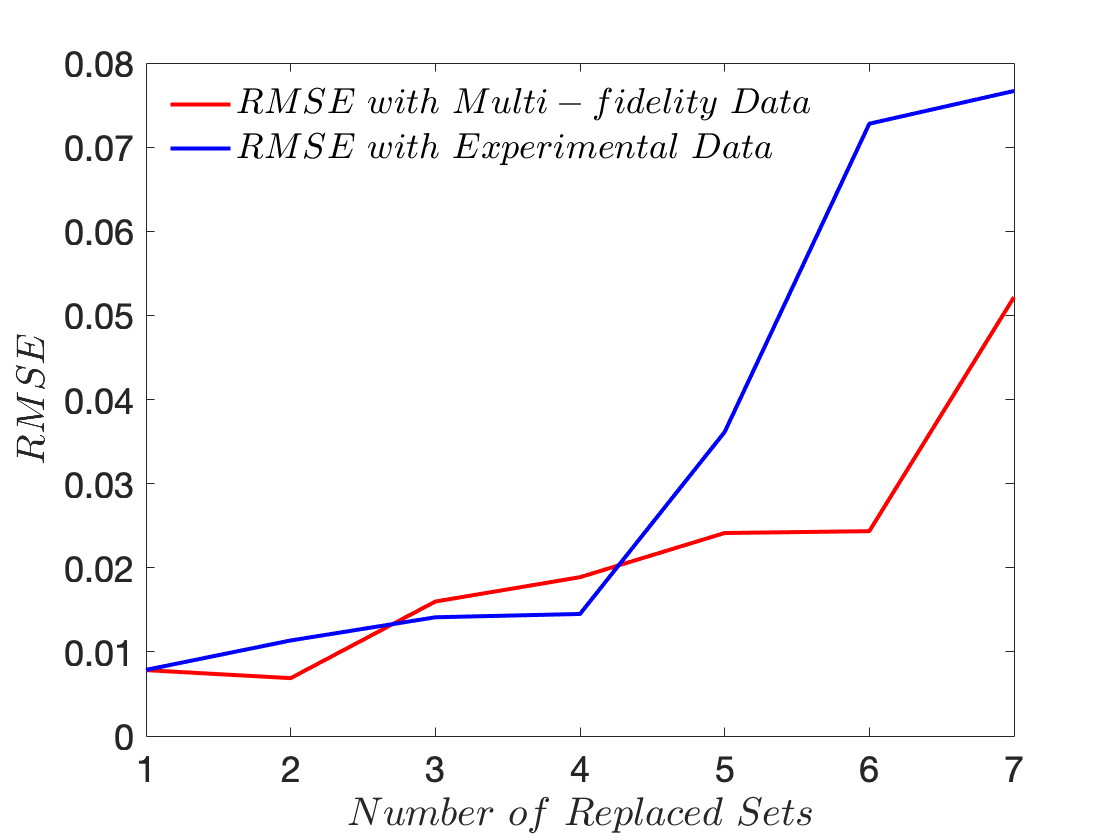}}
    \put(10,0){\includegraphics[width=0.41\textwidth]{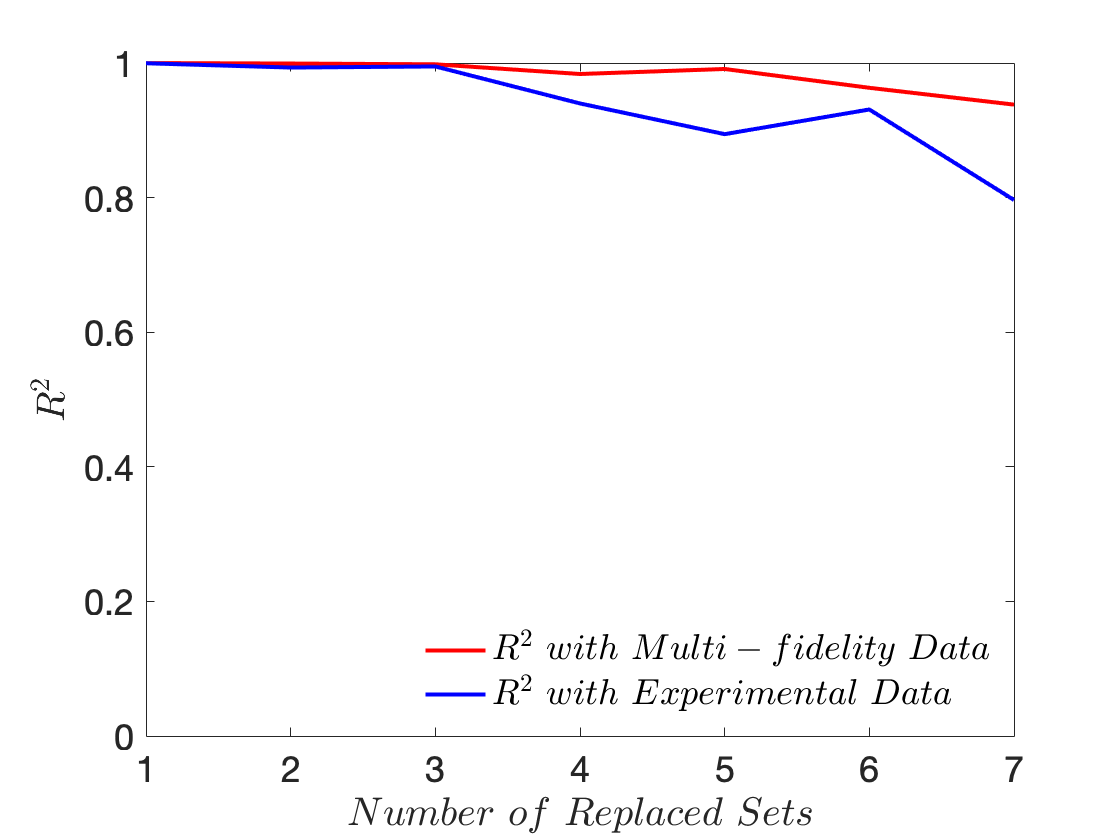}}
    \put(224,0){\includegraphics[width=0.41\textwidth]{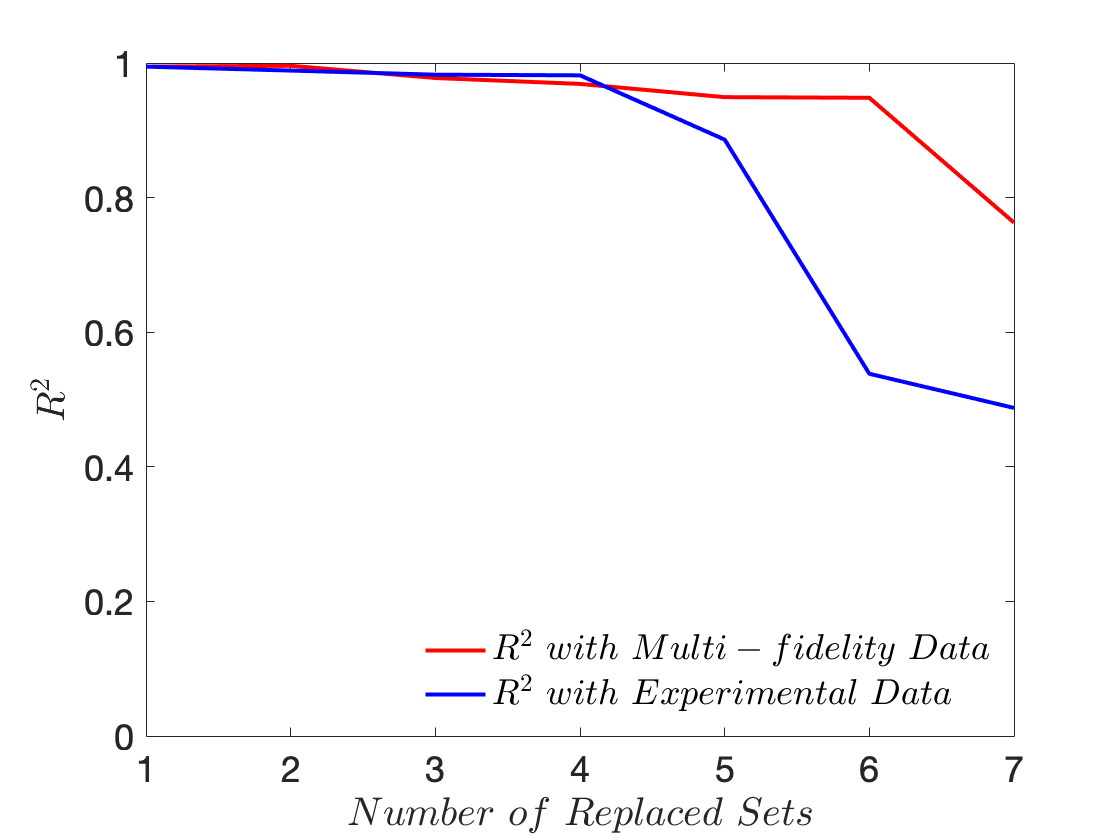}}
    \put(196,170){\color{black} \large {\fontfamily{phv}\selectfont \textbf{a}}}
    \put(410,170){\large {\fontfamily{phv}\selectfont \textbf{b}}}
   \put(196,16){\large {\fontfamily{phv}\selectfont \textbf{c}}} 
   \put(410,16){\large {\fontfamily{phv}\selectfont \textbf{d}}} 
    \end{picture} 
    \caption{RMSE and ${R}^2$ comparison between GPRM and multi-fidelity GPRM for path 2-6 and 3-5.}
\label{fig:test1_6} 
\end{figure}

\subsubsection{Task 2: Constant Total Number of States}
%
% In the second task, the performance of multi-fidelity GPRMs was evaluated by introducing simulated DIs at locations where experimental DI sets were unavailable. This scenario mimics situations where conducting experiments is challenging, yet lower-fidelity data can be generated to fill in the gaps. Paths 2-6 and 3-5 were selected to represent this condition. The damage size ranges from 0 to 20 mm with 2 mm increments, resulting in 11 total states. As previously discussed, the experimental data always includes the 0 mm and 20 mm cases, leaving seven experimental DI sets within the 2-14 mm range that can be substituted with simulated data.

In this test, the experimental DIs that originally covered all 11 states were incrementally replaced by simulated DIs. The results were then compared to standard GPRMs with identical amount of experimental data. Figure \ref{fig:test1_4}, panels (a) and (b), show the performance of the standard GPRM and the multi-fidelity GPRM using five experimental sets, while panels (c) and (d) display the results with four experimental sets from path 2-6. By comparing the first two plots, it can be observed that rather than providing a very rough estimate as standard GPRM, the mean estimates reflect the actual trend and the ground truth data which are the black dots almost all lie between the confidence bonds while using multi-fidelity GPRM. 
Although the confidence bounds fail to encompass most of the ground truth values, the mean estimates show significant improvement in plots (c) and (d). The narrow confidence range is attributed to the small variance in the experimental data. Figure \ref{fig:test1_5} provides a similar comparison for path 3-5. Three different baseline cases from GPRM are shown in the top three plots while the bottom three plots exhibit the results from multi-fidelity GPRM. Both figure \ref{fig:test1_4} and \ref{fig:test1_5} demonstrate that incorporating data from multiple sources allows the predictive mean to better approximate reality, while the $95\%$ confidence intervals are also significantly narrowed. 

Figure \ref{fig:test1_6} presents a compact comparison of the two models as the number of replaced experimental sets increases. Although both models exhibit reduced accuracy as the amount of high-fidelity data decreases, the RMSE of the multi-fidelity GPRM increases at a much slower rate, reflecting its robustness. A similar conclusion is drawn by observing the more gradual decline in ${R}^2$.

\section{Second Test Case: Al Coupon Under Varying Loads} \label{Sec:test2}

In the second test case, experiments were conducted on a different aluminum plate with a similar sensor arrangement. However, the damage states were based on applied loads rather than crack sizes. Two tasks, analogous to those in the previous test case, were performed using the same models to assess the universal superiority of multi-fidelity GPRMs across different datasets.

It was observed that lower-fidelity data could be utilized more efficiently, prompting the integration of the multi-fidelity model with both random selection and active-learning strategies. The results from these two approaches were compared to highlight a data-efficient strategy for real-world applications. With active learning, after each iteration, the location corresponding to the maximum confidence interval was identified, and the nearest load-matching data point from the low-fidelity source was added to the input array. The results demonstrat the effectiveness of combining multi-fidelity GPRMs with active learning, achieving higher model accuracy while requiring a limited amount of data.

\subsection{Experimental Setup}

\begin{figure}[t!]
\centering
\includegraphics[scale=0.42]{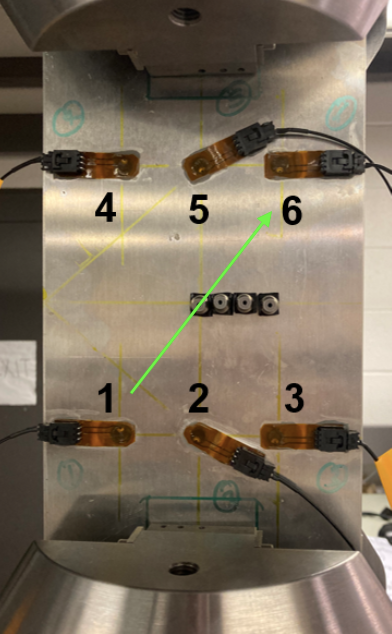}
\caption{The Al coupon used in the second test case with four 3-gm weights and 6 PZT sensors.}
\label{fig:2nd_coupon} 
% \vspace{10pt}
\end{figure}

% \begin{figure}[t!]
% \centering
% \includegraphics[scale=0.4]{Figures/2ndtest/case2/task2_plate.png}
% \caption{Discretization of the wave propagation path to address phase-shift due to non-uniform strain distribution.
% }
% \label{fig:2nd_plate} 
% % \vspace{10pt}
% \end{figure}

The second test case was conducted on a 6061 Aluminum coupon with the dimension of 152.4 x 304.8 x 2.36 mm. 6 PZT sensors were attached on the coupon similarly as in the first test case, as shown in Figure \ref{fig:2nd_coupon}. To ensure stability during the experiments, the adhesive was cured for 24 hours under vacuum. Unlike the first test case, where structural states were determined by varying crack sizes, this test generated different states by applying various loads using a tensile machine (Instron, Inc.). Five static loads—0, 5, 10, 15, and 20 kN—were applied to produce the experimental signals.

For each realization, 5-peak tone burst signals were generated by each actuator sensor, while the remaining sensors acted as receivers. The signals were captured using a ScanGenie III data acquisition system at a sampling frequency of 24 MHz. Under the loads of 0, 5, 10, and 15 kN, each experiment was repeated 20 times, with 15 realizations assigned to the training set and the remaining 5 to the testing set. For the 20 kN case, the data comprised only 2 realizations, evenly divided between the training and testing sets.

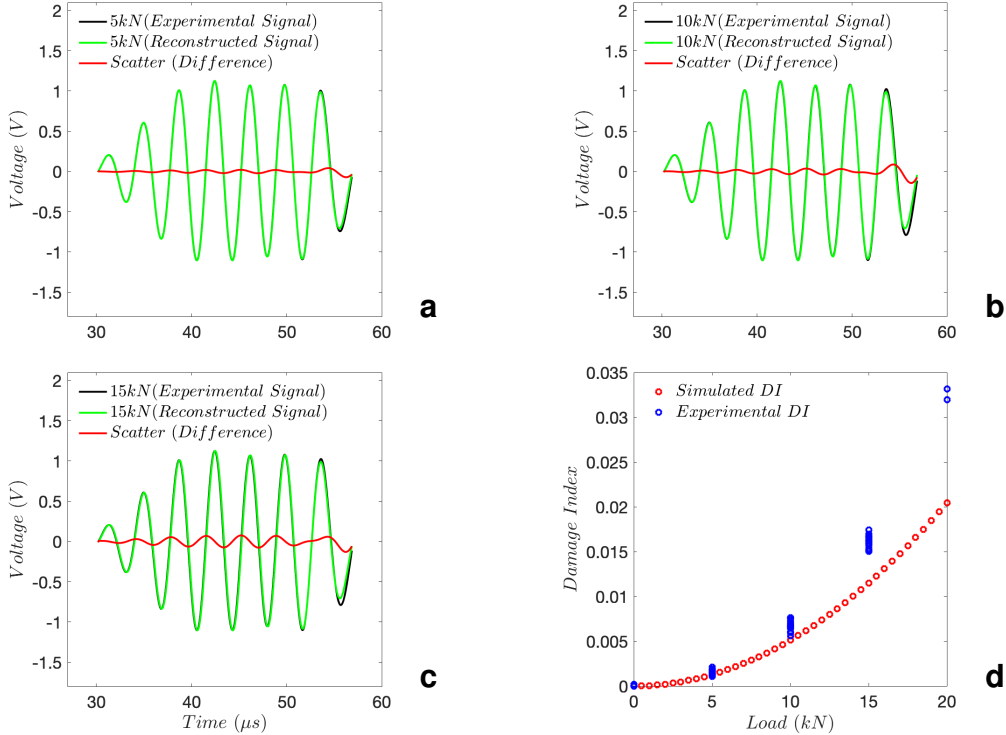
\begin{figure}[b!]
    % \centering
    \begin{picture}(500,270)
    \put(10,140){\includegraphics[width=0.42\textwidth]{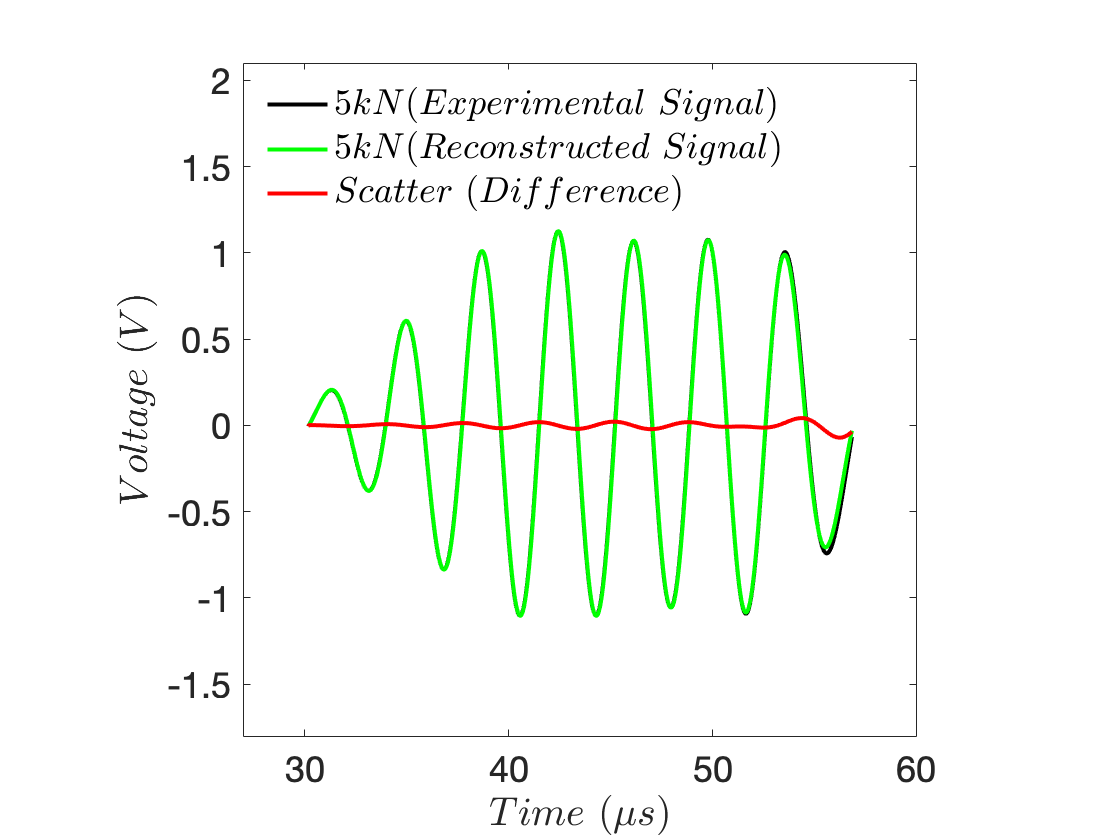}}
    \put(224,140){\includegraphics[width=0.42\textwidth]{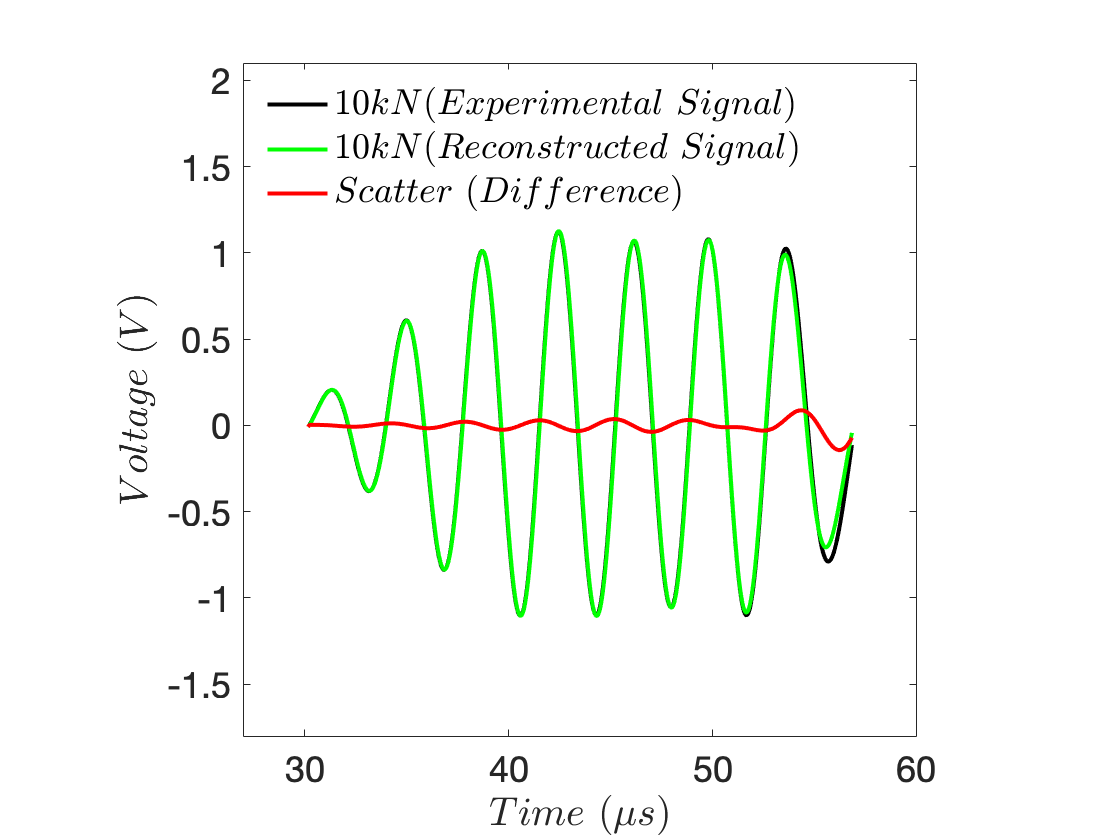}}
    \put(10,0){\includegraphics[width=0.42\textwidth]{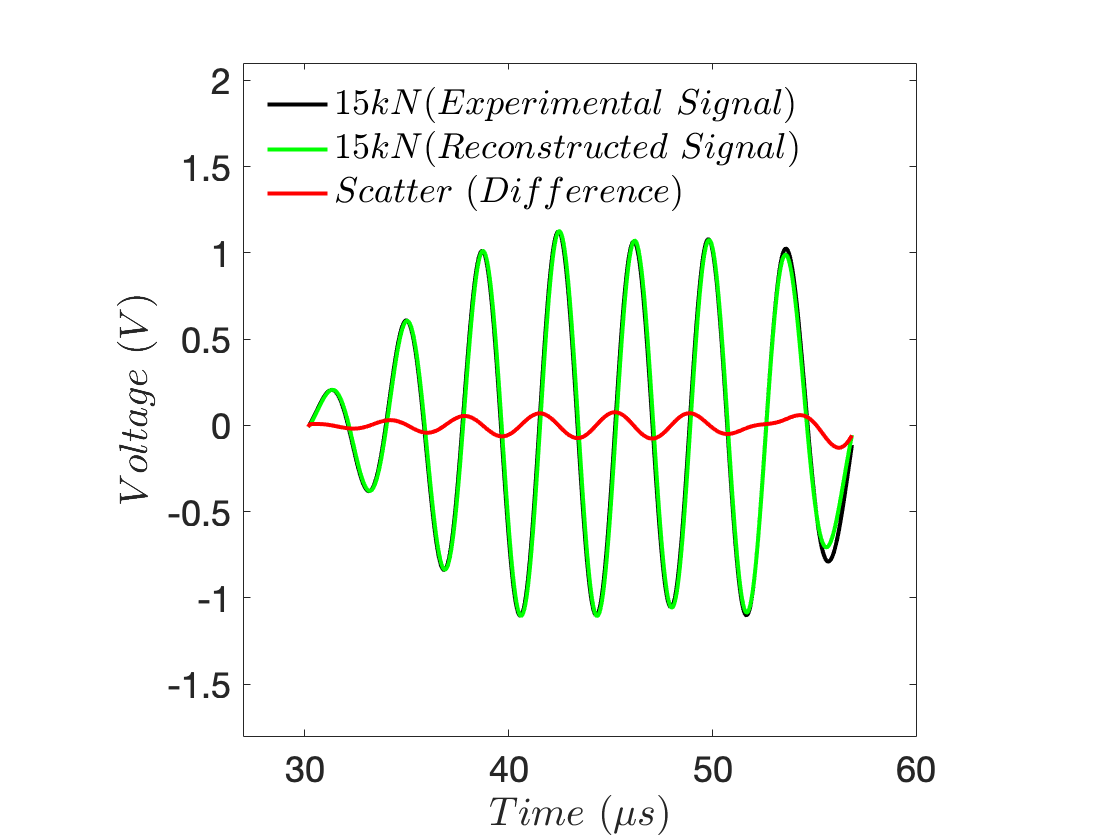}}
    \put(224,0){\includegraphics[width=0.42\textwidth]{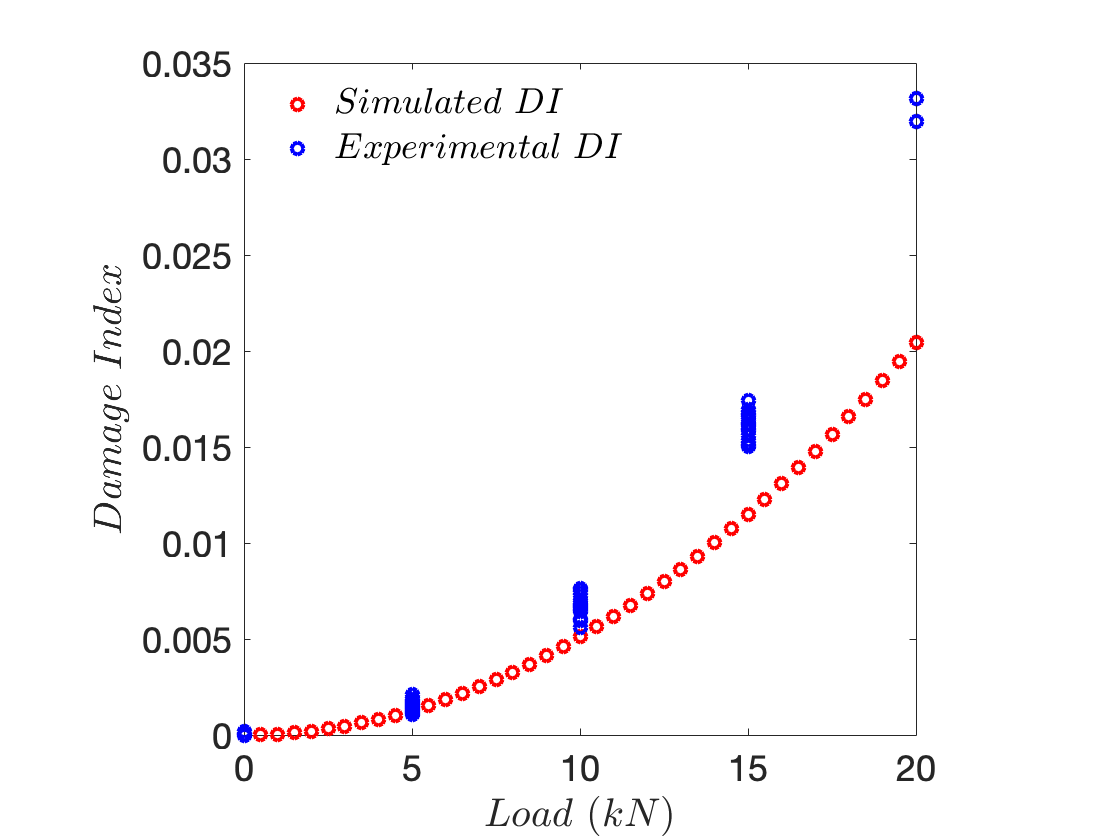}}
    \put(186,158){\color{black} \large {\fontfamily{phv}\selectfont \textbf{a}}}
    \put(400,158){\large {\fontfamily{phv}\selectfont \textbf{b}}}
   \put(186,18){\large {\fontfamily{phv}\selectfont \textbf{c}}} 
   \put(400,18){\large {\fontfamily{phv}\selectfont \textbf{d}}} 
    \end{picture} 
    \caption{The first wave package of both experimental and the physics-based reconstructed signals for path 1-6 as well as their differences under different loads respectively (a) 5kN; (b) 10kN; (c) 15kN; (d) 20kN.}
\label{fig:2nd_sig} 
\end{figure}
\subsection{Physics-based Load Compensation Model}

Directional changes in ultrasonic guided waves occur as the surrounding strains vary due to changing loads. Roy \etal \cite{roy2015load} introduced a model to reconstruct signals by capturing the variations in signal amplitude and phase, based on the property that guided wave propagation velocity increases monotonically with local stress. Ahmad \etal \cite{amer2021gaussian} successfully applied this model to generate baseline signals at different loads along various wave propagation paths by accounting for signal variations under different loading conditions. In our second test case, this model was similarly employed to reconstruct signals and build the simulated dataset.

A brief summary of the underlying theory for the physics-based model from \cite{roy2015load} is provided here. The load compensation model reconstructs signals based on the structure's in-situ strain and temperature distribution, which are obtained from experimental signals. The change in signal amplitude, which is proportional to the local strain, can be expressed as:
\begin{equation}
    \frac{\Delta V_{out}^{(\epsilon)}}{V_{out}} \approx A\epsilon_{path}^{(act)} + B\epsilon_{path}^{(sen)}
\end{equation}
where $V_{out}$ is the voltage output, $\epsilon_{path}^{(act)}$ and $\epsilon_{path}^{(sen)}$ are the local strain captured by actuators and sensors along the signal paths respectively. A and B are unknown model constants which can be calculated from experimental signals collected at various loading conditions. Other than merely affecting the actuating and sensing, mechanical loads also influence deformation of wave propagation path as well as propagation velocity which then lead to change of time of arrival, the net value of which can be expressed by:
\begin{equation}
    \Delta ToA = K_{phase}\sum_{t=1}^{N}(d_{i}\epsilon_{path}^{(i)})
\end{equation}
where $K_{phase}$ is an unknown constant and can be estimated from experimental measurements along with strain distribution. Wave propagation $d$ is divided into small segments with uniform length and averaged strain. The first wave package can then be reconstructed with the calculated signal amplitude change and net time-of-arrival change. 

In this test case, the simulated signals were generated using this physics-based model. Notably, unlike the fixed increments used in the first test case, the DIs derived from the simulated first wave packets can have flexible load increments. This flexibility allows for a denser data distribution within the target region. For this test, an increment of 0.5N, which is one-tenth of the experimental state increment, was selected to ensure sufficient resolution.

\subsection{Results}

Figure \ref{fig:2nd_sig}, panels (a-c), presents the first wave packets of both experimental and physics-based reconstructed signals for path 1-6, along with their differences under varying loads. The negligible scatter in the amplitude differences suggests that the reconstructed signals closely approximate the real values, making them suitable as a distinct dataset. The DIs from both the experimental and reconstructed signals were then extracted and used as inputs for the proposed model.

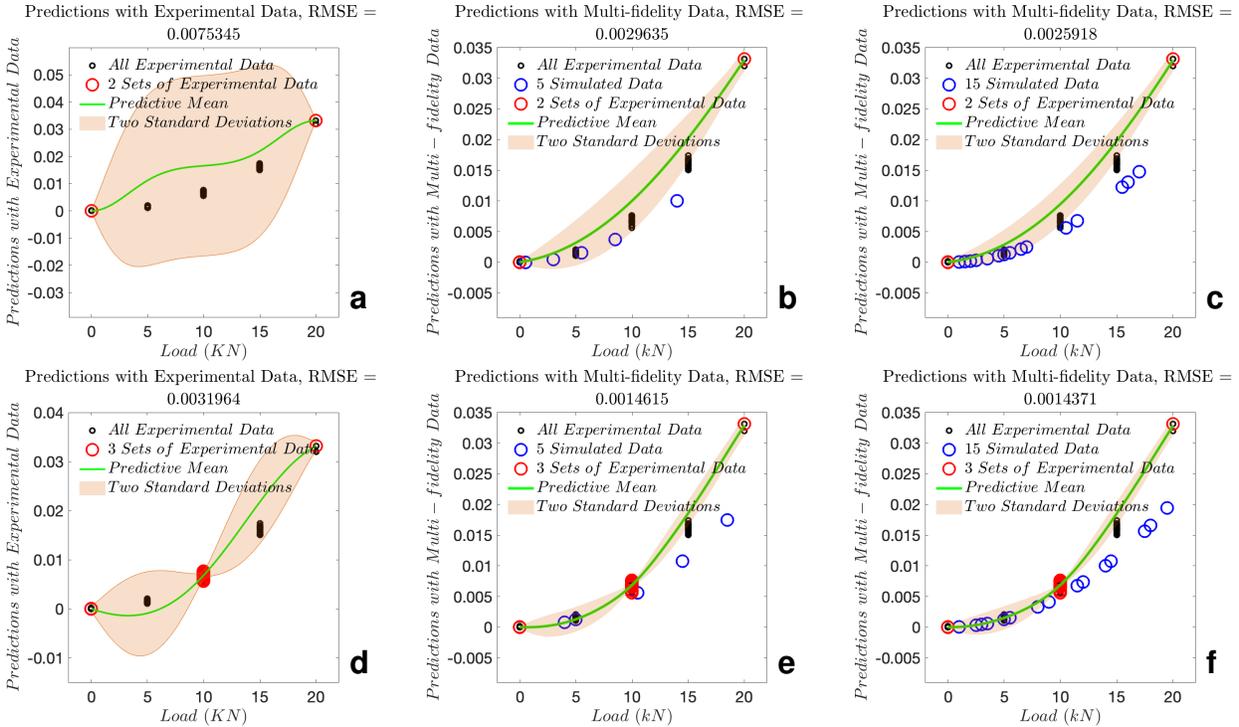
\begin{figure}[b!]
    % \centering
    \begin{picture}(500,260)
    \put(-20,138){\includegraphics[width=0.39\textwidth]{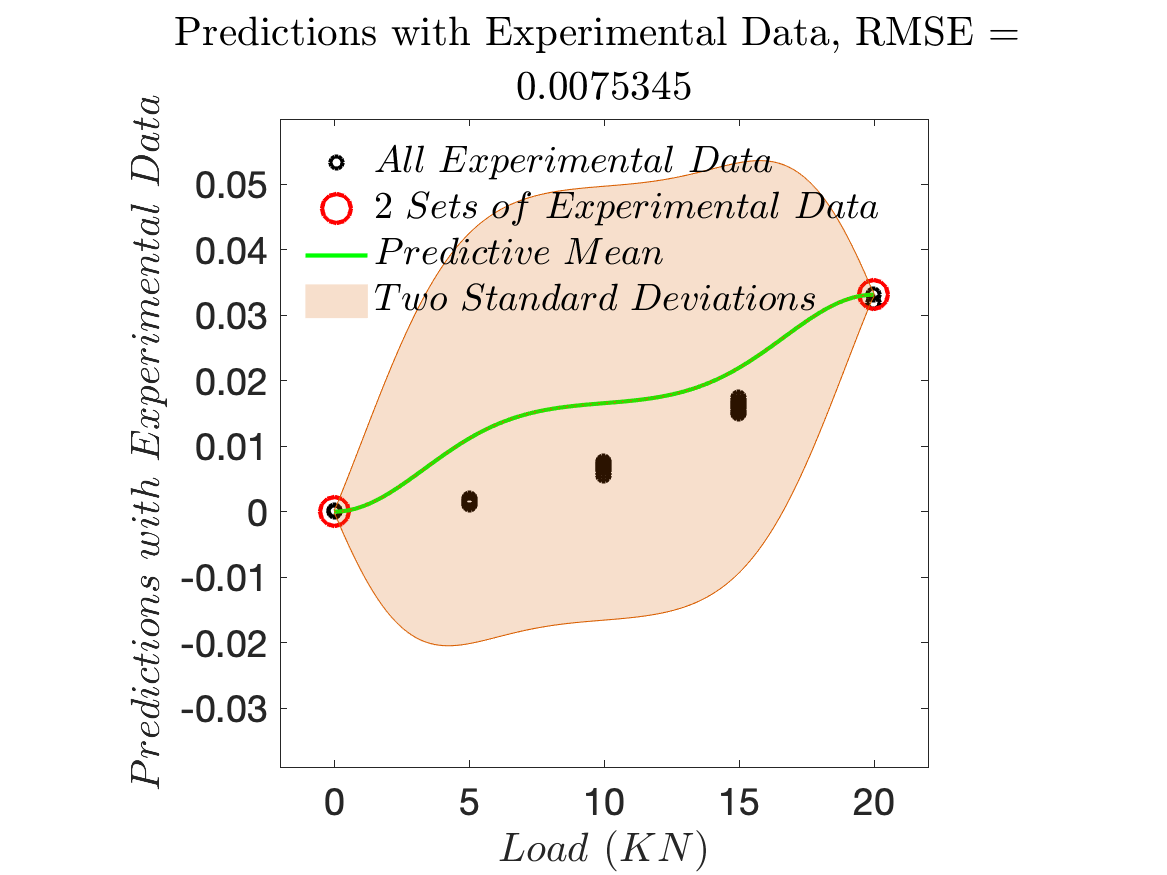}}
    \put(142,138){\includegraphics[width=0.39\textwidth]{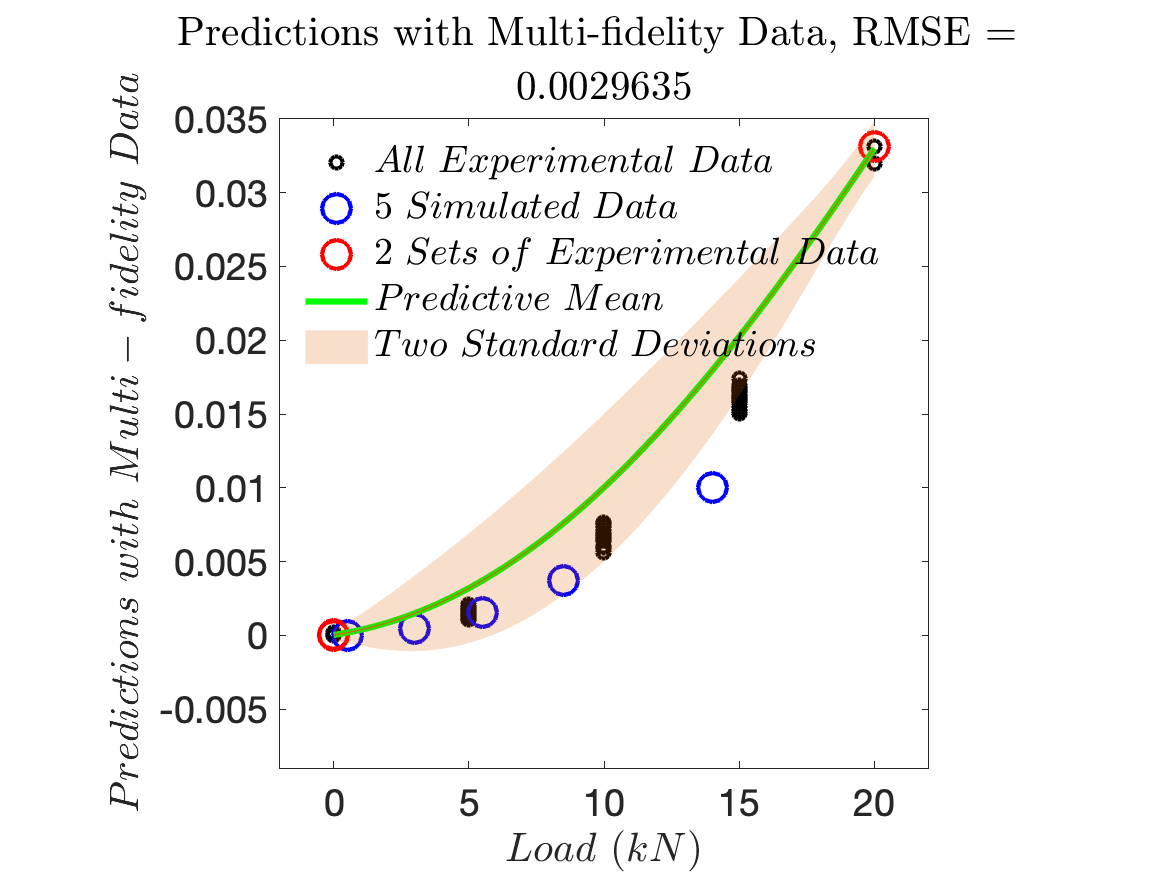}}
    \put(304,138){\includegraphics[width=0.39\textwidth]{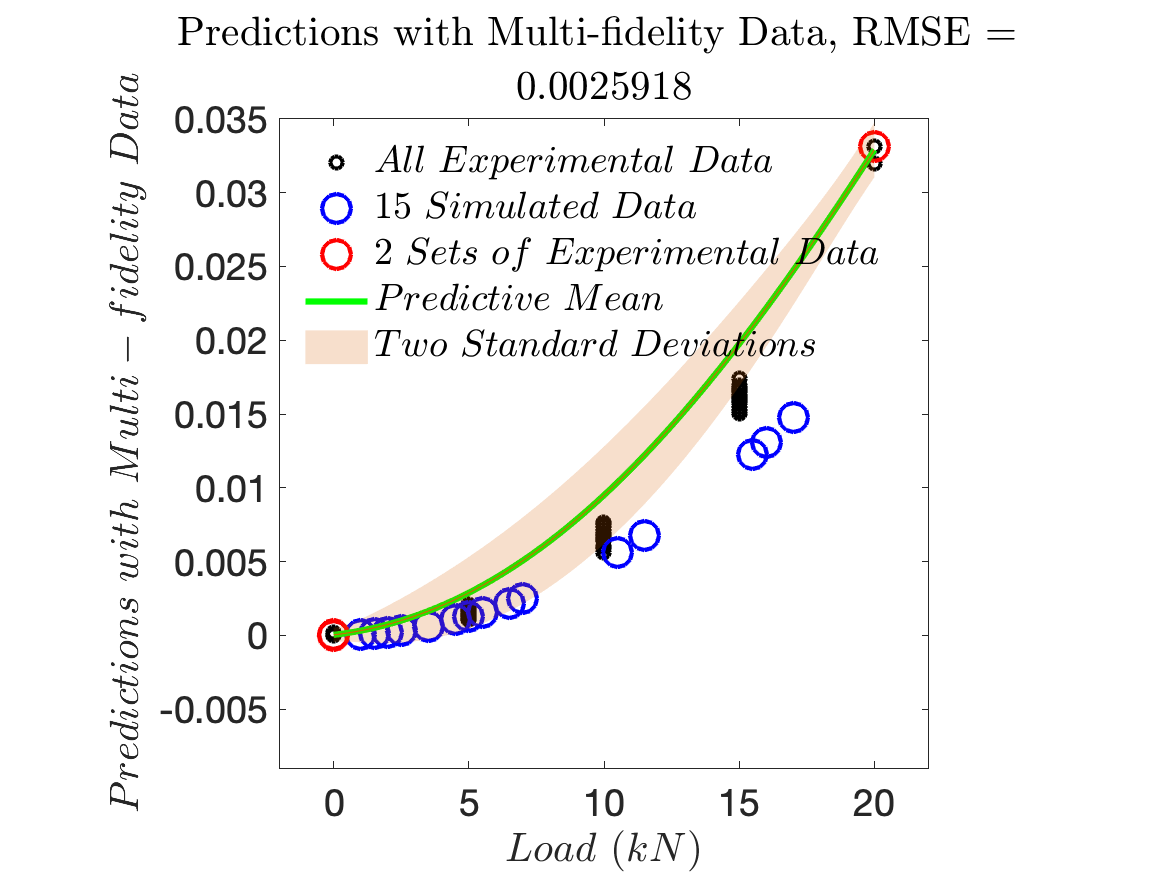}}
    \put(-20,0){\includegraphics[width=0.39\textwidth]{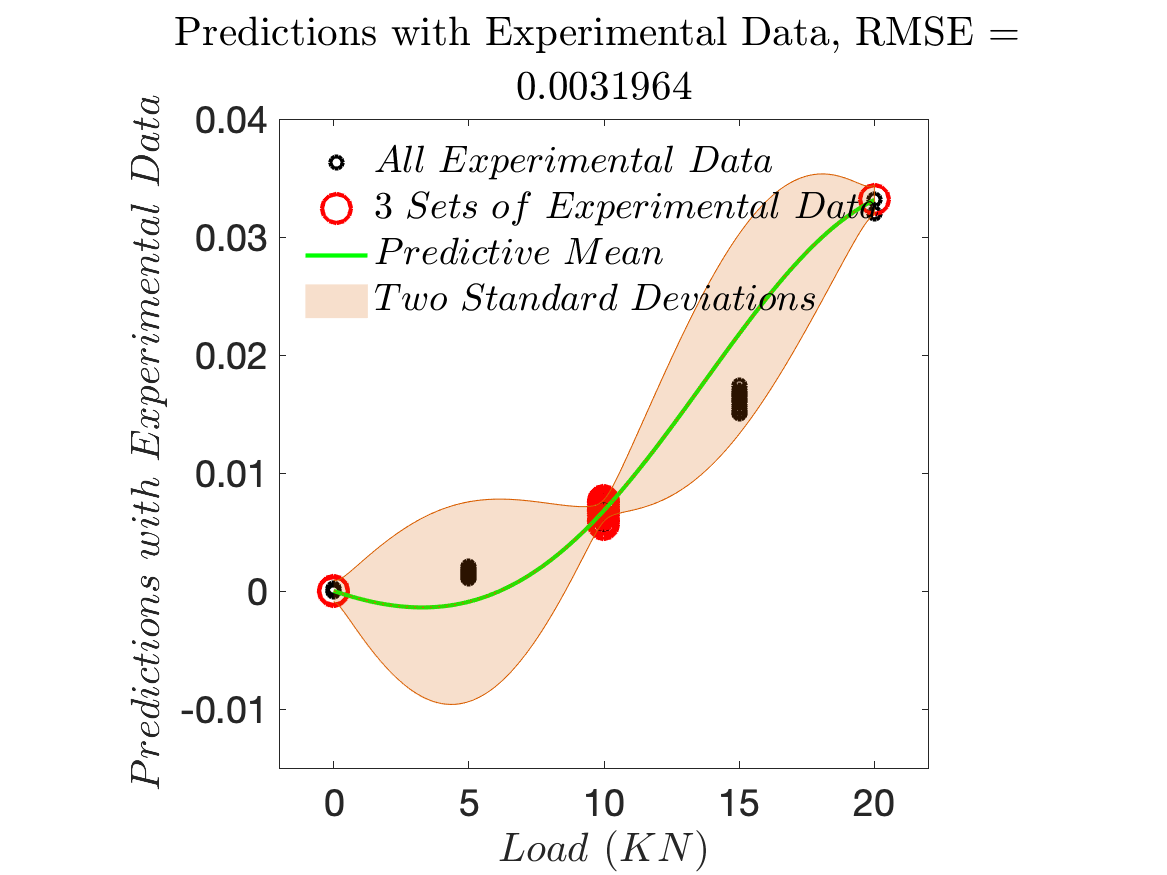}}
    \put(142,0){\includegraphics[width=0.39\textwidth]{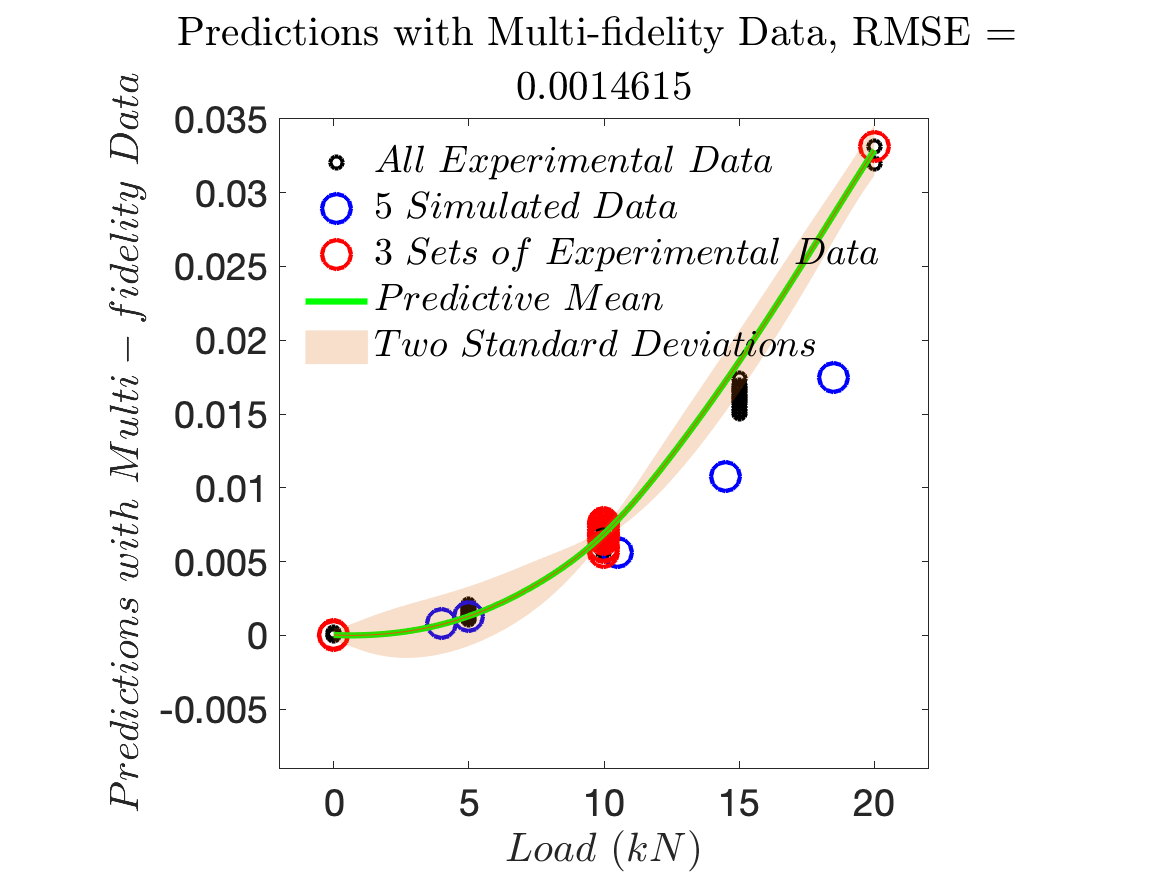}}
    \put(304,0){\includegraphics[width=0.39\textwidth]{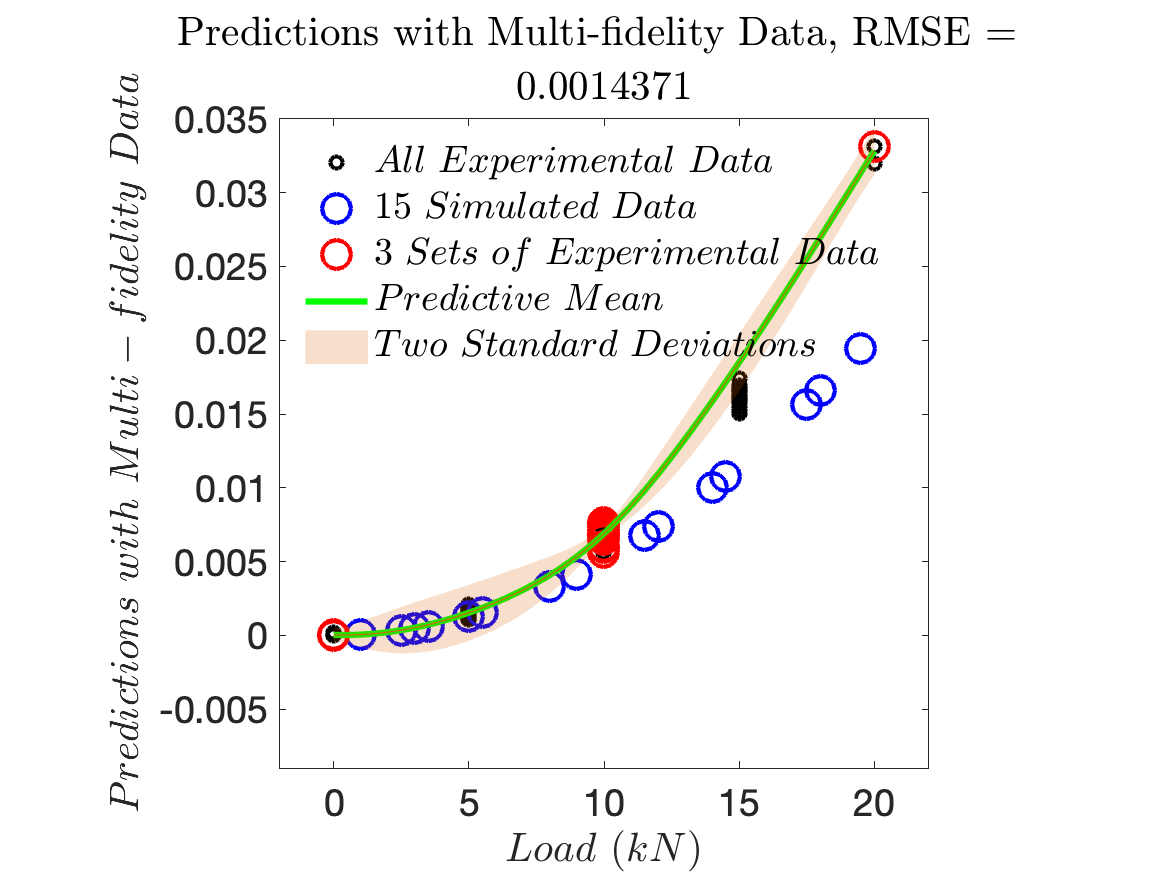}}
    
    \put(130,158){\color{black} \large {\fontfamily{phv}\selectfont \textbf{a}}}
    \put(292,158){\large {\fontfamily{phv}\selectfont \textbf{b}}} 
    \put(454,158){\large {\fontfamily{phv}\selectfont \textbf{c}}} 
    \put(130,20){\large {\fontfamily{phv}\selectfont \textbf{d}}}
   \put(292,20){\large {\fontfamily{phv}\selectfont \textbf{e}}} 
   \put(454,20){\large {\fontfamily{phv}\selectfont \textbf{f}}} 
    \end{picture} 
    \caption{Task 1 results of the 2nd test case. Panels (a) and (d): baseline DI regression for path 1-6 from GPRM; panels (b), (c), (e) and (f): DI regression from multi-fidelity GPRM with batch learning using random selection.}
    
\label{fig:2nd_t1_1} 
\end{figure}

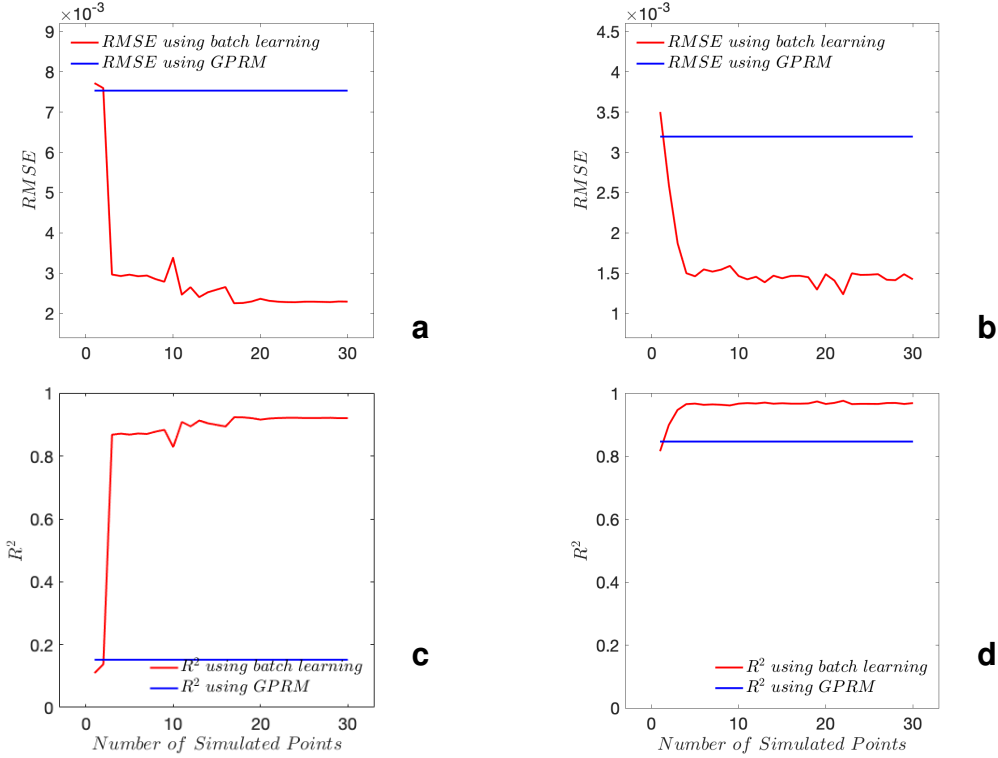
\begin{figure}[t!]
    % \centering
    \begin{picture}(500,278)
    \put(10,140){\includegraphics[width=0.42\textwidth]{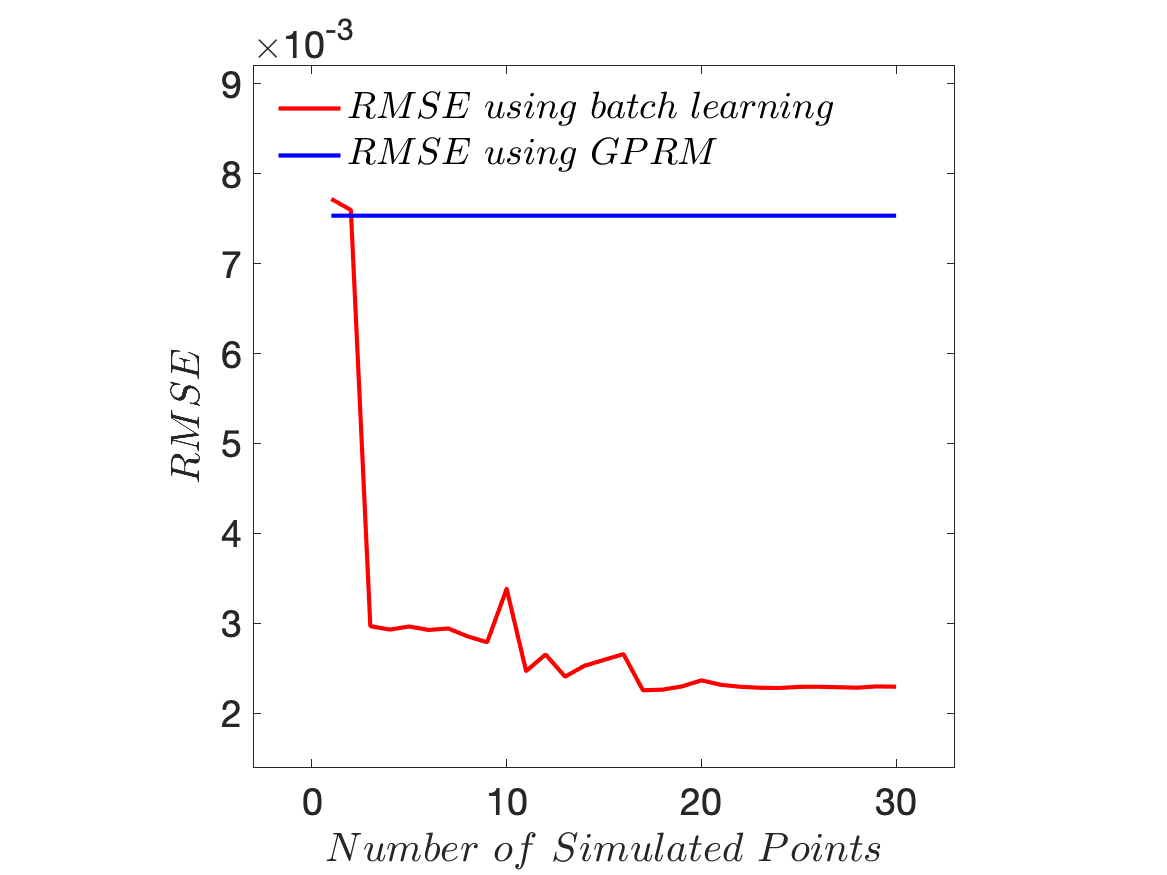}}
    \put(224,140){\includegraphics[width=0.42\textwidth]{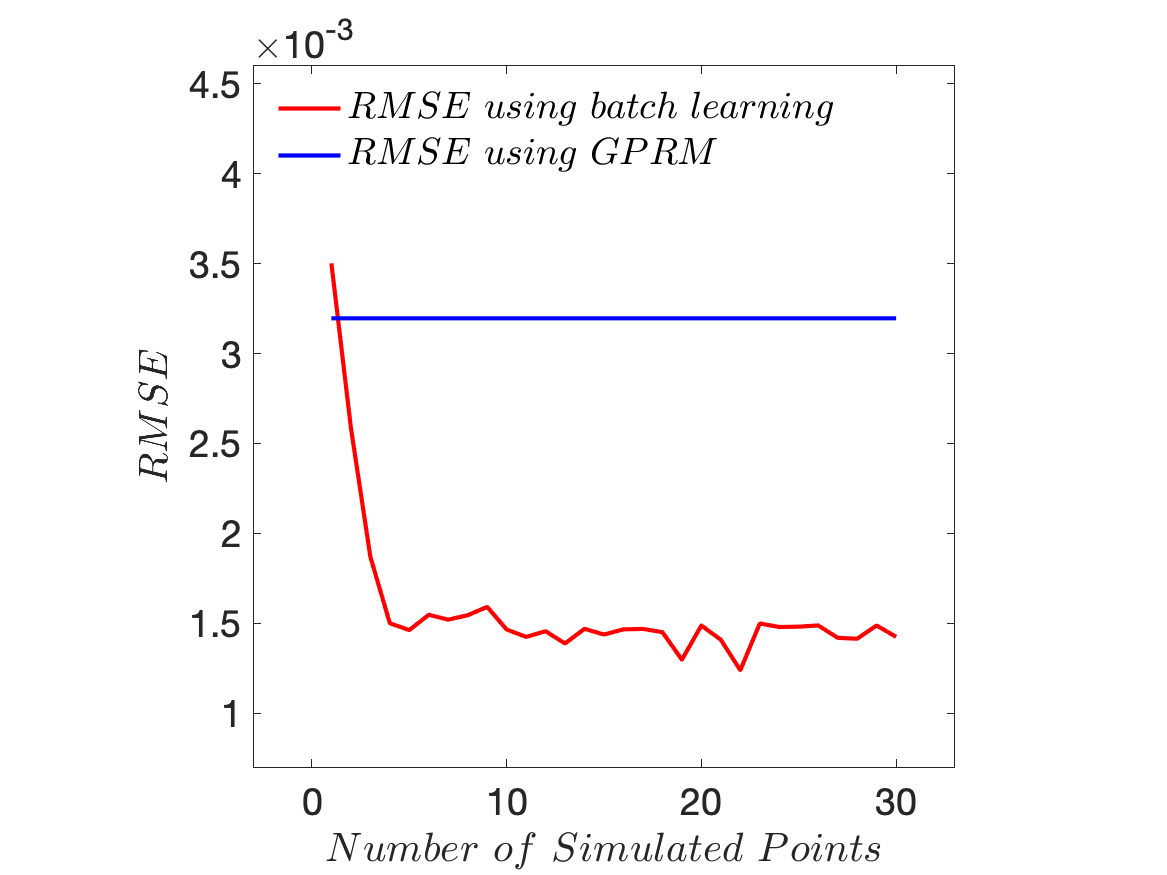}}
    \put(10,0){\includegraphics[width=0.42\textwidth]{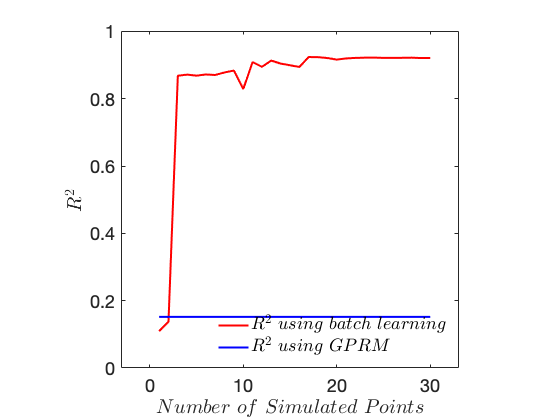}}
    \put(224,0){\includegraphics[width=0.42\textwidth]{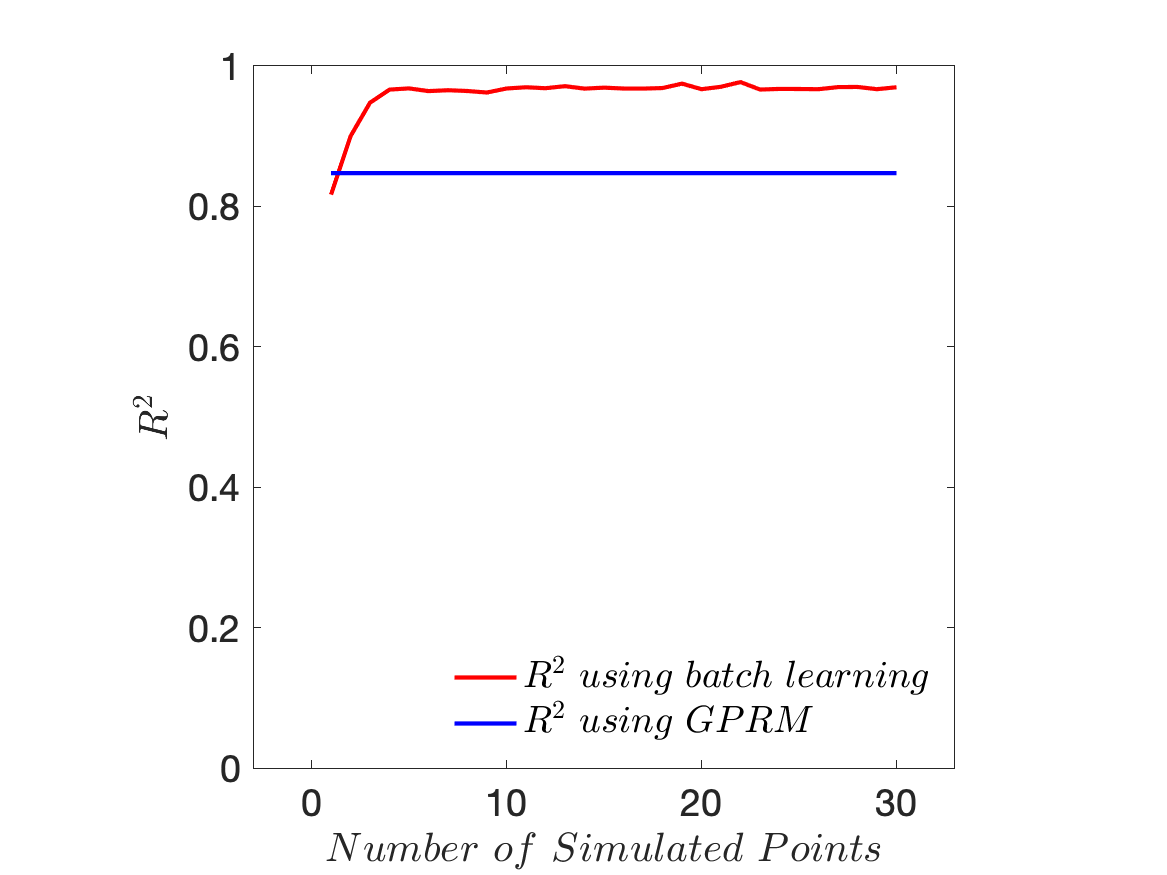}}
    \put(186,158){\color{black} \large {\fontfamily{phv}\selectfont \textbf{a}}}
    \put(400,158){\large {\fontfamily{phv}\selectfont \textbf{b}}}
   \put(186,34){\large {\fontfamily{phv}\selectfont \textbf{c}}} 
   \put(400,34){\large {\fontfamily{phv}\selectfont \textbf{d}}} 
    \end{picture}
    \caption{RMSE and ${R}^2$ comparison between GPRM and multi-fidelity GPRM with batch learning using random selection for path 1-6 : (a) RMSE comparison for path 1-6 with loads at 0 and 20 kN; (b) RMSE comparison for path 1-6 with loads at 0, 10 and 20 kN; (c) ${R}^2$ comparison for path 1-6 with loads at 0 and 20 kN; (d) ${R}^2$ comparison for path 1-6 with loads at 0, 10 and 20 kN.}
\label{fig:2nd_t1_2} 
\end{figure}

\begin{figure}[t!]
    % \centering
    \begin{picture}(500,280)
    \put(-20,138){\includegraphics[width=0.39\textwidth]{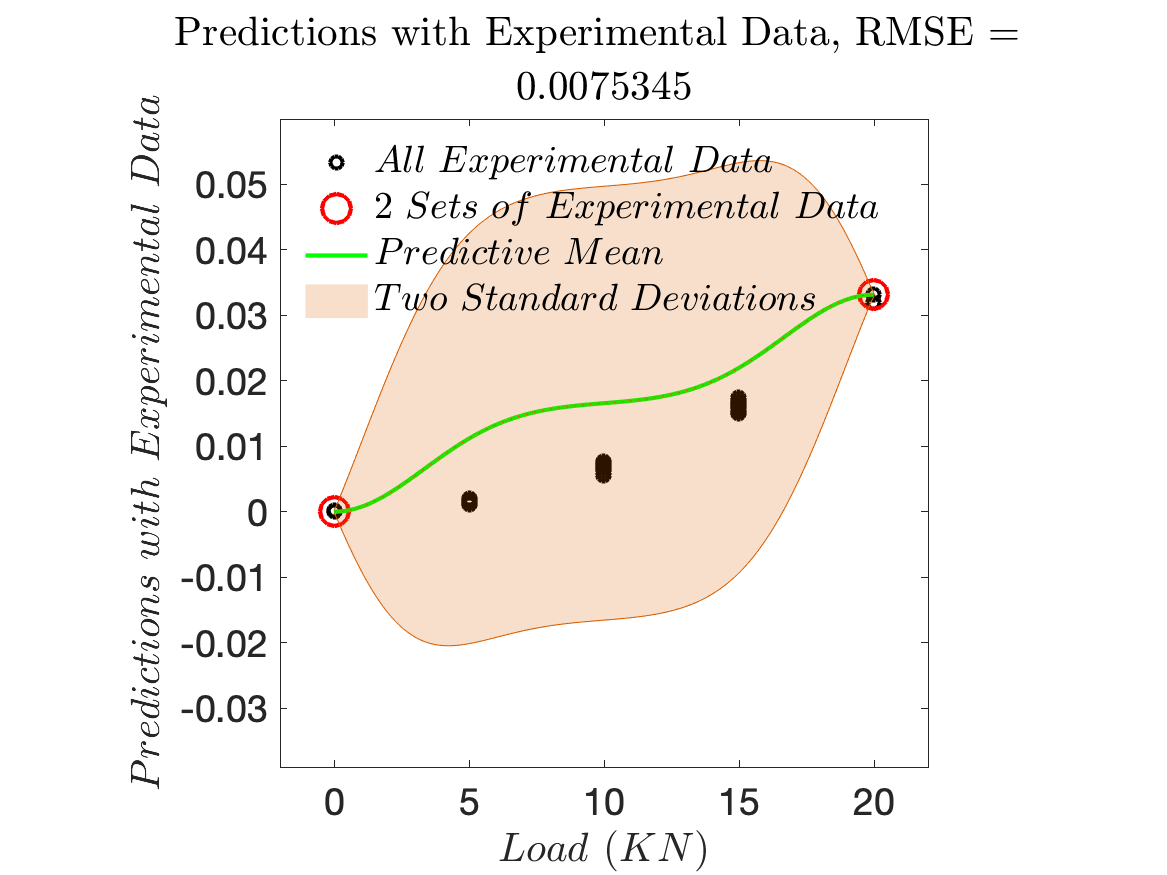}}
    \put(142,138){\includegraphics[width=0.39\textwidth]{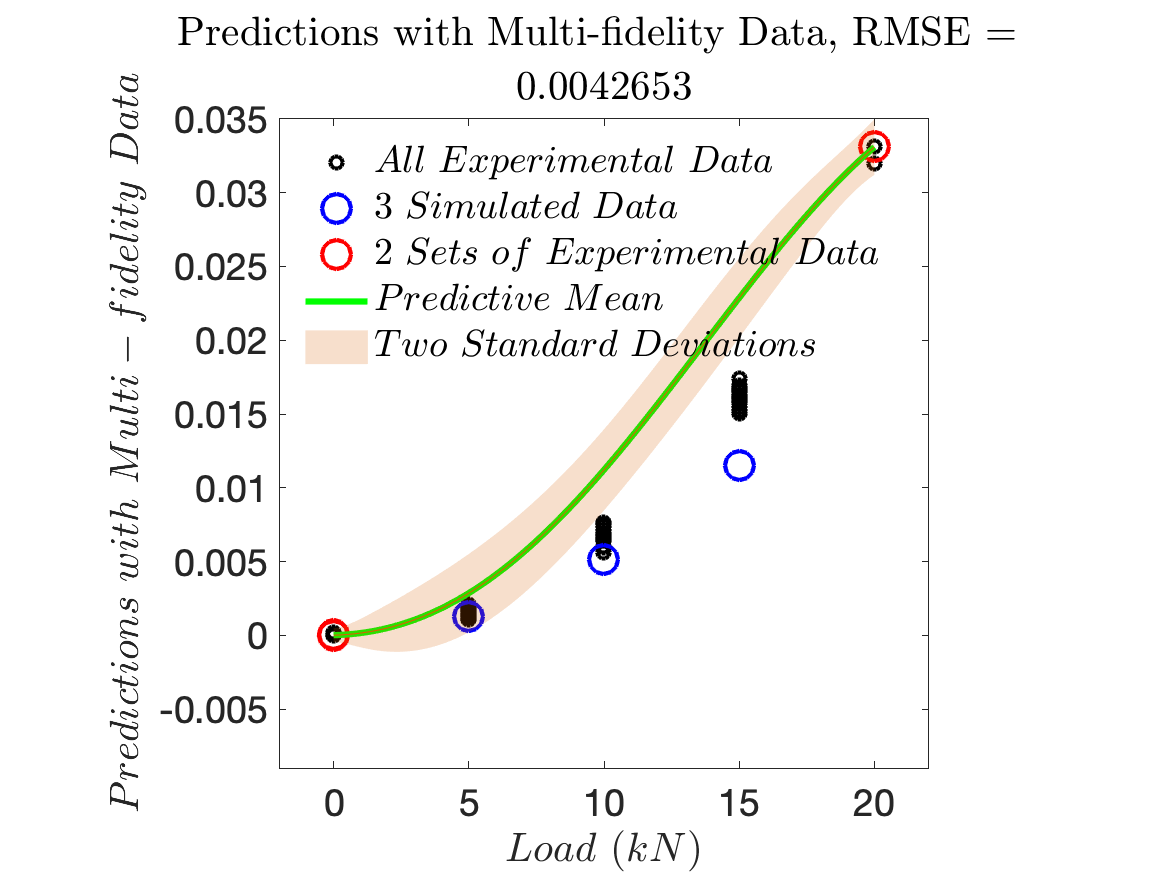}}
    \put(304,138){\includegraphics[width=0.39\textwidth]{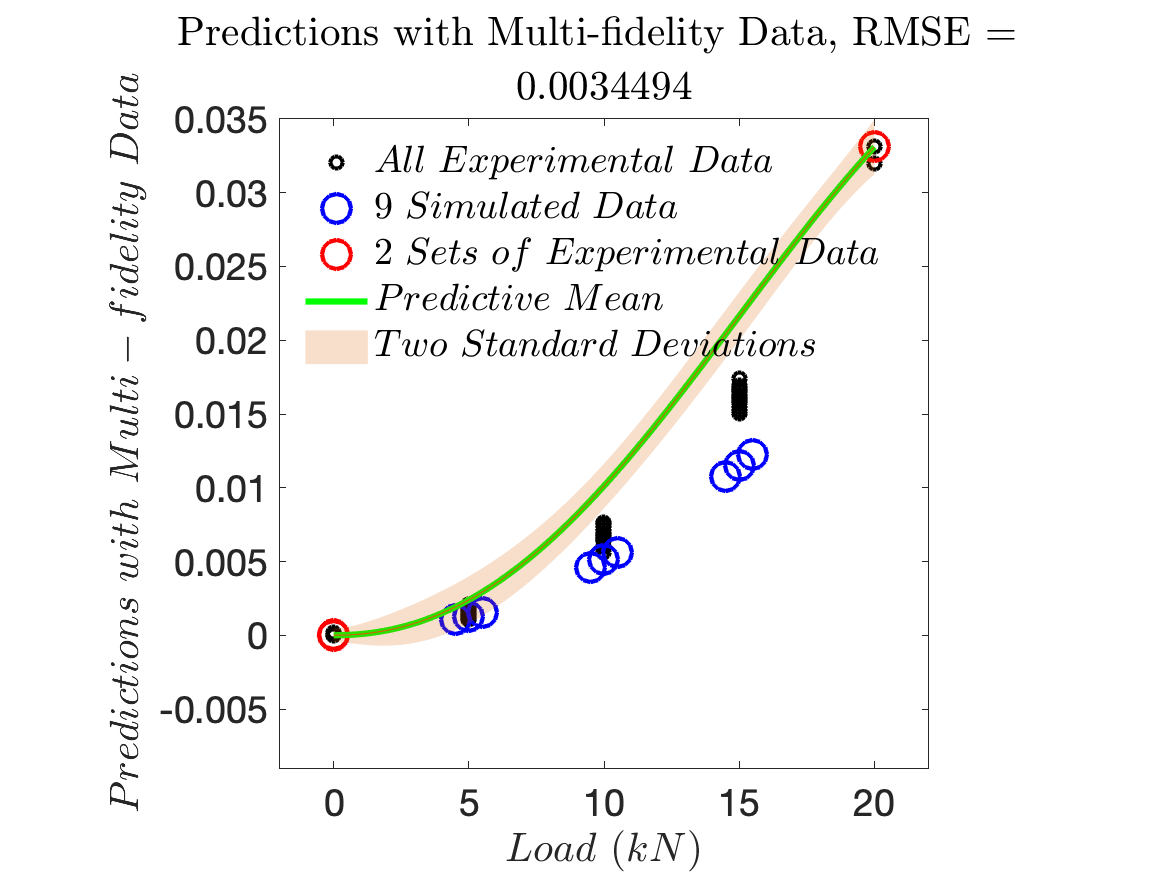}}
    \put(-20,0){\includegraphics[width=0.39\textwidth]{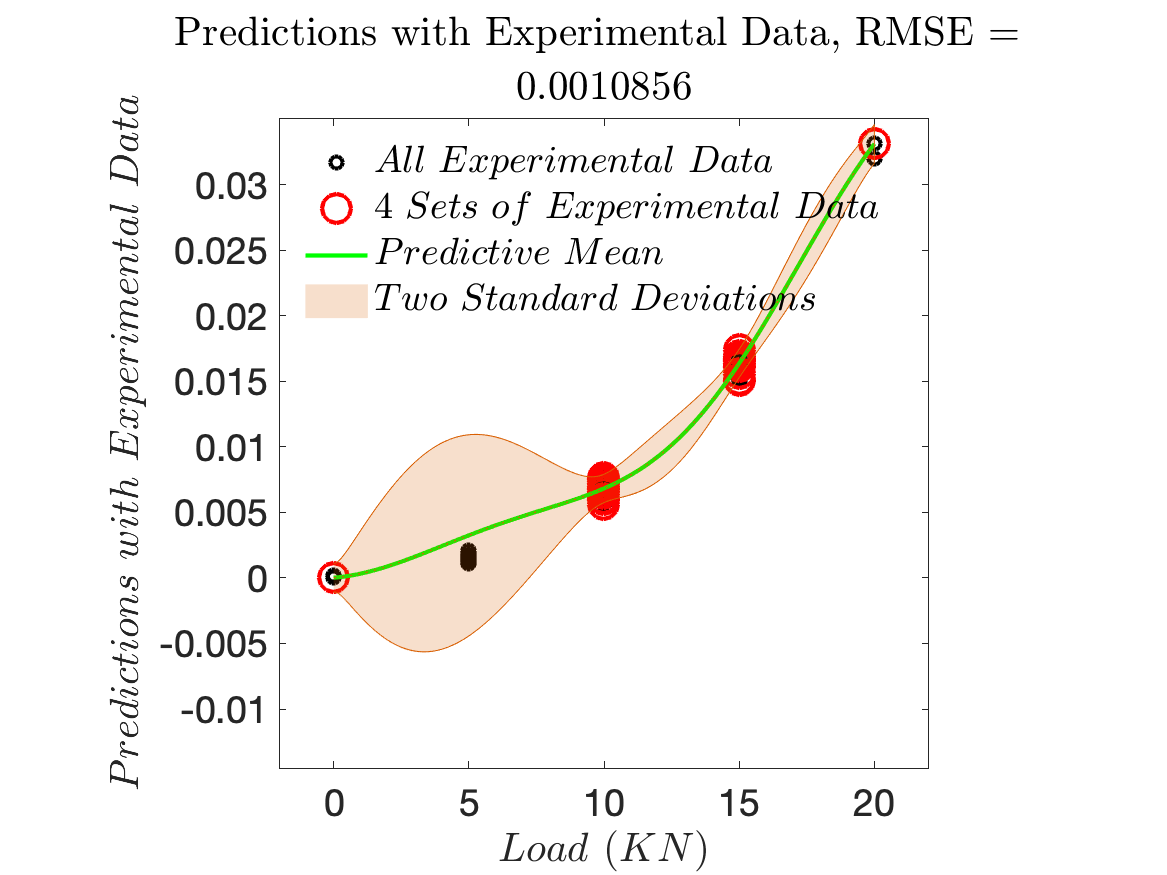}}
    \put(142,0){\includegraphics[width=0.39\textwidth]{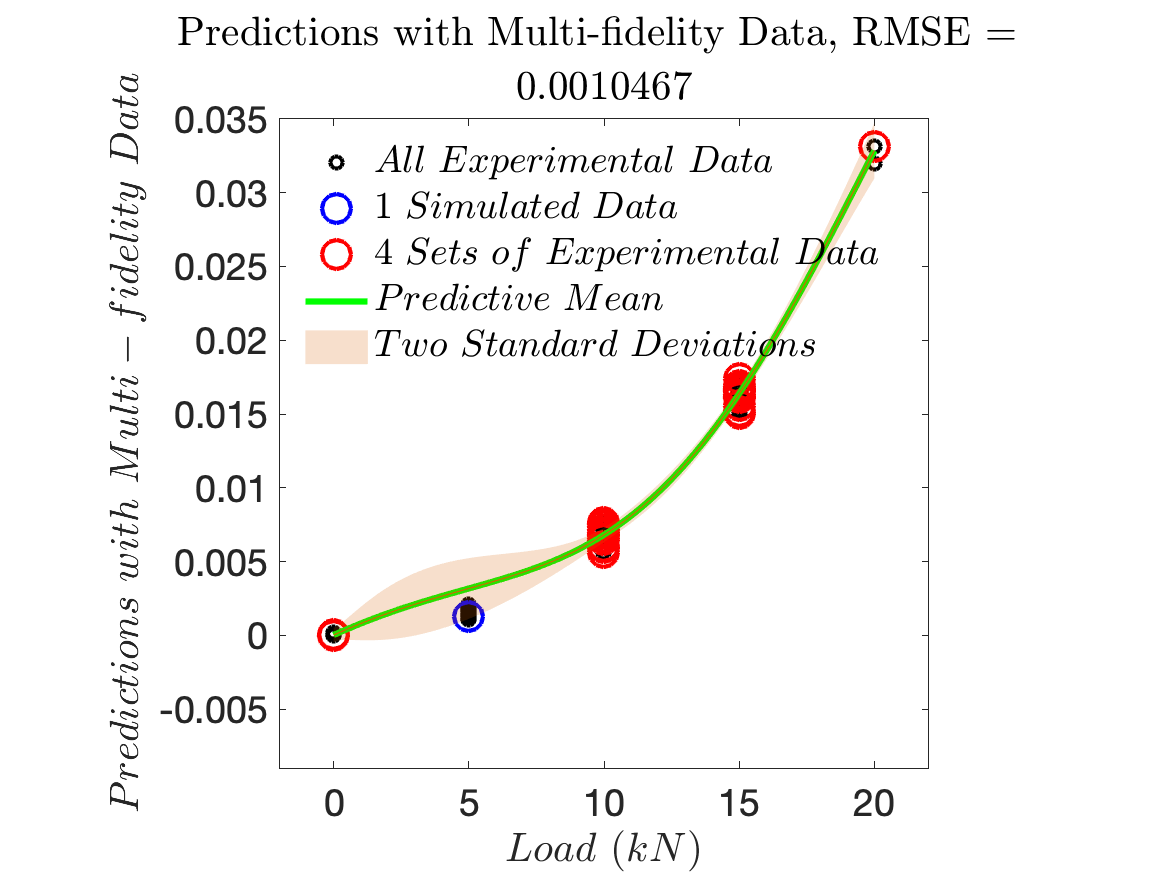}}
    \put(304,0){\includegraphics[width=0.39\textwidth]{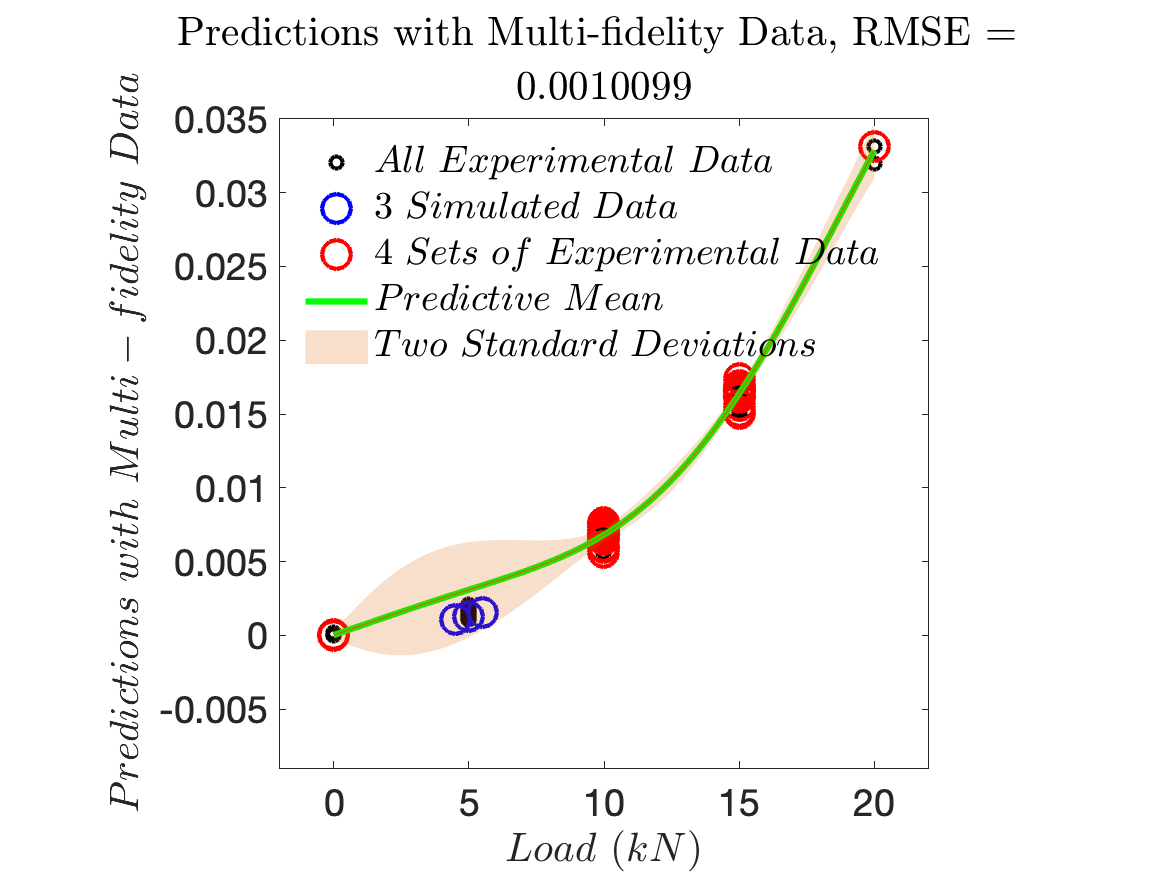}}
    
    \put(130,158){\color{black} \large {\fontfamily{phv}\selectfont \textbf{a}}}
    \put(292,158){\large {\fontfamily{phv}\selectfont \textbf{b}}} 
    \put(454,158){\large {\fontfamily{phv}\selectfont \textbf{c}}} 
    \put(130,0){\large {\fontfamily{phv}\selectfont \textbf{d}}}
   \put(292,0){\large {\fontfamily{phv}\selectfont \textbf{e}}} 
   \put(454,0){\large {\fontfamily{phv}\selectfont \textbf{f}}} 
    \end{picture} 
    \caption{Task 2 of the 2nd test case. DI regression for path 1-6 from GPRM and multi-fidelity GPRM with batch learning using random selection: (a) prediction  using 2 experimental sets at 0 and 20 mm; (b) prediction  using 2 experimental sets and 5 random selected simulated data points; (c) prediction  using 2 experimental sets and 15 random selected simulated data points; (d) prediction  using 3 experimental sets at 0, 10 and 20 mm; (e) prediction  using 3 experimental sets and 5 random selected simulated data points; (f) prediction  using 3 experimental sets and 15 random selected simulated data points.}
\label{fig:2nd_t2_1} 
\end{figure}
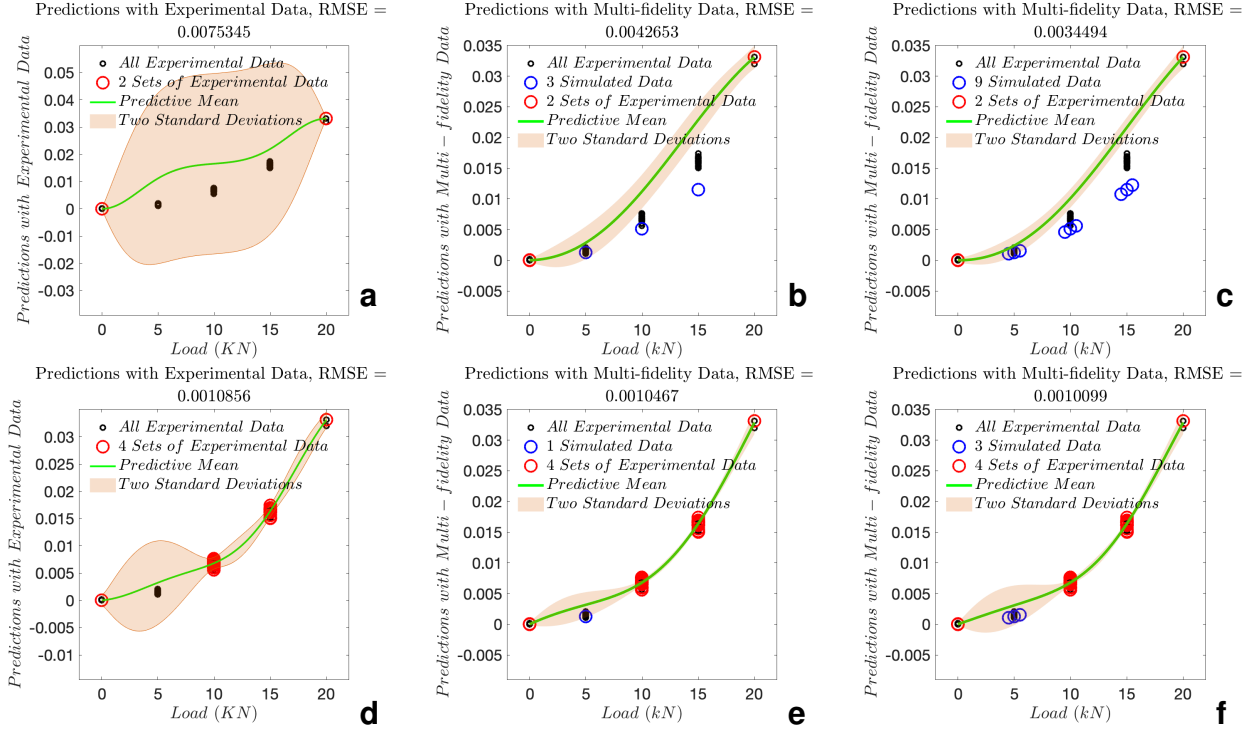
% active learning
% 19
\begin{figure}[t!]
    % \centering
    \begin{picture}(500,160)
    \put(10,0){\includegraphics[width=0.48\textwidth]{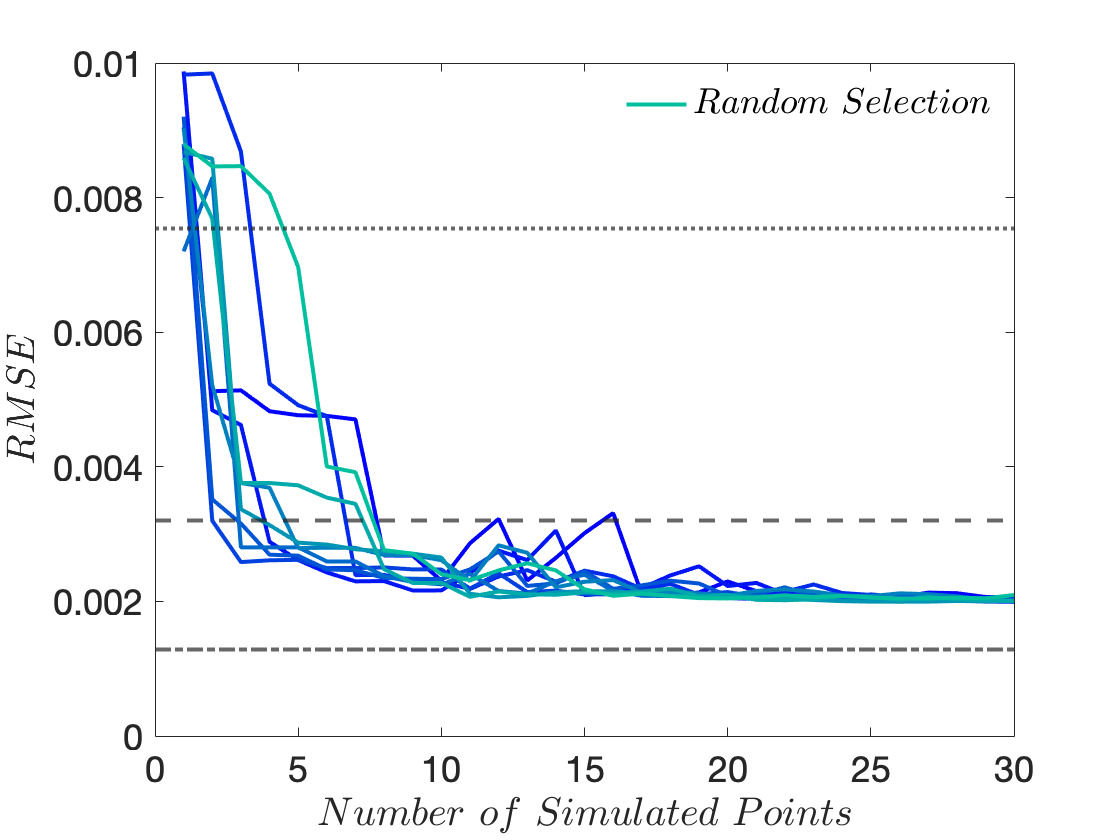}}
    \put(224,0){\includegraphics[width=0.48\textwidth]{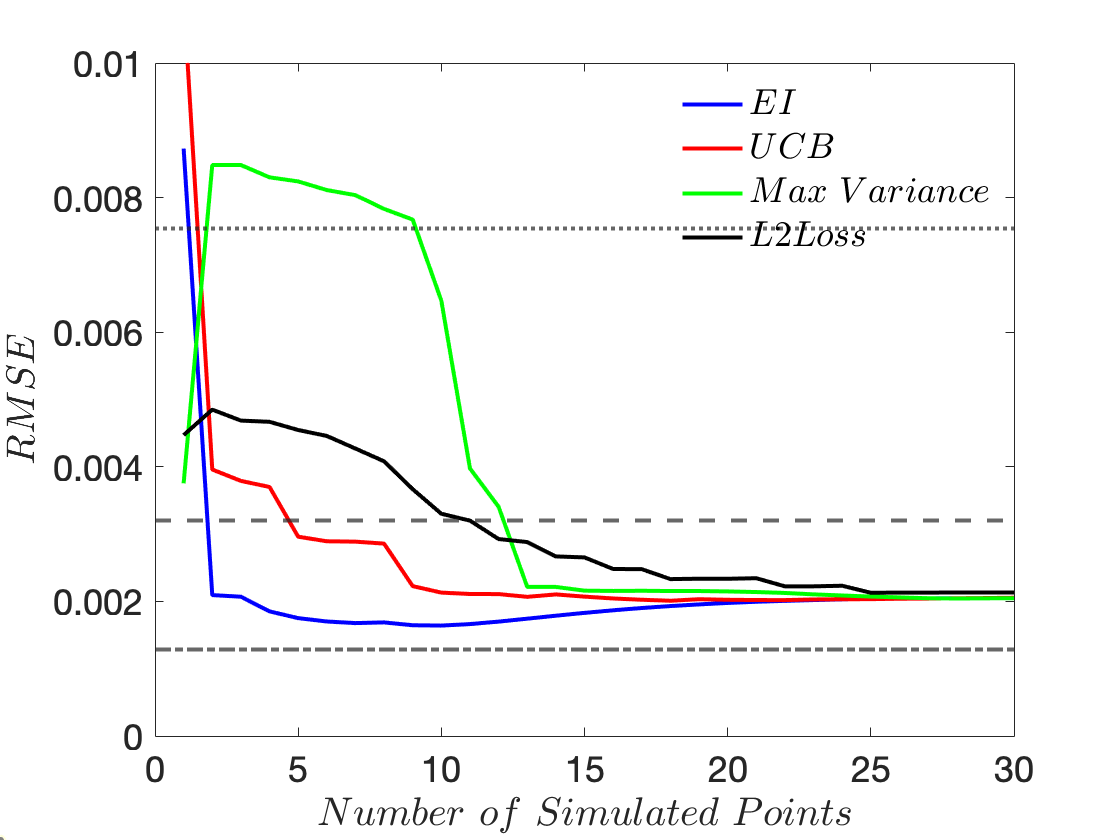}}
    % \put(10,0){\includegraphics[width=0.48\textwidth]{Figures/2ndtest/r2_2d_exp2_2.png}}
    % \put(224,0){\includegraphics[width=0.48\textwidth]{Figures/2ndtest/r2_2d_exp3_2.png}}
    \put(186,26){\color{black} \large {\fontfamily{phv}\selectfont \textbf{a}}}
    \put(396,26){\large {\fontfamily{phv}\selectfont \textbf{b}}}
%   \put(186,130){\large {\fontfamily{phv}\selectfont \textbf{c}}} 
%   \put(396,130){\large {\fontfamily{phv}\selectfont \textbf{d}}} 
    \end{picture} 
    \caption{RMSE trend curves from multi-fidelity models using (a) random selection and (b) four different criteria with active learning methods. In both figures,the model inputs contain 2 sets of experimental data and the number of simulated points increases after each iteration. The dotted line indicates the RMSE from GPRM using 2 experimental sets; the dashed line indicates the RMSE from GPRM using 3 experimental sets; the dash-dotted line indicates the RMSE from GPRM using 4 experimental sets.}
\label{fig:2nd_5} 
\end{figure}
% comparison betw random selection and active learning ends
% UCB example starts
\begin{figure}[t!]
    % \centering
    \begin{picture}(500,410)
    \put(-20,250){\includegraphics[width=0.39\textwidth]{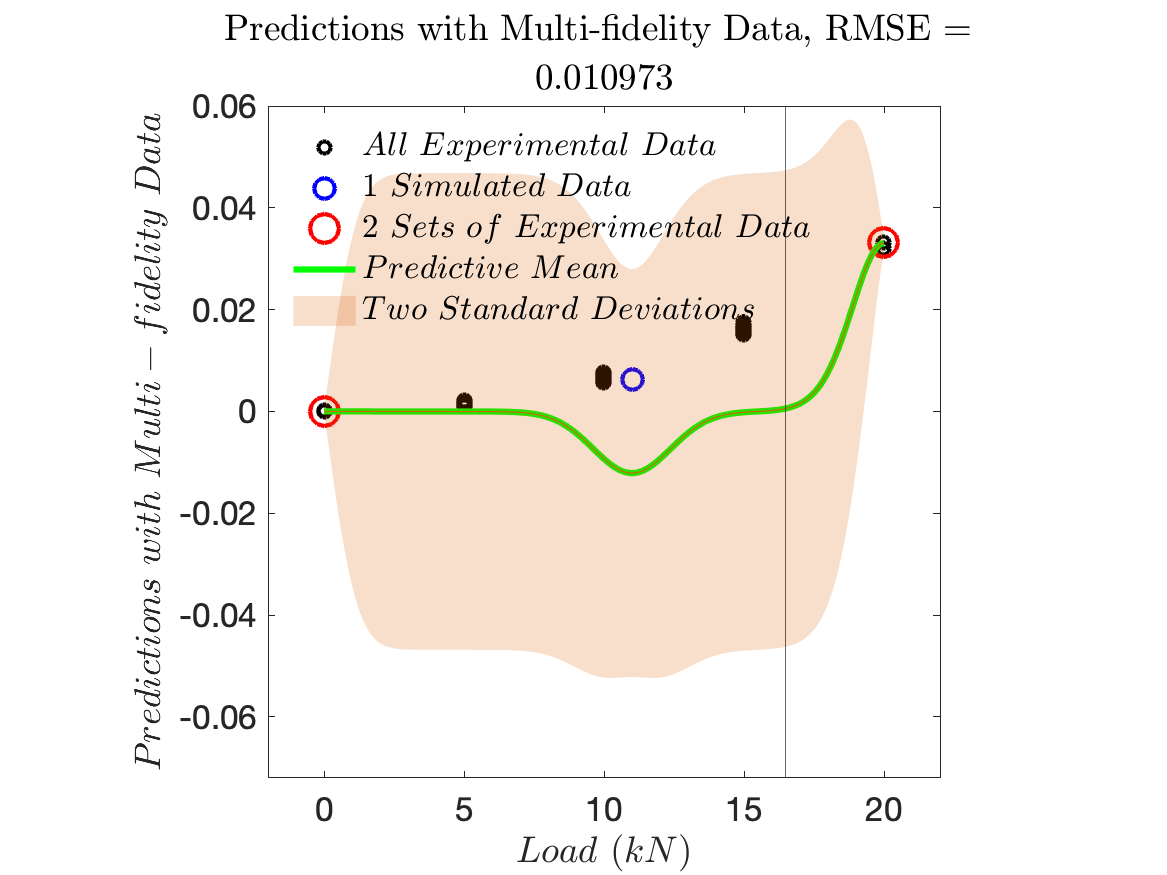}}
    \put(142,250){\includegraphics[width=0.39\textwidth]{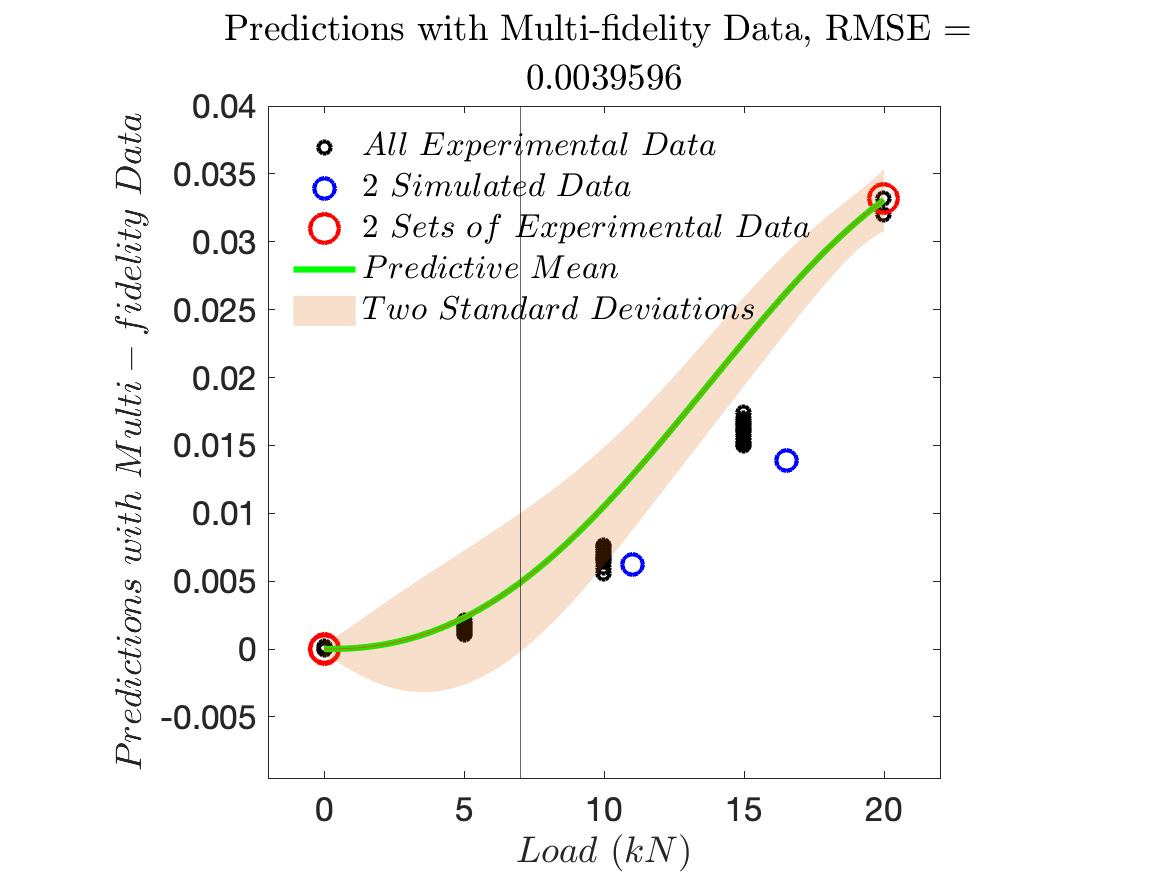}}
    \put(304,250){\includegraphics[width=0.39\textwidth]{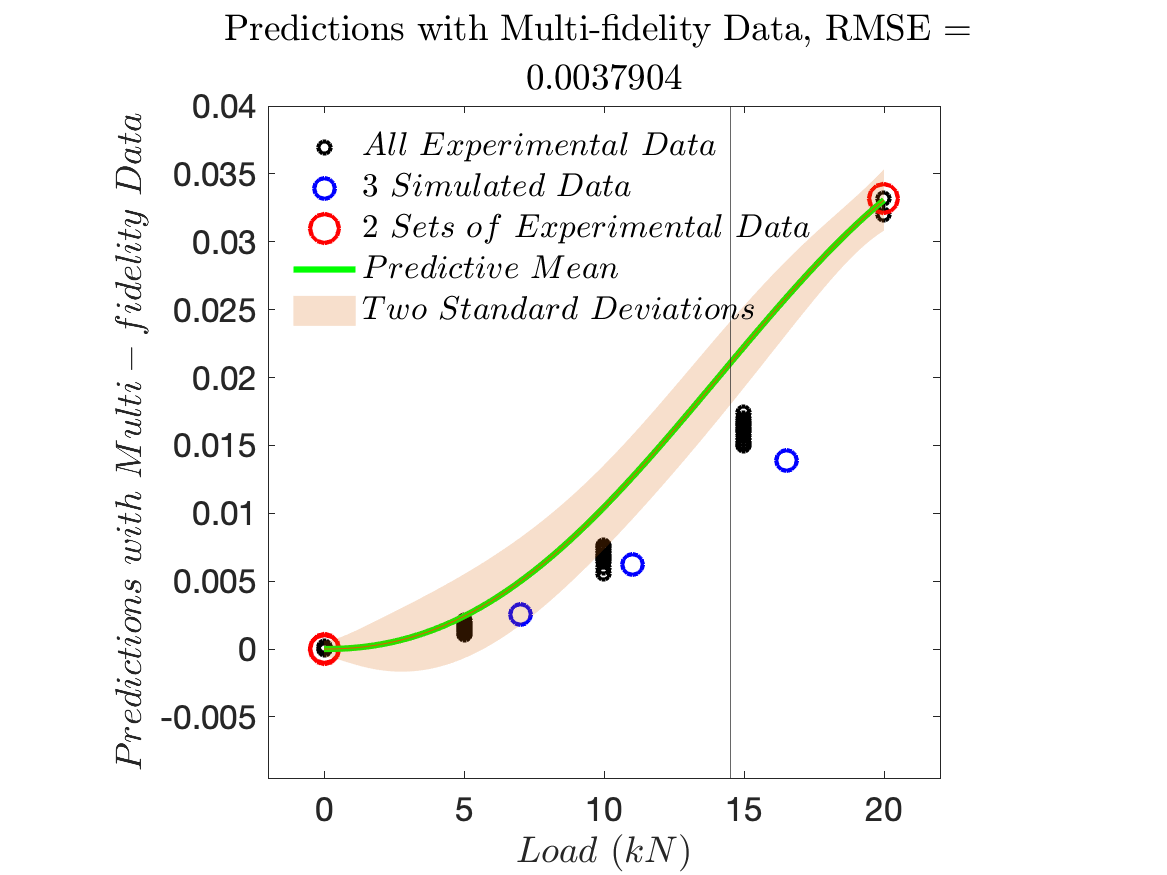}}
    \put(-20,112){\includegraphics[width=0.39\textwidth]{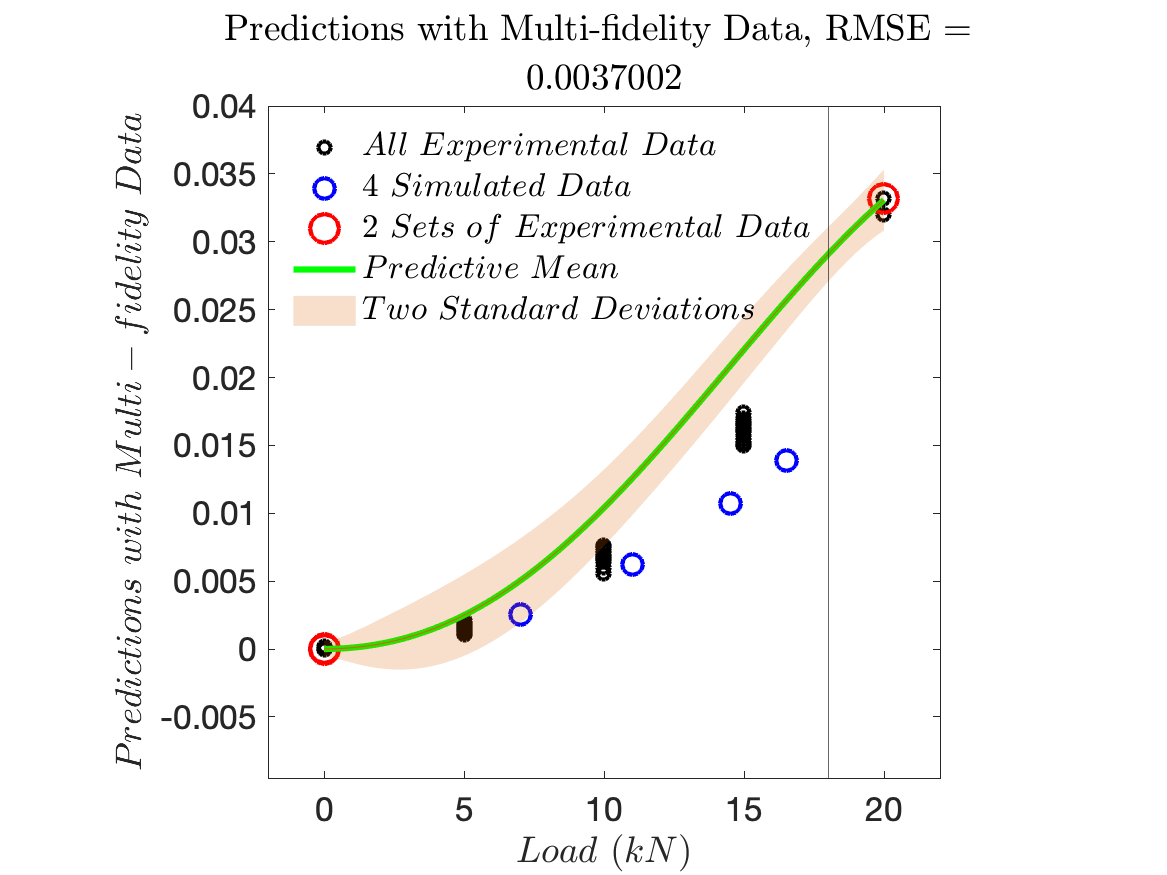}}
    \put(142,112){\includegraphics[width=0.39\textwidth]{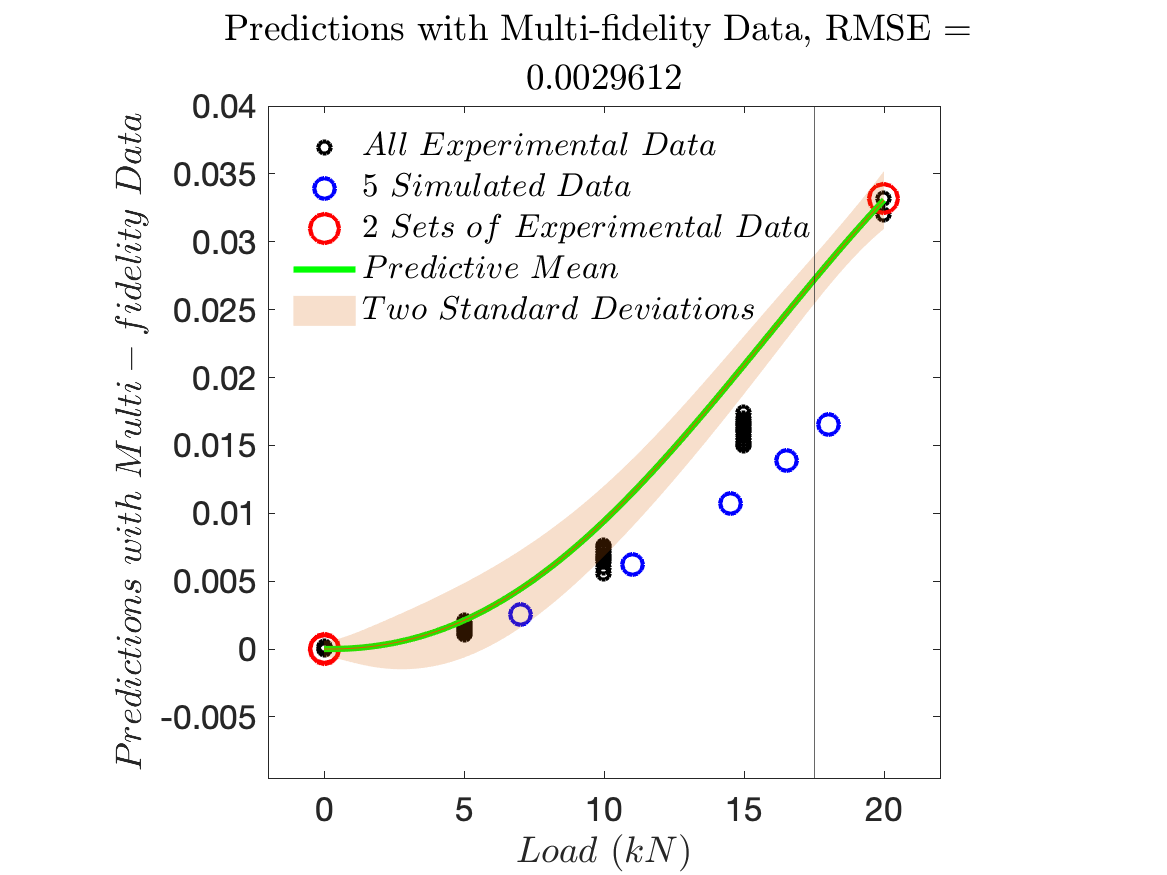}}
    \put(304,112){\includegraphics[width=0.39\textwidth]{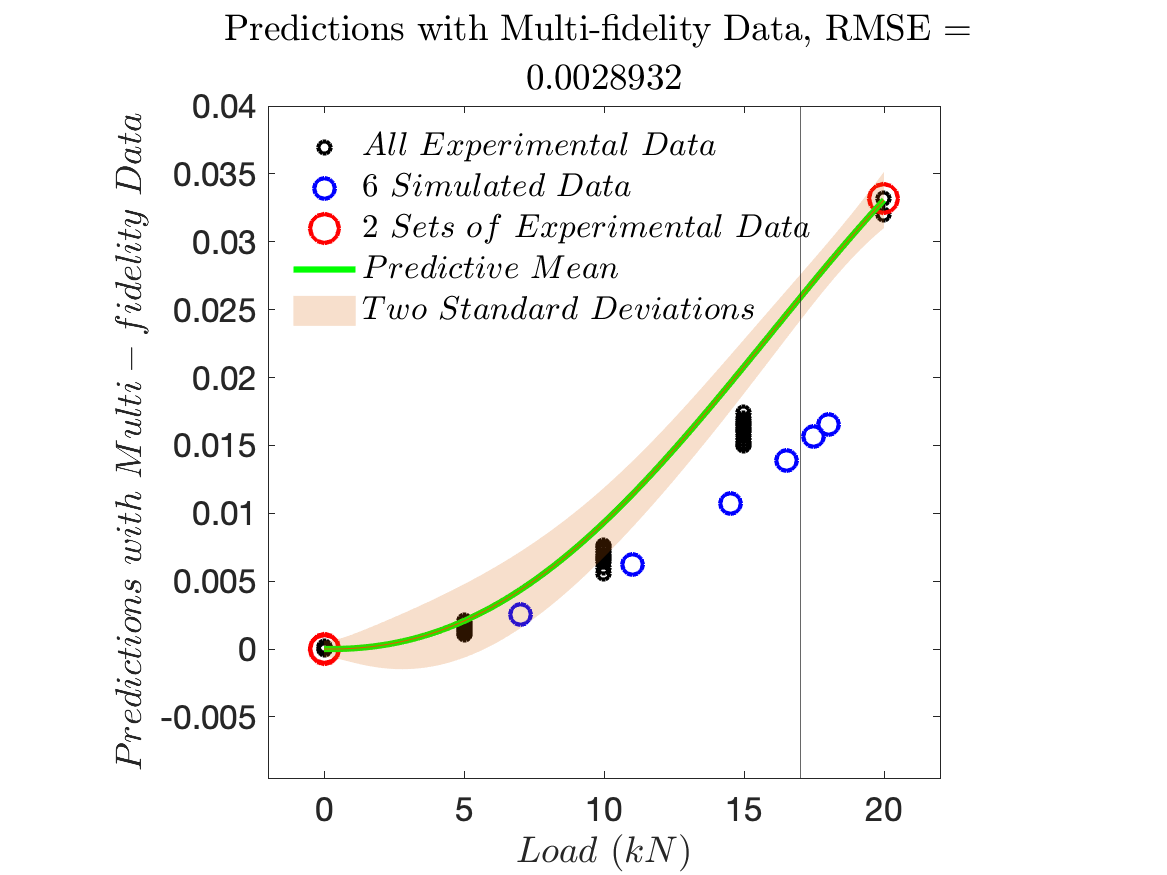}}
    
    \put(110,276){\color{black} \large {\fontfamily{phv}\selectfont \textbf{a}}}
    \put(272,276){\large {\fontfamily{phv}\selectfont \textbf{b}}} 
    \put(434,276){\large {\fontfamily{phv}\selectfont \textbf{c}}} 
    \put(110,138){\large {\fontfamily{phv}\selectfont \textbf{d}}}
   \put(272,138){\large {\fontfamily{phv}\selectfont \textbf{e}}} 
   \put(434,138){\large {\fontfamily{phv}\selectfont \textbf{f}}} 
    \end{picture} \vspace{-130pt}
    \caption{The first 6 iterations of multi-fidelity GPRM results using UCB as acquisition function. The vertical line indicates the location where UCB reaches the maximum.}
\label{fig:2nd_6} 
\vspace{-8pt}
\end{figure}
%
% Task 1 of the 2nd test case. DI regression for path 1-6 from GPRM and multi-fidelity GPRM with batch learning using random selection: (a) prediction  using 2 experimental sets at 0 and 20 mm; (b) prediction  using 2 experimental sets and 5 random selected simulated data points; (c) prediction  using 2 experimental sets and 15 random selected simulated data points; (d) prediction  using 3 experimental sets at 0, 10 and 20 mm; (e) prediction  using 3 experimental sets and 5 random selected simulated data points; (f) prediction  using 3 experimental sets and 15 random selected simulated data points.
\subsubsection{Task 1: Fixed High-Fidelity Data}

Figure \ref{fig:2nd_t1_1}, panels (a) through (c), compare the results of the GPRM model, where only two experimental sets at 0 and 20 kN are available, with the multi-fidelity GPRM prediction that includes an additional 5 and 15 simulated DI points. It is evident that the latter two cases provide superior accuracy and tighter variance bounds. The RMSE decreases from around 0.0075 to 0.0025. Similar trends are observed when an experimental DI set at 10 kN is also included, as shown in panels (d) through (f) where the RMSE goes approximately from 0.0032 to 0.0014. A more compact comparison of the RMSE and ${R}^2$ values for these two cases is provided in Figure \ref{fig:2nd_t1_2}. The blue line indicates the baseline from standard GPRM while the red line is from multi-fidelity GPRM. By comparing the subplots in each row, it is clear that RMSE decreases rapidly with the inclusion of lower-fidelity data, while ${R}^2$ increases sharply with relatively a small amount of simulated points. Although fluctuations may occur, convergence can be achieved within 30 simulated points. It is also evident that the number of high-fidelity data points is crucial for accuracy, as the RMSE for the case with two experimental sets is approximately double that for the case with three sets. While adding more simulated points significantly reduces the error, the RMSE with 30 simulated points in panel (a), around $2.4 \times 10^{-3}$, remains higher than the $1.5 \times 10^{-3}$ shown in panel (b). 
\subsubsection{Task 2: Constant Total Number of States}
Figure \ref{fig:2nd_t2_1} presents the results from task 2, where DIs extracted from physics-based models are used to fill in gaps when experimental DIs are unavailable. Standard GPRMs were first trained using data from 2 and 4 states, as shown in panels (a) and (c), and then compared to multi-fidelity GPRMs with additional data covering the remaining states. In panels (b) and (c), different numbers of simulated data points are added, showing observable improvements as more data are included as inputs. A similar conclusion can be drawn from comparing panels (e) and (f).
% Figure 17
% Figure \ref{fig:2nd_t2_2} shows the corresponding RMSE and R2 values, similar to the analysis in task 2 of test case 1. Although the number of replaceable DIs is limited, the multi-fidelity model consistently outperforms the conventional GPRM. This confirms that the proposed models exhibit robust and satisfactory performance across a range of real-world applications.
%

\begin{table}[b]
\vspace{-16pt}
\centering
% \setlength\tabcolsep{2pt}
% \small
\caption{Summary of results from standard GPRM as baseline}\label{tab:std_gp}
\renewcommand{\arraystretch}{1.3}
{\footnotesize\begin{tabular}{|c|c|c|c|c|c|}
\hline
Number of experimental data & RMSE & ${R}^2$ & Training Time (s) & Prediction Time (s) \\
\hline
2 & 0.0075345 & 0.1521 & 0.011624 & 0.001801 \\
\hline
3 & 0.0031964 & 0.8474 & 0.012901 & 0.000910 \\
\hline
4 & 0.0012905 & 0.9751 & 0.021438 & 0.001241 \\
\hline
5 & 0.0005210 & 0.9959 & 0.024451 & 0.001789 \\
\hline

\end{tabular}} 
% \vspace{+16pt}
\end{table}

\subsubsection{Task 3: Combination with active learning}
Although Figure \ref{fig:2nd_t1_1} demonstrates that a good estimation can be achieved with a sufficient number of simulated points from reconstructed signals, a more efficient implementation approach might be feasible. To address this, we propose a framework that combines the multi-fidelity model with active learning and compare it with a random selection strategy. Same as before, only experimental data was used in standard GPRM to provide a baseline, with results summarized in Table \ref{tab:std_gp}.
To compare with and validate the effectiveness using active learning approach, results using random selection were first generated. Simulated points were chosen randomly within each iteration and the process was repeated 30 times to ensure small error was reached. To guarantee the randomness, 10 different seeds were applied for selecting the added data points.

Then the introduced scheme with active learning methods were applied. The details for implementing active learning with the proposed models introduced in Algorithm \ref{alg:alg1} is as follows: First, a limited amount of experimental data is used to train GPRM, simulating the scenario with scarce data. In this study, data from the two boundary points (0 and 20 kN) are selected initially to capture a rough trend of the regression curve. From the regression results, the location for the next sampling point is determined based on various criteria. In each iteration, an additional data point closest to the recorded location from the lower-fidelity dataset is added for the next round. This process is repeated until a predefined number of iterations is reached. Path 1-6 is used for demonstration, and the framework algorithm is detailed in Algorithm \ref{alg:alg1}.

 Figure \ref{fig:2nd_5} presents the RMSE results of the multi-fidelity GPRM with 2 experimental sets as more simulated points are added, comparing (a) random selection with 10 different seeds and (b) active learning with four different sampling criteria. Plot (a) shows that RMSE convergence rates vary with different seeds, and fluctuations persist after 13 iterations. In contrast, plot (b) demonstrates that while the convergence rates differ among the criteria, the instability significantly diminishes after approximately 13 iterations, indicating more stable performance. The dotted line represents the RMSE for the GPRM with 2 experimental sets, the dashed line for the GPRM with 3 experimental sets, and the dash-dotted line for the GPRM with 4 experimental sets.
It can be concluded that adding more low-fidelity data into the multi-fidelity GPRM can surpass the performance of the GPRM with a higher amount of high-fidelity data. While both methods—random selection and active learning—converge toward a similar RMSE value with enough simulated data, this value remains lower than that of the GPRM with 3 experimental sets but higher than the GPRM with 4 experimental sets. This highlights that adding more low-fidelity data enables the multi-fidelity GPRM to outperform the GPRM with additional high-fidelity data points. Additionally, active learning demonstrates faster and more stable RMSE convergence compared to random selection.
 
 Figure \ref{fig:2nd_6} provides a detailed example of active learning, illustrating the first 6 iterations of the multi-fidelity GPRM with Upper Confidence Bound (UCB) as the acquisition function. The vertical line indicates the location where the UCB reaches its maximum. The nearest simulated data point to this location is added to the model inputs in the next iteration. The overall performance is recorded in Table \ref{tab:mf_gp}.

\begin{table}[t!]
\vspace{-16pt}
\centering
% \setlength\tabcolsep{2pt}
% \small
\caption{Summary of optimal results from multi-fidelity GPRMs with active learning}\label{tab:mf_gp}
\renewcommand{\arraystretch}{1.3}
{\footnotesize\begin{tabular}{|c|c|c|c|c|c|c|}
\hline
Number of experimental & Number of simulated & RMSE & ${R}^2$ & Training & Prediction \\
datasets &data & &  & Time (s) & Time (s) \\
\hline
2 & 8 & 0.0021037 & 0.933899 & 0.729976 & 0.009663 \\
\hline
3 & 6 & 0.0013721 & 0.971881 & 1.029015 & 0.012659 \\
\hline
4 & 4 & 0.0012221 & 0.977691 & 1.104725 & 0.018905 \\
\hline
5 & 5 & 0.0005609 & 0.991698& 1.339533 & 0.017882 \\
\hline

\end{tabular}} 
% \vspace{+16pt}
\end{table}

\section{Conclusion} \label{Sec:conc}

In this work, a damage quantification framework in the realm of active-sensing SHM using a novel variate of Gaussian process regression model was proposed and applied to two test cases. After pre-processing the signals collected from both experiments and simulations, DIs, as the features of signals, have been extracted as inputs to train the proposed models. The multi-fidelity GPRM is then able to provide an estimation of the predictive mean and estimation of the DI evolution along the domain of damage states even at locations where no experiments have been conducted. The performance of the models was compared with standard GPRMs based on the regression results and the values of RMSE and ${R}^2$. In the first test case, two different tasks were conducted to mimic two real world scenarios, i.e., when data with high-fidelity is fixed and when the total number of damage sizes corresponding to data from all available sources is unchanged. In both tasks, the multi-fidelity GPRMs showed more accurate and robust predictions on the DI evolution and exhibited the potential when data with high accuracy is limited. In the second test case, the aforementioned two tasks were repeated and multi-fidelity GPRMs were then combined with random selection and active learning approaches respectively to reach more efficiency in real world applications. The models with random selection, though showed convergence, were not stable enough and the performance was dependent on the sequence of data added. When integrated with active learning, however, the framework showed a rapid convergence rate with robustness and accuracy. 

\vspace{-6pt}

\bibliographystyle{aiaa} 

\bibliography{References} % bibliography data 
%\bibliography{wing_references} % bibliography data 

\vspace{-6pt}

%----------------------------------------------------------------------------------------------------
% % Appendix
% %----------------------------------------------------------------------------------------------------
\clearpage
\appendix
\appendixpage
\addappheadtotoc

\setcounter{table}{0}
\renewcommand\thetable{\Alph{section}.\arabic{table}}

\setcounter{figure}{0}
\renewcommand\thefigure{\Alph{section}.\arabic{figure}}
\section{First Test Case}
\subsection{Task1}
\begin{figure}[H]
    % \centering
    \begin{picture}(500,330)
    \put(10,168){\includegraphics[width=0.48\textwidth]{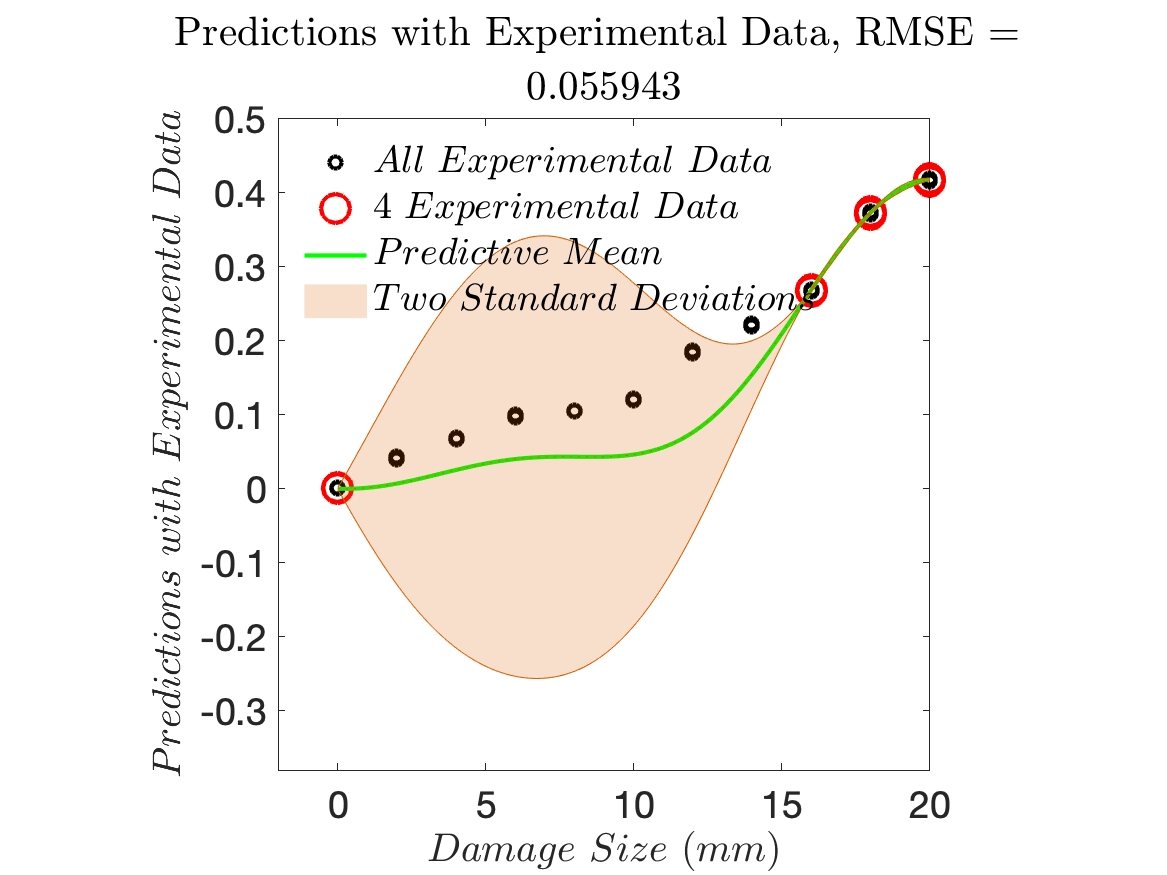}}
    \put(224,168){\includegraphics[width=0.48\textwidth]{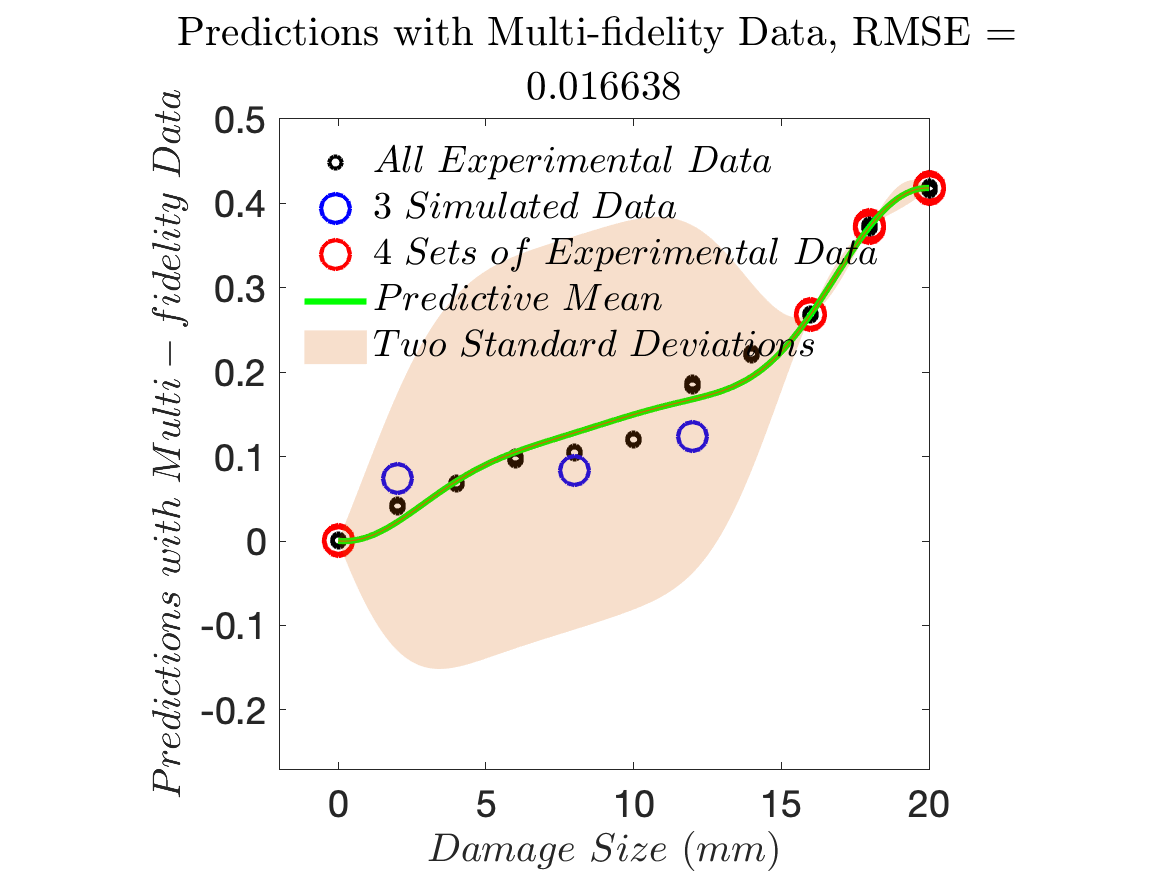}}
    \put(10,0){\includegraphics[width=0.48\textwidth]{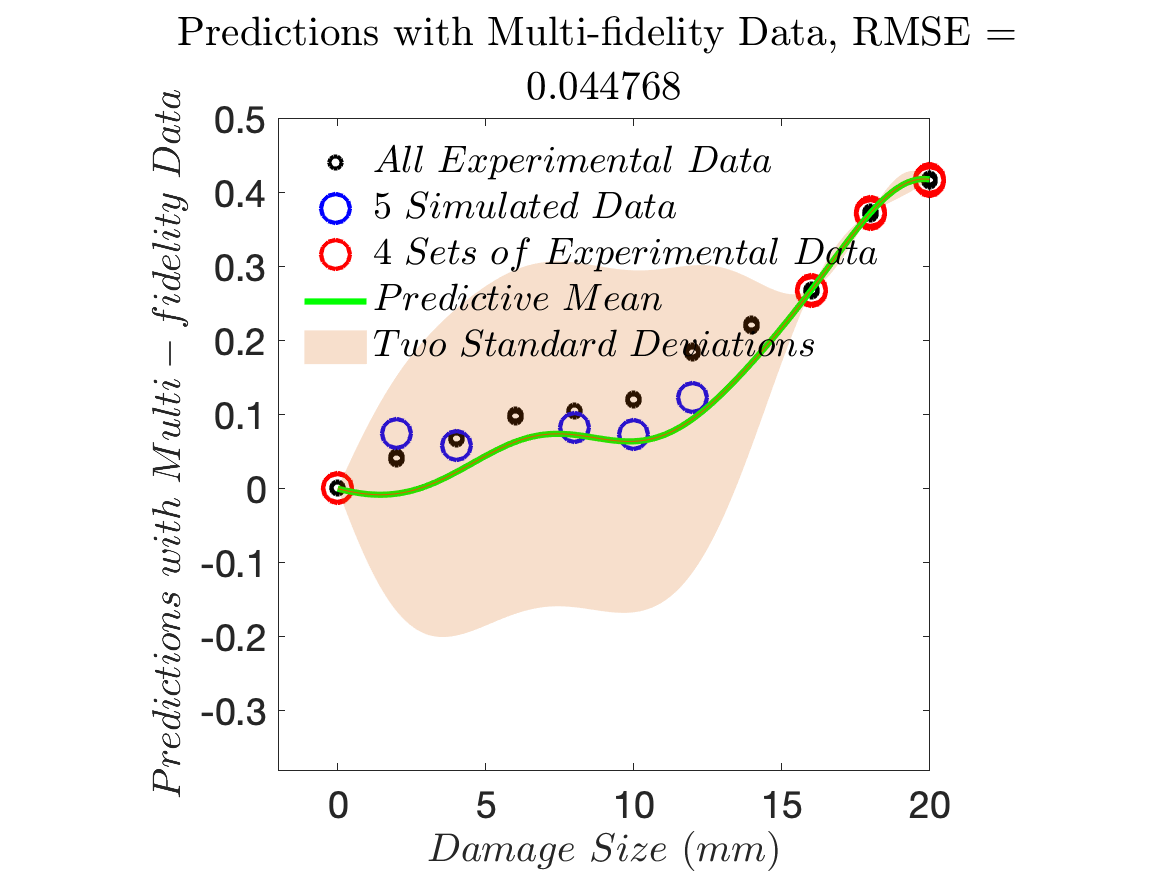}}
    \put(224,0){\includegraphics[width=0.48\textwidth]{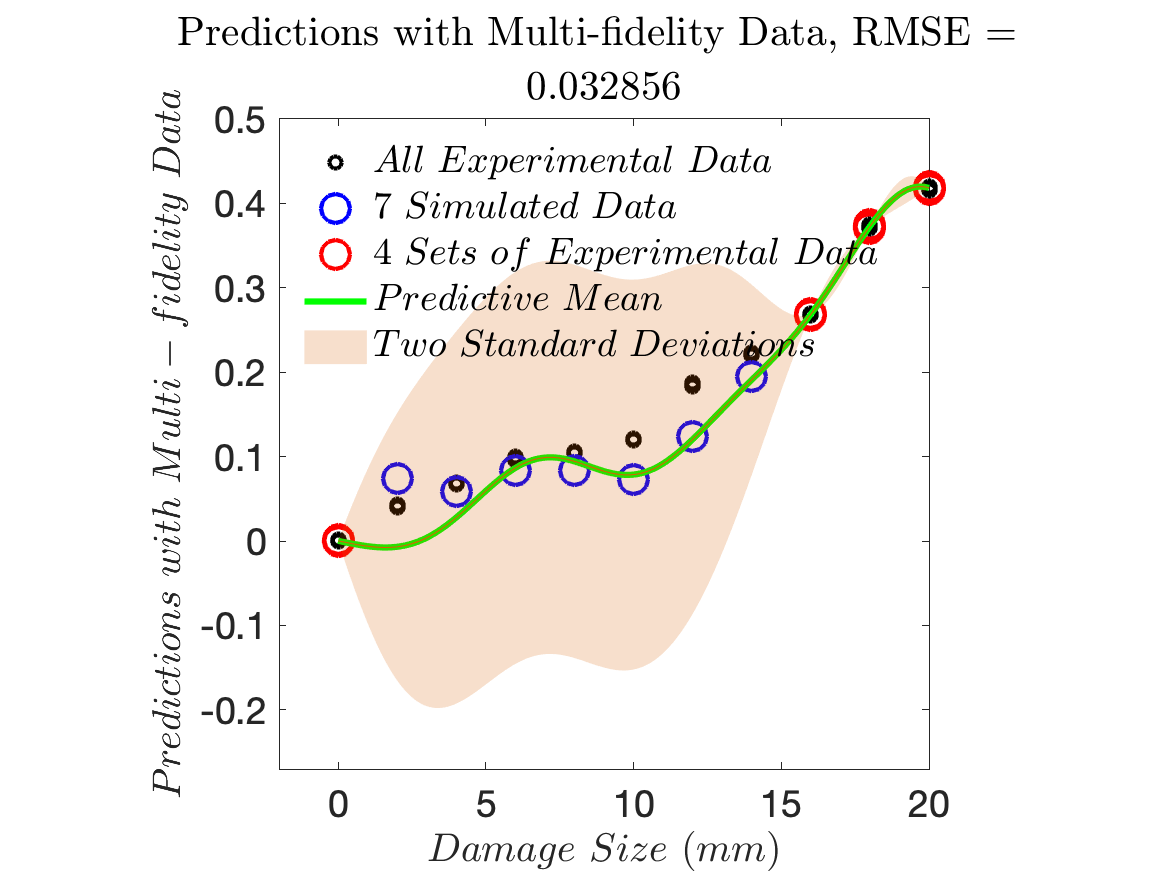}}
    \put(176,200){\color{black} \large {\fontfamily{phv}\selectfont \textbf{a}}}
    \put(390,200){\large {\fontfamily{phv}\selectfont \textbf{b}}}
   \put(176,34){\large {\fontfamily{phv}\selectfont \textbf{c}}} 
   \put(390,34){\large {\fontfamily{phv}\selectfont \textbf{d}}} 
    \end{picture} 
    \caption{DI regression for path 2-6 from GPRM and multi-fidelity GPRM: (a) prediction  using 4 experimental sets; (b) prediction  using 4 experimental sets and 3 simulated data points; (c) prediction  using 4 experimental sets and 5 simulated data points; (d) prediction  using 4 experimental sets and 7 simulated data points.}
\label{fig:app11} 
\end{figure}
\begin{figure}[h!]
    % \centering
    \begin{picture}(500,330)
    \put(10,168){\includegraphics[width=0.48\textwidth]{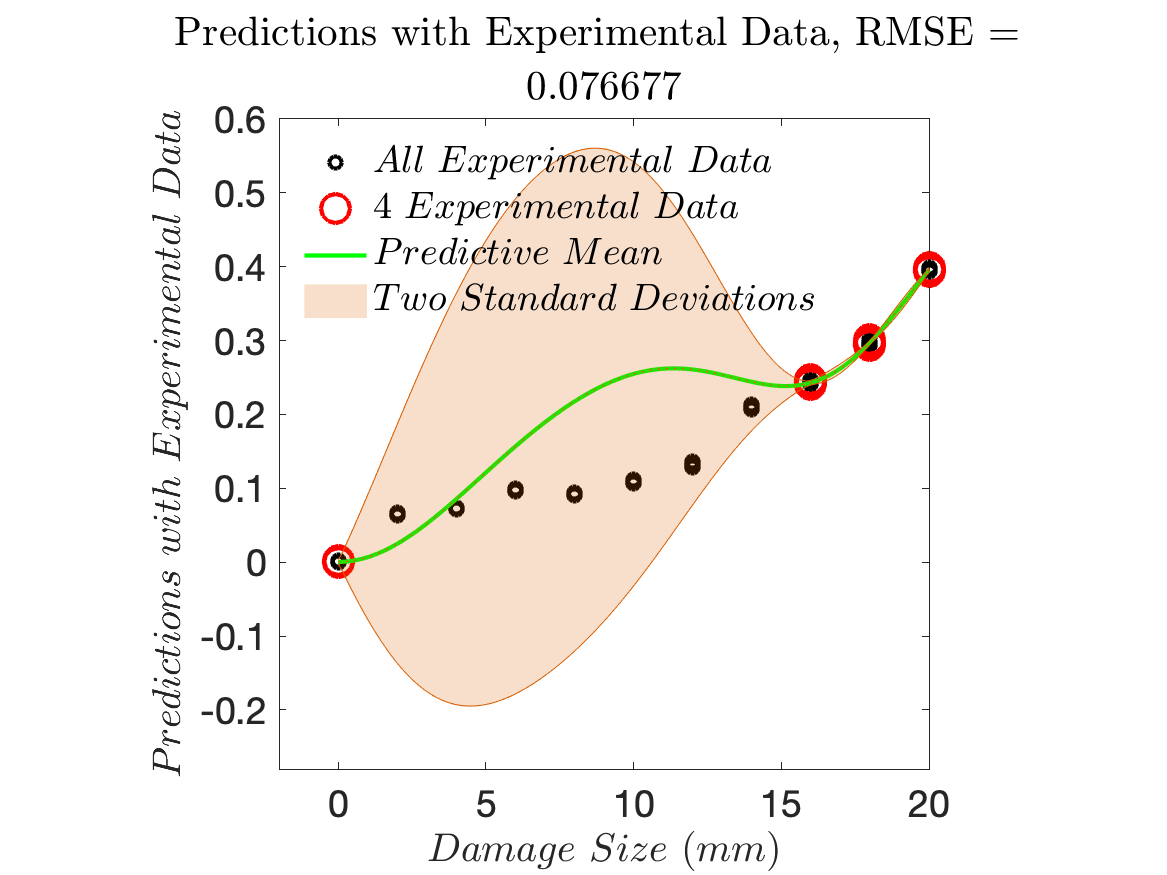}}
    \put(224,168){\includegraphics[width=0.48\textwidth]{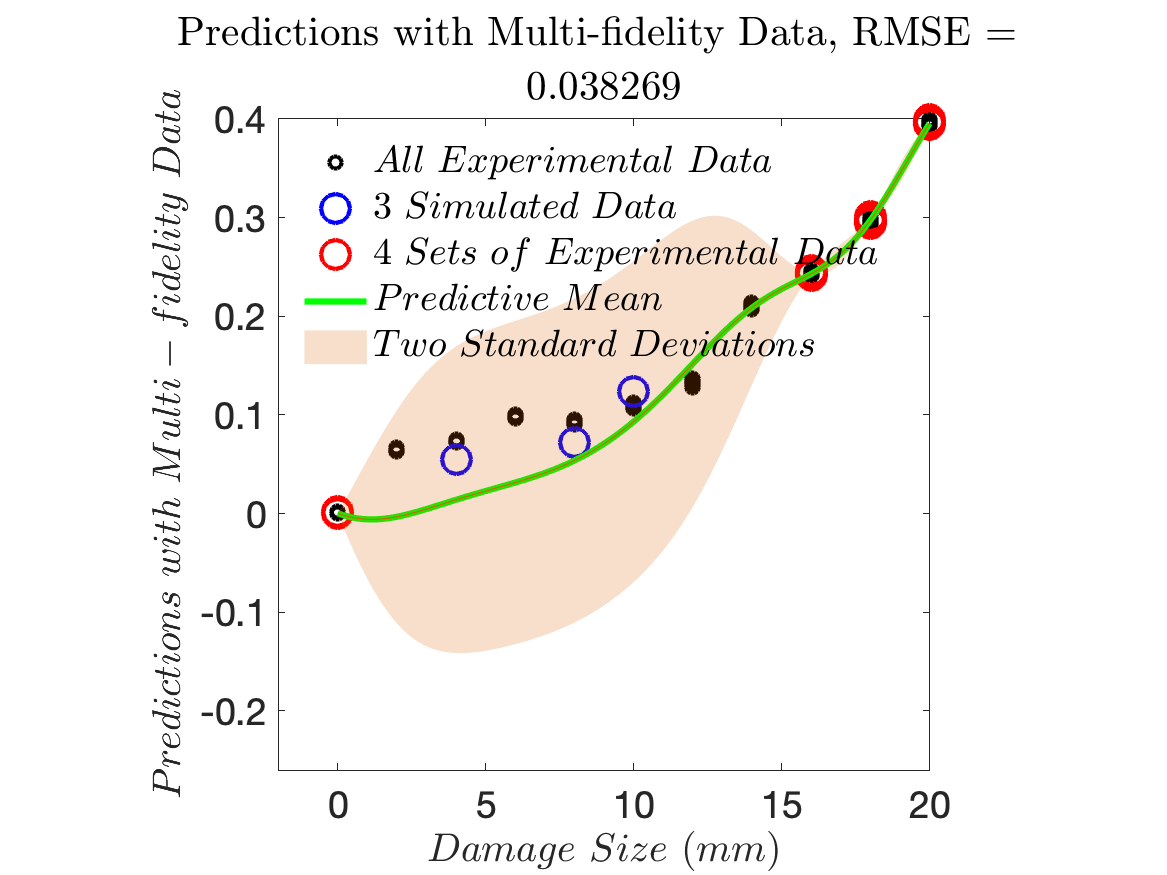}}
    \put(10,0){\includegraphics[width=0.48\textwidth]{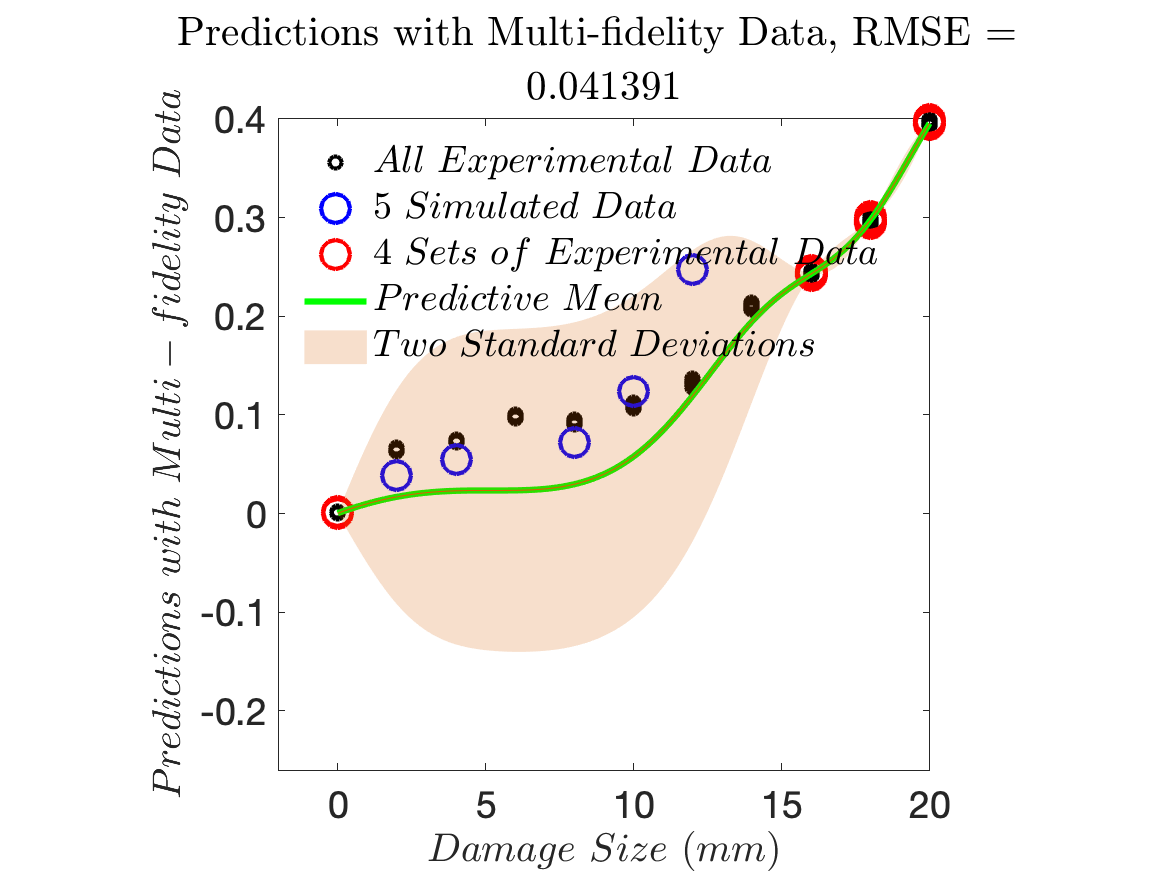}}
    \put(224,0){\includegraphics[width=0.48\textwidth]{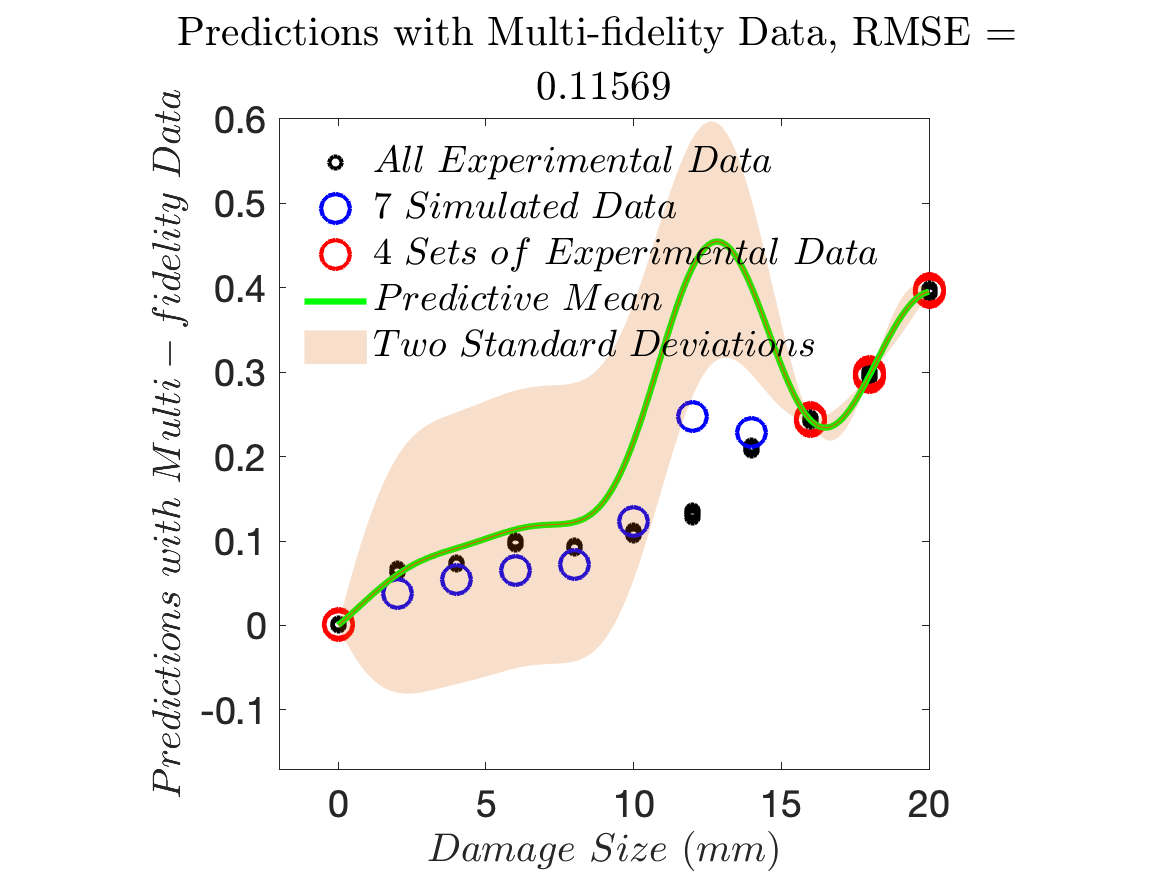}}
    \put(176,200){\color{black} \large {\fontfamily{phv}\selectfont \textbf{a}}}
    \put(390,200){\large {\fontfamily{phv}\selectfont \textbf{b}}}
   \put(176,34){\large {\fontfamily{phv}\selectfont \textbf{c}}} 
   \put(390,34){\large {\fontfamily{phv}\selectfont \textbf{d}}} 
    \end{picture} 
    \caption{DI regression for path 3-5 from GPRM and multi-fidelity GPRM: (a) prediction  using 4 experimental sets; (b) prediction  using 4 experimental sets and 3 simulated data points; (c) prediction  using 4 experimental sets and 5 simulated data points; (d) prediction  using 4 experimental sets and 7 simulated data points.}
\label{fig:app12} 
\end{figure}
\begin{figure}[h!]
    % \centering
    \begin{picture}(500,330)
    \put(10,168){\includegraphics[width=0.48\textwidth]{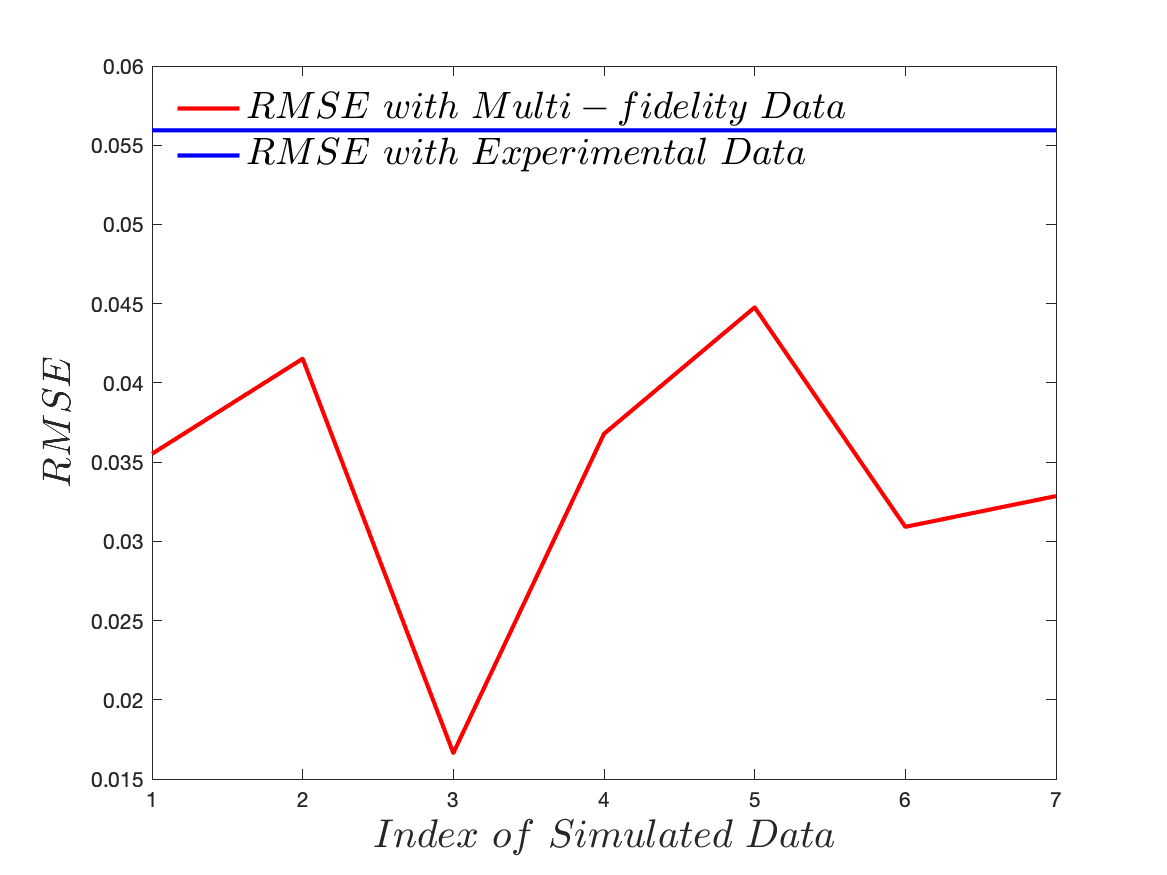}}
    \put(224,168){\includegraphics[width=0.48\textwidth]{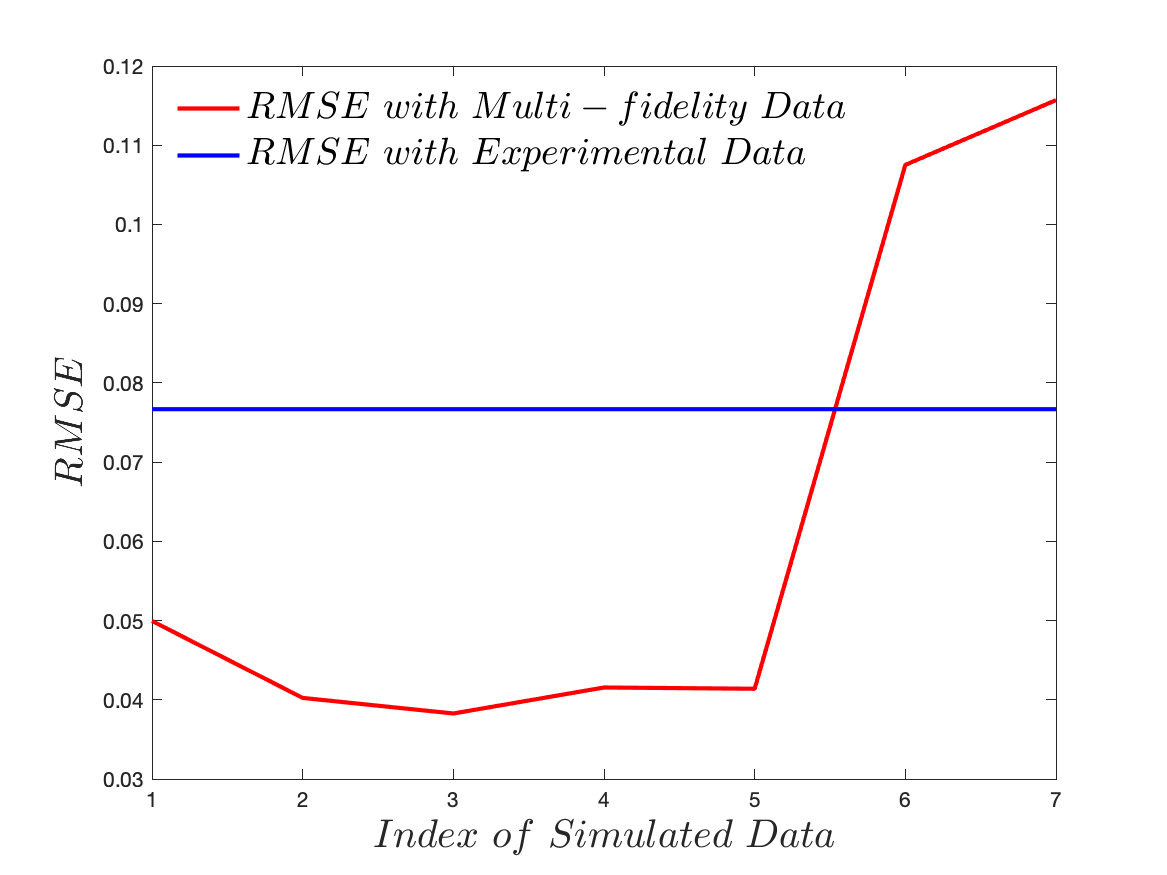}}
    \put(10,0){\includegraphics[width=0.48\textwidth]{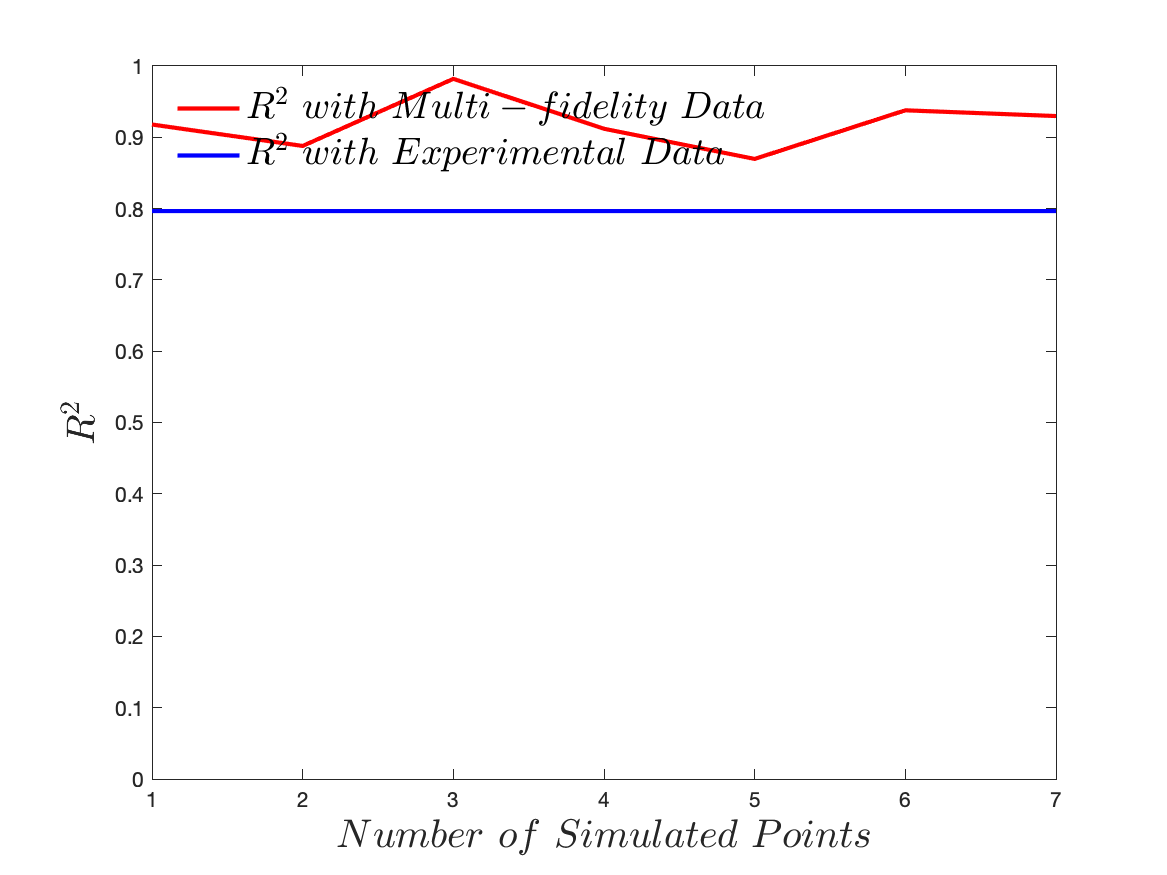}}
    \put(224,0){\includegraphics[width=0.48\textwidth]{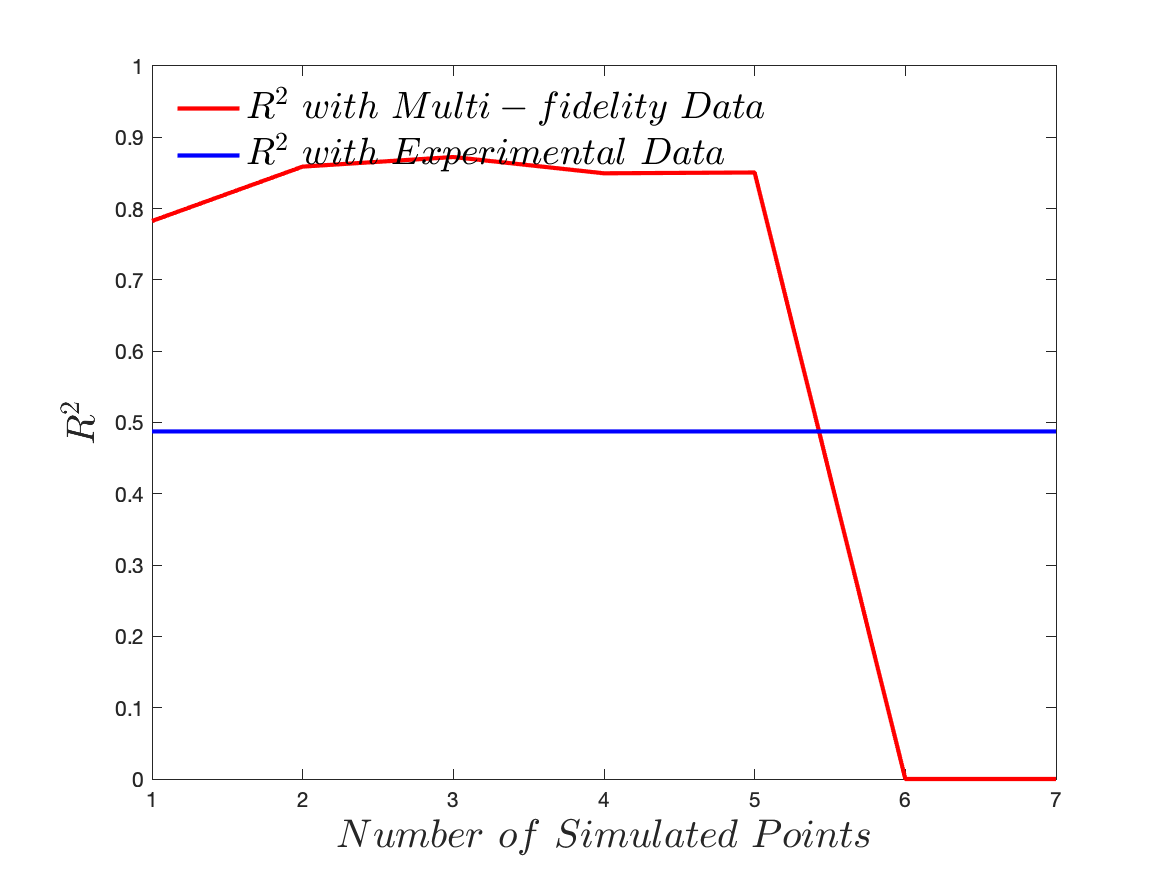}}
    \put(176,200){\color{black} \large {\fontfamily{phv}\selectfont \textbf{a}}}
    \put(390,200){\large {\fontfamily{phv}\selectfont \textbf{b}}}
   \put(176,34){\large {\fontfamily{phv}\selectfont \textbf{c}}} 
   \put(390,34){\large {\fontfamily{phv}\selectfont \textbf{d}}} 
    \end{picture}
    \caption{RMSE and ${R}^2$ comparison between GPRM and multi-fidelity GPRM for path 2-6 and 3-5 : (a) RMSE comparison for path 2-6 (b) RMSE comparison for path 3-5; (c) ${R}^2$ comparison for path 2-6; (d) ${R}^2$ comparison for path 3-5.}
\label{fig:app13} 
\end{figure}
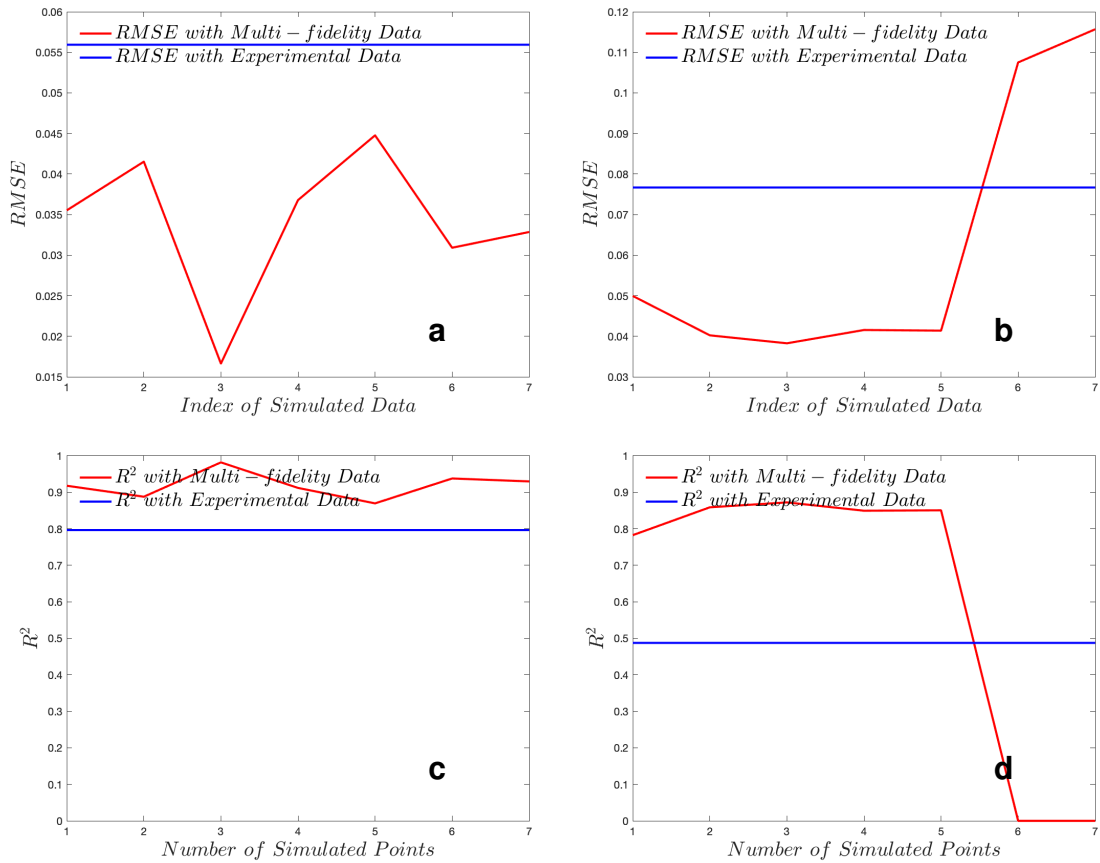
%
% examine effect of different bounds for task 1, 3-4 & 3-6
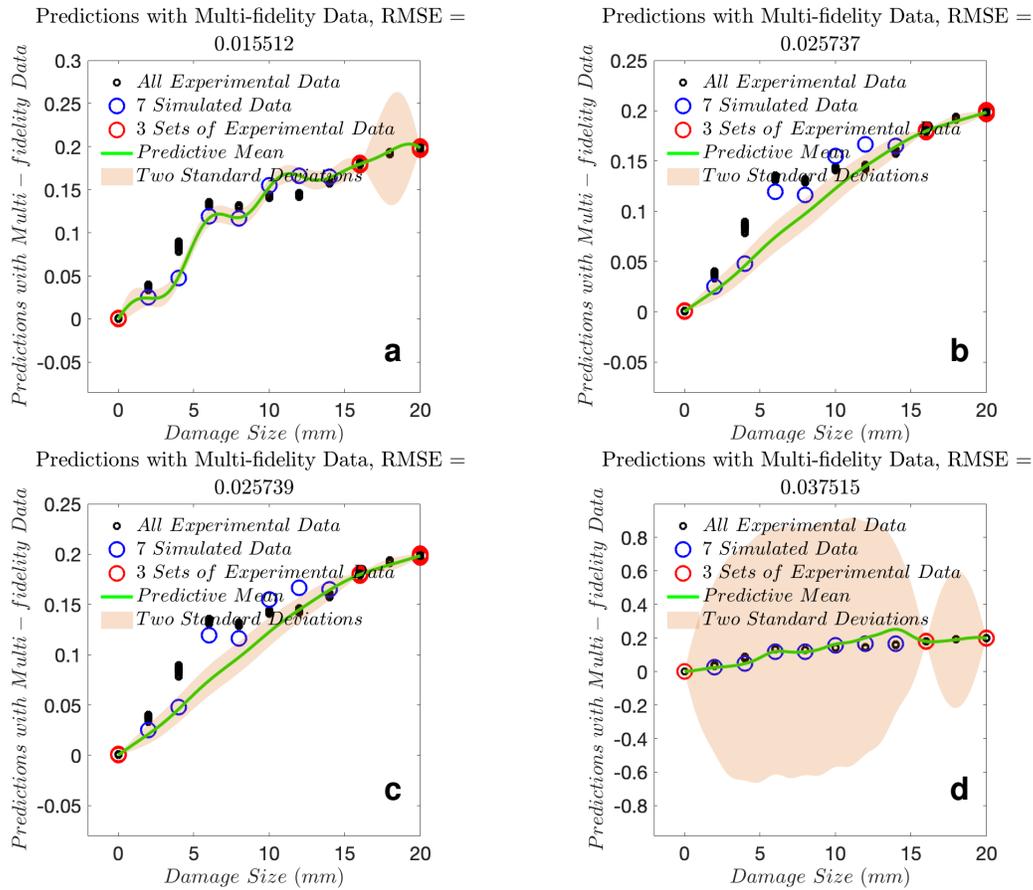
\begin{figure}[h!]
    % \centering
    \begin{picture}(500,330)
    \put(10,168){\includegraphics[width=0.48\textwidth]{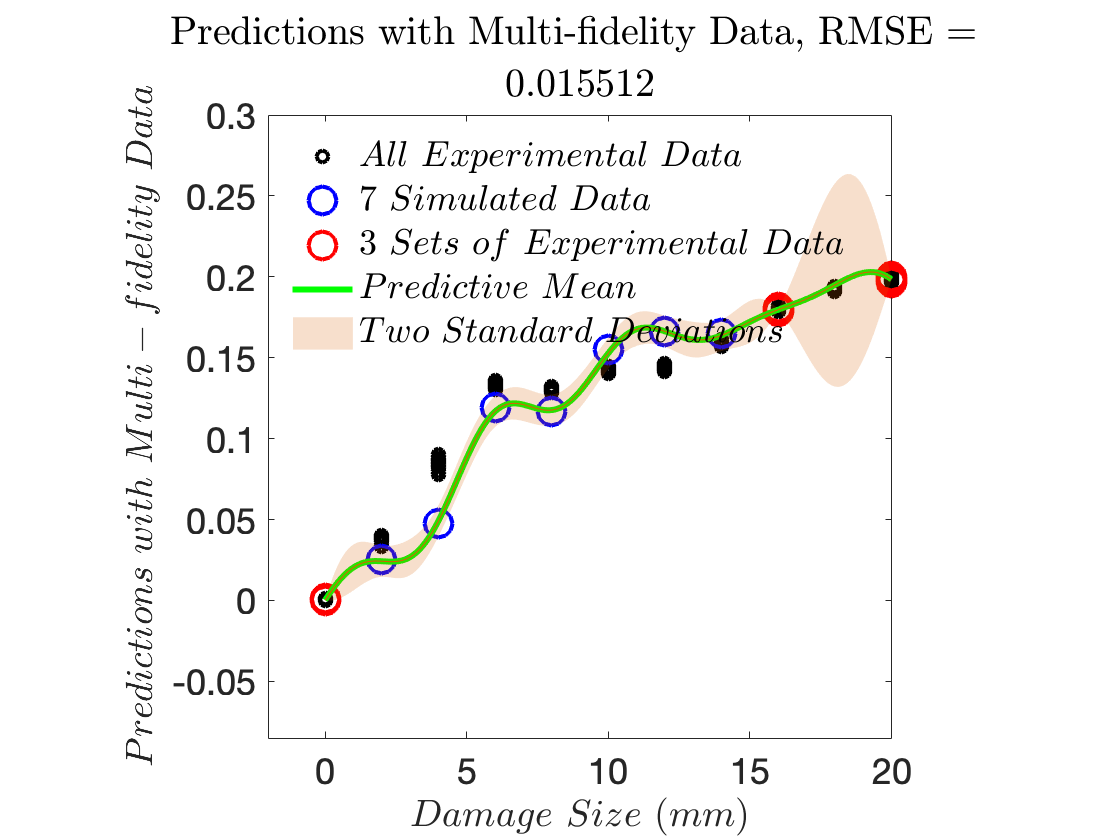}}
    \put(224,168){\includegraphics[width=0.48\textwidth]{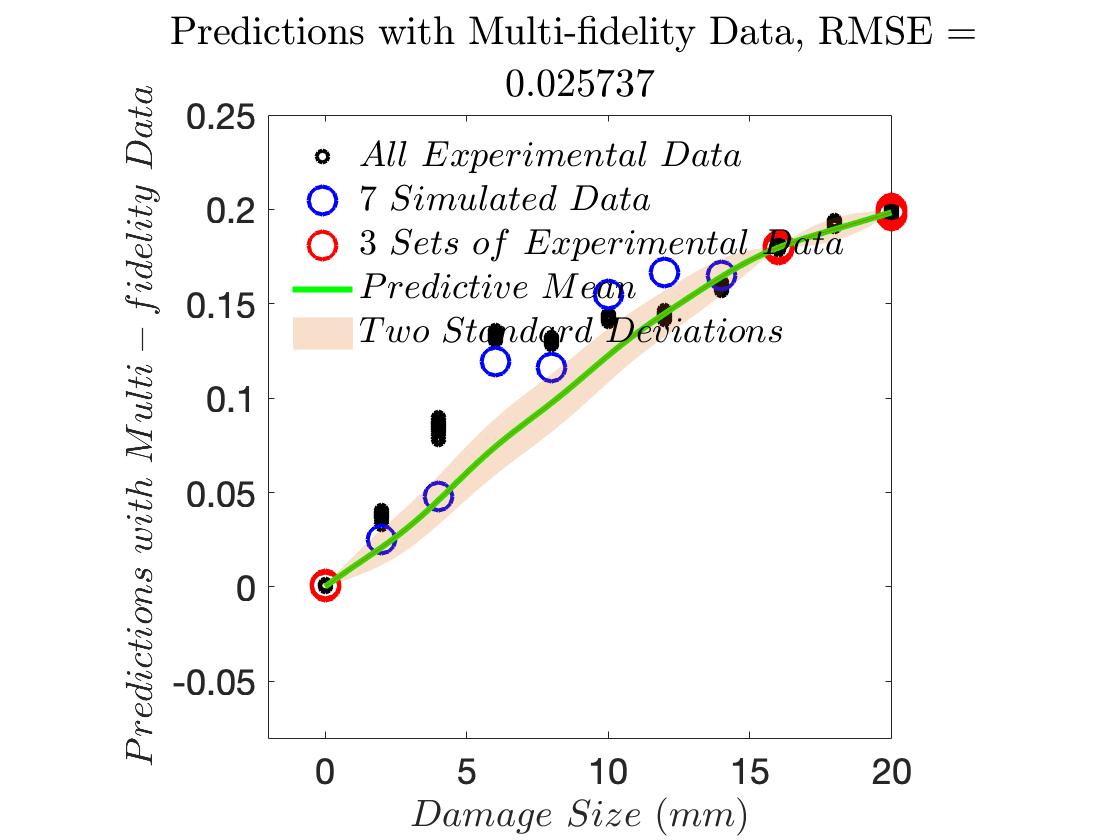}}
    \put(10,0){\includegraphics[width=0.48\textwidth]{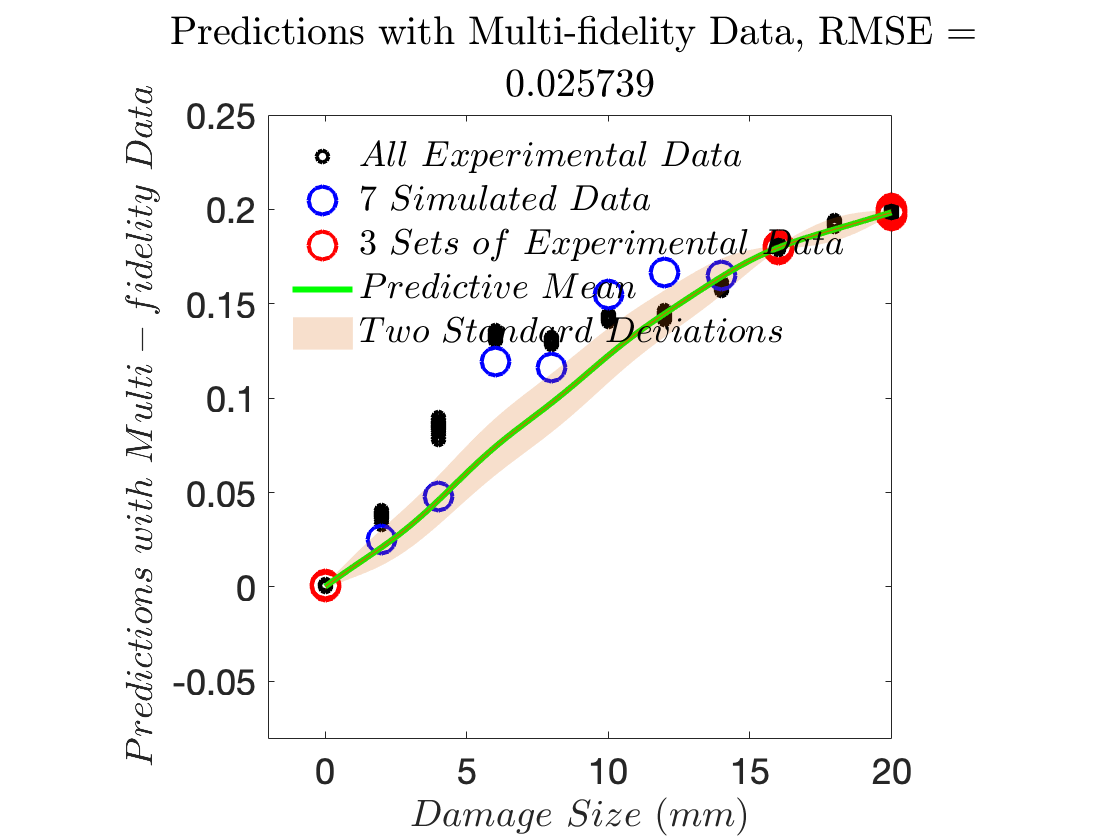}}
    \put(224,0){\includegraphics[width=0.48\textwidth]{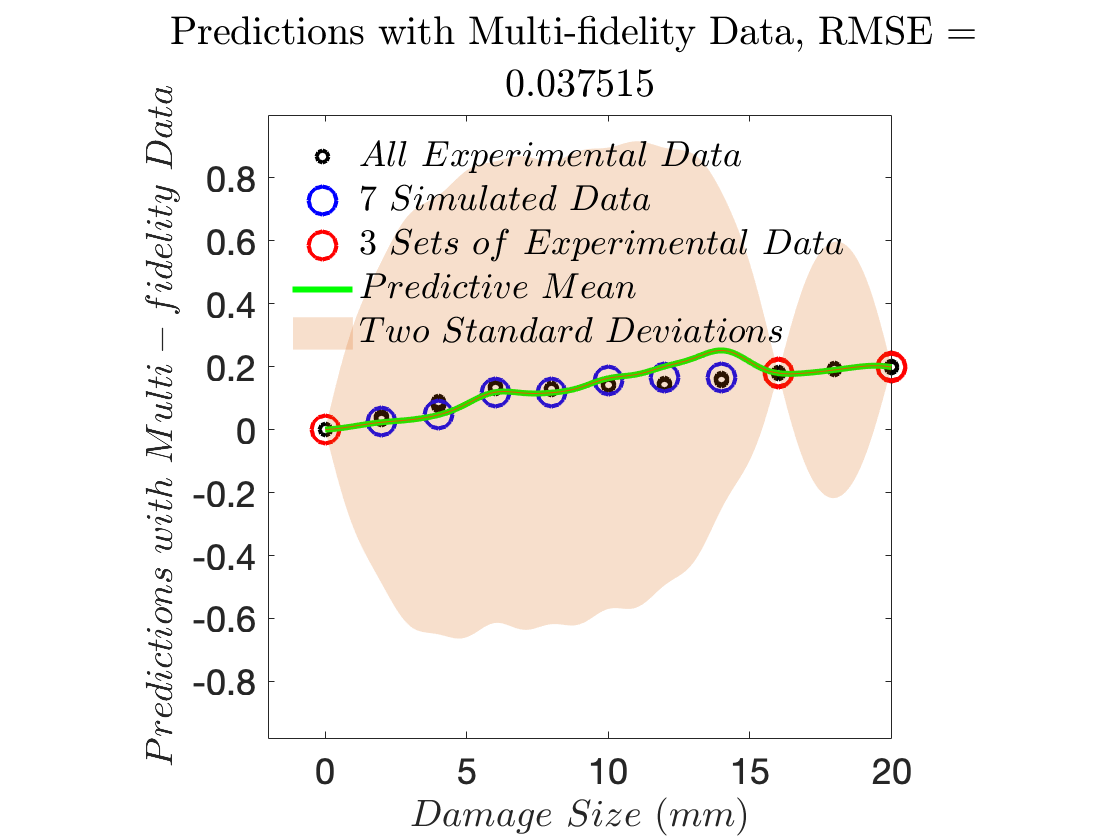}}
    \put(176,200){\color{black} \large {\fontfamily{phv}\selectfont \textbf{a}}}
    \put(390,200){\large {\fontfamily{phv}\selectfont \textbf{b}}}
   \put(176,34){\large {\fontfamily{phv}\selectfont \textbf{c}}} 
   \put(390,34){\large {\fontfamily{phv}\selectfont \textbf{d}}} 
    \end{picture}
    \caption{Results of test case 1 task 1 for path 3-4 using different lower bound constrains during optimization. (a): using the largest variance of experimental data as lower bound; (b): using 5 times the largest variance of experimental data as lower bound; (a): using 10 times the largest variance of experimental data as lower bound; (a): using 15 times the largest variance of experimental data as lower bound.}
\label{fig:app14} 
\end{figure}

\clearpage
\subsection{Examine effects of lower bound of variance in Task1}
% examine effect of different bounds for task 1, 3-4 & 3-6
\begin{figure}[H]
    % \centering
    \begin{picture}(500,330)
    \put(10,168){\includegraphics[width=0.48\textwidth]{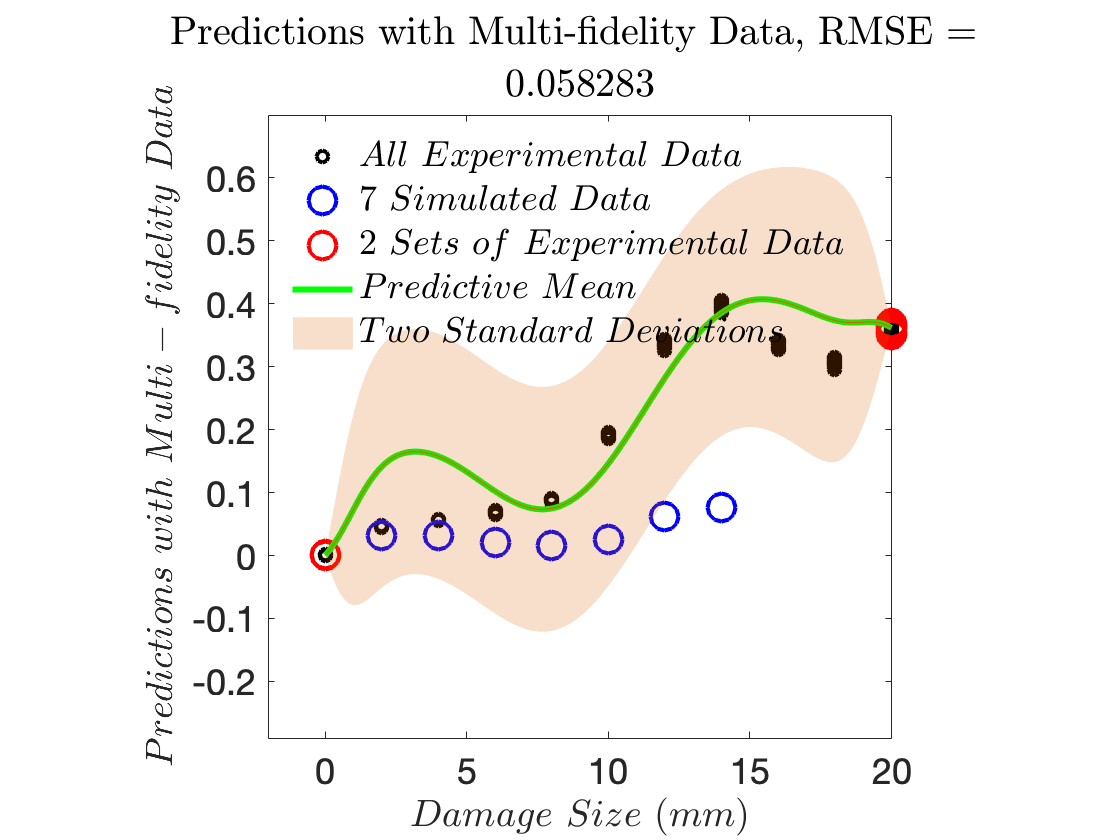}}
    \put(224,168){\includegraphics[width=0.48\textwidth]{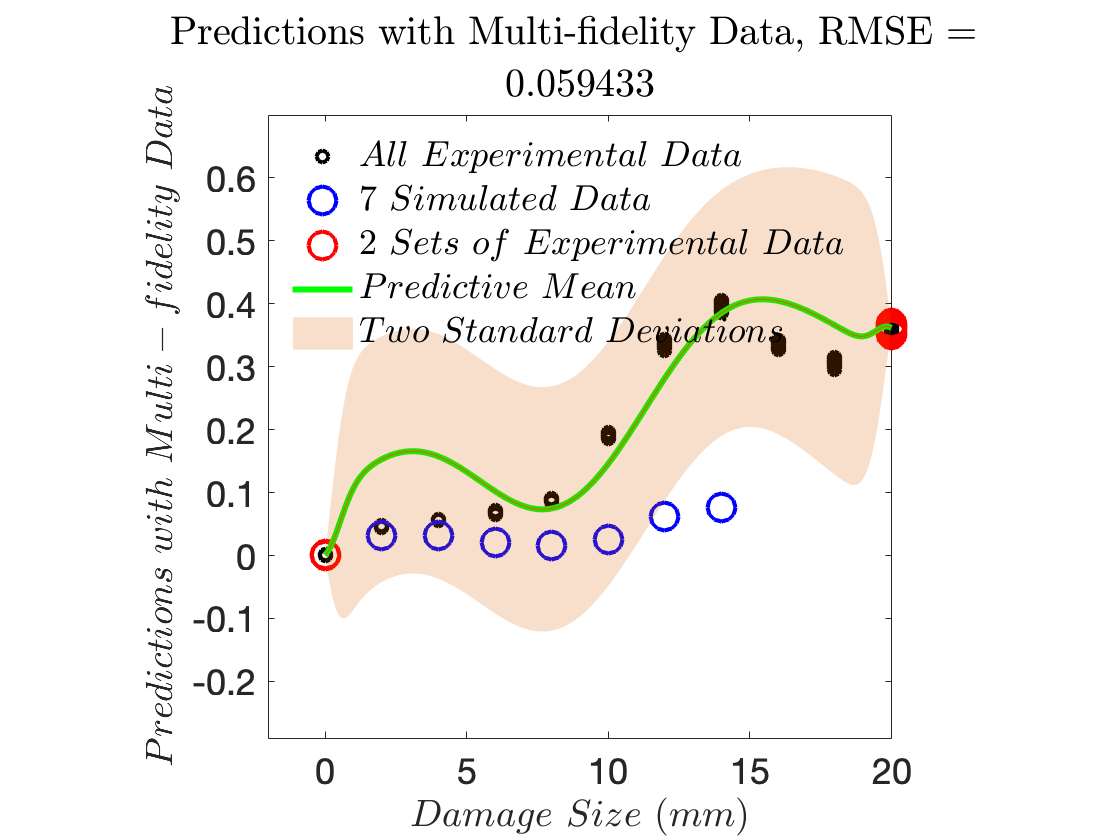}}
    \put(10,0){\includegraphics[width=0.48\textwidth]{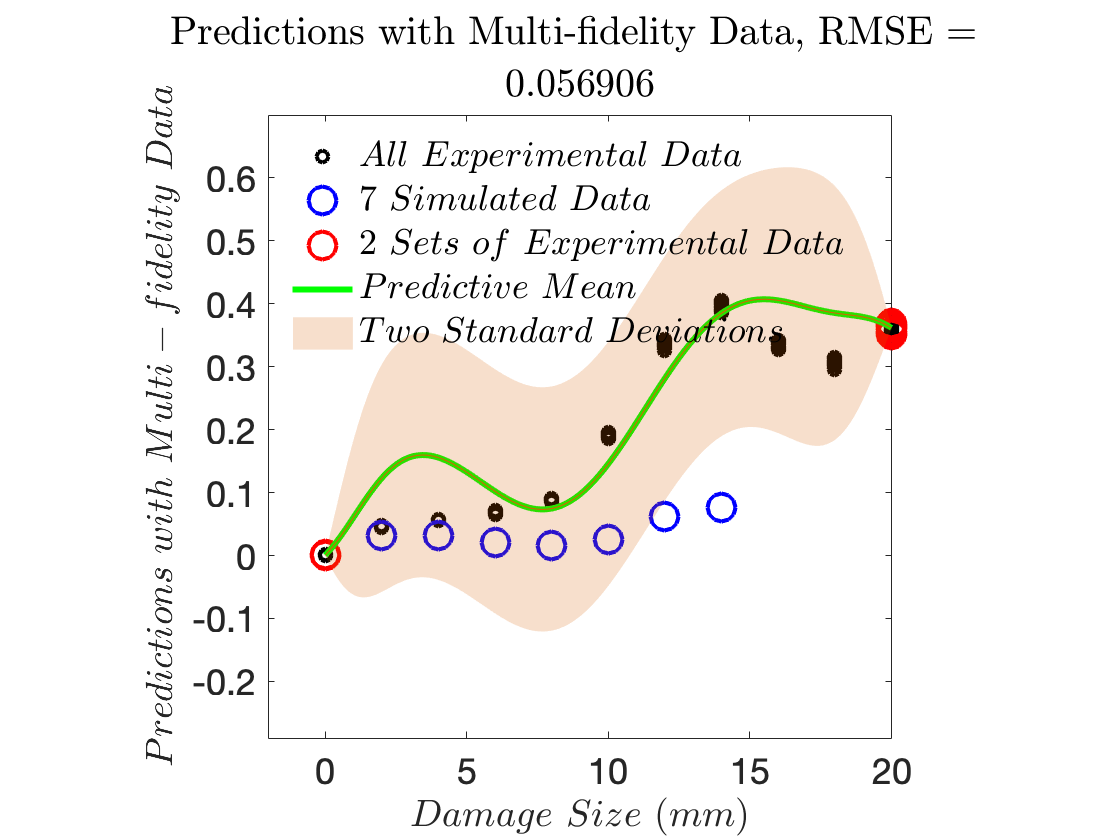}}
    \put(224,0){\includegraphics[width=0.48\textwidth]{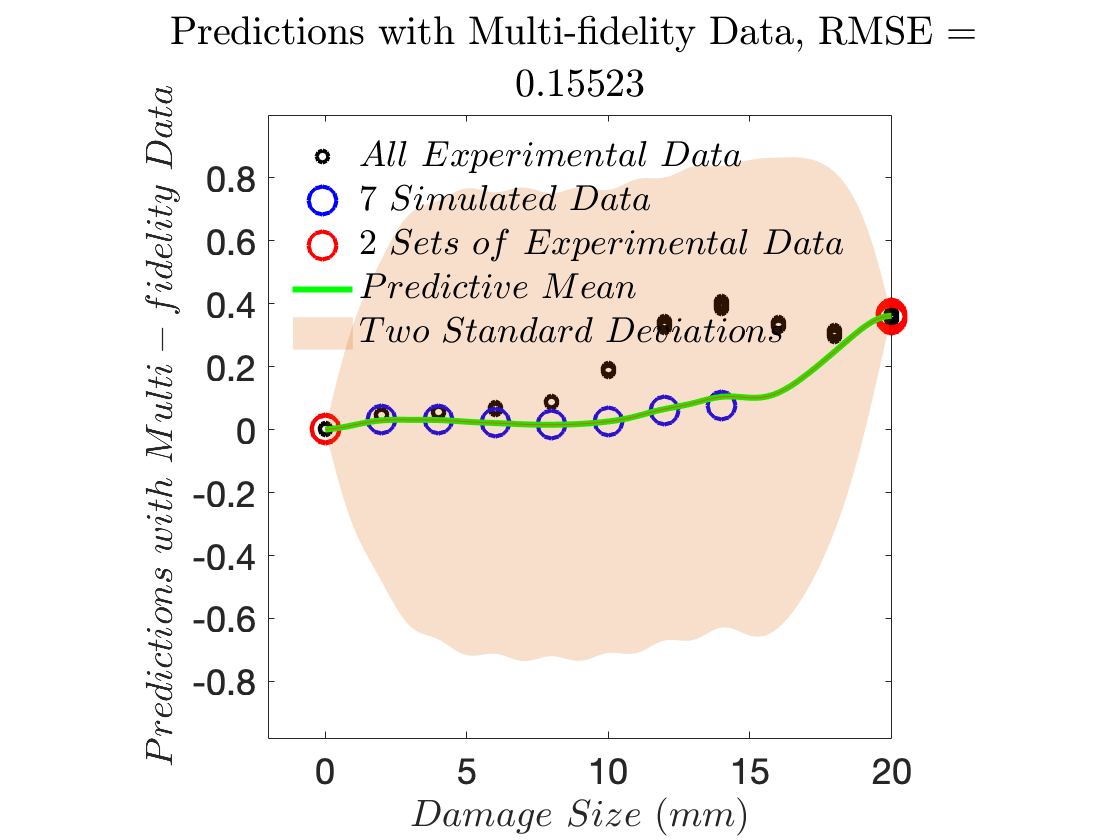}}
    \put(176,200){\color{black} \large {\fontfamily{phv}\selectfont \textbf{a}}}
    \put(390,200){\large {\fontfamily{phv}\selectfont \textbf{b}}}
   \put(176,34){\large {\fontfamily{phv}\selectfont \textbf{c}}} 
   \put(390,34){\large {\fontfamily{phv}\selectfont \textbf{d}}} 
    \end{picture} 
    \caption{Results of test case 1 task 1 for path 3-6 using different lower bound constrains during optimization. (a): using the largest variance of experimental data as lower bound; (b): using 5 times the largest variance of experimental data as lower bound; (a): using 10 times the largest variance of experimental data as lower bound; (a): using 15 times the largest variance of experimental data as lower bound.}
\label{fig:app15} 
\end{figure}
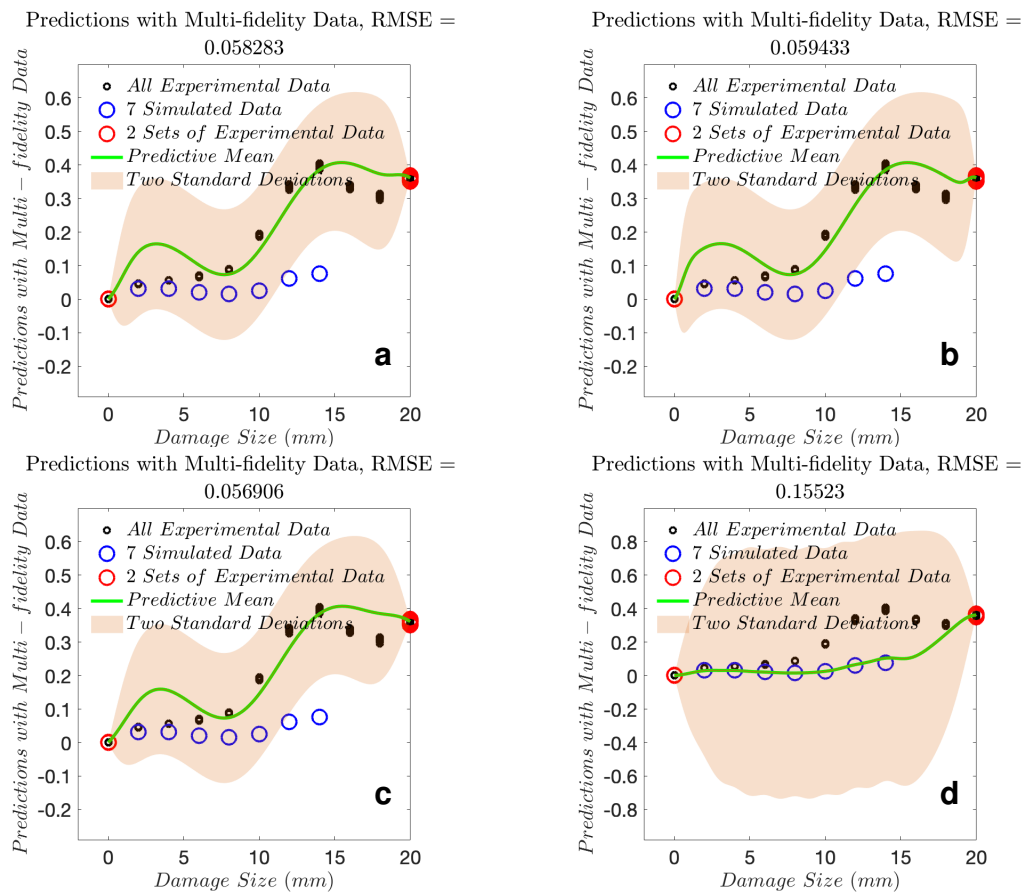

% \clearpage
% \vspace*{\fill}
\newpage
\subsection{Task2}

% \clearpage
% \mbox{}  % force non-empty page
% \newpage
% \subsection{Task2}
% extra for 2-6 task2
% extra at 4/6

\begin{figure}[H]
    % \centering
    \begin{picture}(500,330)
    \put(10,168){\includegraphics[width=0.48\textwidth]{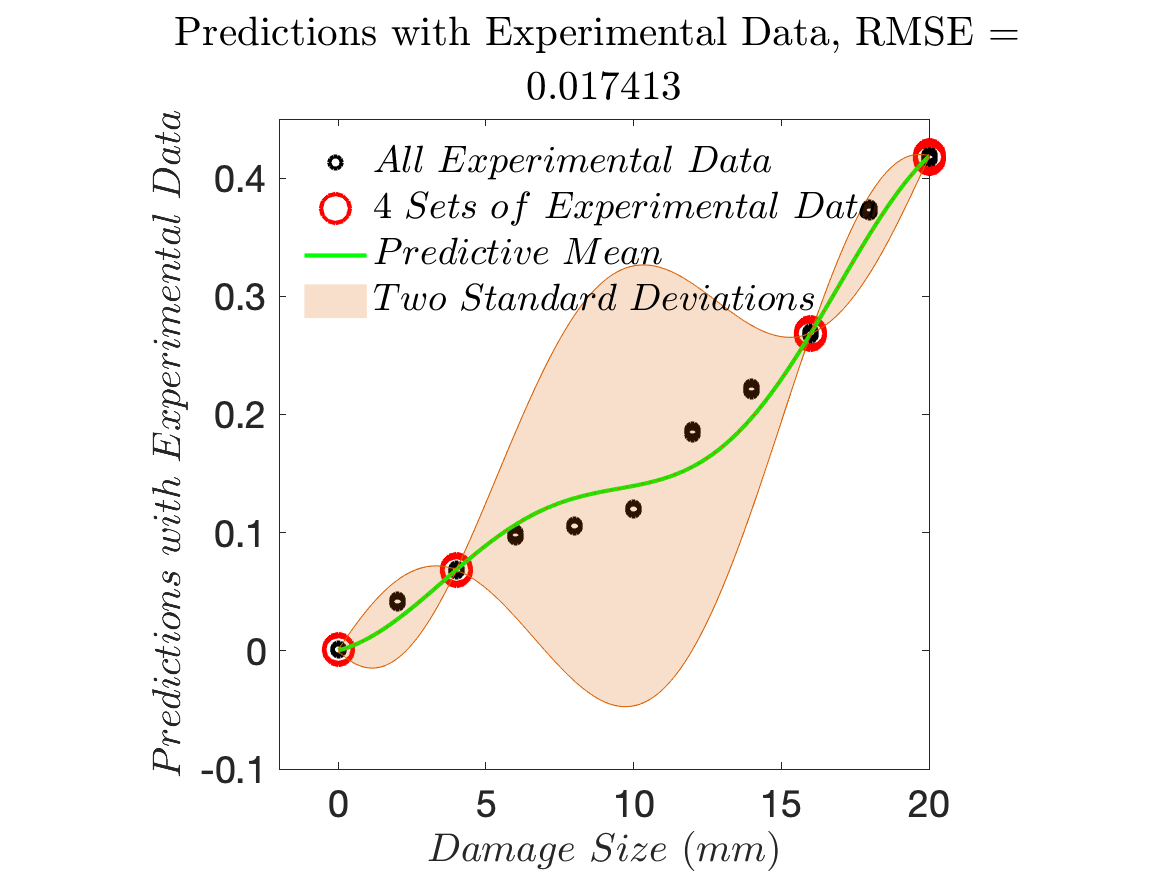}}
    \put(224,168){\includegraphics[width=0.48\textwidth]{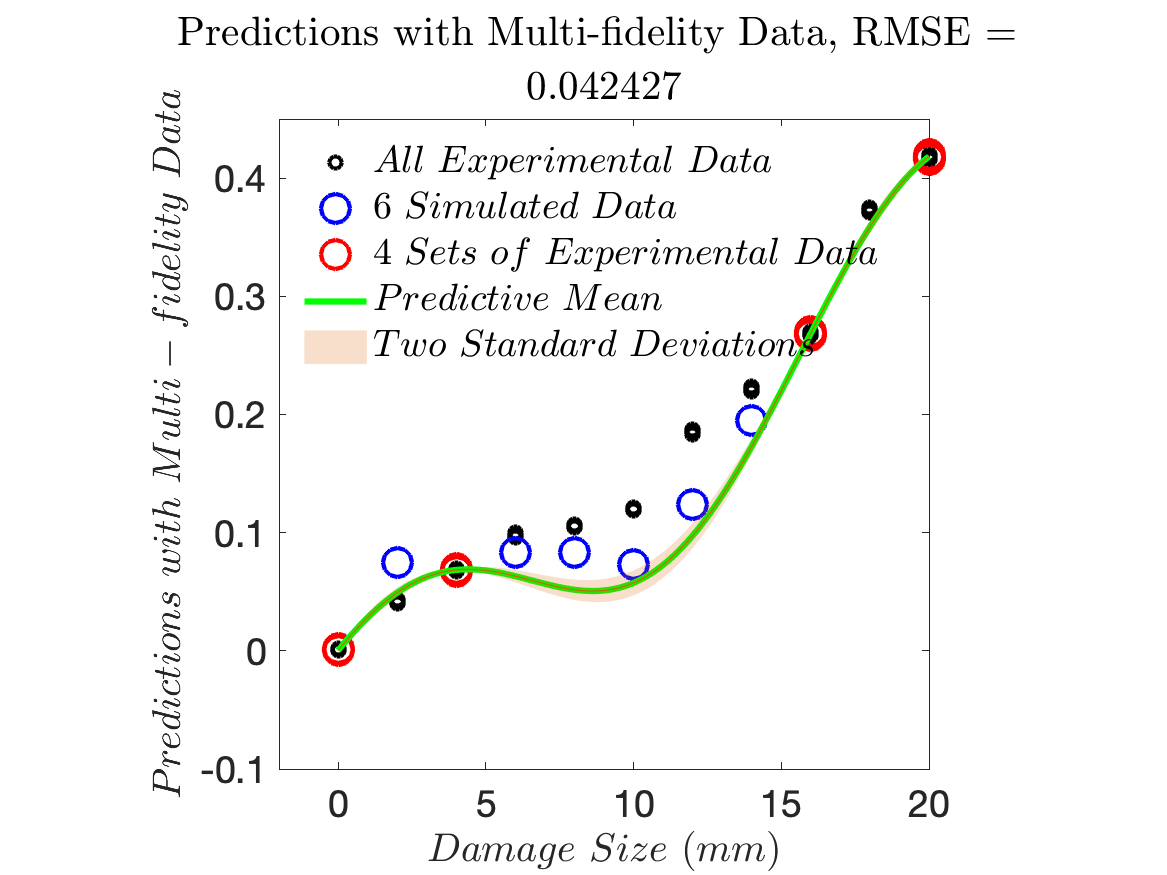}}
    \put(10,0){\includegraphics[width=0.48\textwidth]{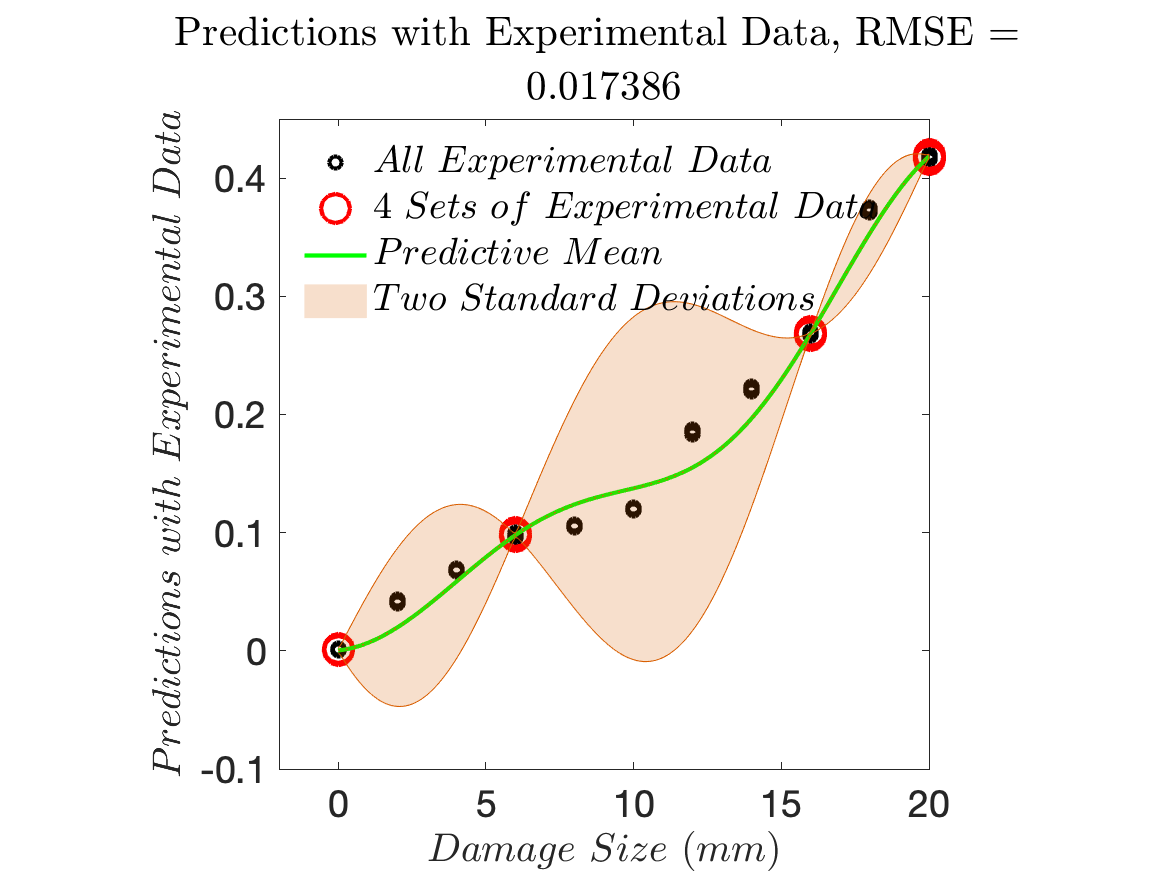}}
    \put(224,0){\includegraphics[width=0.48\textwidth]{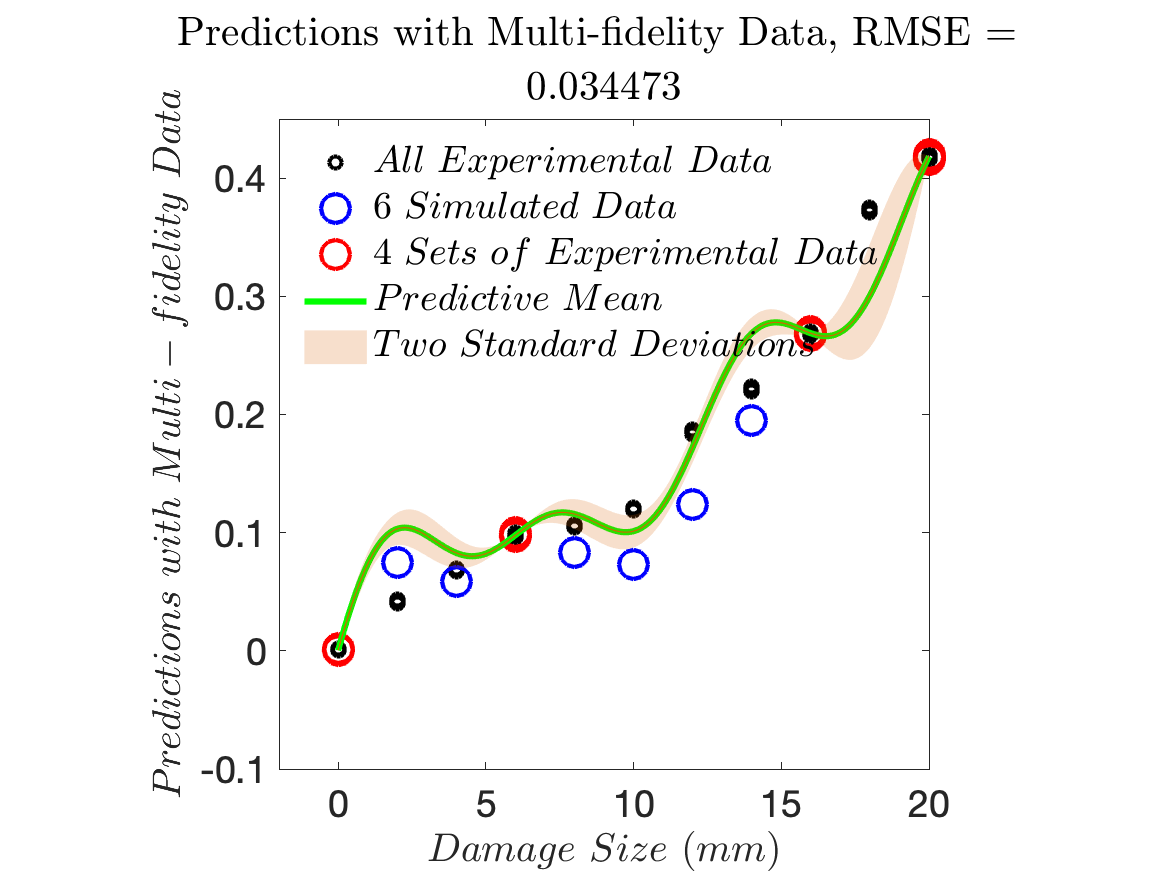}}
    \put(176,200){\color{black} \large {\fontfamily{phv}\selectfont \textbf{a}}}
    \put(380,200){\large {\fontfamily{phv}\selectfont \textbf{b}}}
   \put(176,34){\large {\fontfamily{phv}\selectfont \textbf{c}}} 
   \put(380,34){\large {\fontfamily{phv}\selectfont \textbf{d}}} 
    \end{picture} 
    \caption{DI regression for path 2-6 from GPRM and multi-fidelity GPRM: (a) prediction  using 4 experimental sets at 0, 4, 16 and 20 mm; (b) prediction  using 4 experimental sets and 6 simulated data; (c) prediction  using 4 experimental sets at 0, 6, 16 and 20 mm; (d) prediction  using 4 experimental sets and 6 simulated data.}
\label{fig:test2_extra1} 
\end{figure}
%
% extra at 8/12
\begin{figure}[t]
    % \centering
    \begin{picture}(500,330)
    \put(10,168){\includegraphics[width=0.48\textwidth]{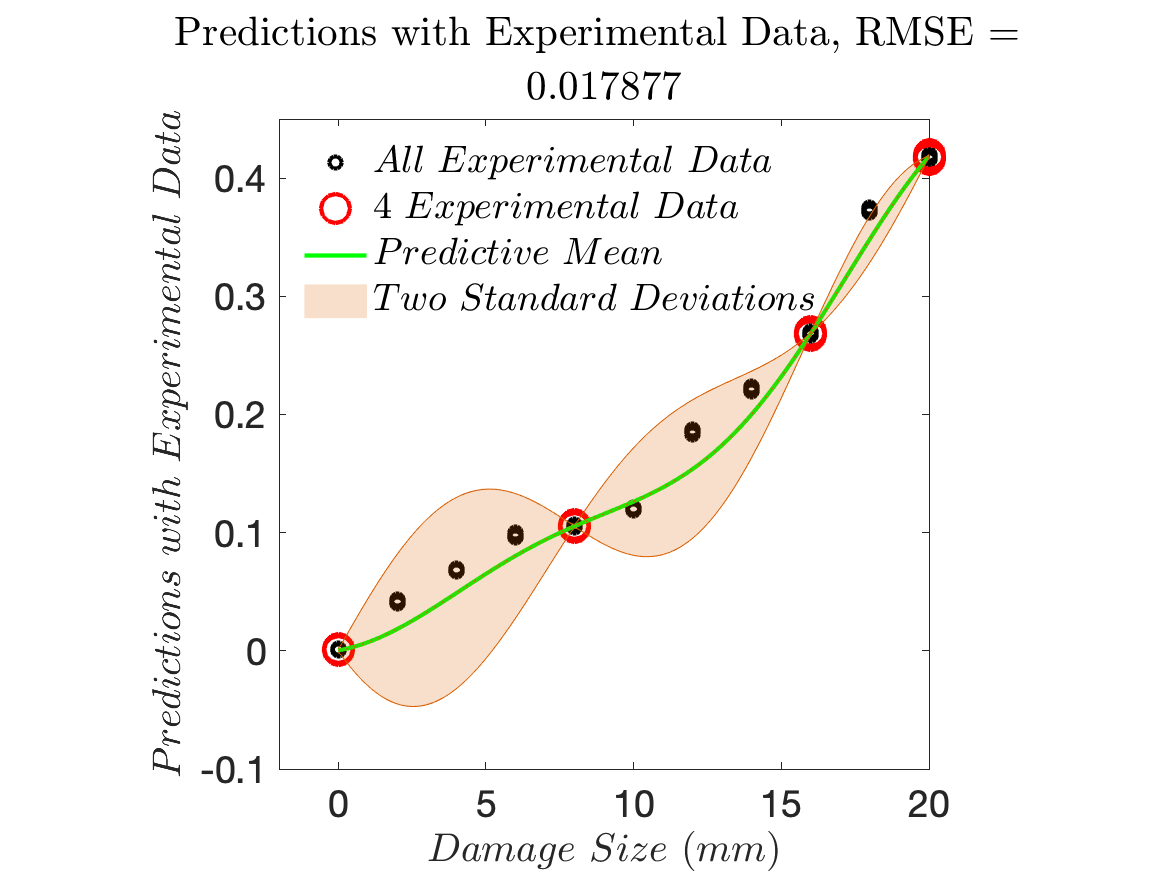}}
    \put(224,168){\includegraphics[width=0.48\textwidth]{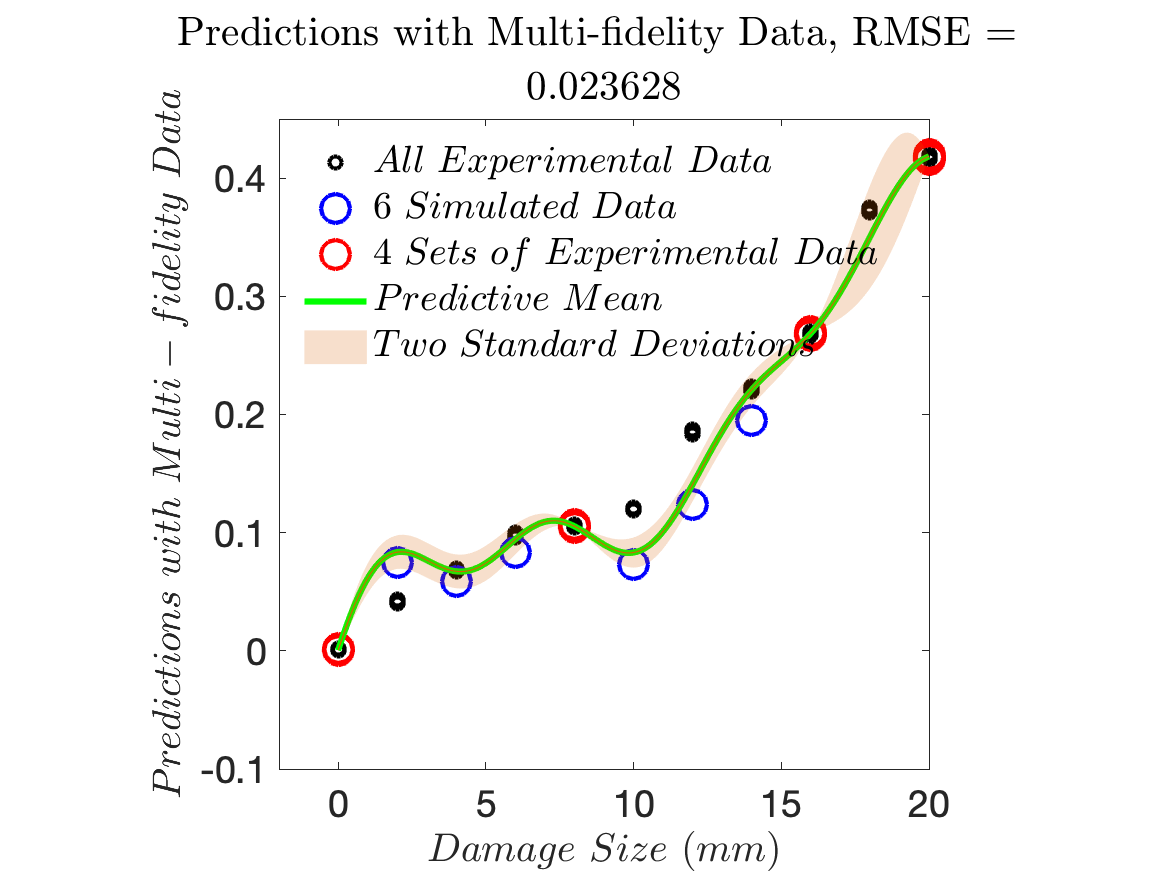}}
    \put(10,0){\includegraphics[width=0.48\textwidth]{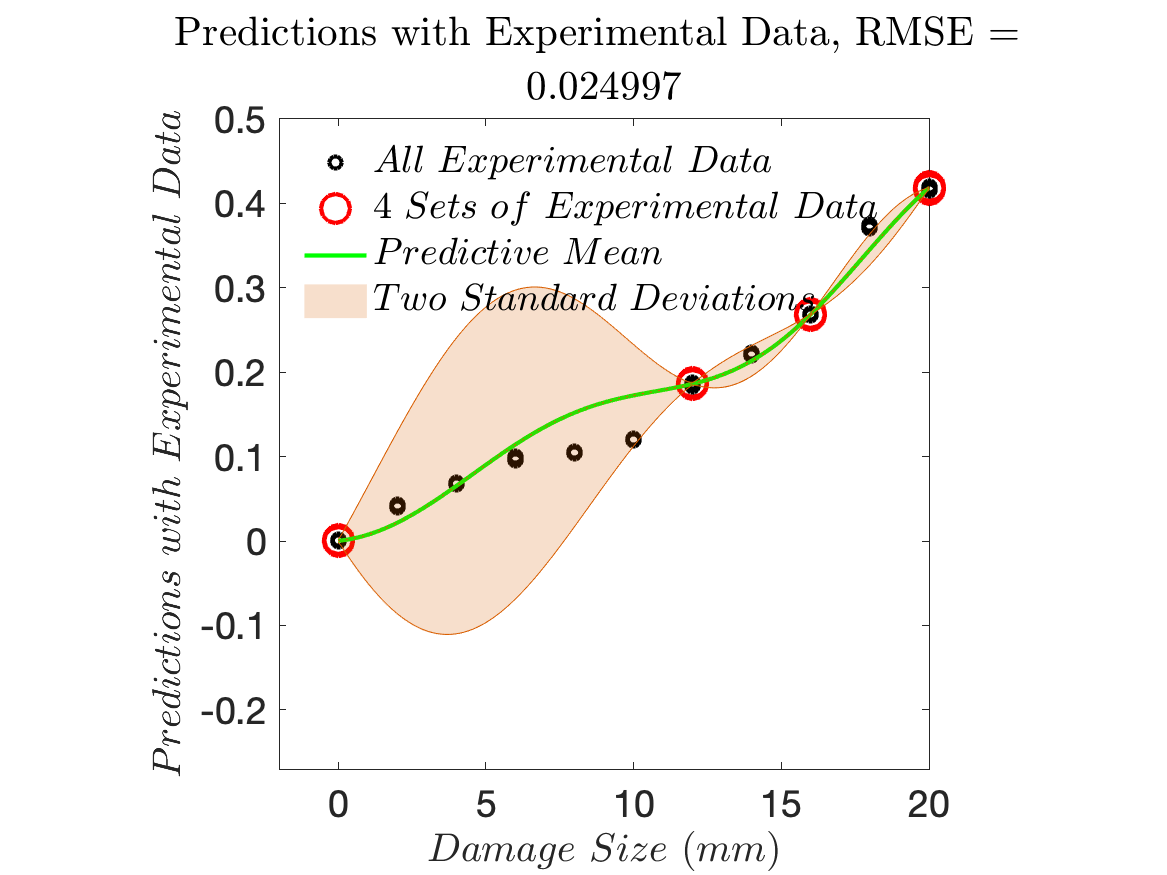}}
    \put(224,0){\includegraphics[width=0.48\textwidth]{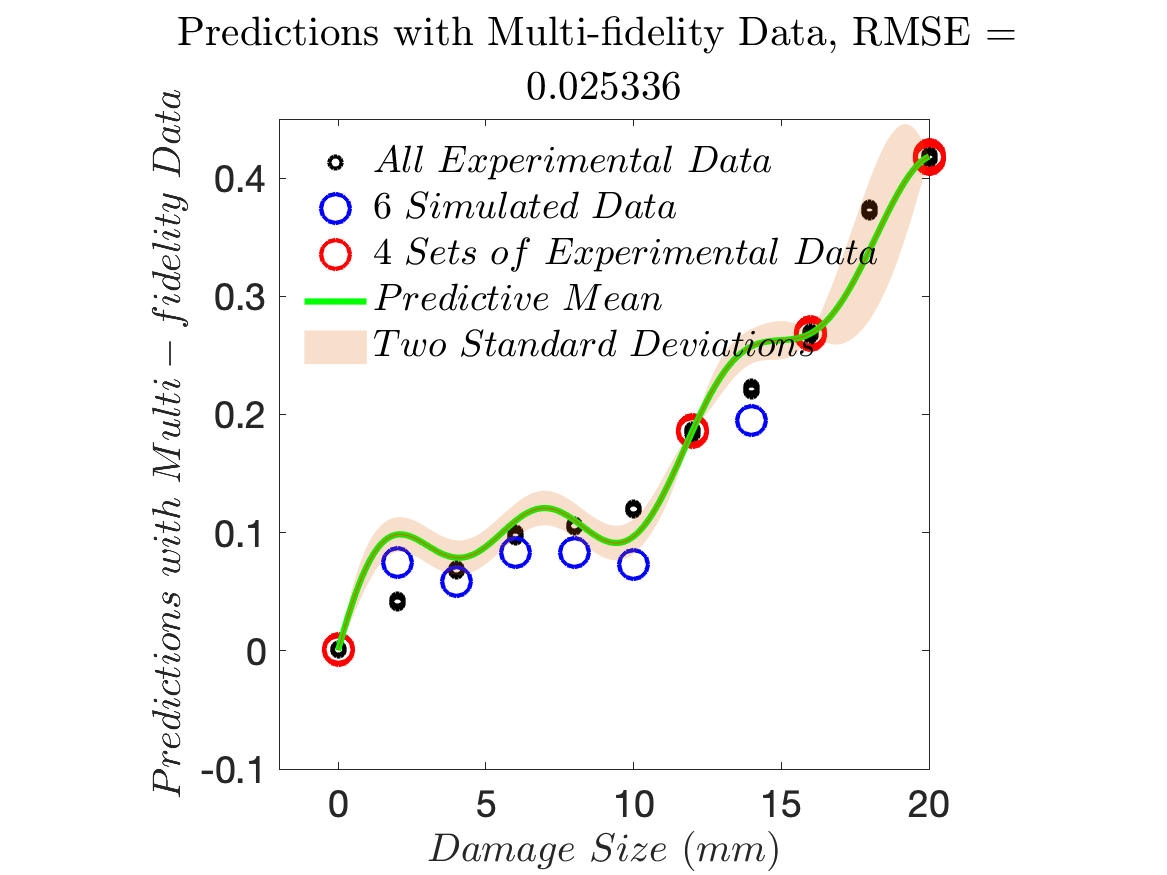}}
    \put(176,200){\color{black} \large {\fontfamily{phv}\selectfont \textbf{a}}}
    \put(380,200){\large {\fontfamily{phv}\selectfont \textbf{b}}}
   \put(176,34){\large {\fontfamily{phv}\selectfont \textbf{c}}} 
   \put(380,34){\large {\fontfamily{phv}\selectfont \textbf{d}}} 
    \end{picture}
    \caption{DI regression for path 2-6 from GPRM and multi-fidelity GPRM: (a) prediction  using 5 experimental sets at 0, 8, 16 and 20 mm; (b) prediction  using 4 experimental sets and 6 simulated data; (c) prediction  using 4 experimental sets at 0, 12, 16 and 20 mm; (d) prediction  using 4 experimental sets and 6 simulated data.}
\label{fig:test2_extra2} 
\end{figure}
%
% extra ends
%
%
\begin{figure}[t!]
    % \centering
    \begin{picture}(500,330)
    \put(10,168){\includegraphics[width=0.48\textwidth]{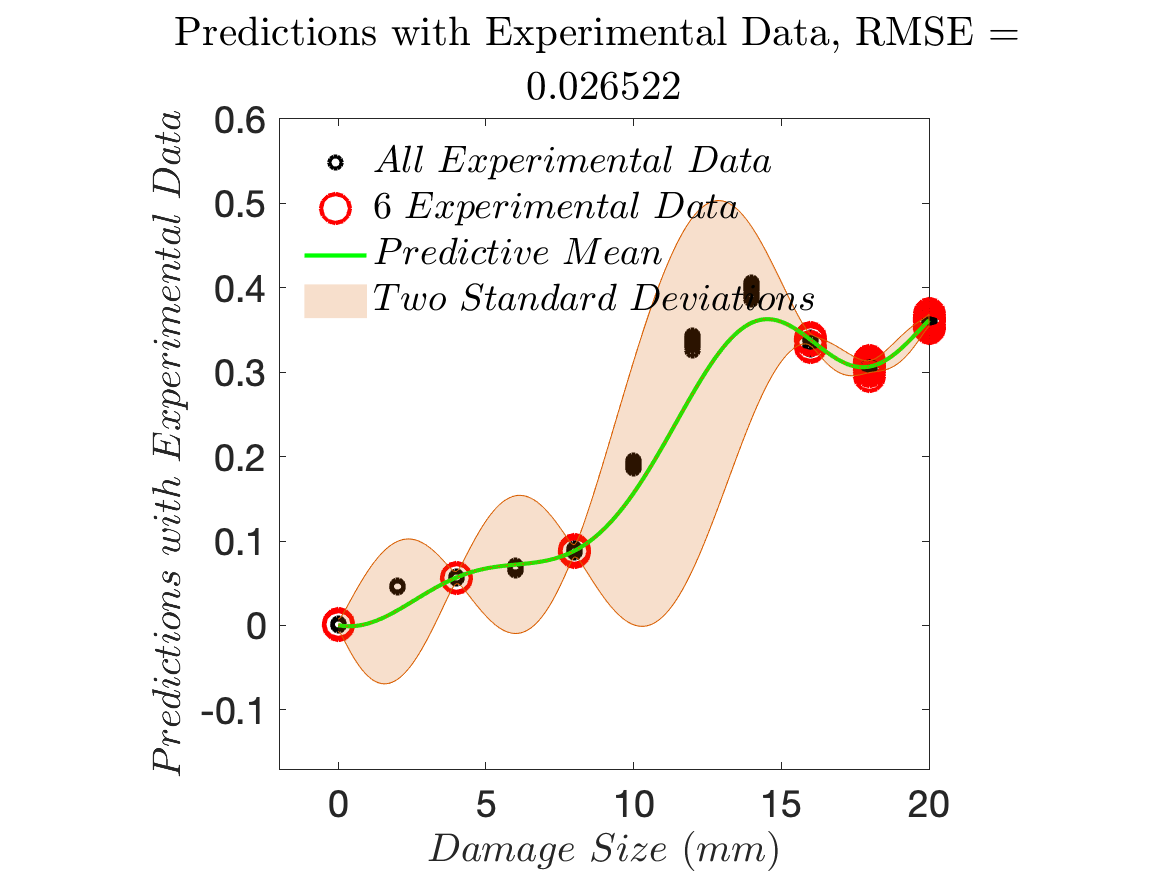}}
    \put(224,168){\includegraphics[width=0.48\textwidth]{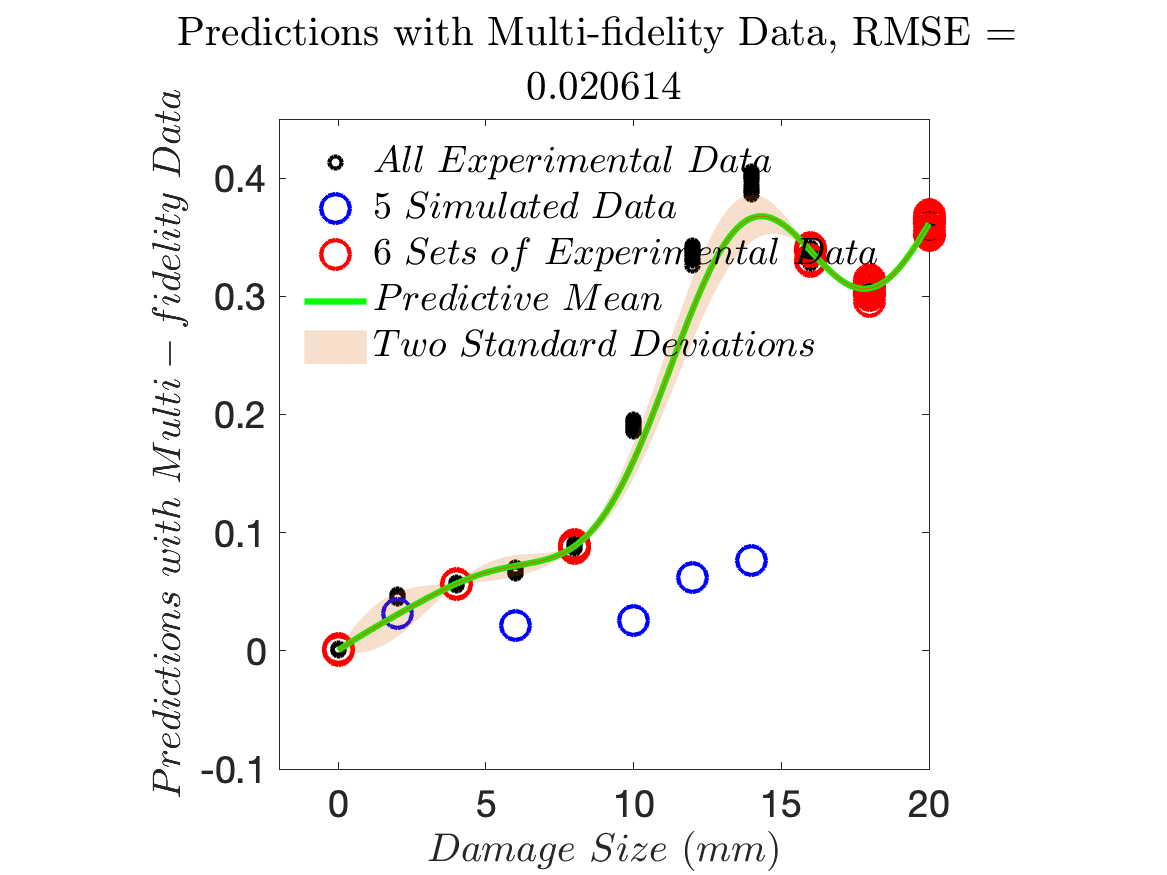}}
    \put(10,0){\includegraphics[width=0.48\textwidth]{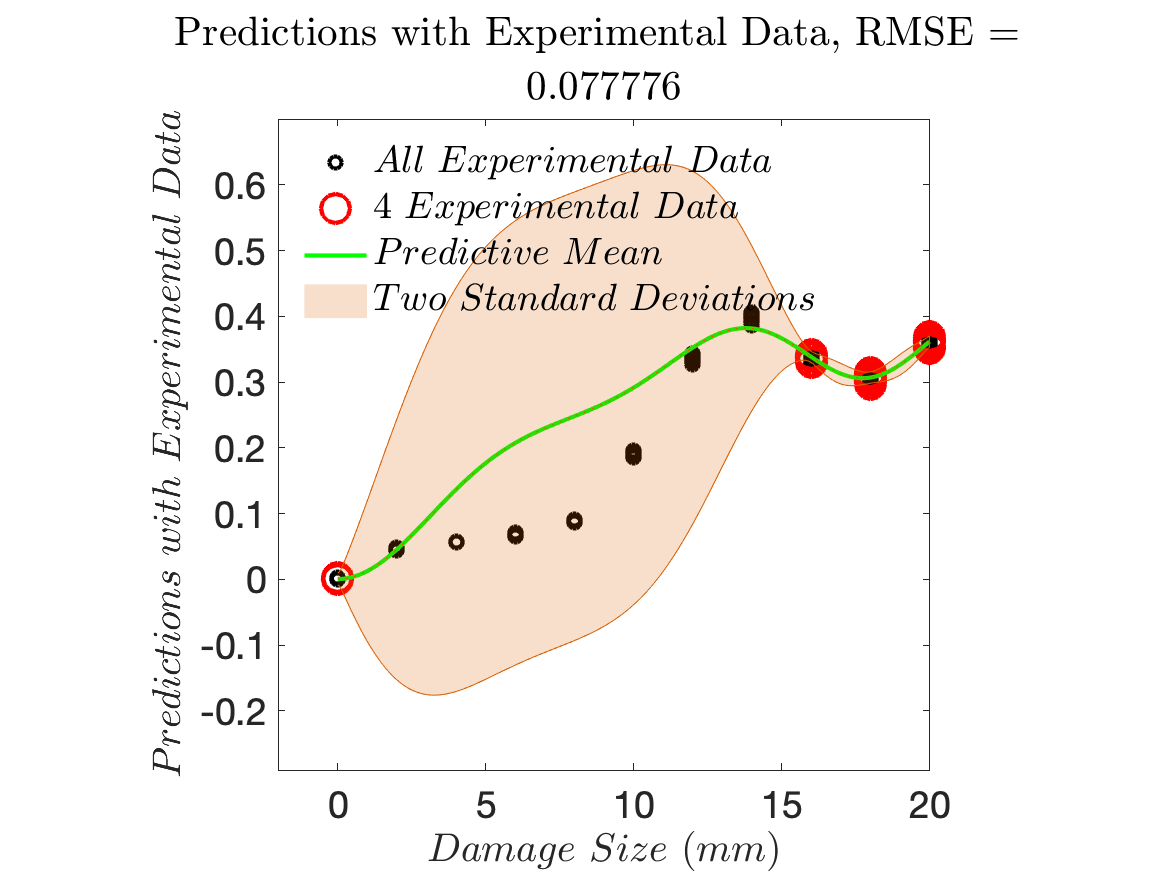}}
    \put(224,0){\includegraphics[width=0.48\textwidth]{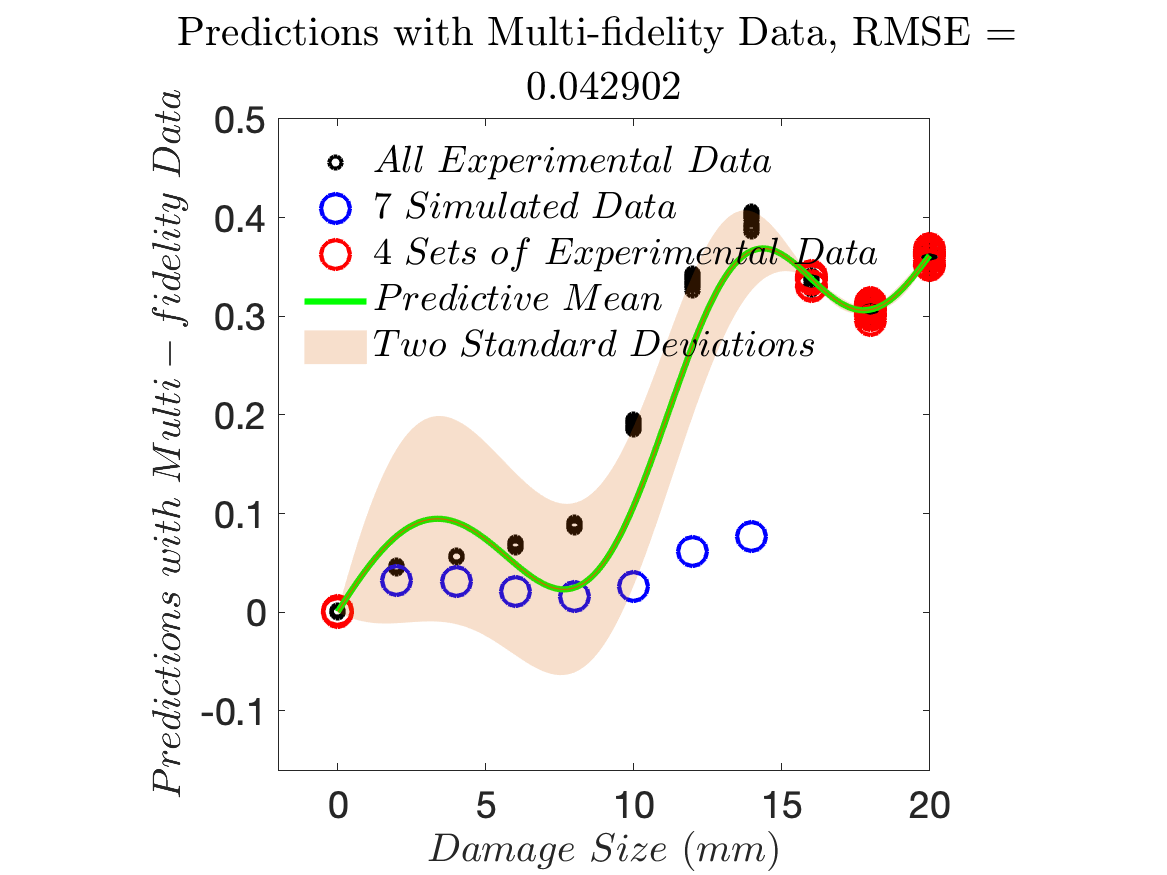}}
    \put(176,200){\color{black} \large {\fontfamily{phv}\selectfont \textbf{a}}}
    \put(390,200){\large {\fontfamily{phv}\selectfont \textbf{b}}}
   \put(176,34){\large {\fontfamily{phv}\selectfont \textbf{c}}} 
   \put(390,34){\large {\fontfamily{phv}\selectfont \textbf{d}}} 
    \end{picture} 
    \caption{DI regression for path 3-6 from GPRM and multi-fidelity GPRM: (a) prediction  using 6 experimental sets; (b) prediction  using 6 experimental sets and 5 simulated data; (c) prediction  using 4 experimental sets; (d) prediction  using 4 experimental sets and 7 simulated data.}
\label{fig:app14} 
\end{figure}
\begin{figure}[t!]
\hspace{-10pt}\includegraphics[width=0.54\textwidth]{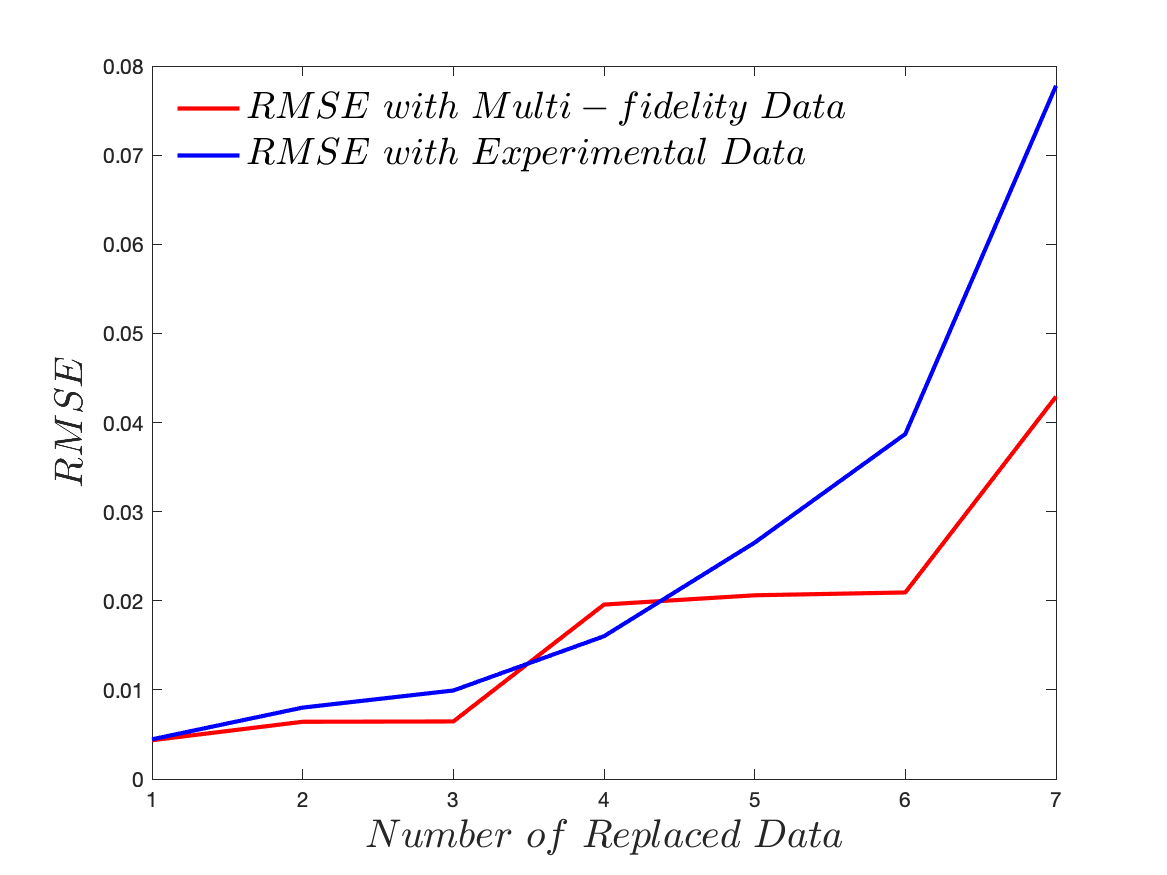}\hspace{-0.5cm}\includegraphics[width=0.54\textwidth]{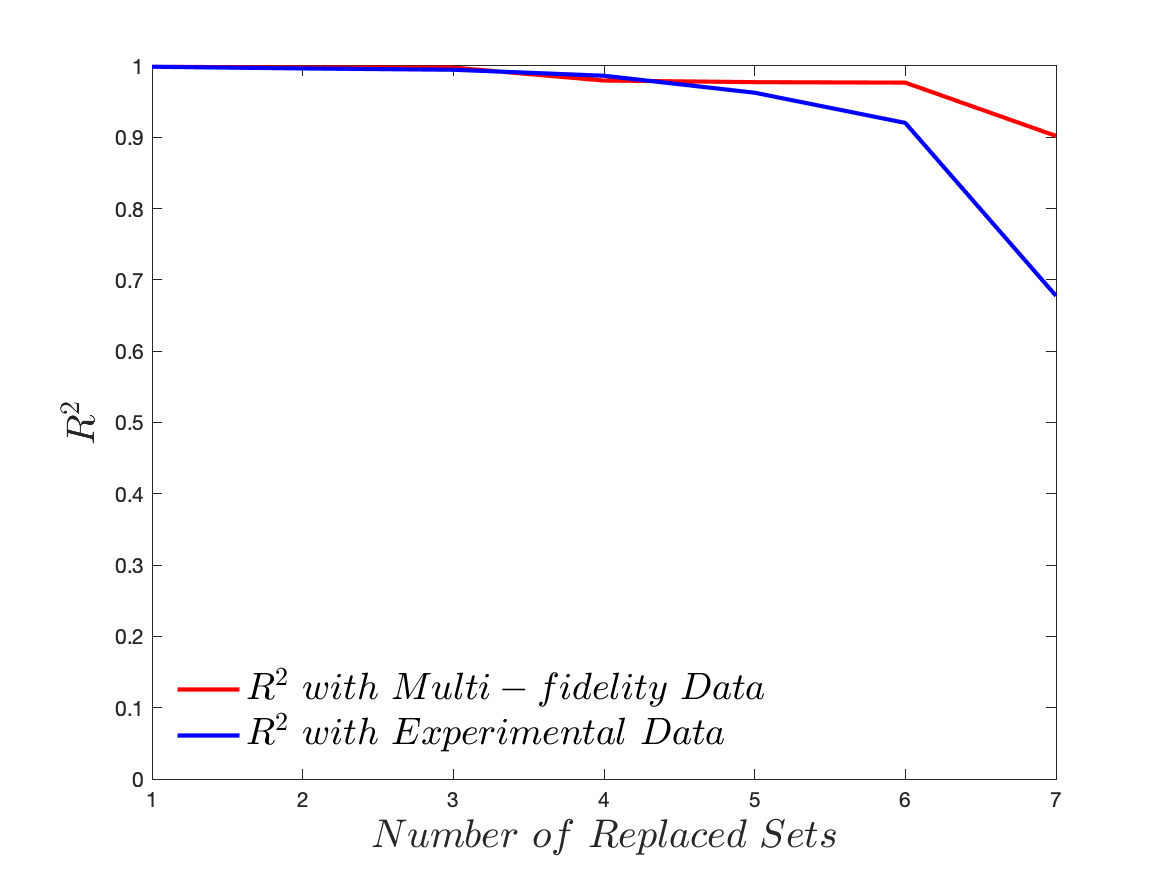}

\caption{RMSE and ${R}^2$ comparison between GPRM and multi-fidelity GPRM for path 3-6: the left panel is the RMSE comparison for path 3-6; the right panel is ${R}^2$ comparison for path 3-6.} 
% \label{fig:mean and std of nets}
\label{fig:app15} 
\end{figure}
% \vspace{+8pt}
%
% \subsection{Task2}
%% R2 starts

%%

\clearpage
% \subsection{GP}
%% GP starts
%%
% 2-5

%
%% GP ends
% \section{Additional results}
\section{Second Test Case}
\subsection{Task1}
% task 1
% h=2
\begin{figure}[H]
    % \centering
    \begin{picture}(500,330)
    \put(10,168){\includegraphics[width=0.48\textwidth]{Figures/2ndtest/case2/c2t1_h2_exp_num_=_2_copy.png}}
    \put(224,168){\includegraphics[width=0.48\textwidth]{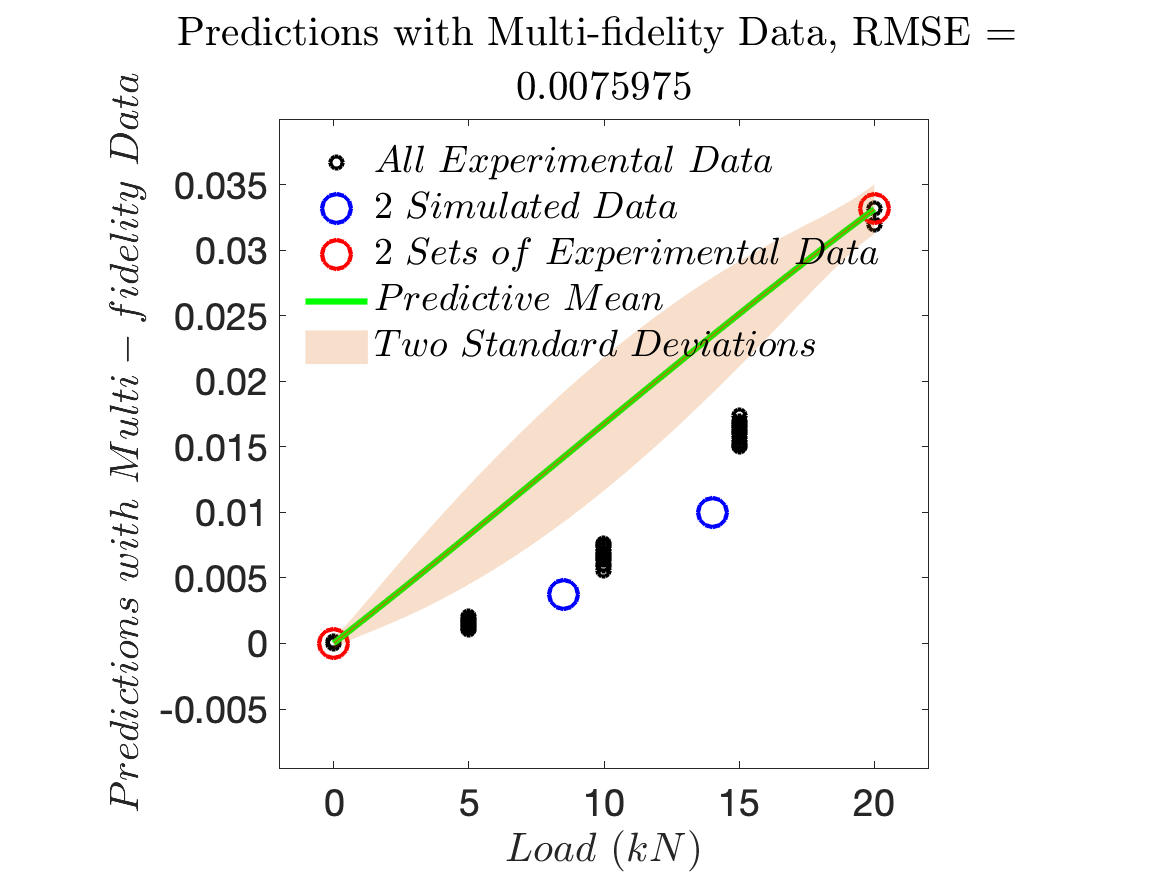}}
    \put(10,0){\includegraphics[width=0.48\textwidth]{Figures/2ndtest/case2/c2t1_h2_itr_is_5,exp_2_2var_copy.png}}
    \put(224,0){\includegraphics[width=0.48\textwidth]{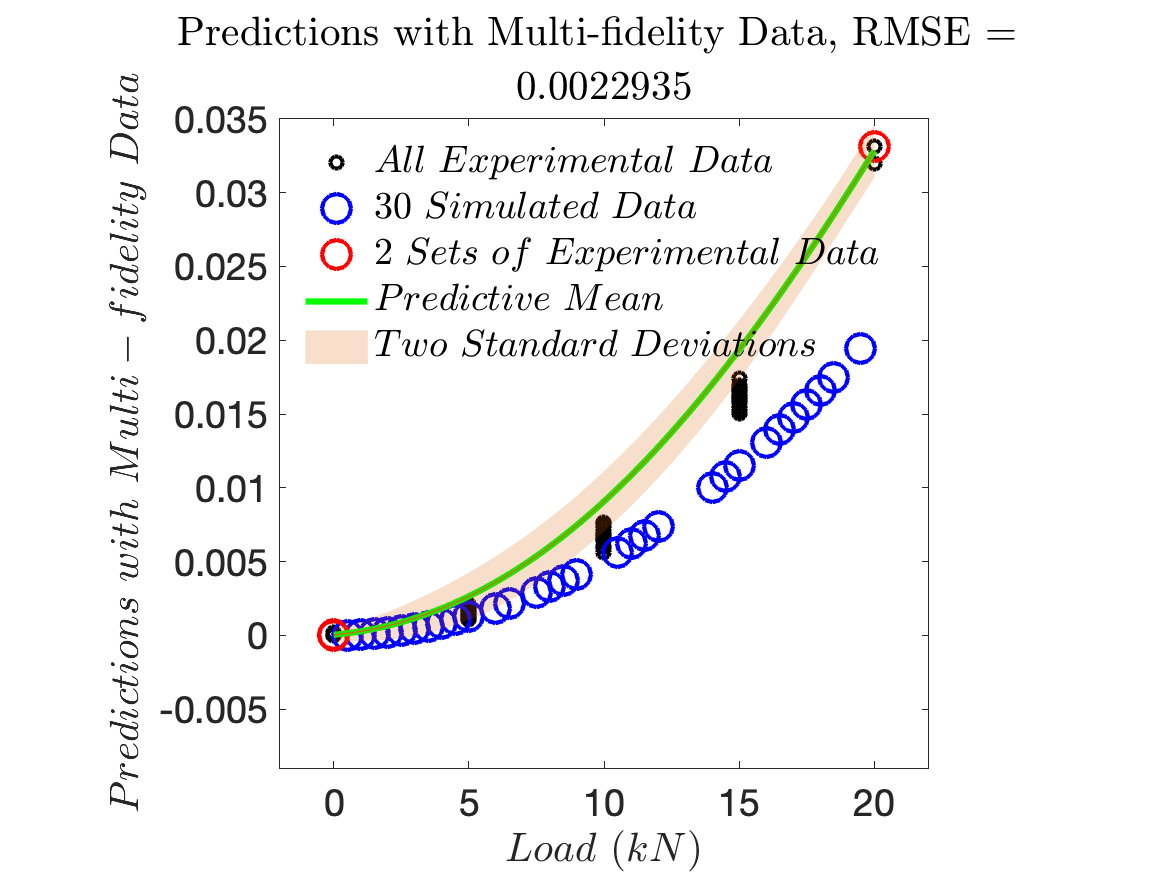}}
    \put(170,200){\color{black} \large {\fontfamily{phv}\selectfont \textbf{a}}}
    \put(380,200){\large {\fontfamily{phv}\selectfont \textbf{b}}}
   \put(170,30){\large {\fontfamily{phv}\selectfont \textbf{c}}} 
   \put(390,30){\large {\fontfamily{phv}\selectfont \textbf{d}}} 
    \end{picture}
    \caption{DI regression for path 1-6 from GPRM and multi-fidelity GPRM with batch learning using random selection: (a) prediction  using 2 experimental sets at 0 and 20 mm; (b) prediction  using 2 experimental sets and 2 random selected simulated data points; (c) prediction  using 2 experimental sets and 5 random selected simulated data points; (d) prediction  using 2 experimental sets and 30 random selected simulated data points.}
\label{fig:fix_mse} 
\end{figure}
%
%h=3
\begin{figure}[h!]
    % \centering
    \begin{picture}(500,330)
    \put(10,168){\includegraphics[width=0.48\textwidth]{Figures/2ndtest/case2/c2t1_h3_exp_num_=_3_copy.png}}
    \put(224,168){\includegraphics[width=0.48\textwidth]{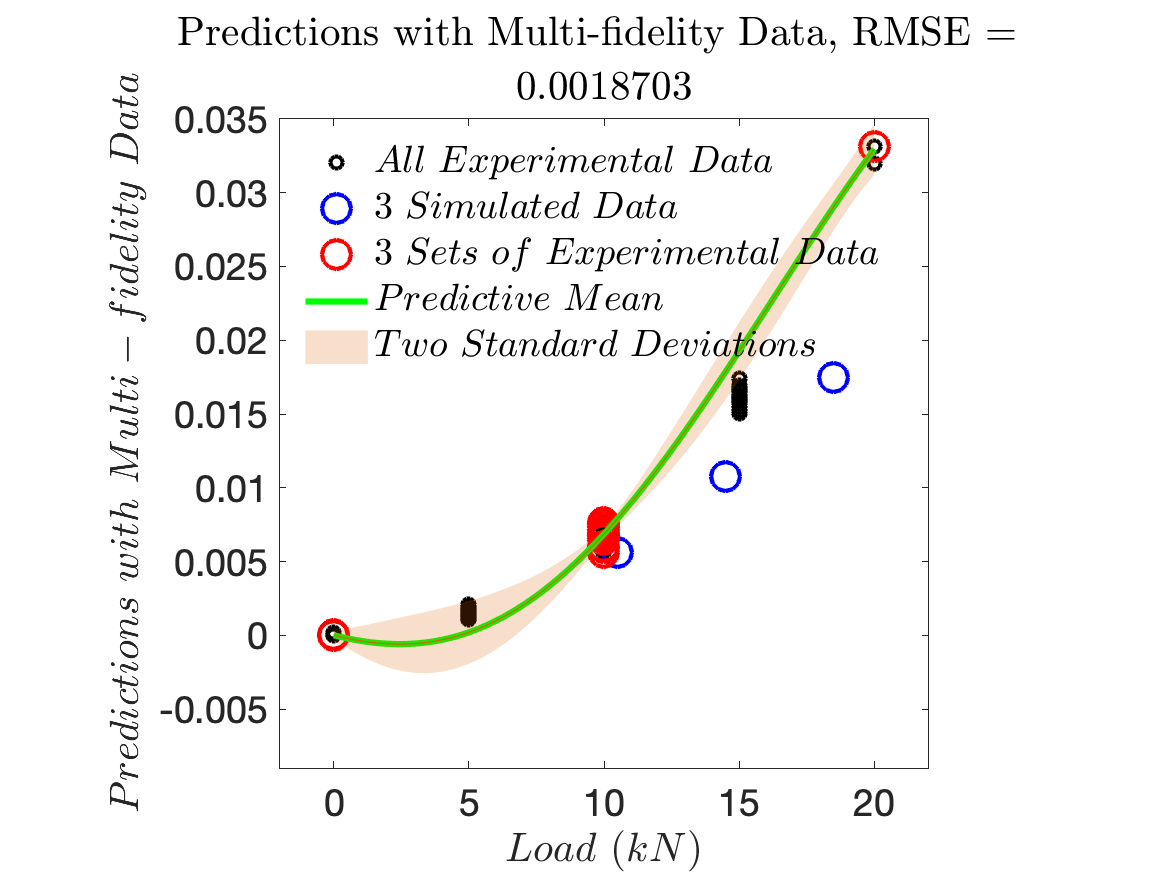}}
    \put(10,0){\includegraphics[width=0.48\textwidth]{Figures/2ndtest/case2/c2t1_h3_itr_is_5,exp_3_2var_copy.png}}
    \put(224,0){\includegraphics[width=0.48\textwidth]{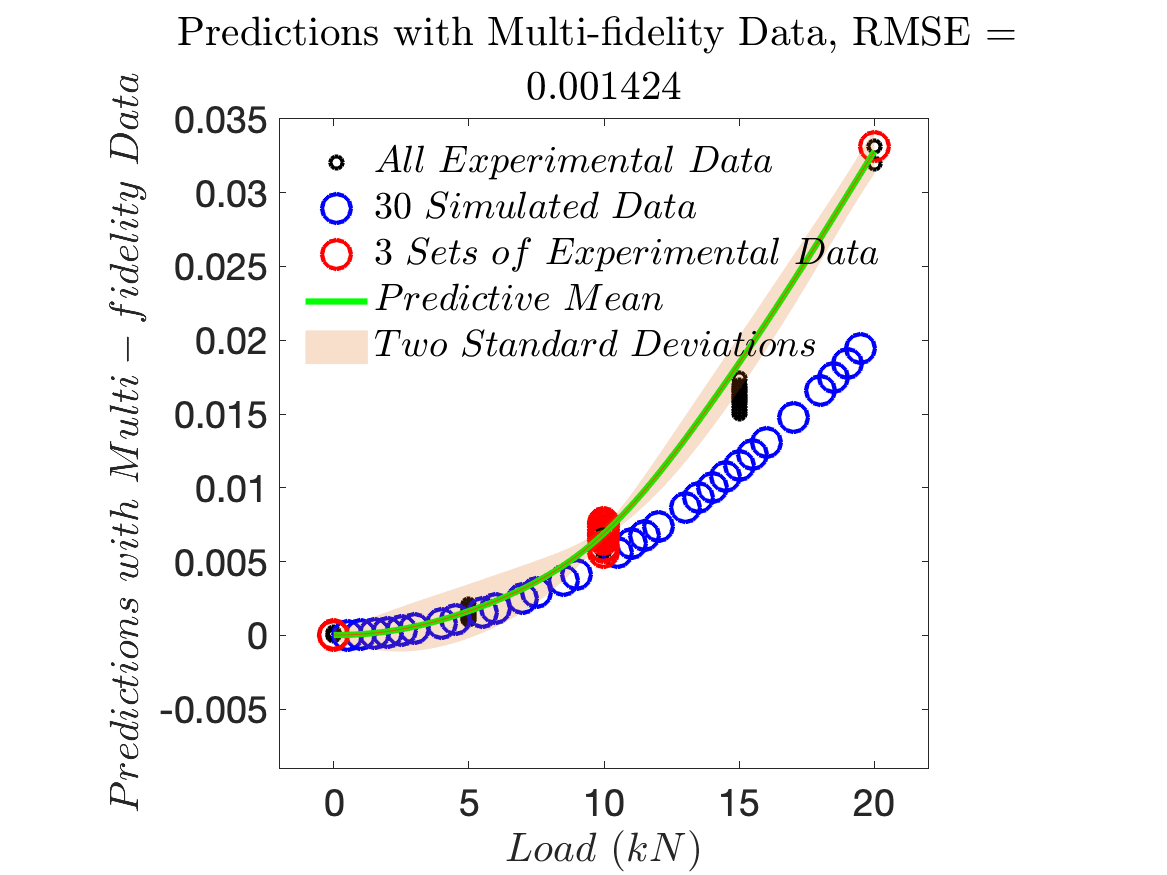}}
    \put(170,200){\color{black} \large {\fontfamily{phv}\selectfont \textbf{a}}}
    \put(380,200){\large {\fontfamily{phv}\selectfont \textbf{b}}}
   \put(170,30){\large {\fontfamily{phv}\selectfont \textbf{c}}} 
   \put(380,30){\large {\fontfamily{phv}\selectfont \textbf{d}}} 
    \end{picture} 
    \caption{DI regression for path 1-6 from GPRM and multi-fidelity GPRM with batch learning using random selection: (a) prediction  using 3 experimental sets at 0, 10 and 20 mm; (b) prediction  using 3 experimental sets and 3 random selected simulated data points; (c) prediction  using 3 experimental sets and 5 random selected simulated data points; (d) prediction  using 3 experimental sets and 30 random selected simulated data points.}
\label{fig:fix_mse} 
\end{figure}
%h=4
\begin{figure}[h!]
    % \centering
    \begin{picture}(500,330)
    \put(10,168){\includegraphics[width=0.48\textwidth]{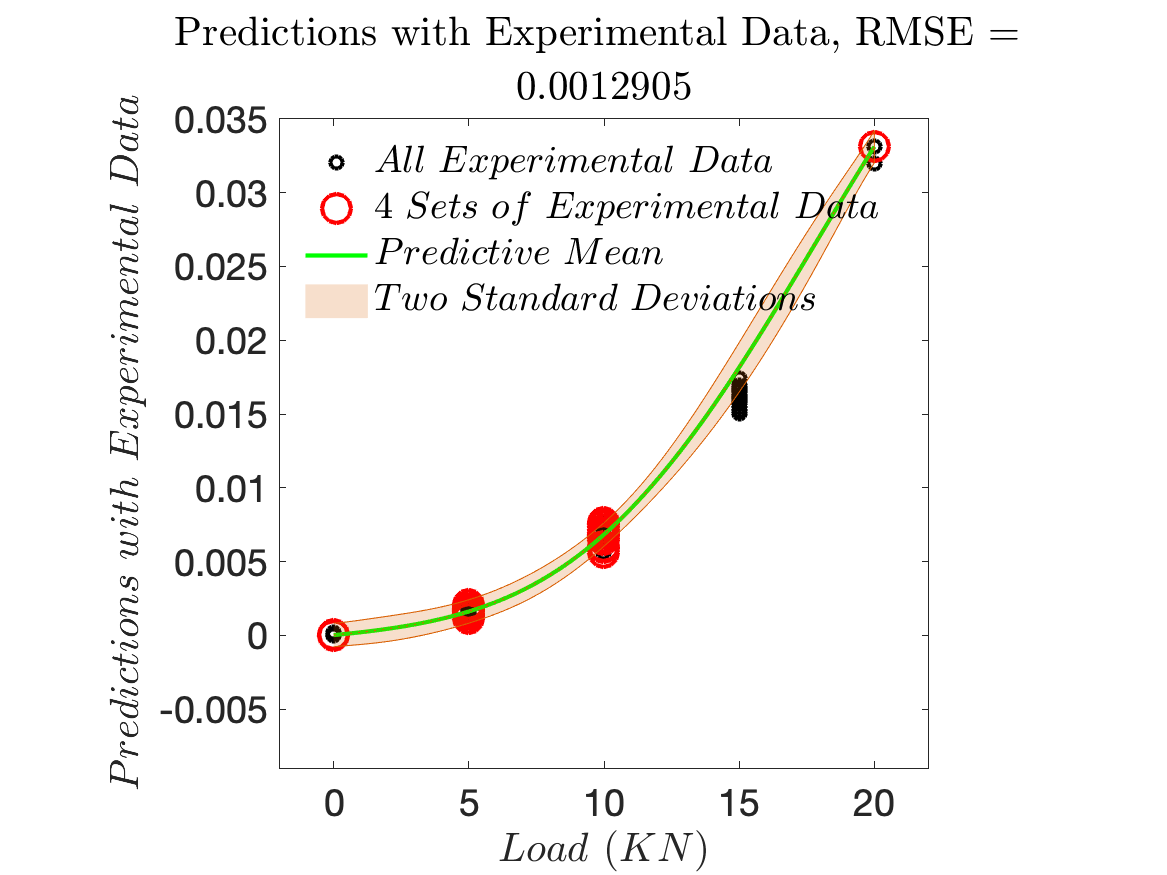}}
    \put(224,168){\includegraphics[width=0.48\textwidth]{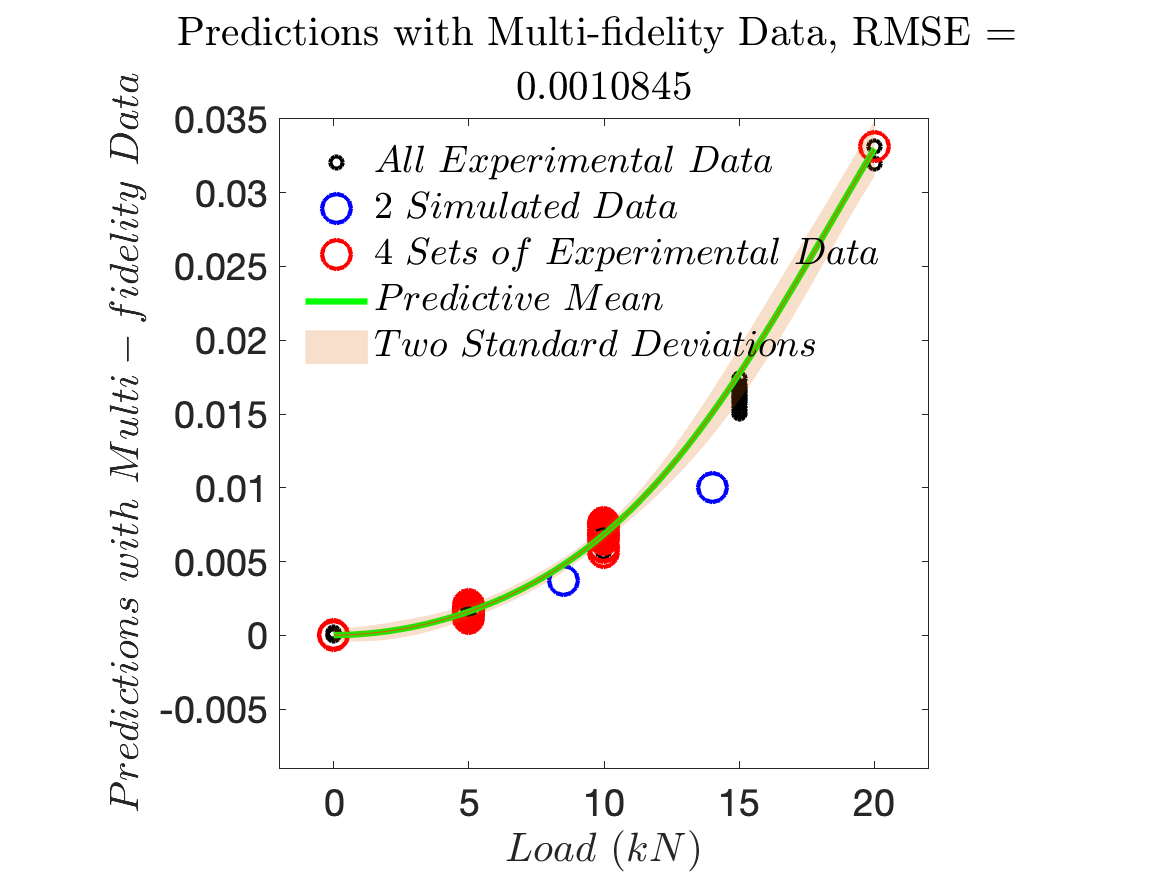}}
    \put(10,0){\includegraphics[width=0.48\textwidth]{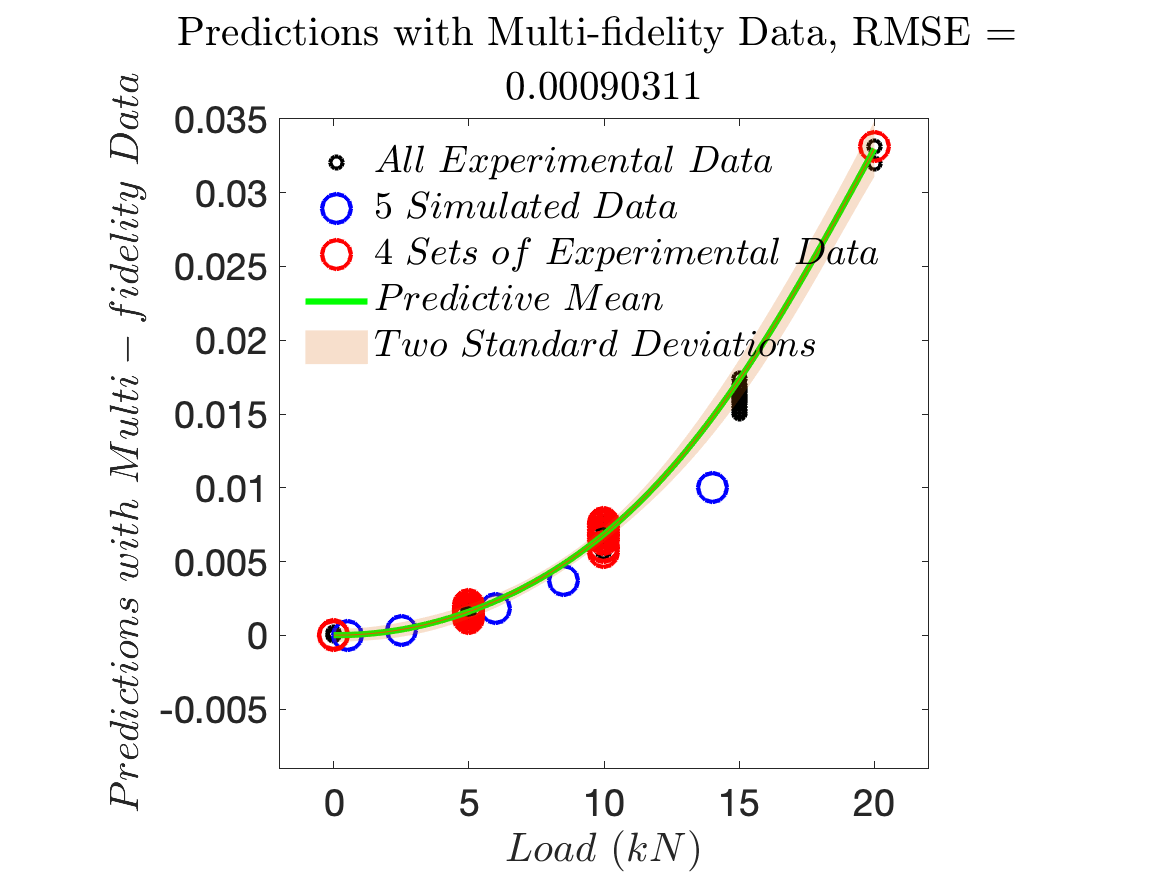}}
    \put(224,0){\includegraphics[width=0.48\textwidth]{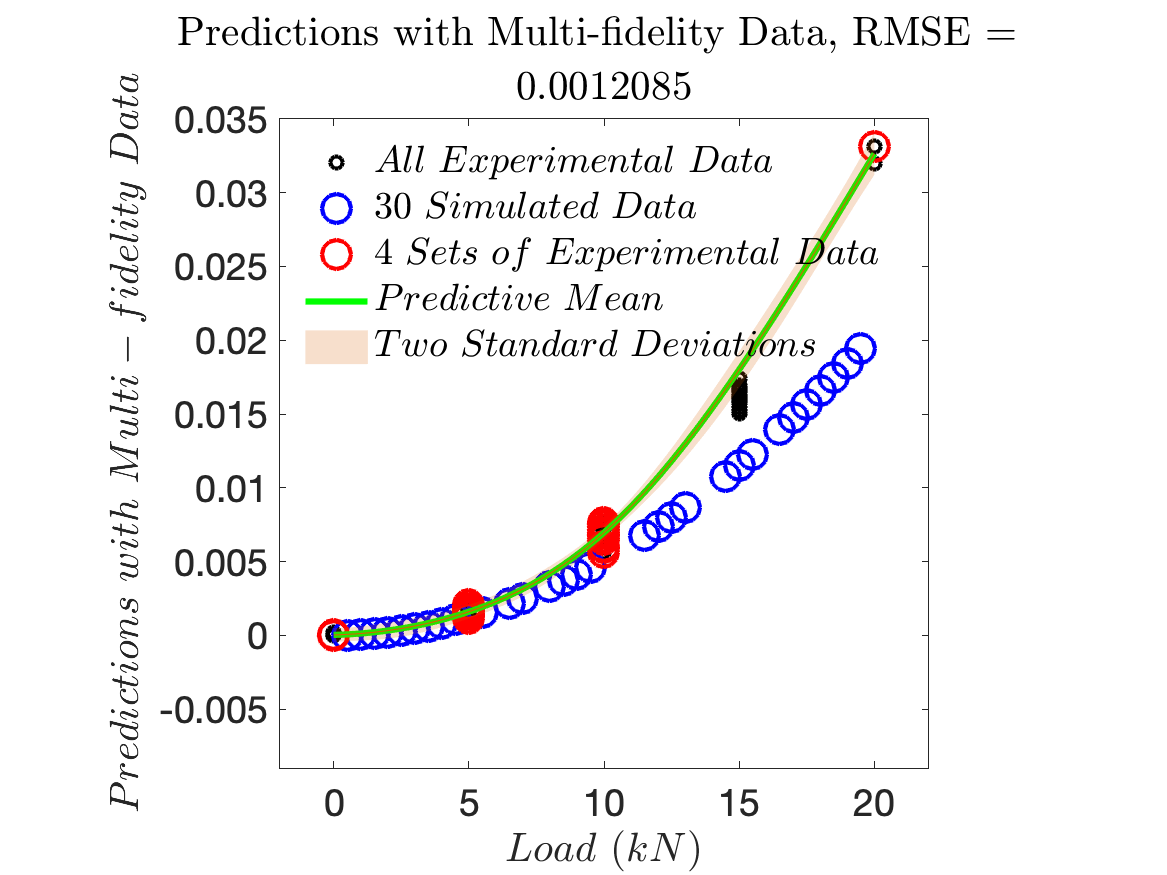}}
    \put(170,200){\color{black} \large {\fontfamily{phv}\selectfont \textbf{a}}}
    \put(380,200){\large {\fontfamily{phv}\selectfont \textbf{b}}}
   \put(170,30){\large {\fontfamily{phv}\selectfont \textbf{c}}} 
   \put(380,30){\large {\fontfamily{phv}\selectfont \textbf{d}}} 
    \end{picture}
    \caption{DI regression for path 1-6 from GPRM and multi-fidelity GPRM with batch learning using random selection: (a) prediction  using 4 experimental sets at 0, 5, 10 and 20 mm; (b) prediction  using 4 experimental sets and 2 random selected simulated data points; (c) prediction  using 4 experimental sets and 5 random selected simulated data points; (d) prediction  using 4 experimental sets and 30 random selected simulated data points.}
\label{fig:fix_mse} 
\end{figure}
%
%h=5
\begin{figure}[h!]
    % \centering
    \begin{picture}(500,330)
    \put(10,168){\includegraphics[width=0.48\textwidth]{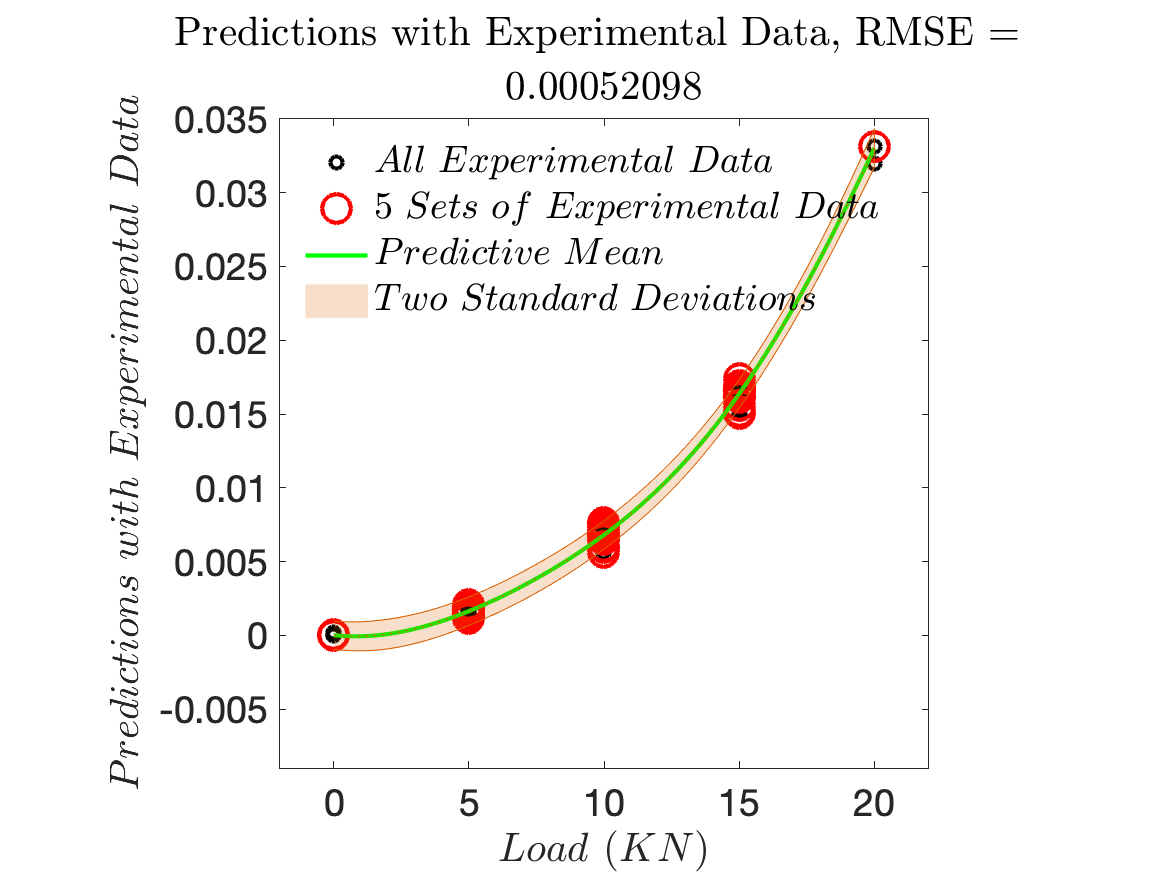}}
    \put(224,168){\includegraphics[width=0.48\textwidth]{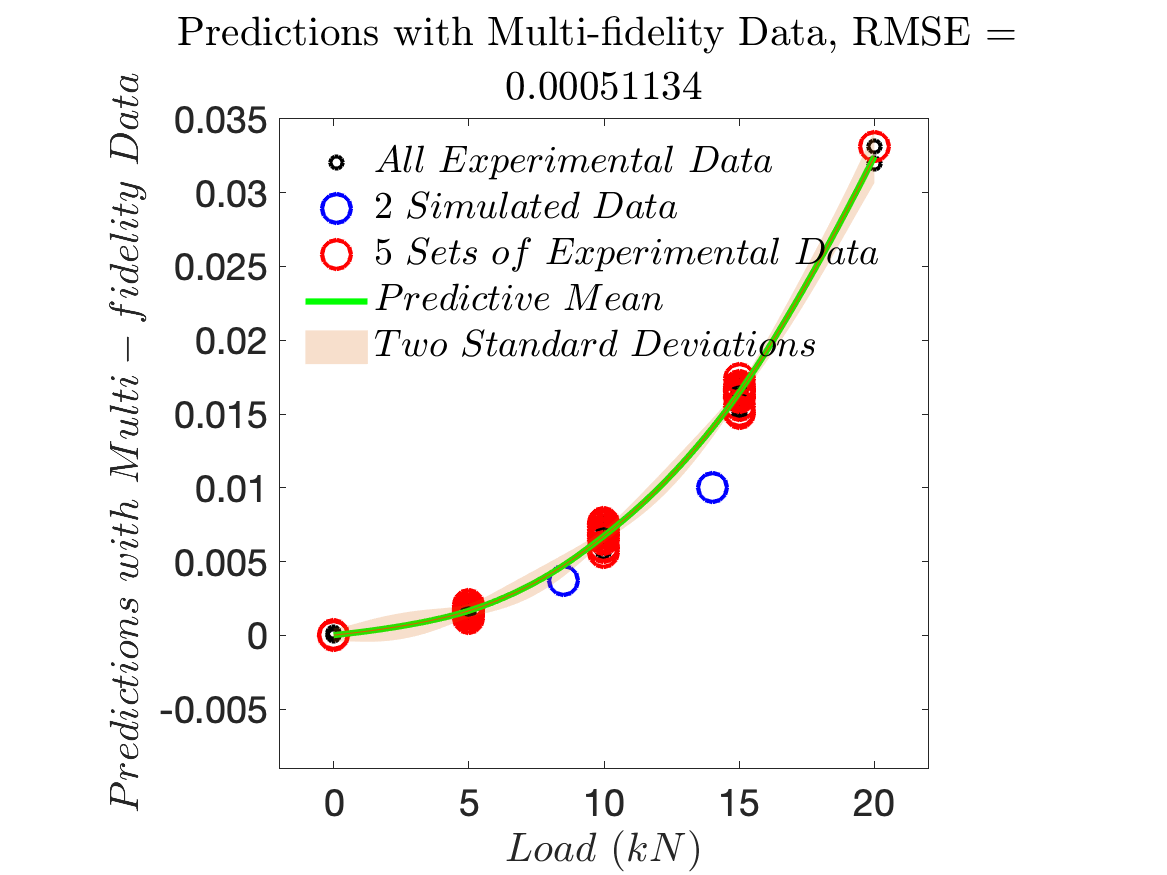}}
    \put(10,0){\includegraphics[width=0.48\textwidth]{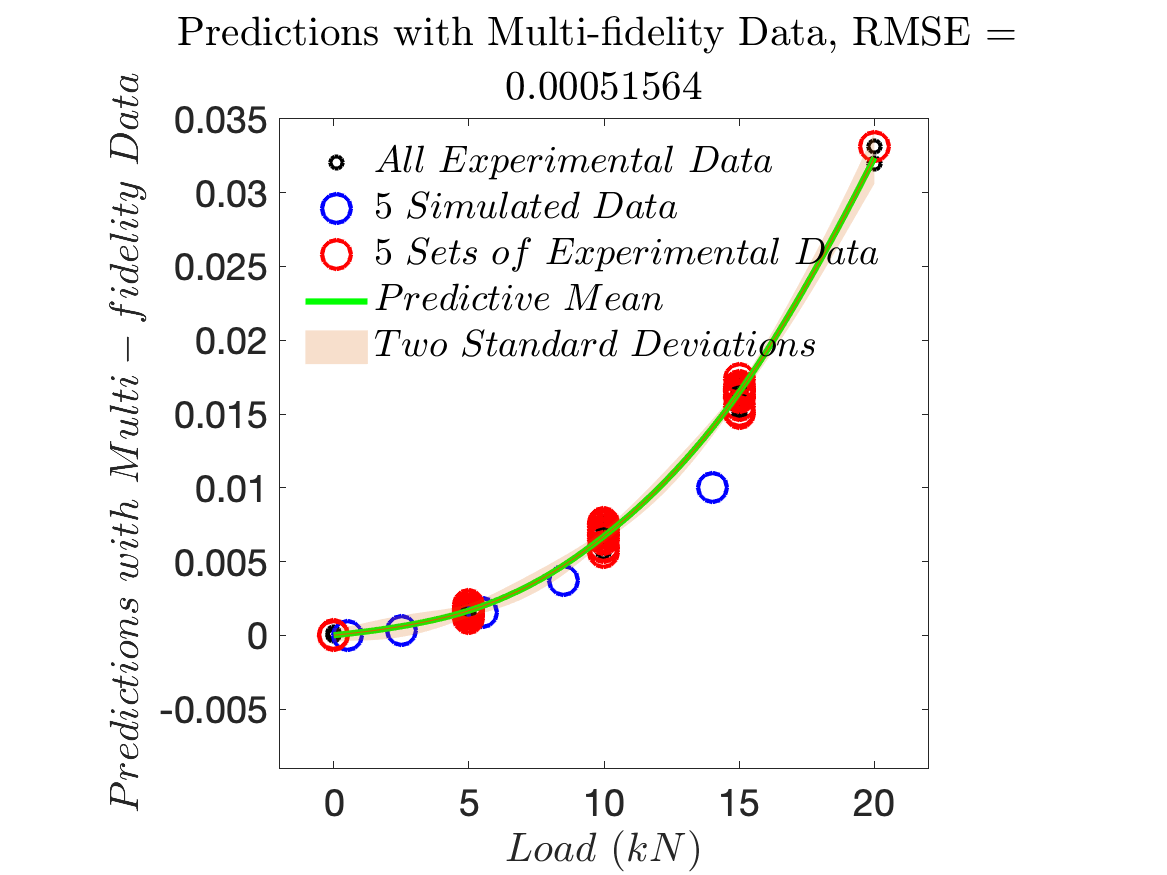}}
    \put(224,0){\includegraphics[width=0.48\textwidth]{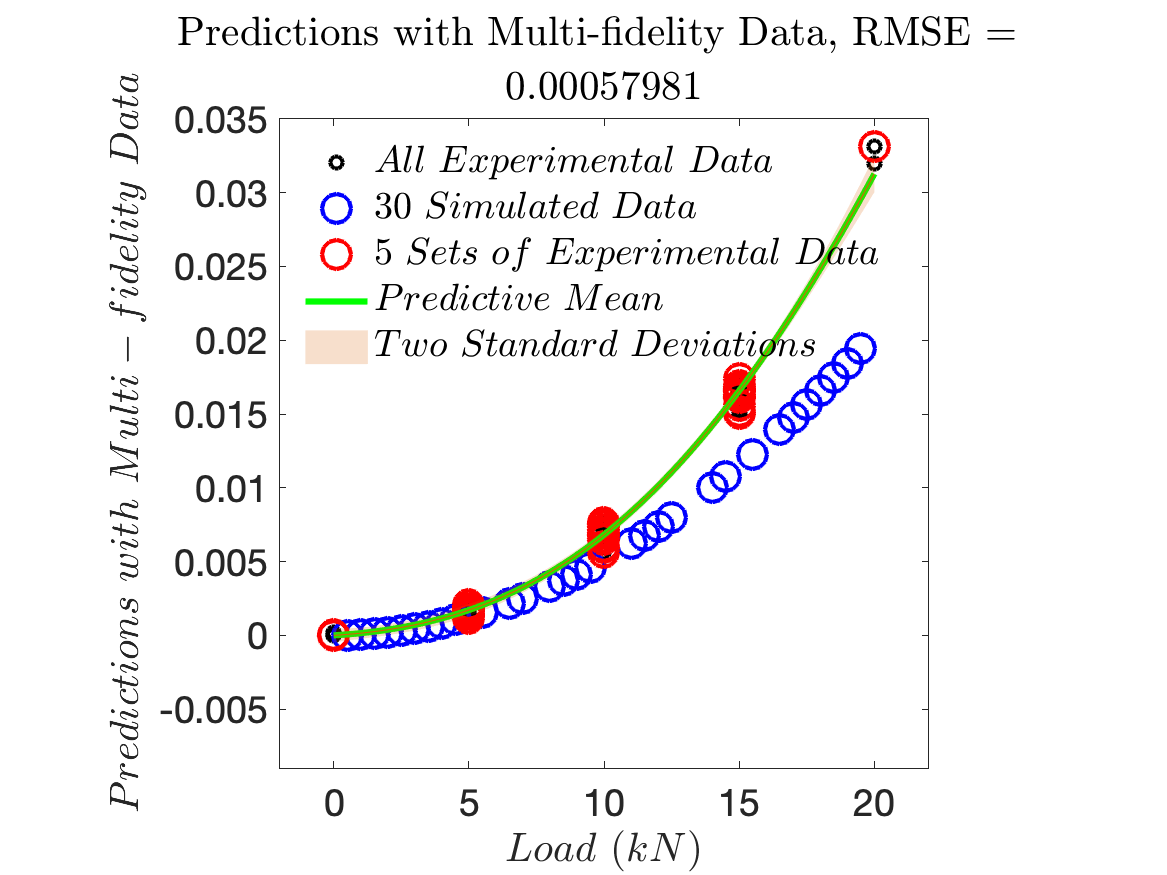}}
    \put(170,200){\color{black} \large {\fontfamily{phv}\selectfont \textbf{a}}}
    \put(380,200){\large {\fontfamily{phv}\selectfont \textbf{b}}}
   \put(170,30){\large {\fontfamily{phv}\selectfont \textbf{c}}} 
   \put(380,30){\large {\fontfamily{phv}\selectfont \textbf{d}}} 
    \end{picture} 
    \caption{DI regression for path 1-6 from GPRM and multi-fidelity GPRM with batch learning using random selection: (a) prediction  using 5 experimental sets at 0, 5, 10, 15 and 20 mm; (b) prediction  using 5 experimental sets and 2 random selected simulated data points; (c) prediction  using 5 experimental sets and 5 random selected simulated data points; (d) prediction  using 5 experimental sets and 30 random selected simulated data points.}
\label{fig:fix_mse} 
\end{figure}

\end{document}